\documentclass[10pt,aps,twocolumn,superscriptaddress,pra]{revtex4-2}
\usepackage{amsmath,amssymb,amsfonts,comment,ulem}
\usepackage[dvipsnames]{xcolor}
\usepackage{graphicx}
\definecolor{myred}{rgb}{1,0.,0.3}
\definecolor{myblue}{rgb}{0.2,0.2,0.8}
\usepackage[colorlinks=true,citecolor=myblue,linkcolor=myred]{hyperref}
\usepackage{color}

 \def\be{\begin{equation}}
 \def\ee{\end{equation}}
\def\ba{\begin{eqnarray}}
\def\ea{\end{eqnarray}}
\def\rang{\rangle}

\definecolor{BRICKRED}{rgb}{0.796, 0.255, 0.329}

\usepackage{tikz}

\newcommand{\ket}[1]{|#1\rangle}
\newcommand{\bra}[1]{\langle #1|}

\begin{document}

\allowdisplaybreaks

\title{Cavity-mediated cross-cross-resonance gate}
\author{Alexey V.~Gorshkov}
\affiliation{AWS Center for Quantum Computing, Pasadena, California 91125, USA}
\author{Or Golan}
\affiliation{AWS Center for Quantum Computing, Pasadena, California 91125, USA}
\author{Daniel Cohen}
\affiliation{AWS Center for Quantum Computing, Pasadena, California 91125, USA}
\author{Arbel Haim}
\affiliation{AWS Center for Quantum Computing, Pasadena, California 91125, USA}
\author{Amit Rotem}
\affiliation{AWS Center for Quantum Computing, Pasadena, California 91125, USA}
\author{Gihwan Kim}
\affiliation{AWS Center for Quantum Computing, Pasadena, California 91125, USA}
\affiliation{Kavli Nanoscience Institute and Thomas J. Watson, Sr., Laboratory of Applied Physics, California Institute of Technology, Pasadena, California 91125, USA.}
\affiliation{Institute for Quantum Information and Matter, California Institute of Technology, Pasadena, California 91125, USA.}
\author{Andreas~Butler}
\affiliation{AWS Center for Quantum Computing, Pasadena, California 91125, USA}
\affiliation{Kavli Nanoscience Institute and Thomas J. Watson, Sr., Laboratory of Applied Physics, California Institute of Technology, Pasadena, California 91125, USA.}
\affiliation{Institute for Quantum Information and Matter, California Institute of Technology, Pasadena, California 91125, USA.}
\author{Connor T.~Hann}
\affiliation{AWS Center for Quantum Computing, Pasadena, California 91125, USA}
\author{Oskar Painter}
\affiliation{AWS Center for Quantum Computing, Pasadena, California 91125, USA}
\author{Fernando G.S.L.~Brand\~ao}
\affiliation{AWS Center for Quantum Computing, Pasadena, California 91125, USA}
\author{Alex Retzker}
\affiliation{AWS Center for Quantum Computing, Pasadena, California 91125, USA}
\affiliation{Racah Institute of Physics, The Hebrew University of Jerusalem, Jerusalem 91904, Givat Ram, Israel}

\begin{abstract}
We propose a cavity-mediated gate between two transmon qubits or other nonlinear superconducting elements. The gate is realized by driving both qubits at a frequency that is near-resonant with the frequency of the cavity. Since both qubits are subject to a cross-resonant drive, we call this gate a cross-cross-resonance gate. In close analogy with gates between trapped-ion qubits, in phase space, the state of the cavity makes a circle whose area depends on the state of the two qubits, realizing a controlled-phase gate. We propose two schemes for canceling the dominant error, which is the  {qubit-cavity} dispersive coupling. We also show that this cross-cross-resonance gate allows one to realize simultaneous gates between multiple pairs of qubits coupled via the same metamaterial composed of an array of coupled cavities or other linear mediators.
\end{abstract}
\maketitle

\tableofcontents

\section{Introduction \label{sec:intro}}

Superconducting systems are among the most promising architectures for quantum information processing \cite{krantz19,kjaergaard20,blais21}. A fundamental ingredient used to manipulate quantum information is a two-qubit gate. A large variety of two-qubit gates have been explored both theoretically and experimentally. The conceptually simplest gate 
 {resonantly couples two-transmon states 01 and 10} via direct capacitive coupling that is turned on and off by tuning transmon frequencies  \cite{blais03,bialczak10,dewes12,barends14} or using tunable couplers \cite{chen14,yan18,arute19}. A similar gate can be implemented when the two qubits are coupled via a virtually populated resonator bus, and the gate can be turned on and off again either by tuning the transmon frequencies \cite{majer07} or by controlling qubit-cavity couplings \cite{gambetta11,srinivasan11}. Another gate is the 11-02 controlled phase gate achieved by tuning the two-transmon states 11 and 02 to resonance with each other \cite{dicarlo09,strauch03,dicarlo10,yamamoto10,barends14,chen14,negirneac21}. A variety of all-microwave gates include {the cross-resonance gate \cite{paraoanu06,rigetti10},}  the resonator-induced phase (RIP) gate \cite{cross15,puri16,paik16,malekakhlagh22a,kumph24a,huang2024fast},   the sideband iSWAP gate \cite{leek09}, the bSWAP gate \cite{poletto12}, the microwave-activated CPHASE gate \cite{chow13}, and the fg-ge gate \cite{zeytinoglu15,egger19}. Finally, in parametric gates, a system parameter is varied in time with a frequency matching the detuning between the two qubits \cite{bertet06,niskanen06,liu07a,niskanen07,beaudoin12,strand13,kapit15,sirois15,mckay16,naik17,caldwell18,didier18,reagor18} {or the energy difference between two-transmon states 11 and 02 \cite{caldwell18}}. 

{ The quality of a quantum gate is determined by several factors, but one major contributor is the amount and rate of parasitic transitions that may occur during gate operation. These unwanted transitions limit both the fidelity and the speed of the gate. This issue becomes even more critical for gates mediated by a metamaterial (an array of coupled cavities), the focus of this work, where the number of parasitic transitions increases dramatically due to the large number of degrees of freedom in the system.} 
{For example, baseband flux-controlled gates \cite{sung2021realization} require reaching a specific resonance. To do so, the qubit frequency may have to cross several metamaterial modes. Moreover, another parasitic  
transition can appear as the 1-2 transition may become resonant with another mode.}
{An alternative is a parametric-type gate \cite{reagor18}, which could avoid crossing physical metamaterial resonances. However, the higher harmonics associated with this approach can still hit unwanted resonances involving other metamaterial modes and possibly the 1-2 transition.}
{While transverse adiabatic gates \cite{stehlik2021tunable} have many advantages, they cannot be easily used to target a pair of qubits out of many qubits coupled via the same metamaterial. Cross-resonance gates, on the other hand, enable such targeting by being frequency-specific: one qubit is driven at the frequency of the target qubit. This also means that the drive frequency corresponds to the target qubit rather than an energy gap in the system. As a result, the number of unwanted resonances is reduced, which is another crucial advantage for metamaterial architectures. For these reasons, we chose to focus on this type of gate in this manuscript.}

In the standard realization of {the cross resonance gate}  \cite{paraoanu06,rigetti10},  
one qubit is driven at the frequency of the second qubit, resulting in an effective ZX coupling between the two. One advantage of this gate is that it works with fixed-frequency transmons, which maintain coherence better than their tunable counterparts. Another advantage is that the gate can be realized using the same controls as those driving single-qubit gates.  
To the best of our knowledge, all prior theoretical and experimental studies of the cross-resonance gate \cite{paraoanu06,rigetti10,degroot10,chow11,kandala21,wei21,chow12,corcoles13,takita16,sheldon16,patterson19,hazra20,dogan22,degroot12,kirchhoff18,tripathi19,magesan20,sundaresan20,malekakhlagh20,malekakhlagh22,nesterov22,heya21,petrescu21} considered the case where the two qubits are coupled directly (or via an auxiliary system that is populated only virtually during the operation of the gate) and where one of them is driven at the frequency of the other. In contrast, we propose a cross-resonance-based gate for two qubits  
that are coupled through a cavity (which we will  interchangeably call an oscillator or a resonator) or an array of coupled cavities (a metamaterial) and where the cavity plays a significant role during the gate operation and is actively and significantly populated. Specifically, in our implementation, the two qubits are driven at a frequency that is near-resonant with the cavity. Since both qubits are subject to a cross-resonant drive, we call this gate a {cavity-mediated} cross-cross-resonance gate. As a result, in close analogy with the M{\o}lmer-S{\o}rensen gate in trapped ions \cite{sorensen00}, the state of the cavity in phase space makes a circle whose area depends on the state of the two qubits, realizing a controlled-phase gate. Such near-resonant gates can be faster than gates mediated by adiabatically eliminated cavity modes. Furthermore, thanks to the near-resonant drive, when multiple qubits are coupled via an array of coupled cavities, we can simultaneously implement multiple long-range two-qubit gates (for any desired pairing of qubits), each mediated by a different eigenmode of the cavity array. Such simultaneous gates work even if all qubits have approximately the same frequency. {We will make a detailed comparison between our gate and the resonator-induced phase (RIP) gate \cite{cross15,puri16,paik16,malekakhlagh22a,kumph24a} at the end of Sec.\ \ref{sec:qubits}, after we explain the operation of our gate in more detail.}

{The term cross-cross resonance gate has been introduced before in Ref.\ \cite{heya21} to describe a gate in which qubit 1 is driven at the frequency of qubit 2, while qubit 2 is simultaneously driven at the frequency of qubit 1. To differentiate our gate from the gate of Ref.\ \cite{heya21}, we call our gate cavity-mediated cross-cross-resonance gate. However, since we will not be discussing Ref.\ \cite{heya21} in the rest of the manuscript, we will often refer to our gate for brevity as cross-cross-resonance gate.} 

{Our} fast parallelizable long-range {cavity-mediated cross-cross-resonance} gates can be used to realize better quantum error correcting codes with superconducting qubits. In particular, quantum low-density parity-check (LDPC) codes can offer large rates (ratios of logical qubits to physical qubits) and large code distances at the expense of requiring non-local error syndrome measurements \cite{breuckmann21,gottesman14,tremblay22,panteleev22,leverrier22a,bravyi24b,cain2026shor,zhao2026towards,kasai2026breaking}. By enabling fast measurements of such non-local syndromes, our fast parallelizable long-range gates can make superconducting-qubit architectures more scalable, which is currently one of the main challenges facing these systems. Our scheme for enabling fast measurements of non-local syndromes is complementary to the recently demonstrated scheme \cite{wang2026demonstration}, where each long-range connection is hardwired. This contrasts with our scheme, where each metamaterial can mediate multiple simultaneous two-qubit gates for any desired pairing of qubits coupled to the metamaterial. We note that a detailed investigation of correlated errors induced by implementing simultaneous gates is outside the scope of this manuscript.

The proposed architecture also offers advantages in horizontal scaling between different chips \cite{norris2025performance,smith2022scaling,gold2021entanglement,field2024modular,deng2025long,wu2024modular}. One option for horizontal scaling is to use coaxial cables as in Ref.~\cite{deng2025long}. If we use the cross-cross-resonance gate, then the two extra LC circuits on each chip from  Ref.~\cite{deng2025long} are not needed, as the transmons could be coupled directly to the coaxial oscillator. Horizontal scaling, which is enabled by a coaxial cable, could also be realized via a standard cross-resonance gate mediated by a virtually populated resonator, as in Ref.~\cite{song2024realization}. Since, in our cross-cross-resonance gate, the resonator is not adiabatically eliminated and is populated with real photons, we have the potential to realize faster gates. In a 3D configuration of horizontal scaling \cite{norris2025performance}, the metamaterial could serve as a long-range connector between distant qubits in two different chips, and our approach allows using the same connector for multiple two-qubit gates simultaneously.

The remainder of this manuscript is organized as follows. In Sec.\ \ref{sec:qubits}, we describe the basic operation of the gate when transmons are treated in the qubit approximation. In Sec.\ \ref{sec:quantumstark}, staying in the qubit approximation, we derive the dispersive coupling, which is the dominant correction to the ideal implementation of the gate. In Sec.\ \ref{sec:transmons}, we describe the gate operation and the dispersive coupling for transmons beyond the qubit approximation. Then, in Sec.\ \ref{sec:integers} and Sec.\ \ref{sec:flowers}, we propose two approaches, called the ``integers'' approach and the ``flowers'' approach,  respectively, for canceling the dispersive coupling. We then demonstrate how these two approaches  can be used for the case of coupling two transmons via two oscillators (Sec.\ \ref{sec:2oscillators}), to cancel coupling to adjacent cavities (Sec.\ \ref{sec:adjacent}), to cancel always-on ZZ interactions (Sec.\ \ref{sec:ZZcancel}), and for the case of coupling via a collection of oscillators, i.e.\ a metamaterial (Sec.\ \ref{sec:meta}). {In Sec.\ \ref{sec:imperfections}, we discuss gate parameters and the expected effects of decoherence. In Sec.\ \ref{sec:simul}, we present numerical simulations to validate the performance of our gate.}  We present our conclusions and outlook in Sec.\ \ref{sec:conclusions}. In Appendices, we present details omitted from the main text.

\section{Cavity-mediated cross-cross-resonance gate \label{sec:gate}} 

In this section, we first describe the operation of the cavity-mediated cross-cross-resonance gate (Sec.\ \ref{sec:qubits}) and the dominant correction (Sec.\ \ref{sec:quantumstark}), which is the dispersive coupling, when transmons are treated in the qubit approximation. We then present the generalization of both the gate operation and the dispersive coupling to transmons beyond the qubit approximation (Sec.\ \ref{sec:transmons}). 

\subsection{Gate operation for qubits \label{sec:qubits}}

In this subsection, we describe the operation of the cavity-mediated cross-cross-resonance gate when transmons are treated within a two-level (qubit) approximation {corresponding to infinite anharmonicity}. 
\begin{figure}[t!]
\centering
\includegraphics[width=0.99 \linewidth]{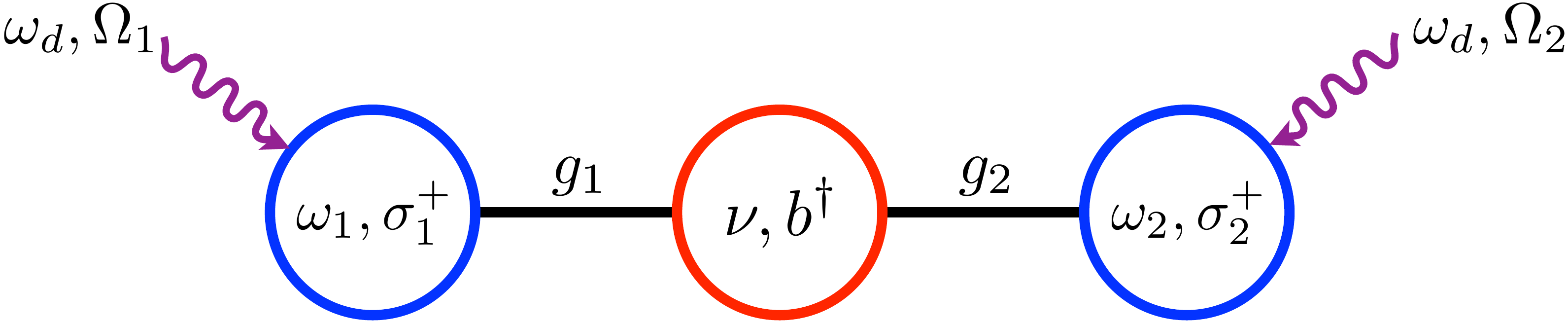} 
\caption{Cavity-mediated cross-cross-resonance gate. Two qubits with frequencies $\omega_1$ and $\omega_2$ and raising operators $\sigma_1^+$ and $\sigma_2^+$ are coupled, with coupling strengths $g_1$ and $g_2$, via a cavity with frequency $\nu$ and creation operator $b^\dagger$. The qubits are driven with Rabi frequencies $\Omega_1$ and $\Omega_2$ at the same frequency $\omega_d$ that is near resonant with the cavity frequency $\nu$.   \label{fig:cr}
}
\end{figure}
As shown in Fig.\ \ref{fig:cr}, consider two qubits 
coupled via a resonator, and both driven at the same frequency near resonance with the resonator. {Postponing the analysis of qubit decoherence and cavity loss until Secs.\ \ref{sec:imperfections} and \ref{sec:simul}, we can describe the system by the following Hamiltonian}  
 (setting $\hbar = 1$ throughout the manuscript) \cite{blais21}
\be
H=\nu b^\dagger b+\!\!\! \sum_{j=1,2} \!\! \left[\frac{\omega_j}{2}  Z_j +\Omega_j\cos(\omega_d t)X_j+g_j (\sigma^+_j b+h.c.)\right].\label{eq:ham1}
\ee
Here $b^\dagger$ creates an excitation in the resonator ($[b,b^\dagger] = 1$), which has frequency $\nu$. $X_j$, $Y_j$, and $Z_j$ are Pauli matrices acting on qubit $j = 1,2$ in the $\{\ket{1},\ket{0}\}$ basis and having frequency $\omega_j$. $\sigma^+_j = (X_j + i Y_j)/2 = \ket{1}_j\bra{0}_j$ is the raising operator of transmon $j$. $\Omega_j$, which we assume to be non-negative, is 
the Rabi frequency driving transmon $j$ at frequency $\omega_d \approx \nu$. $g_j$ is the coupling between qubit $j$ and the resonator. Such cross-resonant driving of a qubit near resonance with a coupled resonator can be used for qubit readout \cite{touzard19} and generation of Schr\"odinger cat states in the resonator \cite{solano03,ayyash24}. Cross-resonant drives in this qubit-resonant setting have also been considered for the cases of multi-photon resonances \cite{ayyash24a} and of multiple resonators/modes  \cite{ayyash25}. 

Moving into the interaction picture with respect to $H_0=\omega_d b^\dagger b+\frac{\omega_d}{2} (Z_1+Z_2)$ and making the rotating-wave approximation (RWA), we arrive at
\begin{align}
H &= \epsilon b^\dagger b + \sum_{j=1,2} \left[\frac{\Delta_j}{2}  Z_j +\frac{\Omega_j}{2} X_j+g_j(\sigma^+_j b +h.c.)\right], \label{eq:ham1b}
\end{align}
where $\epsilon = \nu - \omega_d$ is the 
detuning between the drive frequency and the resonator and $\Delta_j = \omega_j - \omega_d$ is the  
detuning between the drive frequency and the qubit \footnote{Technically, $\epsilon$ and $\Delta_j$ are negatives of the corresponding detunings since they are negative when the drive is blue-detuned. However, throughout the paper, we will abuse the terminology and refer to them simply as detunings.}. 
To avoid putting absolute values around $\Delta_j$ and $\epsilon$ in inequalities throughout the manuscript, we will assume that they are positive, but all the results apply to negative $\Delta_j$ and $\epsilon$, as well.

To understand how the cross-cross-resonance gate works, let us change the local basis to diagonalize the qubit Hamiltonian in the absence of interactions. To do this, we rotate the Hamiltonian into the dressed basis of each qubit using the unitary
\ba
U=\exp\left(-i \sum_j Y_j \theta_j/2\right), \label{eq:dressingU}
\ea
where
\ba
\cos\theta_j &=& \frac{\Delta_j}{\sqrt{\Delta_j^2+\Omega_j^2}},\\
\sin\theta_j &=& \frac{\Omega_j}{\sqrt{\Delta_j^2+\Omega_j^2}},
\ea
so that the new Hamiltonian is
\ba
H & \rightarrow & U^\dagger H U\\
  &=& \epsilon b^\dagger b + \sum_{j=1,2}\Bigg\{\frac{\sqrt{\Delta_j^2+\Omega_j^2}}{2} Z_j \nonumber \\
  &&+\frac{g_j}{2} [(X_j \cos\theta_j + Z_j \sin\theta_j + i Y_j )b+h.c.]\Bigg\}.\label{eq:hamfull}
\ea
In an experiment, one can go between the bare basis and the dressed basis by adiabatically ramping $\Omega_j$ up at the beginning of the gate and adiabatically ramping it down at the end of the gate. To simplify the presentation, we will suppress this detail in the remainder of the paper.

Assuming $\sqrt{\Delta_j^2+\Omega_j^2}\gg g_j, \epsilon$ and moving into the interaction picture with respect to $H_0= \sum_{j=1,2} \frac{\sqrt{\Delta_j^2+\Omega_j^2}}{2} Z_j$, we arrive at 
\be
H = \epsilon b^\dagger b + \frac{1}{2} \sum_{j = 1,2} g_j Z_j \sin \theta_j (b + b^\dagger) = \epsilon b^\dagger b + M (b + b^\dagger), \label{eq:noquantumstark}
\ee
where we defined $M = \frac{1}{2} \sum_{j = 1,2} g_j Z_j \sin \theta_j$.

Notice the simple structure of the resulting coupling: each qubit participates in the coupling only through its Pauli matrix $Z_j$. This allows for a very simple exact solution of the resulting evolution: one can think of $M$ as a number (rather than an operator) that takes one of four possible values depending on what eigenstates of $Z_j$ we are in. In this sense, the cross-cross-resonance gate enables the implementation with superconducting qubits of what can be done directly with trapped ions \cite{bruzewicz19, sorensen00, leibfried03}. With trapped ions, where the harmonic oscillator is a collective motional mode, one separately engineers the flip-flop coupling $\sigma^+ b + \textrm{h.c.}$ (by driving the red sideband) and the flip-flip coupling $\sigma^+ b^\dagger + \textrm{h.c.}$ (by driving the blue sideband), resulting in an overall coupling of the form $X (b + b^\dagger)$. 
While one could consider engineering the flip-flop and flip-flip couplings with superconducting qubits, our cross-cross-resonance gate instead relies on engineering  
the $Z (b + b^\dagger)$ coupling---the same coupling as in Leibfried's trapped ion gate \cite{leibfried03}---instead of $X (b + b^\dagger)$.

Treating, therefore, $M$ as a real number, time evolution under the Hamiltonian in Eq.\ (\ref{eq:noquantumstark}) can be easily solved analytically as follows. It is convenient to go into an interaction picture with respect to $\epsilon b^\dagger b$, which gives
\ba
H = M (b e^{- i \epsilon t} + b^\dagger e^{i \epsilon t}). \label{eq:hamM}
\ea
This Hamiltonian is a special case of
\ba
H = f(t) x + g(t) p, \label{eq:generalSM}
\ea
where $f(t)$ and $g(t)$ are real functions, $x = (b^\dagger + b)/\sqrt{2}$, $p = i (b^\dagger - b)/\sqrt{2}$, and $[x,p]=i$. The propagator for Eq.\ (\ref{eq:generalSM}) is  \cite{sorensen00}
\ba
U(t) = e^{-i x F(t)} e^{-i p G(t)} e^{- i A(t) }, \label{eq:UMS}
\ea
where
\ba
F(t) &=& \int_0^t f(t') d t', \\
G(t) &=& \int_0^t g(t') d t', \\
A(t) &=& - \int_0^t F(t') g(t') dt'. \label{eq:Aoft}
\ea
One can easily check that this propagator is correct by verifying that $U(0)$ is identity and that $i \dot U = H U$. 

Heisenberg evolution of $x$ and $p$ is then given by $U^\dagger x U = e^{i p G} x e^{-i p G} = x + G$ and $U^\dagger p U = e^{i x F} p e^{-i x F} = p - F$. In other words, 
$U(t)$ translates operators $x$ and $p$  
according to 
\ba
\{x, p\} \rightarrow \{x  + G(t),  p  - F(t)\},
\ea
which is equivalent to
\ba
b \rightarrow b + \frac{G(t) - i F(t)}{\sqrt{2}}.\label{eq:btob}
\ea
If $f(t)$ and $g(t)$ are designed in such a way that, at some particular time $\tau$, the oscillator returns to its initial state, i.e.\ $F(\tau) = G(\tau) = 0$, then the propagator reduces to
\ba
U(\tau) = e^{- i A(\tau) }, \label{eq:eiA}
\ea
where $A(\tau)$ is equal to the area enclosed by the path $(G(t),-F(t))$ in the $(x,p)$ phase space if the path is clockwise (and negative of the area if the path is counterclockwise) \cite{sorensen00}. To see this, it is sufficient to note that the area of the rectangle under the $[t,t+dt]$ segment of the curve $(G(t),-F(t))$ is given by the product of the height of the rectangle, $-F(t)$, and its width, $G'(t) dt = g(t) dt$. The resulting product, $-F(t) g(t) d t$, is exactly the infinitesimal element that is summed in the integral defining $A(t)$ in Eq.\ (\ref{eq:Aoft}).

For the specific case of the Hamiltonian in Eq.\ (\ref{eq:hamM}), we have
\ba
f(t) &=& \sqrt{2} M \cos (\epsilon t),\label{eq:f}\\
g(t) &=& \sqrt{2} M \sin(\epsilon t),\label{eq:g}\\
F(t) &=& \sqrt{2} M \sin(\epsilon t)/\epsilon, \label{eq:F}\\
G(t) &=& \sqrt{2} M (1 - \cos( \epsilon t))/\epsilon, \label{eq:G}\\
A(t) &=& - \frac{M^2}{\epsilon^2}\left( \epsilon t - \frac{1}{2} \sin (2 \epsilon t)\right).\label{eq:A}
\ea
To avoid the factors of $\sqrt{2}$ that show up in Eqs.\ (\ref{eq:F},\ref{eq:G}), we rescale the axes of phase space for the remainder of the paper. Specifically, we will make all plots in the $\{\langle x\rangle,\langle p \rangle\}/\sqrt{2} = \{\textrm{Re}[\langle b \rangle], \textrm{Im}[\langle b \rangle]\}$ plane, which we can also think of as the complex plane for $\langle b \rangle = \textrm{Re}[\langle b \rangle] + i \textrm{Im}[\langle b \rangle]$. In this plane, the initial state evolves according to  
\ba
\{\textrm{Re}[\langle b(0) \rangle], \textrm{Im}[\langle b(0) \rangle]\} + \{G(t),- F(t)\}/\sqrt{2}.
\ea 
Although we will often choose to take the expectation value of $b$ (to make the meaning of the axes in our plots more concrete), we emphasize that the full Heisenberg evolution of $b$ (and not just of its expectation value) is given by the same formula, as shown in Eq.\ (\ref{eq:btob}).

\begin{figure}[tb]
\centering
\includegraphics[width=0.6 \linewidth]{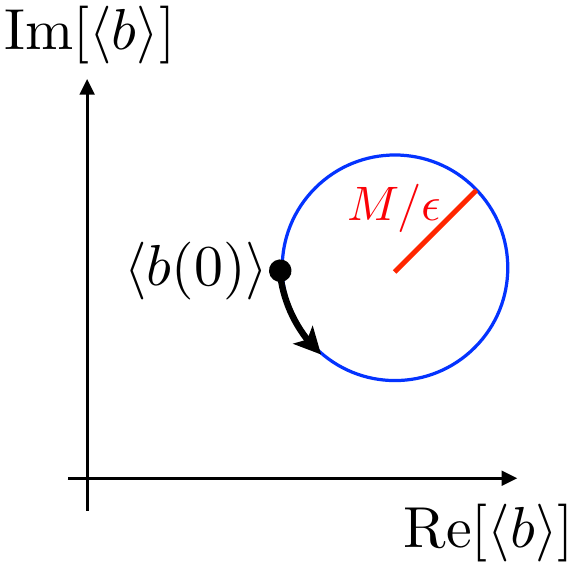} 
\caption{Time evolution, given in Eq.\ (\ref{eq:brot}), of the real and imaginary parts of $\langle b \rangle$ under the Hamiltonian in Eq.\ (\ref{eq:hamM}). The figure assumes $\epsilon > 0$ and $M > 0$. In the complex plane, in time $t = 2 \pi/\epsilon$, $\langle b \rangle$ makes a complete counterclockwise circle of radius $M/\epsilon$ whose center is $M/\epsilon$ to the right of the initial point $\langle b(0)\rangle$.    \label{fig:circle}
}
\end{figure}

Combining Eqs.\ (\ref{eq:btob},\ref{eq:F},\ref{eq:G}), we find that the Heisenberg evolution of $b$ under $H$ in Eq.\ (\ref{eq:hamM}) is given by 
\ba
b(t) = b(0) + \frac{M}{\epsilon} \left(1 - e^{i \epsilon t}\right).\label{eq:brot}
\ea
Taking the expectation value of both sides in some initial state,  
we arrive at the evolution of $\langle b(t) \rangle$ shown in Fig.~\ref{fig:circle}.
The evolution corresponds to completing counterclockwise circles of radius $M/\epsilon$ with period $2 \pi/\epsilon$. Therefore, if we choose the final time $\tau$ such that
\ba
\epsilon \tau = 2 \pi n,\label{eq:epsilon}
\ea
where $n$ is an integer, then $F(\tau) = G(\tau) = 0$, the evolution corresponds to $n$ complete circles, and the unitary simplifies [by plugging Eq.\ (\ref{eq:A}) into Eq.\ (\ref{eq:eiA})] to
\ba
U(\tau) = e^{i \frac{M^2}{\epsilon} \tau}.\label{eq:phase}
\ea
We can equivalently rewrite $U(\tau)$ as
\ba
U(\tau) = e^{i 2 \textrm{(enclosed area)}}, \label{eq:area}
\ea
where the enclosed area of $n \pi M^2/\epsilon^2$ corresponds to the area of $n$ circles in Fig.\ \ref{fig:circle}. The minus sign in Eq.\ (\ref{eq:eiA}) cancels with the negative sign due to counterclockwise rotation of the circle to give a positive phase in Eq.\ (\ref{eq:area}). 
The factor of $2$ in Eq.\ (\ref{eq:area}) comes from the fact that we rescaled each of our axes by $\sqrt{2}$, so that the enclosed area in Fig.\ \ref{fig:circle} is a factor of 2 smaller than in the $\{x,p\}$ plane.

Up to an irrelevant global phase,   Eq.\ (\ref{eq:phase}) is equivalent to evolution for time $\tau$ under
\ba
H_{ZZ} = - Z_1 Z_2 \frac{g_1 g_2}{2 \epsilon} \sin \theta_1 \sin \theta_2. \label{eq:hz1z2}
\ea 
{[In the limit of large detuning---$\epsilon \gg |M|$ for all assignments of $Z_1$ and $Z_2$---the oscillator can be adiabatically eliminated, so that Eq.\ (\ref{eq:phase}) and the effective Hamiltonian description in Eq.\ (\ref{eq:hz1z2}) apply at all times. On the other hand, at small detunings, Eqs.\ (\ref{eq:phase},\ref{eq:hz1z2})  apply only at discrete times satisfying Eq.\ (\ref{eq:epsilon}); outside of these times, the qubits are entangled with the oscillator.]}
Since $Z_j = 2 \ket{1}_j \bra{1}_j - 1$, if we further choose our parameters such that 
\ba
2 \tau \frac{g_1 g_2}{\epsilon} \sin \theta_1 \sin \theta_2  = \pi,\label{eq:pi}
\ea
we obtain a gate that, up to single-qubit $Z$ rotations, is equivalent to a CZ gate. Eliminating $\epsilon$ from the two constraints [Eqs.\ (\ref{eq:epsilon},\ref{eq:pi})] and solving for $\tau$, we obtain
\ba
\tau = \pi \sqrt{\frac{n}{g_1 g_2 \sin \theta_1 \sin \theta_2}}. \label{eq:rootn}
\ea
The fastest gate occurs for $n=1$ and takes time 
\ba
\tau = \pi \frac{1}{\sqrt{g_1 g_2 \sin \theta_1 \sin \theta_2}}.\label{eq:fastest}
\ea

Notice that, under the conditions of Eq.\ (\ref{eq:epsilon}), the unitary that takes us from the rotating frame to the lab frame is
\ba
V(\tau) = e^{- i \tau \epsilon b^\dagger b} = e^{- i 2 \pi n b^\dagger b} = 1,
\ea
i.e.\ it is an identity, so, at time $\tau$, the rotating frame evolution $U(\tau)$ and the lab frame evolution $V(\tau) U(\tau)$ coincide.

For readers more familiar with qubits than with oscillators, it may be instructive to consider the limit where the resonator starts in the vacuum state and $\epsilon \gg M$ in Eq.\ (\ref{eq:noquantumstark}). In that case, the oscillator is barely excited during the evolution and can be approximated with an effective qubit corresponding to the oscillator's lowest Fock states $\ket{0}$ and $\ket{1}$. In that case, Eq.\ (\ref{eq:noquantumstark}) becomes
\ba
H = \epsilon \ket{1} \bra{1} + M X,
\ea
where $X$ is the $X$ Pauli matrix of the effective qubit. Starting in $\ket{0}$, the evolution under this Hamiltonian is given by
\ba
&& e^{- i \epsilon t/2} \Bigg[\frac{- i M }{M'} \sin (M' t) \ket{1} + \nonumber \\
&& \left(\cos(M' t) + \frac{i \epsilon}{2 M'} \sin (M' t) \right) \ket{0}\Bigg],\label{eq:threequbits}
\ea
where $M' = \sqrt{M^2 + (\epsilon/2)^2}$ is twice the generalized Rabi frequency. This qubit undergoes off-resonant Rabi oscillations, making small circles near the $\ket{0}$ pole [with 3D coordinates $(0,0,-1)$] of the Bloch sphere  around the axis pointing along $-(M,0,\epsilon/2)$. 
The qubit comes back to the initial state $\ket{0}$ after time $\tau = \pi/M' \approx 2 \pi/\epsilon$, and the picked up phase is $\pi - \epsilon \tau/2 = \pi (1 - \epsilon/(2 M')) \approx 2 \pi M^2/\epsilon^2$, in agreement with Eq.\ (\ref{eq:phase}) evaluated at $\tau = 2 \pi/\epsilon$. Unlike in the cavity case, using  a qubit to mediate the interactions requires knowing the initial state of the qubit. Similarly to the cavity case, using a qubit to mediate the interactions works also in the limit of small detuning. Assuming that $g_1 \sin \theta_1 = g_2 \sin \theta_2$, {and explicitly introducing the arguments $Z_1$ and $Z_2$ into $M(Z_1,Z_2)$, we have $M(1, -1) = M(-1, 1) = 0$ and $M(1,1) = - M(-1,-1)$.}  Therefore, after time {$\pi/\sqrt{|M(1,1)|^2 + (\epsilon/2)^2}$,}  we obtain a two-qubit gate that is diagonal in the computational basis and is given by $\textrm{diag}(e^{i \phi}, 1, 1, e^{i \phi})$, where {$\phi = \pi (1 - \epsilon/\sqrt{\epsilon^2 + 4 |M(1,1)|^2})$.}  This is an entangling gate unless $\epsilon = 0$ or {$|M(1,1)| = 0$.}  {By driving the two data qubits and the qubit mediating the interactions at the frequency of the latter, a three-qubit $i$Toffoli gate was realized in Ref.\ \cite{kim22}.} 

 It is worth mentioning the importance of the relative phase between the two cross-resonance drives in Eq.\ (\ref{eq:ham1}). Suppose that the cross-resonance drive on qubit 2 is $\xi$ out of phase relative to the cross-resonance drive on qubit 1. Let us absorb this phase into the definition of state $\ket{1}$ of qubit 2, so that Eq.\ (\ref{eq:ham1b}) becomes
\begin{align}
H & =  \epsilon b^\dagger b + \sum_{j=1,2} \left[\frac{\Delta_j}{2}  Z_j +\frac{\Omega_j}{2} X_j\right] \nonumber \\
& \quad + g_1(\sigma^+_1 b +h.c.)+ g_2(\sigma^+_2 b e^{i \xi} +h.c.).
\end{align}
Rotating this Hamiltonian into the dressed basis using Eq.\ (\ref{eq:dressingU}), Eq.\ (\ref{eq:noquantumstark}) becomes 
\be
H = \epsilon b^\dagger b + M^\dagger b + M b^\dagger,
\ee
where 
\be
M = \frac{1}{2} g_1 Z_1 \sin \theta_1 + \frac{1}{2} g_2 Z_2 \sin \theta_2 e^{- i \xi}. \label{eq:complexM}
\ee
Going into the interaction picture with respect to $\epsilon b^\dagger b$, Eq.\ (\ref{eq:hamM}) becomes
\ba
H = M^\dagger b e^{- i \epsilon t} + M b^\dagger e^{i \epsilon t}, 
\ea
where we can think of $M = |M| e^{i \arg(M)}$ as a complex number and of $M^\dagger$ as its conjugate. This Hamiltonian is a special case of Eq.\ (\ref{eq:generalSM}) with
\ba
f(t) &=& \sqrt{2} |M| \cos[\epsilon t + \arg(M)],\\
g(t) &=& \sqrt{2} |M| \sin[\epsilon t + \arg(M)],
\ea
so that
\begin{align}
F(t) &= \sqrt{2} |M| \left(\sin[\epsilon t + \arg(M)]-\sin[\arg(M)]\right)/\epsilon,\\
G(t) &=  \sqrt{2} |M| \left(\cos[\arg(M)] - \cos[\epsilon t + \arg(M)]\right)/\epsilon, \\
A(t) &= - \frac{|M|^2}{2 \epsilon^2} \big( 2 \epsilon t + 2 \sin[2 \arg(M) + \epsilon t] - \sin[2 \arg(M)] \nonumber \\
& \quad - 2 \sin[\epsilon t] - \sin[2 \epsilon t+2 \arg(M)]\big).
\end{align}
If we now choose, as before, the final time $\tau$ such that $\epsilon \tau = 2 \pi n$ with integer $n$, the evolution unitary reduces to 
\ba
U(\tau) = e^{-i A(\tau)} = e^{i \frac{|M|^2}{\epsilon} \tau},
\ea
which is the generalization of Eq.\ (\ref{eq:phase}) to complex $M$. Inserting $M$ from Eq.\ (\ref{eq:complexM}), we see that the effective Hamiltonian in Eq.\ (\ref{eq:hz1z2}) generalizes to
\ba
H_{ZZ} = - Z_1 Z_2 \frac{g_1 g_2}{2 \epsilon} \sin \theta_1 \sin \theta_2 \cos \xi. \label{eq:hz1z2cosxi}
\ea 
In other words, we pick up an extra factor of $\cos \xi$. The intuition for vanishing effective interaction at $\xi = \pi/2$ is that, in this case, $|M|$, which determines the area in phase space, is the same for all four choices of $(Z_1, Z_2)$. It is therefore important to know the relative phase $\xi$ between the two cross-resonance drives and ideally set it to a multiple of $\pi$ so as to minimize the gate time. Alternatively, by setting $\xi = \pi/2$, we can set the interaction between the two qubits to zero. For example, if four qubits are coupled to the same resonator, and qubits 3 and 4 are driven $\xi = \pi/2$ out of phase relative to qubits 1 and 2, then we can get $M \propto Z_1 + Z_2 + i (Z_3 + Z_4)$, in which case the effective Hamiltonian has only $Z_1 Z_2$ coupling and $Z_3 Z_4$ coupling. This means that the same resonator can mediate two independent simultaneous two-qubit gates (or multi-qubit gates, as we discuss in Sec.\ \ref{sec:conclusions}). {Analyzing this scheme in a bit more detail, the nonlinear phase picked up by qubits 1 and 2 and the nonlinear phase picked up by qubits 3 and 4 can both be equal to $\pi$ at the same time if one sets $g_1 g_2 \sin \theta_1 \sin \theta_2 = g_3 g_4 \sin \theta_3 \sin \theta_4$ [see Eq.\ (\ref{eq:pi})]. If any of the $g_i$ or $\sin \theta_j$ has a fractional deviation of $\sim f$ from its nominal value, one of the nonlinear phases will be $\sim \pi (1 \pm f)$ when the other one is exactly at $\pi$. However, such deviations can easily be compensated to achieve a $\pi$ phase on both pairs of qubits at the same time: one can simply tune the cross-resonance-drive Rabi frequencies, which determine $\sin \theta_j$. }

Now that we have described our cavity-mediated cross-cross-resonance gate, it is worth comparing it to another cavity-mediated gate called the resonator-induced phase (RIP) gate \cite{cross15,puri16,paik16,malekakhlagh22a,kumph24a}. The RIP gate works in the dispersive-coupling regime (see Sec.\ \ref{sec:quantumstark}) where the effective cavity frequency depends on which of the four computational basis states the two transmons are in. The cavity is typically (but not always \cite{cross15,puri16}) driven off-resonantly and, as a result, sent on a small excursion in phase space (see also Ref.\ \cite{jandura24}, where a driven cavity mediates a gate between qubits outside of the dispersive-coupling regime). At the end of the gate, the cavity returns to the original state in phase space but with a phase that depends on its frequency, which, in turn, depends on the state of the two transmons, resulting in a diagonal controlled-phase gate. The main difference between the RIP gate and our cavity-mediated cross-cross-resonance gate is that we drive the transmons, while the RIP gate is implemented by driving the cavity. Another difference is that, in the RIP gate, the cavity is usually 
driven off-resonantly and is therefore populated only weakly, while, in our gate---and in certain implementations of the RIP gate \cite{cross15,puri16}---the drives are nearly resonant with the cavity, resulting in significant cavity population and enabling fast gates. As we describe in Sec.\ \ref{sec:meta}, the fact that we drive the qubits near resonance with the cavity makes our gate particularly well-suited for applying two-qubit gates in parallel on multiple pairs of qubits interacting via an array of coupled oscillators (i.e.\ a metamaterial): each pair of qubits can be driven at a frequency close to one of the eigenmodes of the metamaterial resulting in a simultaneous independent operation of multiple cavity-mediated cross-cross-resonance gates via different metamaterial eigenmodes. 
On the other hand, 
it is challenging to use the RIP gate in a metamaterial setting where driving a given mode will address all transmons coupled to the mode, implementing a complicated multi-qubit gate instead of one or more independent two-qubit gates.

\subsection{Dispersive coupling for qubits \label{sec:quantumstark}}

In this subsection, we derive the dominant correction to the Hamiltonian in Eq.\ (\ref{eq:noquantumstark}), which is the dispersive coupling.

Suppose $\Delta_i \gg \Omega_i$. This implies $\theta_i \ll 1$, i.e.\ the dressing is weak. In this case, the term proportional to $g_i$ in the dressed-picture Hamiltonian [Eq.\ (\ref{eq:hamfull})] is approximately equal to the original interaction term in Eq.\ (\ref{eq:ham1}):
\ba
\approx \sum_i g_i (\sigma^+_i b+h.c.).
\ea
The effect of these terms is suppressed because $\Delta_i \gg g_i$, which makes them highly off-resonant. The lowest-order correction can be obtained in second-order perturbation theory (in $g$) \cite{blais21}:
\ba
H_\textrm{disp} = \sum_i f_i Z_i b^\dagger b, \label{eq:starkqubits}
\ea
where $f_i$ are coefficients that  
scale as $f_i \sim g_i^2/\Delta_i$,
where $\Delta_i$ can be equivalently thought of as $\omega_i - \nu$ or $\omega_i - \omega_d$ (the two are approximately equal since $\epsilon \ll \Delta_i$). 
$H_\textrm{disp}$ has such a simple diagonal form only provided that  $|\omega_1 - \omega_2| \gg g_1 g_2/|\omega_1-\nu|, g_1 g_2/|\omega_2-\nu|$, a condition we assume for the rest of the paper. If this condition were not satisfied, the resonator would have passively mediated flip-flop interactions $\sigma^+_1 \sigma^-_2 + \textrm{h.c.}$ between the two qubits. $H_\textrm{disp}$ {can be thought of as a qubit-resonator cross-Kerr term and} is the usual qubit-resonator dispersive coupling in circuit quantum electrodynamics \cite{blais21}. 
One can also think of $H_\textrm{disp}$ as a quantum Stark shift because, thanks to $H_\textrm{disp}$, the qubit experiences a Stark-like shift that is proportional to the number of photons in the oscillator, much like the regular Stark shift, which is also proportional to  intensity.   

Adding the dispersive coupling to our Hamiltonian in Eq.\ (\ref{eq:noquantumstark}), we obtain
\begin{align}
H &= \left[\epsilon + \sum_i f_i Z_i\right] b^\dagger b + \frac{1}{2} \sum_{i = 1,2} g_i Z_i \sin \theta_i (b + b^\dagger) \\
&= \epsilon' b^\dagger b + M (b + b^\dagger),\label{eq:HeppM}
\end{align}
where we defined $\epsilon' = \epsilon + \sum_i f_i Z_i$ and where, as before, $M = \frac{1}{2} \sum_{i = 1,2} g_i Z_i \sin \theta_i$.
The evolution under this Hamiltonian, much like under Eq.\ (\ref{eq:noquantumstark}), is again exactly solvable using the same approach, i.e.~treating $Z_i$ as numbers. The only difference is that, not only $M$, but also the frequency of the oscillator $\epsilon'$ now depends on $Z_i$. Going into a rotating frame relative to $\epsilon' b^\dagger b$, we get exactly the same unitary $U(t)$ as before [see Eq.\ (\ref{eq:UMS}) and Eqs.\ (\ref{eq:f}-\ref{eq:A})], 
except $\epsilon$ is replaced with $\epsilon'$:
\ba
U(t) &=& e^{-i x \sqrt{2} M \sin (\epsilon ' t)/\epsilon'} e^{-i p \sqrt{2} M (1- \cos (\epsilon' t))/\epsilon'} \nonumber \\
&& \times e^{i \frac{M^2}{\epsilon'^2} \left( \epsilon' t - \frac{1}{2} \sin (2 \epsilon' t)\right)} \label{eq:U} \\  
&=& e^{\frac{1- e^{i \epsilon' t}}{\epsilon'} M b^\dagger - \textrm{h.c.}} e^{i \frac{M^2}{\epsilon'^2} (\epsilon' t - \sin(\epsilon' t))}, \label{eq:U2}
\ea
where in the last line we used the fact that $D(\alpha) D(\beta) = e^{(\alpha \beta^* - \alpha^* \beta)/2} D(\alpha + \beta)$ with $D(\alpha) = e^{\alpha b^\dagger - \alpha^* b}$ being the displacement operator.
The lab-frame evolution is now given by $V(t) U(t)$, where 
\ba
V(t) = e^{-i \epsilon' t b^\dagger b}, \label{eq:V}
\ea
{so that the full gate at time $\tau$ is
\begin{align}
V(\tau) U(\tau) = e^{-i \epsilon' \tau b^\dagger b} e^{\frac{1- e^{i \epsilon' \tau}}{\epsilon'} M b^\dagger - \textrm{h.c.}} e^{i \frac{M^2}{\epsilon'^2} (\epsilon' \tau - \sin(\epsilon' \tau))}.\label{eq:VU}
\end{align}}

{We see from Eq.\ (\ref{eq:VU}) that the dispersive coupling is problematic for the implementation of a two-qubit gate. Assuming we do only one circle in phase space, the gate takes time $\tau \sim 1/(g \sin \theta)$, and we have $\epsilon_0 = 2 \pi/\tau \sim g \sin \theta$. While $\epsilon_0 \tau = 2 \pi$ is chosen such that 
the photons come back to their initial state in phase space in the absence of the dispersive coupling, due to the second factor in Eq.\ (\ref{eq:VU}), they now end up a distance $\sim (M/\epsilon') ((\epsilon'-\epsilon_0)/\epsilon_0)$ away from where they started, thus producing entanglement between the photons on the one hand and the two qubits (whose state determines the value of $M$) on the other. Assuming that the oscillator starts in vacuum, the overlap between the initial vacuum state and the vacuum state displaced by  
$\sim (M/\epsilon') ((\epsilon'-\epsilon_0)/\epsilon_0)$) is given by 1 minus roughly the square of this displacement. The resulting infidelity is
\be
\sim (M/\epsilon')^2 ((\epsilon'-\epsilon_0)/\epsilon_0)^2 \sim (M/\epsilon_0)^2 (f_{i,0}/\epsilon_0)^2 \sim  \left(\frac{g^2}{\Delta \epsilon_0}\right)^2,
\ee
where we assumed $f_{i,0} \sim g^2/\Delta$ for some characteristic $g$ and $\Delta$. So the resulting infidelity is negligible provided
\ba
\frac{g^2}{\Delta \epsilon_0} \ll 1.
\ea
Since $\epsilon_0 \sim g \Omega/\Delta$, this condition can also be written as
\ba
g \ll \Omega,
\ea
which, in turn, is equivalent to the intuitive condition that the dispersive coupling $f \sim g^2/\Delta$ is negligible compared to the strength $M \sim g \Omega/\Delta$ of the cross-resonance interaction.}

{The first factor in Eq.\ (\ref{eq:VU}) can also produce entanglement between the photons and the two qubits because $\epsilon'$ depends on the state of the qubits. However, the amount of entanglement depends on how populated the oscillator is:  
the photonic state gets rotated around the origin by a small angle $\tau \sum_i f_{i,0} Z_i$, which needs to be multiplied by the distance from the origin to get the displacement in phase space. If the oscillator starts in vacuum, this factor doesn't contribute to the infidelity.}

{Finally, the third factor in Eq.\ (\ref{eq:VU}) is modified by the presence of the dispersive coupling (i.e.\ when $\epsilon' \neq \epsilon$), but, for small dispersive coupling, $\tau$ can be slightly adjusted so that this term gives the correct gate up to single-qubit rotations.}

{In the above analysis, we are neglecting always-on resonator-mediated ZZ coupling between the transmons since this coupling occurs only at fourth order in $g$ (we note that the strength of this ZZ coupling is affected by the cross-cross-resonance gate, which populates the resonator).}  {We are also neglecting always-on} ZZ coupling between the two transmons mediated by direct transmon-transmon coupling that we also neglect. 
{Both of these ZZ couplings are further reduced when the interaction is mediated by two oscillators coupled in series (see Sec.\ \ref{sec:2oscillators}) or by more than two oscillators coupled in series (as can be the case with a metamaterial discussed in Sec.\ \ref{sec:meta}).}
 {Furthermore, always-on transmon-transmon} ZZ coupling commutes with the effective qubit-resonator coupling [Eq.\ (\ref{eq:noquantumstark})] and, at the end, simply adds to the cross-resonance-mediated effective ZZ Hamiltonian [Eq.\ (\ref{eq:hz1z2})]. Therefore, the equation for setting the phase to $\pi$ in the CZ gate [Eq.\ (\ref{eq:pi})] will now simply have two terms on the left-hand side, making the CZ gate just as easy to implement in the presence of  {always-on} ZZ coupling between the transmons. In Sec.\ \ref{sec:ZZcancel}, we discuss approaches for canceling unwanted always-on ZZ interactions.

In Sec.\ \ref{sec:integers} and Sec.\ \ref{sec:flowers}, we will show how to make the gate work perfectly despite the dispersive coupling using two different approaches: the ``integers'' approach in Sec.\ \ref{sec:integers} and the ``flowers'' approach in Sec.\ \ref{sec:flowers}. Before doing this, we will discuss in the next section (Sec.\ \ref{sec:transmons}) how the operation of the cavity-mediated cross-cross-resonance gate and the dispersive coupling are modified for transmons, i.e.\ in the presence of finite (as opposed to infinite) anharmonicity. 

\subsection{Gate operation and dispersive coupling for transmons \label{sec:transmons}}

So far, we have been discussing the case of a cavity-mediated cross-cross-resonance gate between qubits. In this subsection, we generalize this treatment to the cavity-mediated cross-cross-resonance gate between transmons, which feature a finite (i.e.~not infinite) anharmonicity. The Hamiltonian in Eq.\ (\ref{eq:ham1}) is then generalized to  
\ba
H &=& \nu b^\dagger b + \sum_{i = 1,2} \Big[ \omega_i c^\dagger_i c_i - \frac{\eta_i}{2} c^\dagger_i c^\dagger_i c_i c_i \nonumber \\
&& + \Omega_i \cos(\omega_d t) (c_i + c^\dagger_i) + g_i (c^\dagger_i b + h.c.)\Big],
\ea
where $c^\dagger_i$ is the bosonic creation operator describing transmon $i$ and $\eta_i > 0$ is the anharmonicity of transmon $i$.
We now make the rotating-wave approximation and go to a rotating frame with respect to 
\ba
H_0 = \omega_d (b^\dagger b + c^\dagger_1 c_1 + c^\dagger_2 c_2)
\ea
to get 
\ba
H &=& \epsilon b^\dagger b + \sum_{i = 1,2} \Big[ \Delta_i c^\dagger_i c_i - \frac{\eta_i}{2} c^\dagger_i c^\dagger_i c_i c_i \nonumber \\
&& + \frac{\Omega_i}{2} (c_i + c^\dagger_i) + g_i (c^\dagger_i b + h.c.)\Big].
\label{eq:CR1}
\ea 
Let us now diagonalize each transmon assuming that $\Omega_i \ll \Delta_i,|\Delta_i-\eta_i|$. The lowest two new (dressed) eigenstates are then 
\ba
|\tilde 0\rangle_i &\approx& |0\rangle_i - \frac{\Omega_i}{2 \Delta_i} |1\rangle_i,\\
|\tilde 1\rangle_i &\approx& |1\rangle_i + \frac{\Omega_i}{2 \Delta_i} |0\rangle_i- \frac{\Omega_i}{\sqrt{2} (\Delta_i-\eta_i)} |2\rangle_i.\label{eq:1tilde}
\ea 
 In the dressed basis, we therefore get a cross-cross-resonance gate driven by the Hamiltonian {\cite{malekakhlagh20}}
{\begin{align}
H &= \epsilon b^\dagger b +  \sum_{i = 1,2} \frac{g_i}{2} (Z_i A_i + B_i) (b + b^\dagger) \\& \equiv \epsilon b^\dagger b + (M+M_0) (b + b^\dagger),
\label{eq:Hcc} 
\end{align}
where 
\begin{align}
A_i &= {}_i\langle \tilde 1| c_i |\tilde 1\rangle_i - {}_i\langle \tilde 0| c_i |\tilde 0\rangle_i \nonumber \\
&= \left(\frac{\Omega_i}{2 \Delta_i} - \frac{\Omega_i}{\Delta_i - \eta_i} \right) + \frac{\Omega_i}{2 \Delta_i}  \nonumber \\
&= -\frac{\Omega_i \eta_i}{\Delta_i (\Delta_i - \eta_i)},  \label{eq:Ai}\\
B_i &=  {}_i\langle \tilde 1| c_i |\tilde 1\rangle_i + {}_i\langle \tilde 0| c_i |\tilde 0\rangle_i = - \frac{\Omega_i}{\Delta_i - \eta_i}, \label{eq:Bi}\\
M &= \sum_{i = 1,2} \frac{g_i}{2} Z_i A_i, \\
M_0 &= \sum_{i = 1,2} \frac{g_i}{2} B_i.\label{eq:M0}
\end{align}}
As a sanity check, at vanishing anharmonicity {(when the transmon becomes a perfectly harmonic oscillator) the presence of state $\ket{2}$ causes the proposed gate mechanism to fail}  ($A_i = 0$), while at infinite anharmonicity {(when the transmon becomes a two-level system)}, we recover $A_i = \Omega_i/\Delta_i$ {and $B_i = 0$}, as in Eq.\ (\ref{eq:noquantumstark}). 

{The presence of the $M_0$ term does not materially affect the closed-system operation of the gate. Indeed, for transmons, $M^2$ in Eq.\ (\ref{eq:phase}) is replaced with $(M + M_0)^2 = M^2 + 2 M M_0 + M_0^2$, so that we get the desired gate up to single-qubit $Z$ rotations. If desired, the $M_0$ term can also be canceled by driving the cavity.}

We now turn to computing the dispersive coupling for transmons. As in the case of qubits, in the limit of weak dressing, the dispersive coupling can be computed in the absence of the cross-resonance drive. As in the case of qubits,  
the dispersive coupling arises at second order in $g_i$ \cite{blais21}. Since the dispersive coupling is computed independently for each transmon, we will focus on one transmon with index $i$. In the basis $\ket{\textrm{transmon $i$}, \textrm{cavity}}$, the vacuum state $\ket{00}$ is an exact eigenstate and is therefore not shifted. The states $\ket{10}$ and $\ket{01}$ couple to each other and therefore experience equal and opposite shifts that cancel each other when computing the strength of the $Z_i b^\dagger b$ interaction. Finally, the state $\ket{11}$ couples to states $\ket{02}$ and $\ket{20}$, giving rise to the shift
\ba
\frac{2 g_i^2}{\Delta_i} - \frac{2 g_i^2}{\Delta_i - \eta_i} = - \frac{2 g_i^2 \eta_i}{\Delta_i (\Delta_i - \eta_i)}, \label{eq:disp}
\ea
where, as in the qubit approximation, $\Delta_i$ can be equivalently thought of as $\omega_i - \nu$ or $\omega_i - \omega_d$ (the two are approximately equal since $\epsilon \ll \Delta_i$). 
We can therefore write the dispersive coupling as \cite{blais21}
\ba
H_\textrm{disp} = \sum_i f_i Z_i b^\dagger b,
\ea
where 
\ba
f_i \sim \frac{g_i^2 \eta_i}{\Delta_i (\Delta_i - \eta_i)}.
\ea
This again reduces to zero in the case of zero anharmonicity and is $\propto g_i^2/\Delta_i$ in the regime of infinite anharmonicity in agreement with Eq.\ (\ref{eq:starkqubits}). 

To conclude, in the case of transmons, the desired (``good'') interaction and the (``bad'') dispersive coupling have strength $-\frac{g_i \Omega_i \eta_i}{2 \Delta_i (\Delta_i - \eta_i)}$ and $f_i \sim \frac{g_i^2 \eta_i}{\Delta_i (\Delta_i - \eta_i)}$, respectively, as compared to their qubit versions (i.e.\ $\eta_i \rightarrow \infty$ versions) $g_i \Omega_i/(2 \Delta_i)$ and $f_i \sim g_i^2/\Delta_i$, respectively.

An alternative derivation goes as follows. Starting with Eq.~(\ref{eq:CR1}) and focusing on one of the transmons, we begin by diagonalizing the linear term, i.e., 
\ba
H_0 = \epsilon b^\dagger b +   \Delta c^\dagger c +  g (c^\dagger b + c b^\dagger) + \frac{\Omega}{2}(c + c^\dagger).
\ea 
Focusing on the regime $\epsilon, g, \Omega \ll \Delta,$ {we first do the unitary transformation $c \rightarrow c - \frac{\Omega}{2 \Delta}$, which gives (up to an additive constant)
\ba
H_0 = \epsilon b^\dagger b +   \Delta c^\dagger c +  g (c^\dagger b + c b^\dagger) - \frac{g \Omega}{2 \Delta}(b + b^\dagger),
\ea
where the last term recovers $B_i$ from Eq.\ (\ref{eq:Bi}) in the limit $\eta = 0$. We then do a second unitary transformation to diagonalize the $g$ term: 
\begin{align}
c &\rightarrow c - \frac{g}{\Delta} b + O\left(\frac{1}{\Delta^2}\right),\\
b &\rightarrow b + \frac{g}{\Delta} c + O\left(\frac{1}{\Delta^2}\right).
\end{align}}
{The composition of the two unitary transformations corresponds, at $O\left(\frac{1}{\Delta^2}\right)$, to $c \rightarrow c - \frac{g}{\Delta} b - \frac{\Omega}{2 \Delta}$, which we insert into the anharmonicity term to obtain  
\begin{align}
&- \frac{\eta}{2} \left(c - \frac{g}{\Delta} b - \frac{\Omega}{2 \Delta}\right)^\dagger \left(c - \frac{g}{\Delta} b - \frac{\Omega}{2 \Delta}\right)^\dagger \nonumber \\
& \times 
\left(c - \frac{g}{\Delta} b - \frac{\Omega}{2 \Delta}\right) 
\left(c - \frac{g}{\Delta} b - \frac{\Omega}{2 \Delta}\right).
\end{align}}
To quadratic order in $1/\Delta$, the terms proportional to $\eta$ that conserve $c^\dagger c$ are (dropping the $c^\dagger c$ term itself)
\be
- \frac{\eta}{2}  c^\dagger c^\dagger c c   - 2 \eta   \left( \frac{g}{\Delta} \right)^2 b^\dagger b c^\dagger c - \eta \frac{g \Omega}{\Delta^2} c^\dagger c (b +b^\dagger), 
\ee
where the second term is the ``bad'' dispersive shift (which agrees with Eq.\ (\ref{eq:disp}) to first order in $\eta$), while the last term {can be written as
\be
- \eta \frac{g \Omega}{\Delta^2} \frac{Z + 1}{2} (b +b^\dagger)
\ee
and therefore reproduces the desired ``good'' interaction in Eq.\ (\ref{eq:Ai}) to first order in $\eta$, as well as the first-order-in-$\eta$ contribution to $B_i$ in Eq.\ (\ref{eq:Bi}).}

\section{Two approaches for canceling the dispersive coupling}

\subsection{The ``integers'' approach \label{sec:integers}}

In this subsection, we describe one approach, which we call the ``integers'' approach, to canceling the dispersive coupling. In the following subsection (Sec.\ \ref{sec:flowers}), we will describe the second approach.

To make sure the unitary $U(t = \tau)$ in Eq.\ (\ref{eq:U}) doesn't entangle qubits with the oscillator, we now need to make sure that
\ba
\epsilon' \tau = 2 \pi * (\textrm{nonzero integer}) \label{eq:epsprime}
\ea
for all four choices of $\{Z_1, Z_2\}$ (recall that $\epsilon' = \epsilon + \sum_i f_i Z_i$). [The reason $\epsilon' = 0$ has to be avoided is because, when $\epsilon'= 0$, the evolution under $H$ in Eq.\ (\ref{eq:HeppM}) never returns the photons back to the initial state and never disentangles the photons from the qubits.] Notice that the condition in Eq.\ (\ref{eq:epsprime}) not only dramatically simplifies $U(\tau)$, but also ensures that $V(\tau) = 1$ [see Eq.\ (\ref{eq:V})], so that the rotating frame and the lab frame coincide at the final time $\tau$, which is essential because $V(t)$ is in general entangling. 

The condition in Eq.\ (\ref{eq:epsprime}) is equivalent to requiring that
\ba
\epsilon \tau &=& \pi n_\epsilon,\\
f_1 \tau &=& \pi n_1,\\
f_2 \tau &=& \pi n_2,
\ea
where $n_\epsilon$, $n_1$, and $n_2$ are integers [so that $\epsilon' \tau = \pi (n_\epsilon + n_1 Z_1 + n_2 Z_2)$] satisfying two additional conditions: (1) that $n_\epsilon + n_1 + n_2$ is even and (2) that $n_\epsilon \pm n_1 \pm n_2 \neq 0$ for all 4 combinations of signs. [A special case of condition (1) is when $n_\epsilon$, $n_1$, and $n_2$ are all even. Another special case of condition (1) is when $f_1 = f_2 = 0$, in which case $n_1 = n_2 = 0$, so $n_\epsilon$ must be even, recovering Eq.\ (\ref{eq:epsilon}).] 

Once Eq.\ (\ref{eq:epsprime}) is satisfied, both the rotating-frame unitary in Eq.\ (\ref{eq:U}) and the lab-frame unitary $V(\tau) U(\tau)$ [see Eq.\ (\ref{eq:V}) for the definition of $V(t)$] 
simplify to
\ba
U(\tau) = e^{i \frac{M^2}{\epsilon'} \tau}. \label{eq:Uintegers} 
\ea
Defining four phases $\phi(Z_1,Z_2) = \frac{M^2}{\epsilon'} \tau$ (where $Z_i = \pm 1$ and $M = \frac{1}{2} \sum_{i = 1,2} g_i Z_i \sin \theta_i$), the condition for implementing a gate that is, up to single-qubit rotations, equivalent to a CZ gate is
\ba
|\phi(1,1) + \phi(-1,-1) - \phi(1,-1) - \phi(-1,1)| = \pi. \label{eq:piphase}
\ea
While any odd multiple of $\pi$ would work on the right-hand side of Eq.\ (\ref{eq:piphase}), having $\pi$ on the right-hand side gives the fastest gate. {The combination of Eq.\ (\ref{eq:epsprime}) and Eq.\ (\ref{eq:piphase}) is conceptually similar to synchronizing two-qubit-gate entangling parameters with minima in the leakage channel in Refs.\  \cite{strauch03,barends19}.}

An important question is whether we have enough tuning knobs to satisfy Eq.\ (\ref{eq:piphase}) while simultaneously obtaining integer values for $n_\epsilon$, $n_1$, and $n_2$. If $g_i$, $\omega_i$, and $\nu$ are not tunable, then we only have two knobs---$\tau$ and $\omega_d$---to obtain three integer values $n_\epsilon$, $n_1$, and $n_2$. Therefore, we need an additional knob, which can be realized by making $g_1$, $g_2$, $\omega_1$, $\omega_2$, or $\nu$ tunable (potentially at the expense of reduced coherence time of the qubit and/or the resonator). Finally, Eq.\ (\ref{eq:piphase}) is then easy to satisfy since this equation has two additional tuning knobs $\Omega_1$ and $\Omega_2$.

Much like in the construction without dispersive coupling, where it is important to have stability to ensure that $\epsilon \tau/(2 \pi)$ remains very close to an integer despite all fluctuations, we will now similarly need to have stability to ensure that $\epsilon \tau/\pi$, $f_1 \tau/\pi$, and $f_2 \tau/\pi$ remain very close to integer values.

It is also important to point out that, typically, $f_i \ll \epsilon$, which means that $n_\epsilon \gg 1$, which in turn means that this approach will typically lead to a  slowdown of the gate by a factor of $\sqrt{n_\epsilon} \gtrsim \sqrt{\epsilon/f_i}$ [see Eq.\ (\ref{eq:rootn})].

{We note that the $M_0$ term in Eq.\ (\ref{eq:Hcc}), which arises for transmons in addition to the desired $M$ term, does not affect the performance of the integers approach. In the presence of the $M_0$ term, Eq.\ (\ref{eq:Uintegers}) becomes
\be
U(\tau) = e^{i \frac{(M+M_0)^2}{\epsilon'} \tau}.
\ee
The contribution of the term $(2 M_0 M + M_0^2)/\epsilon'$ to $\phi(1,1) + \phi(-1,-1) - \phi(1,-1) - \phi(-1,1)$ in Eq.\ (\ref{eq:piphase}) is small because this contribution vanishes when $\epsilon' \rightarrow \epsilon$. We also point out that the $M_0$ term in Eq.\ (\ref{eq:Hcc}) can be canceled by driving the cavity.}

\subsection{The “flowers” approach \label{sec:flowers}}

In this subsection, we describe the second approach, which we call the ``flowers'' approach, for canceling the dispersive coupling.

Recall that the Hamiltonian in the presence of the dispersive coupling is [Eq.\ (\ref{eq:HeppM})]
\ba
H = \epsilon' b^\dagger b  + M (b + b^\dagger),
\ea
where $\epsilon' = \epsilon + \sum_i f_i Z_i$ and $M = \frac{1}{2} \sum_{i = 1,2} g_i Z_i \sin \theta_i$. Since the dispersive coupling is linear in $Z_i$, a natural approach for canceling it is to apply dynamical decoupling in the form of $\pi$ pulses that change $Z_i \rightarrow -Z_i$ periodically and simultaneously for both $i = 1$ and $i=2$. Unfortunately, a naive application of this approach would in general fail since the desired $M$ term is also linear in $Z_i$ and will also be affected. In particular, if we apply the $\pi$ pulses rapidly, both the dispersive coupling and the desired $M$ term would vanish. However, the dispersive coupling and the desired $M$ term couple differently to the resonator ($b^\dagger b$ versus $b^\dagger + b$). Therefore, if we apply the $\pi$ pulses less rapidly, it  {is} plausible{---and is indeed the case, as we will show in this section---}that  
{we can} refocus the dispersive coupling but not the desired $M$ term. As another intuition for why such dynamical decoupling may work, recall that, in the limit of large detuning $\epsilon \gg |M|$, one can adiabatically eliminate the resonator to obtain an interaction between the qubits $\propto Z_1 Z_2/\epsilon$ [Eq.\ (\ref{eq:hz1z2})]. Assuming that $\epsilon \gg |\epsilon - \epsilon'|$, the effective $Z_1 Z_2$ interaction is approximately the same in the presence of the dispersive coupling. Dynamical decoupling can then cancel the dispersive coupling in the original $\epsilon' b^\dagger b$ term (which is linear in $Z_i$) without affecting the product $Z_1 Z_2$ (which is quadratic in $Z_i$). 
What we show in the ``flowers'' approach described in this section is that, by strategically timing the application of simultaneous $\pi$ pulses on both qubits, one can cancel the dispersive coupling exactly and do so even when the detuning $\epsilon$ is too small for the adiabatic elimination to work.

The ``flowers'' approach presented in this section is identical to the dynamical decoupling approach 
proposed and demonstrated on trapped ions in Ref.\  \cite{manovitz17} and inspired by ultrafast gate schemes \cite{mizrahi2013ultrafast,garcia2003speed}. However, in Ref.\ \cite{manovitz17}, this  approach was used to cancel slowly varying $Z$ noise acting on the qubits during a M{\o}lmer-S{\o}rensen gate. Here we show that this approach also perfectly cancels the dispersive coupling, which can also be thought of as $Z$ noise, but with a coefficient proportional to $b^\dagger b$.

Let us show how this approach for canceling the dispersive coupling works in the case where we apply $P=4$ pairs of $\pi$ pulses ({we show in App.\ \ref{sec:Pneq4} that} this is the smallest $P$ that works). 
Defining $x = \sum_i f_i Z_i/\epsilon$ and $\epsilon_+ = \epsilon (1+x)$, 
the Hamiltonian with the dispersive coupling [Eq.\ (\ref{eq:HeppM})] can be written as 
\ba
H = \epsilon_+ b^\dagger b  + M (b + b^\dagger),\label{eq:H1px}
\ea
where $M = \frac{1}{2} \sum_{i = 1,2} g_i Z_i \sin \theta_i$, as before. 
\begin{figure}[tb!]
\centering
\includegraphics[width= \linewidth]{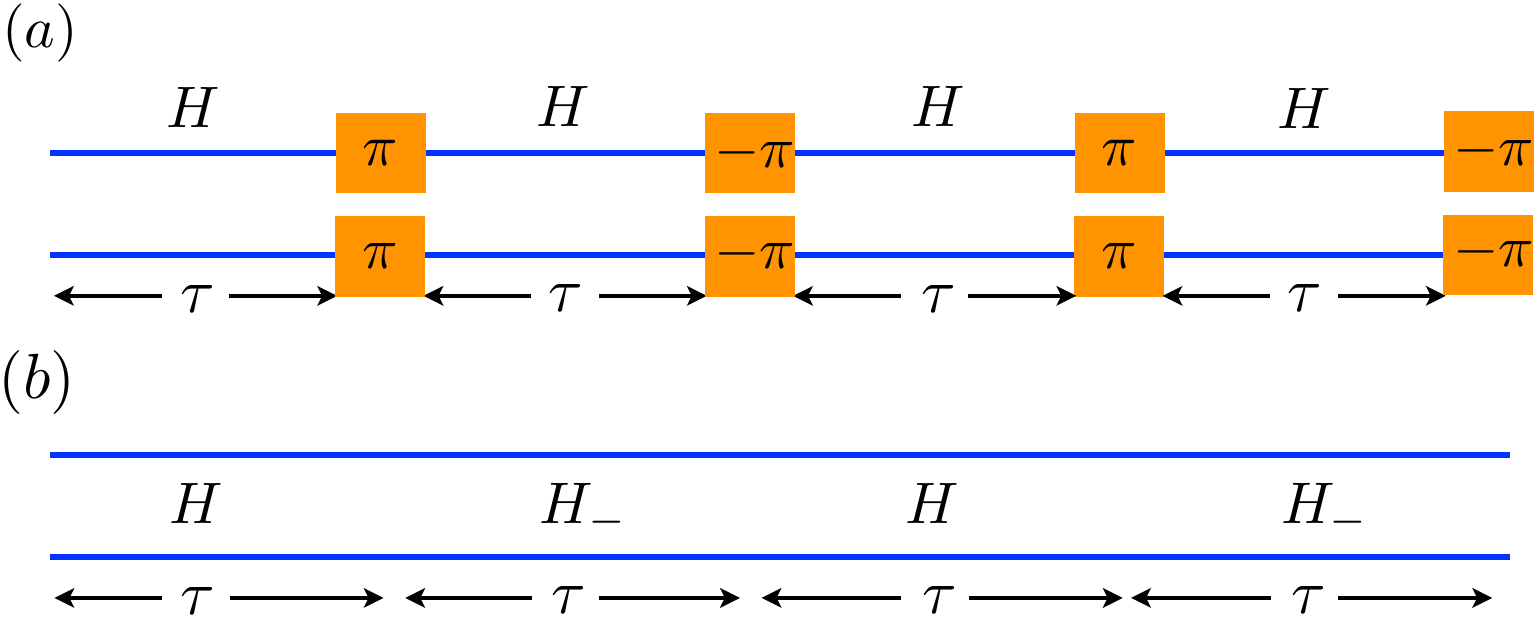} 
\caption{(a) The pulse sequence used in the $P = 4$ ``flowers'' approach for canceling the dispersive coupling. The two blue lines represent the qubits, and time goes from left to right. The system, including both qubits and the resonator, evolves under Hamiltonian $H$ in Eq.\ (\ref{eq:H1px}) for time $\tau = 3 \pi/(2 \epsilon)$. Then a $\pi$ pulse $\exp(-i X_j \pi/2)$ is applied on both qubits $j = 1,2$. Then the system evolves again for time $\tau$. Then a $-\pi$ pulse $\exp(i X_j \pi/2)$ is applied on both qubits. Then the entire process is repeated one more time. (b) Since conjugating $H$ by a pair of $\pi$ pulses gives $H_-$ [Eq.\ (\ref{eq:Hminus})], the evolution in (a) is equivalent to the evolution in (b).  \label{fig:N4pulses}
}
\end{figure}
As shown in Fig.\ \ref{fig:N4pulses}, we start by applying this Hamiltonian for time $\tau = 3 \pi/(2 \epsilon)$. We then apply the Hamiltonian again for time $\tau$, but this time conjugated by a pair of $\pi$ pulses on both qubits, which changes $Z_j \rightarrow -Z_j$ for both $j$. This effectively changes the signs of $x$ and $M$ to give
\ba
H_- = \epsilon_- b^\dagger b  - M (b + b^\dagger), \label{eq:Hminus}
\ea
where $\epsilon_- = \epsilon (1-x)$.
We then repeat evolution under $H$ for time $\tau$, followed again by an evolution under $H_-$ for time $\tau$. 

It is important to emphasize that the $\pi$ pulses, which we take to be of duration $\tau_\pi$, must be performed slowly enough to respect the dressing [see Eq.\ (\ref{eq:dressingU})], but fast enough to take negligible time compared to the dynamics in Eq.\ (\ref{eq:H1px}). The latter condition is $(\tau_\pi)^{-1} \gg \epsilon, g \Omega/\Delta$. To derive the condition for respecting the dressing, we write---in the rotating-wave approximation and in the rotating frame---the single-qubit Hamiltonian subject to the cross-resonance drive with Rabi frequency $\Omega$ and detuning $\Delta$ and to the $\pi$-pulse drive with Rabi frequency $\Omega'$ and frequency shift $\tilde \Delta = \sqrt{\Omega^2 + \Delta^2}$ relative to the cross-resonance drive:
 \ba
 H = \frac{\Delta}{2} Z + \left(\frac{\Omega + \Omega' e^{i \tilde \Delta t}}{2} \ket{0}\bra{1} + h.c.\right). \label{eq:pipulseH}
\ea
As a sanity check, in the absence of the cross-resonance drive ($\Omega = 0$), we have $\tilde \Delta = \Delta$ and obtain a perfect $\pi$ pulse on the bare states after time $\pi/\Omega'$. In the presence of the cross-resonance drive, we go into the dressed basis $\ket{\tilde 0}$ and $\ket{\tilde 1}$ that diagonalizes the $\Omega' = 0$ Hamiltonian [see Eq.\ (\ref{eq:dressingU})]. Re-expressing $\Omega' e^{i \tilde \Delta t} \ket{0} \bra{1}$ in Eq.\ (\ref{eq:pipulseH}) in terms of $\ket{\tilde 0}$ and $\ket{\tilde 1}$, the dominant correction to the desired $\Omega' e^{i \tilde \Delta t} \ket{\tilde 0} \bra{\tilde 1}$ term is $\sim \frac{\Omega' \Omega}{\Delta} (\ket{\tilde 1} \bra{\tilde 1} - \ket{\tilde 0} \bra{\tilde 0}) \cos(\tilde \Delta t)$. We see that this term is suppressed in amplitude by $\Omega/\Delta$ and oscillates at frequency $\tilde \Delta \approx \Delta$. We therefore expect this term to be negligible provided $\frac{\Omega' \Omega}{\Delta} \ll \Delta$. Since the duration of the $\pi$ pulse scales as $\tau_\pi \sim 1/\Omega'$, this condition corresponds to $\Delta^2/\Omega \gg (\tau_\pi)^{-1}$. An alternative approach would be to temporarily ramp down the cross-resonance drives to apply the $\pi$ pulses in the bare basis. Optimal control can be used to reduce  errors associated with $\pi$-pulse imperfections.

We will now show that this sequence can implement a perfect CZ gate between the two qubits. 
It is convenient to work in the rotating frame where both $H$ and $H_-$ have the $b^\dagger b$ term rotated away:
\ba
H &=& M be^{-i \epsilon_+ t} + h.c., \label{eq:Hprf}\\
H_- &=& - M be^{- i \epsilon_- t} + h.c., \label{eq:Hmrf}.
\ea

It is {also} convenient to track the evolution in phase space using Heisenberg equation of motion for $b$.   {We show in App.\ \ref{sec:P4derivation} that} 
our time evolution from $t = 0$ to $t = 4 \tau$ is as follows. With $t$ running from $t = 0$ to $t = \tau$ in each of the following four equations, we have
\ba
b(t) &=& b(0) + \frac{M}{\epsilon_+} \left(1 - e^{i \epsilon_+ t}\right), \label{eq:tau1}\\ 
b(\tau + t) &=& b(\tau) - \frac{M e^{i \epsilon_+ \tau}}{\epsilon_-} \left(1 - e^{i \epsilon_- t}\right), \label{eq:tau2}\\ 
b(2 \tau + t) &=& b(2 \tau) + \frac{M e^{i (\epsilon_++\epsilon_-) \tau}}{\epsilon_+} \left(1 - e^{i \epsilon_+ t}\right),\label{eq:tau3}\\ 
b(3 \tau + t) &=& b(3 \tau) - \frac{M e^{i (2 \epsilon_++\epsilon_-) \tau}}{\epsilon_-} \left(1 - e^{i \epsilon_- t}\right).\label{eq:tau4}
\ea

\begin{figure}[t!]
\centering
\includegraphics[width= \linewidth]{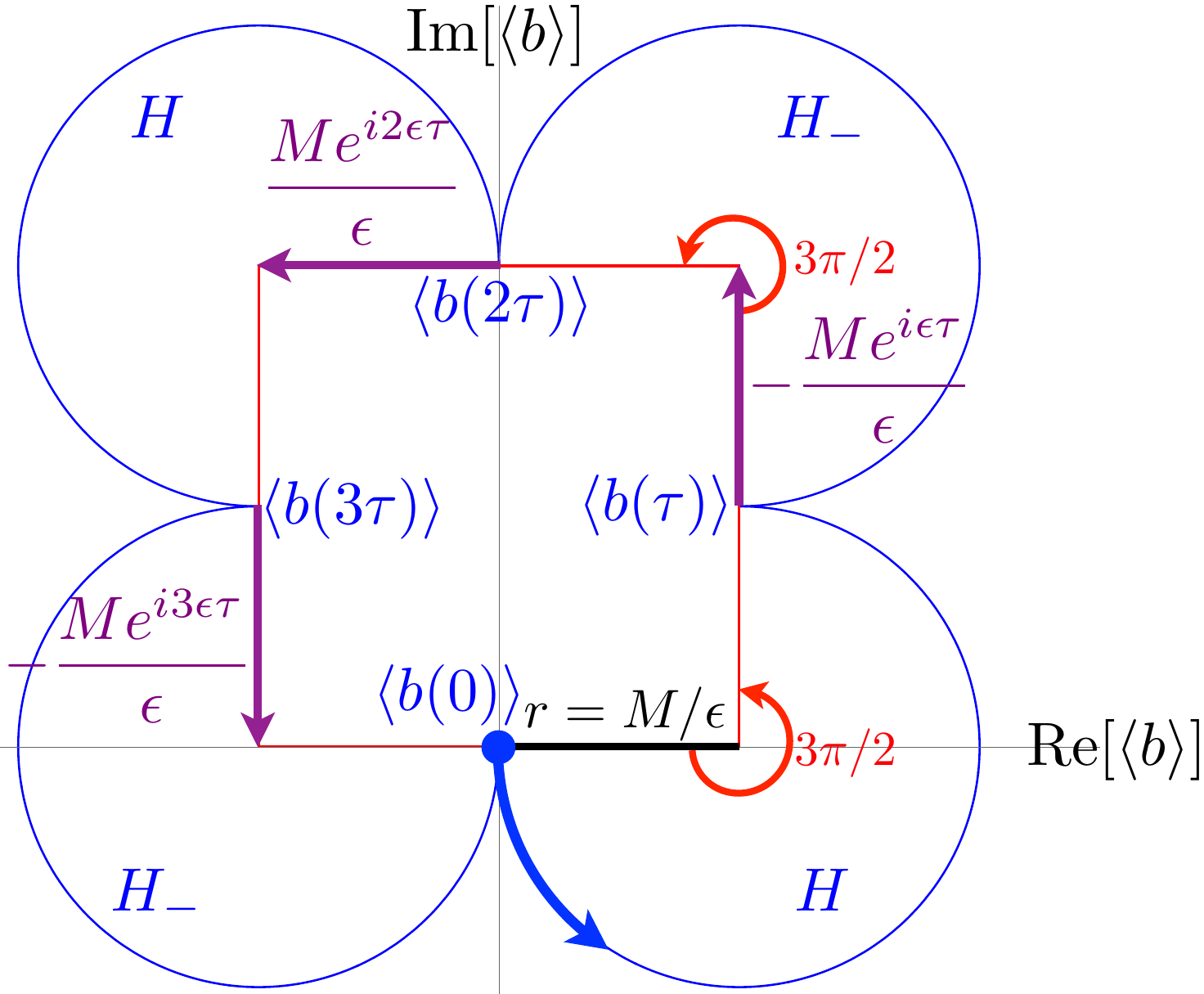} 
\caption{The $P = 4$ flower for the case $x = 0$. Starting at the origin in phase space, $\langle b \rangle$  evolves counterclockwise for angle $3 \pi/2$ around four circles in the following order: lower right circle, upper right circle, upper left circle, lower left circle. The phase-space trajectory comes back to the origin and looks like a 4-petal flower. \label{fig:N4x0}
}
\end{figure}

In Fig.\ \ref{fig:N4x0}, we show what this evolution looks like for the case $x = 0$ considered in Ref.\ \cite{manovitz17}. {We remind the reader that $x = \sum_i f_i Z_i/\epsilon$ is a (fractional) dispersive coupling, so $x=0$ corresponds to a vanishing dispersive coupling.}  In this case, $\epsilon_+ = \epsilon_- = \epsilon$. 
For concreteness, let us assume that we start at the origin, but if we started at a different location, the entire picture would be the same except translated to the new location. Equation (\ref{eq:tau1}) gives us counterclockwise rotation by an angle $\epsilon \tau = 3 \pi/2$ along the  
lower right circle  of radius $r = M/\epsilon$, whose center is shifted by $M/\epsilon$ from the origin in the complex plane. Equation (\ref{eq:tau2}) gives us counterclockwise rotation by an angle $3 \pi/2$ along the top right circle, whose center is shifted by $-M e^{i \epsilon \tau}/\epsilon = -M e^{i 3 \pi/2}/\epsilon = i M/\epsilon$ from $\langle b(\tau)\rangle$ in the complex plane. Equation (\ref{eq:tau3}) gives us counterclockwise rotation by an angle $3 \pi/2$ along the top left circle, whose center is shifted by $M e^{i 2 \epsilon \tau}/\epsilon = M e^{i 3 \pi}/\epsilon = -M/\epsilon$ from $\langle b(2 \tau)\rangle$ in the complex plane. Finally, Eq.~(\ref{eq:tau4}) gives us counterclockwise rotation by an angle $3 \pi/2$ along the bottom left circle, whose center is shifted by $-M e^{i 3 \epsilon \tau}/\epsilon = -M e^{i 9 \pi/2}/\epsilon = -i M/\epsilon$ from $\langle b(3 \tau)\rangle$ in the complex plane.  
We call this approach the ``flowers'' approach because Fig.\ \ref{fig:N4x0} looks like a 4-petal flower. 

\begin{figure}[t!]
\centering
\includegraphics[width=\linewidth]{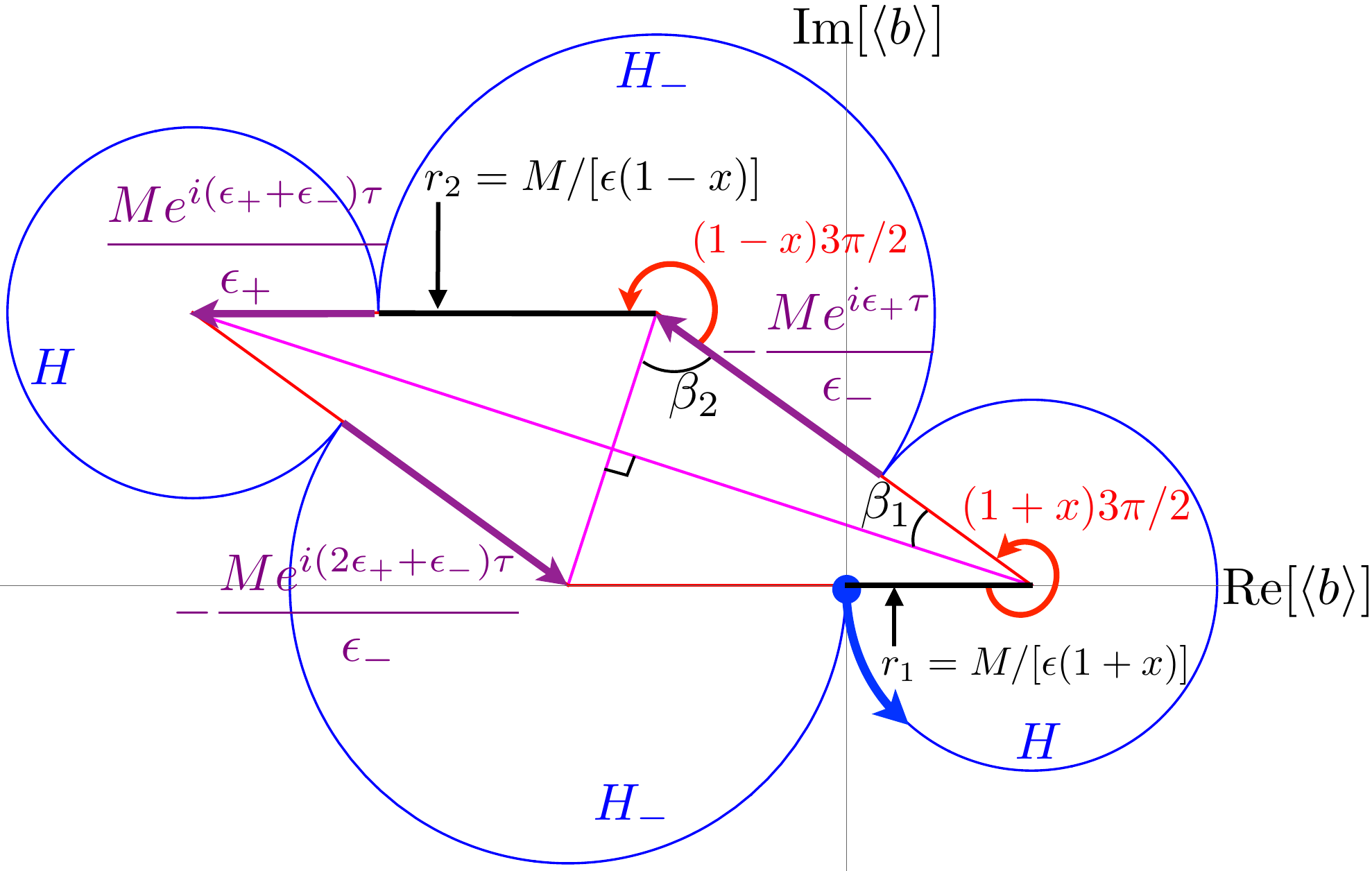} 
\caption{The $P = 4$ flower for the case $x = 0.2$. Starting at the origin in phase space, $\langle b \rangle$  evolves counterclockwise around four circles: for angle $(1+x) 3 \pi/2$ around the right small circle, for angle $(1-x) 3 \pi/2$ around the top large circle, for angle $(1+x) 3 \pi/2$ around left small circle, and for angle $(1-x) 3 \pi/2$ around the bottom large circle. The phase-space trajectory comes back to the origin. \label{fig:N4}
}
\end{figure}

One might worry that, when {the dispersive coupling is present (i.e.}~$x \neq 0${)}, the photonic state does not come back to the origin. However, we show in Fig.\ \ref{fig:N4} that the photonic state does come back to where it started even for $x \neq 0$.
Equation (\ref{eq:tau1}) gives us counterclockwise rotation by an angle $\epsilon (1+x) \tau = (1+x) 3 \pi/2$ along the   
right small circle  of radius $r_1 = M/[\epsilon(1+x)] = M/\epsilon_+$, whose center is shifted by $r_1$ from the origin in the complex plane. Equation (\ref{eq:tau2}) gives us counterclockwise rotation by an angle $\epsilon (1-x) \tau = (1-x) 3 \pi/2$ along the top large circle of radius $r_2 = M/\epsilon_-$, whose center is shifted by $-M e^{i \epsilon_+ \tau}/\epsilon_- = - r_2 e^{i (1+x) 3 \pi/2} = i e^{i x 3 \pi/2} r_2$ from $\langle b(\tau)\rangle$ in the complex plane. Equation (\ref{eq:tau3}) gives us counterclockwise rotation by an angle $(1+x) 3 \pi/2$ along the left small circle, whose center is shifted by $M e^{i (\epsilon_+ + \epsilon_-) \tau}/\epsilon_+ = r_1 e^{i 3 \pi}/\epsilon = -r_1$ from $\langle b(2 \tau)\rangle$ in the complex plane. Finally, Eq.~(\ref{eq:tau4}) gives us counterclockwise rotation by an angle $(1-x) 3 \pi/2$ along the bottom large circle, whose center is shifted by $-M e^{i (2 \epsilon_+ + \epsilon_-) \tau}/\epsilon_- = - i r_2 e^{i x 3 \pi/2}$ from $\langle b(3 \tau)\rangle$ in the complex plane.  
Intuitively, the reason the photonic state comes back to where it started is because the over-rotation by $x 3 \pi/2$ in $(1 + x) 3 \pi/2$ cancels with the under-rotation by $x 3 \pi/2$ in $(1 - x) 3 \pi/2$. This is the reason why we need an even number $P$ of petals in our flowers.

We can also explain this evolution as follows (where we will assume that $x < 1$, although the scheme also works for $x > 1$). The effect of switching the sign of $M$ (as we switch from $H$ to $H_-$ and back) is to flip the instantaneous direction of travel, but keep the counterclockwise sense in which the circle is traced. This means that neighboring circles are tangent to each other at the point where the $\pi$ pulse induces the switch from one circle to the other, and the line connecting the centers of these circles also goes through this point.  
Since we go along a small circle for angle $(1+x) 3 \pi/2$ and then along a big circle for angle $(1-x) 3 \pi/2$, we make an overall counterclockwise rotation by $6 \pi/2 = 3 \pi$, which means that the line connecting the top two circles is horizontal. Similarly, as we now go along the left small circle and then the bottom large circle, we again make an overall rotation by $3 \pi$. This means that the line from the center of the bottom circle to the final point is also horizontal, meaning that we arrive back at the origin, as desired.

As a result, the centers of the four circles form a rhombus. It is convenient to define the angle $\beta_1$ as half of the angle left after going around the small circle: 
\ba
\beta_1 = \pi - (1+x) 3 \pi/4 = (1 - 3 x) \pi/4,
\ea
and $\beta_2$ as half of the angle left after going around a large circle:
\ba
\beta_2 = \pi - (1-x) 3 \pi/4 = (1 + 3 x) \pi/4.\label{eq:beta2}
\ea
As expected, $\beta_1+\beta_2 = \pi/2$, so the corresponding triangle is a right triangle, as must be the case for a rhombus.

{In App.\ \ref{sec:Pneq4}, we show that all flowers with even $P \geq 4$ successfully cancel the dispersive coupling during a cavity-mediated cross-cross-resonance gate. One motivation for considering flowers with larger values of $P$ is their ability to better cancel slowly varying $Z$ noise \cite{manovitz17}, which may be present in our system in addition to the dispersive coupling. Appendix \ref{sec:Pneq4} also provides the formula for the area of the flowers for all even $P \geq 4$. This area determines the nonlinear phase of the resulting controlled-phase gate, and $\epsilon$---which will therefore be $P$-dependent---is tuned to set this phase to $\pi$,  corresponding to a CZ gate.} 

\begin{figure}[t!]
\centering
\includegraphics[width=\linewidth]{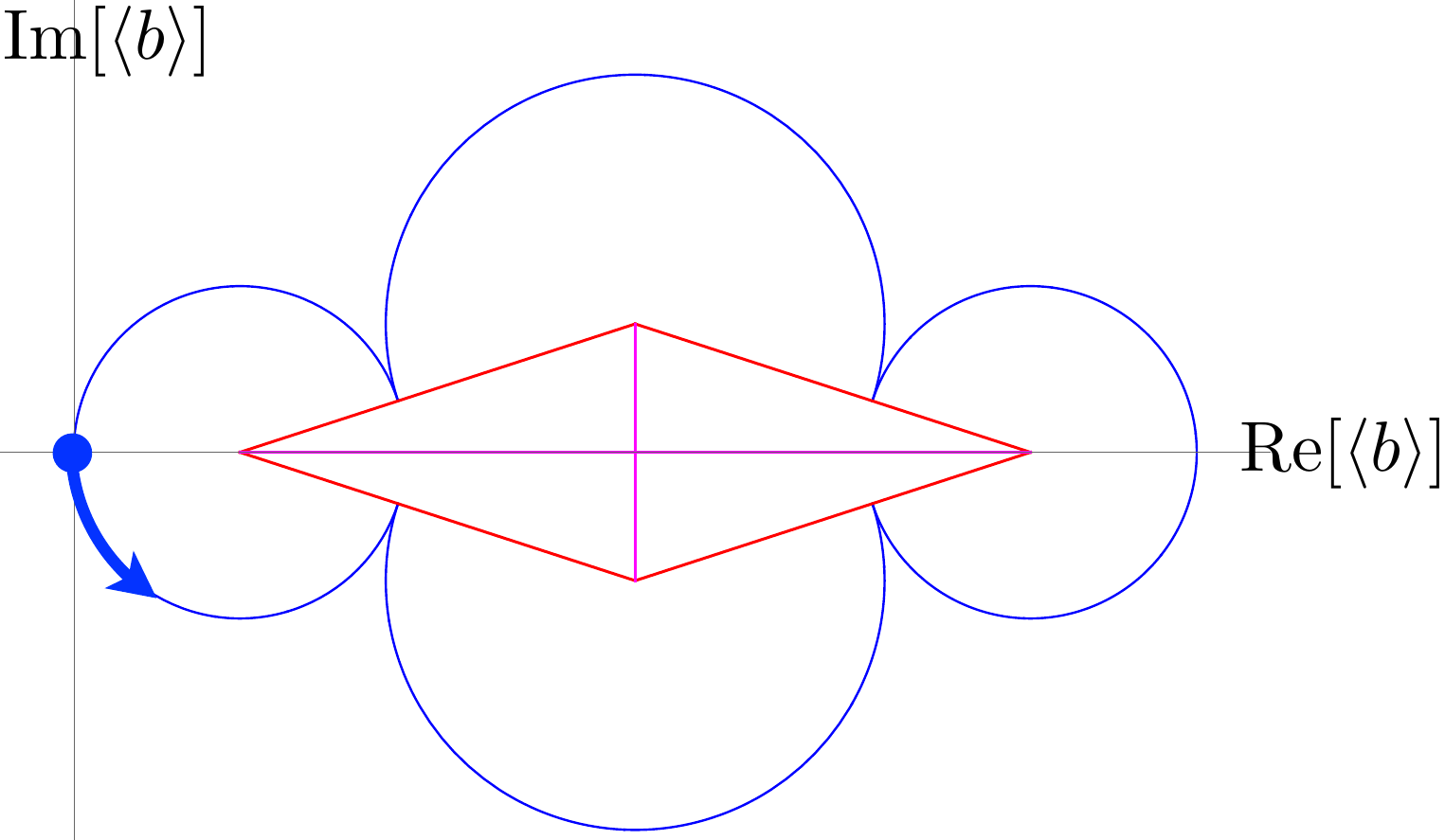} 
\caption{A modification of the $P=4$ flower from Fig.\ \ref{fig:N4}: the first time interval is $\tilde \tau = \tau/2$ [where $\tau = 3 \pi/(4 \epsilon)$], while the last time interval is $\tau - \tilde \tau$. This modification simply rotates the original flower in phase space. \label{fig:modifiedN4}
}
\end{figure}

{In App.\ \ref{sec:Pneq4}, we also slightly generalize the flowers approach to pulse sequences where the first time interval (i.e.~wait time before the first pair of pulses) is shorter than the time intervals between the following pairs of pulses. A special case of this generalization is the Carr-Purcell-Meiboom-Gill-like (CMPG-like) \cite{carr54,meiboom58} pulse sequence where the first time interval is half of the other time intervals. For $P = 4$, the corresponding flower is shown in Fig.\ \ref{fig:modifiedN4}.}

{In App.\ \ref{sec:Pneq4}, we also show that the $P=2$ flowers (i.e.\ a flowers with 2 petals) and flowers with odd $P$ do not work for canceling the dispersive coupling.}

{In App.\ \ref{sec:flowerstransmons}, we show that the flowers approach works for transmons despite the presence of the $M_0$ term in Eq.\ (\ref{eq:Hcc}).}

It would be interesting to find out more generally what pulse sequences can and cannot be used to implement the flowers. As a step in this direction, in App.\ \ref{sec:uhrig}, we show that, while the famous Uhrig pulse sequence \cite{uhrig07} works to cancel classical slowly varying $Z$ noise in the context of Ref.\ \cite{manovitz17}, it cannot be used to cancel the dispersive coupling. We also show in Ref.\ \cite{golan25}  that the strength of  the (``bad'') dispersive coupling can be suppressed relative to the desired (``good'') interaction by spin-locking the qubits, an approach that can be thought of as the continuous version of the flowers approach where discrete $\pi$ pulses are replaced by continuous resonant driving.

\section{Canceling the dispersive coupling when two oscillators mediate the coupling \label{sec:2oscillators}}

In this section, we show how to use the flowers approach to cancel the dispersive coupling in the case where the cross-cross-resonance gate is mediated by two oscillators. {This section is a particularly simple and pedagogical  special case of Sec.\ \ref{sec:meta}, which considers the case of $N$ qubits coupled via a metamaterial composed of $N$ coupled oscillators.} 

\begin{figure}[t!]
\centering
\includegraphics[width=\linewidth]{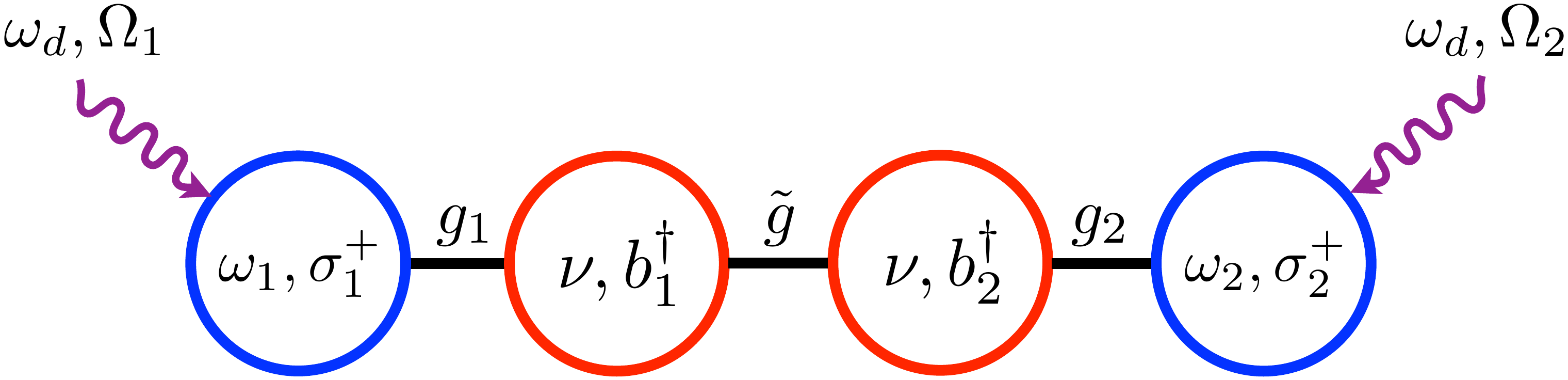} 
\caption{Cross-cross-resonance gate mediated by two oscillators. Two qubits with frequencies $\omega_1$ and $\omega_2$ and raising operators $\sigma_1^\dagger$ and $\sigma_2^\dagger$ are coupled, with coupling strengths $g_1$ and $g_2$, to cavities with frequency $\nu$ and creation operators $b_1^\dagger$ and $b_2^\dagger$. The two cavities are themselves coupled with coupling strength $\tilde g$. The qubits are driven with Rabi frequencies $\Omega_1$ and $\Omega_2$ at the same frequency $\omega_d$. \label{fig:two-cavities}
}
\end{figure}
As shown in Fig.\ \ref{fig:two-cavities}, consider two qubits $i = 1, 2$, each coupled to its own oscillator $b_i$, and assume that the two oscillators are coupled to each other. The Hamiltonian is 
\ba
H &=& \nu \sum_i b_i^\dagger b_i + \sum_i \frac{\omega_i}{2} Z_i + \sum_i g_i (\sigma_i^+ b_i + h.c.) \nonumber \\
&& + \tilde g (b_1^\dagger b_2 + h.c.).
\ea

{We} consider the situation where $g_1$, $g_2$, and $\tilde g$ can all be treated as small parameters proportional to $g$ (in particular, we assume $g_1, g_2, \tilde g \ll |\omega_1-\nu|, |\omega_2 - \nu|$).  {In this} case, the always-on ZZ interaction between the two qubits appears at sixth order in $g$ (see Appendix \ref{sec:ZZ2} and Ref.\ \cite{kumph24a}), making this configuration attractive for suppressing the always-on ZZ interactions, {i.e.\ the qubit-qubit cross-Kerr interactions. In the remainder of this section, we will show that the flowers approach works elegantly to cancel the qubit-oscillator dispersive coupling in this setup.}

We define annihilation operators $b_s = (b_1 + b_2)/\sqrt{2}$ and $b_a = (b_1 - b_2)/\sqrt{2}$ for the symmetric and antisymmetic modes, respectively.  
The Hamiltonian then becomes
\ba
H &=& (\nu + \tilde g) b_s^\dagger b_s + (\nu - \tilde g) b_a^\dagger b_a + \sum_i \frac{\omega_i}{2} Z_i \nonumber \\
&& + \frac{1}{\sqrt{2}} \left[(g_1 \sigma_1^+ + g_2 \sigma_2^+) b_s  + h.c.\right] \nonumber \\
&& + \frac{1}{\sqrt{2}} \left[(g_1 \sigma_1^+-g_2 \sigma_2^+)  b_a + h.c.\right].
\ea
We now add the cross-resonance drive 
\ba
H_d = \sum_i \Omega_i \cos(\omega_d t) X_i.
\ea
We do the rotating-wave approximation and go into the rotating frame to obtain the combined Hamiltonian
\ba
H &=& \epsilon_s b^\dagger_s b_s  + \epsilon_a b_a^\dagger b_a + \sum_i \left(\frac{\Delta_i}{2}Z_i + \frac{\Omega_i}{2} X_i \right) \nonumber \\
&& + \frac{1}{\sqrt{2}} \left[(g_1 \sigma_1^+ + g_2 \sigma_2^+) b_s + h.c.\right] \nonumber \\
&& + \frac{1}{\sqrt{2}} \left[(g_1 \sigma_1^+-g_2 \sigma_2^+)  b_a + h.c.\right],
\ea
where $\epsilon_s = \nu + \tilde g - \omega_d$, $\epsilon_a = \nu - \tilde g - \omega_d$, and $\Delta_i = \omega_i - \omega_d$.
We now diagonalize each transmon and drop the off-resonant terms, which gives the Hamiltonian in the dressed basis:
\ba
H &=& \epsilon_s b^\dagger_s b_s  + \epsilon_a b_a^\dagger b_a  + \sum_i \sqrt{\Delta^2_i + \Omega^2_i} \frac{Z_i}{2} \nonumber \\
&& + \frac{1}{2\sqrt{2}} \left[(g_1 \sin \theta_1 Z_1+g_2 \sin \theta_2 Z_2) b_s + h.c.\right] \nonumber \\
&& + \frac{1}{2 \sqrt{2}} \left[(g_1 \sin \theta_1 Z_1-g_2 \sin \theta_2 Z_2)  b_a + h.c.\right],
\ea
where the angle $\theta_i$ satisfies $\sin \theta_i = \Omega_i /\sqrt{\Omega_i^2 + \Delta_i^2}$.  We can now drop the single-qubit terms since they commute with the rest  of the Hamiltonian and change the rotating frame to obtain 
\ba
H = H_s + H_a,
\ea
where 
\ba
H_s &=& M_s (b_s e^{- i \epsilon_s t} + h.c.),\\
H_a &=& M_a (b_a e^{- i \epsilon_a t} + h.c.),
\ea
where
\ba
M_s &=& \frac{1}{2 \sqrt{2}}(g_1 \sin \theta_1 Z_1 + g_2 \sin \theta_2 Z_2) \label{eq:Msformula},\\
M_a &=& \frac{1}{2 \sqrt{2}}(g_1 \sin \theta_1 Z_1 - g_2 \sin \theta_2 Z_2) \label{eq:Maformula}.
\ea
We see that $H_s$ and $H_a$ commute with each other, so we are simply implementing two commuting cross-cross-resonance gates in parallel, one via the symmetric mode and one via the anti-symmetric mode. 

Let us suppose that we tune close to the symmetric mode, so that $0 < \epsilon_s \ll |\epsilon_a|$ (we will confirm below that this condition can indeed be satisfied for the CZ gate). Since we are tuned close to the symmetric mode, we will choose to make only one circle in the corresponding phase space:
\ba
\epsilon_s \tau = 2 \pi, \label{eq:est2pi}
\ea
while allowing  
for a large integer number $n_a$ of circles in the phase space corresponding to the anti-symmetric mode:
\ba
\epsilon_a \tau = 2\pi n_a. \label{eq:epsilonaformula2}
\ea
Once these two conditions are satisfied, the spin unitary takes the form 
\ba
U(\tau) = e^{i \left(\frac{M_s^2}{\epsilon_s} + \frac{M_a^2}{\epsilon_a} \right) \tau}.
\ea
Therefore, in order to get a gate equivalent (up to single-qubit rotations) to a CZ gate, we need
\ba
g_1 g_2\sin \theta_1 \sin \theta_2 \tau  \left(\frac{1}{\epsilon_s} - \frac{1}{\epsilon_a} \right)= \pi. \label{eq:esa}
\ea
As a sanity check, if $\tilde g = 0$ (i.e.~$\epsilon_s = \epsilon_a$), the two qubits are not coupled at all, so we cannot obtain a two-qubit gate, as Eq.\ (\ref{eq:esa}) confirms.

 Since $\epsilon_s \ll |\epsilon_a|$, we can drop the $1/\epsilon_a$ term in Eq.\ (\ref{eq:esa}). Then 
 Eq.\ (\ref{eq:esa}) and the condition $\epsilon_s \tau = 2 \pi$ [Eq.\ (\ref{eq:est2pi})] can be used to eliminate $\epsilon_s$ to obtain (like in the case of a single-mode cross-cross-resonance gate, except the couplings $g_i$ are reduced by $\sqrt{2}$): 
\ba
\tau \approx  \frac{\sqrt{2} \pi}{\sqrt{g_1 g_2\sin \theta_1 \sin \theta_2}} \label{eq:tauformula},
\ea
and 
\ba
\epsilon_s = \frac{2 \pi}{\tau} \approx \sqrt{2 g_1 g_2 \sin \theta_1 \sin \theta_2} \label{eq:epsilonsformula}.
\ea
Assuming for simplicity that $g_1 = g_2 = g$ and $\sin \theta_1 = \sin \theta_2 = \sin \theta$, we have $\epsilon_s \approx \sqrt{2} g \sin \theta \ll g$ since we typically work at $\sin \theta \ll 1$. If we further assume that $\tilde g \sim g$, this means that $\epsilon_a = \epsilon_s - 2 \tilde g \approx - 2 \tilde g$, which means $\epsilon_s \ll |\epsilon_a|$, as desired. By tuning the parameters such that $\epsilon_a = n_a \epsilon_s$ for some integer $n_a$ and further tuning the parameters slightly so that the CZ phase is exactly $\pi$, we get a CZ gate that is only a factor of $\sqrt{2}$ slower than the vanilla single-mode gate. Notice that, because $\epsilon_a$ is negative (assuming $\tilde g > 0$) and comes with a minus sign in the formula for the total phase in Eq.\ (\ref{eq:esa}), the two modes actually contribute constructively.

Will the flowers approach still work to cancel the {qubit-oscillator} dispersive coupling  coming from both modes? 
Remarkably, as we apply the $P=4$ flower to the symmetric mode, the antisymmetric mode will also come back to the origin provided $n_a$ is odd.  To see this, consider the Hamiltonian with the dispersive coupling:
\ba
H = H_s + H_a, \label{eq:Hsa}
\ea
where
\ba
H_s &=& M_s (b_s e^{- i \epsilon_s (1+ x_s)  t} + h.c.),\\
H_a &=& M_a (b_a e^{- i \epsilon_a (1 + x_a) t} + h.c.),
\ea
where
\ba
M_s &=& \frac{1}{2 \sqrt{2}}(g_1 \sin \theta_1 Z_1 + g_2 \sin \theta_2 Z_2),\label{eq:Msformula2}\\
M_a &=& \frac{1}{2 \sqrt{2}}(g_1 \sin \theta_1 Z_1 - g_2 \sin \theta_2 Z_2),\label{eq:Maformula2}\\
\epsilon_s x_s &=& \sum_i f^{(s)}_i Z_i,\\
\epsilon_a x_a &=& \sum_i f_i^{(a)} Z_i,
\ea
for some coefficients $f^{(s/a)}_i \sim g^2/\Delta_i$,
where, as in the single-mode case, $\Delta_i$ can be equivalently thought of as $\omega_i - \nu$ or $\omega_i - \omega_d$ (the two are approximately equal since we assume $\epsilon_s, |\epsilon_a| \ll \Delta_i$). {We will now show that the $P=4$ flower for canceling the dispersive coupling ($\propto x_s$) to the symmetric mode also cancels the dispersive coupling ($ \propto x_a$) to the antisymmetric mode.}

Let us write $n_a = 2 m + 1$, where $m = 0, 1, 2, \dots$. (In reality, for our example, $n_a$ will be negative; however, we will use positive $n_a$ below to keep phase-space rotations counterclockwise for consistency with  the previous sections; the analysis of negative $n_a$ is similar.) Let us first understand why the $P=4$ flower works in the case where $x_a = 0$. Let us assume we are doing the CPMG version of the flower where we first wait for time  
$\tau/2 = 3 \pi/(4 \epsilon_s)$, followed by three time intervals of duration  
$\tau = 3 \pi/(2 \epsilon_s)$, followed by one more time interval of duration $\tau/2$. {We will now separately discuss cases $m=0,1,2,3$ and then argue that, for larger $m$, the flowers repeat based on what $m$ is modulo 4.}  

\begin{figure}[t!]
\centering
\includegraphics[width=\linewidth]{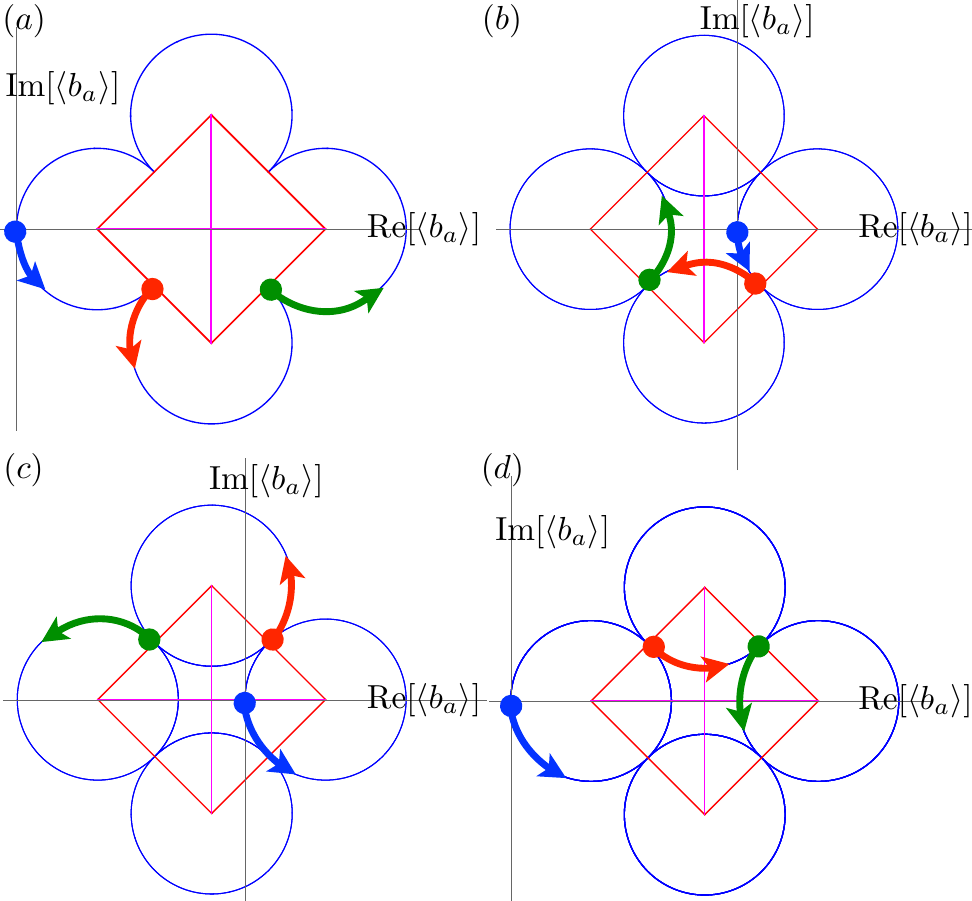} 
\caption{{(a)} For $n_a = 2 m + 1$ with $m=0$ and $x_a = 0$, the antisymmetric mode undergoes the CPMG $P=4$ flower shown in the figure. Starting at the origin, $\langle b_a \rangle$ evolves along the blue arrow, then along the red arrow, then along the green arrow, etc\dots, until it comes back to the origin. {(b)} Same as (a), but now with $m = 1$. Starting at the origin, $\langle b_a \rangle$ evolves along the blue arrow, making a full circle plus $\pi/4$, then along the red arrow, making two full circles plus $\pi/2$, then along the green arrow, making two full circles plus $\pi/2$, etc\dots, until it comes back to the origin. {(c)} Same as (a), but now with $m = 2$. Starting at the origin, $\langle b_a \rangle$ evolves along the blue arrow, making a full circle plus $\pi/2$, then along the red arrow, making 3 full circles plus $3 \pi/2$, then along the green arrow, making 3 full circles plus $3 \pi/2$, etc\dots, until it comes back to the origin. {(d)} Same as (a), but now with $m = 3$. Starting at the origin, $\langle b_a \rangle$ evolves along the blue arrow, making a 2 full circles plus $5 \pi/4$, then along the red arrow, making 5 full circles plus $\pi/2$, then along the green arrow, making 5 full circles plus $\pi/2$, etc\dots, until it comes back to the origin.  \label{fig:m0123}
}
\end{figure}

For $m=0$, the antisymmetric mode undergoes the simple CPMG $P=4$ flower shown in Fig.\ \ref{fig:m0123}(a).
In Fig.\ \ref{fig:m0123}(a), we start at the origin and follow the blue arrow until we get to the square, then we follow the red arrow, then the green arrow, etc\dots, until we come back to the origin.

For $m=1$, the antisymmetric mode undergoes the flower shown in Fig.\ \ref{fig:m0123}(b). The first angle (blue arrow), instead of being $3 \pi/4$ is now $3 \pi/4 + m  3 \pi/2 = 9 \pi/4 = 2 \pi + \pi/4$, so we make a full circle and another $\pi/4$. We then start down the red arrow and go for twice as large of an angle, i.e.~$3 \pi/2 + m 3 \pi = 4 \pi + \pi/2$, i.e.~we make two full circles plus $\pi/2$. We then start down the green arrow and go for the same angle, etc\dots, until we come back to the origin.

For $m=2$, the antisymmetric mode undergoes the flower in Fig.\ \ref{fig:m0123}(c). The first angle (blue arrow) is $3 \pi/4 + m  3 \pi/2 = 2 \pi + 7 \pi/4$, so we make a full circle and another $7 \pi/4$. We then start down the red arrow and go for twice as large of an angle, i.e.~$3 \pi/2 + m 3 \pi = 6 \pi + 3 \pi/2$, i.e.~we make 3 full circles plus $3 \pi/2$. We then start down the green arrow and go for the same angle, etc... until we come back to the origin.

For $m=3$, the antisymmetric mode undergoes the flower in Fig.\ \ref{fig:m0123}(d). The first angle (blue arrow) is $3 \pi/4 + m  3 \pi/2 = 4 \pi + 5 \pi/4$, so we make 2 full circles and another $5 \pi/4$. We then start down the red arrow and go for twice as large of an angle, i.e.~$3 \pi/2 + m 3 \pi = 10 \pi + \pi/2$, i.e.~we make 5 full circles plus $\pi/2$. We then start down the green arrow and go for the same angle, etc\dots, until we come back to the origin.

Finally, for $m=4$, we do exactly the same flower as for $m=0$ except we also go 6 times around each of the four circles. The flowers then repeat based on what $m$ is modulo 4. 

Now let us introduce the imperfection $x_a$. For $m=0$, everything will work exactly as in the regular single-mode $P=4$ flower discussed in Sec.\ \ref{sec:flowers}. For $m=1$, the flower is shown in Fig.\ \ref{fig:m0imperfect}:
\begin{figure}[t!]
\centering
\includegraphics[width=\linewidth]{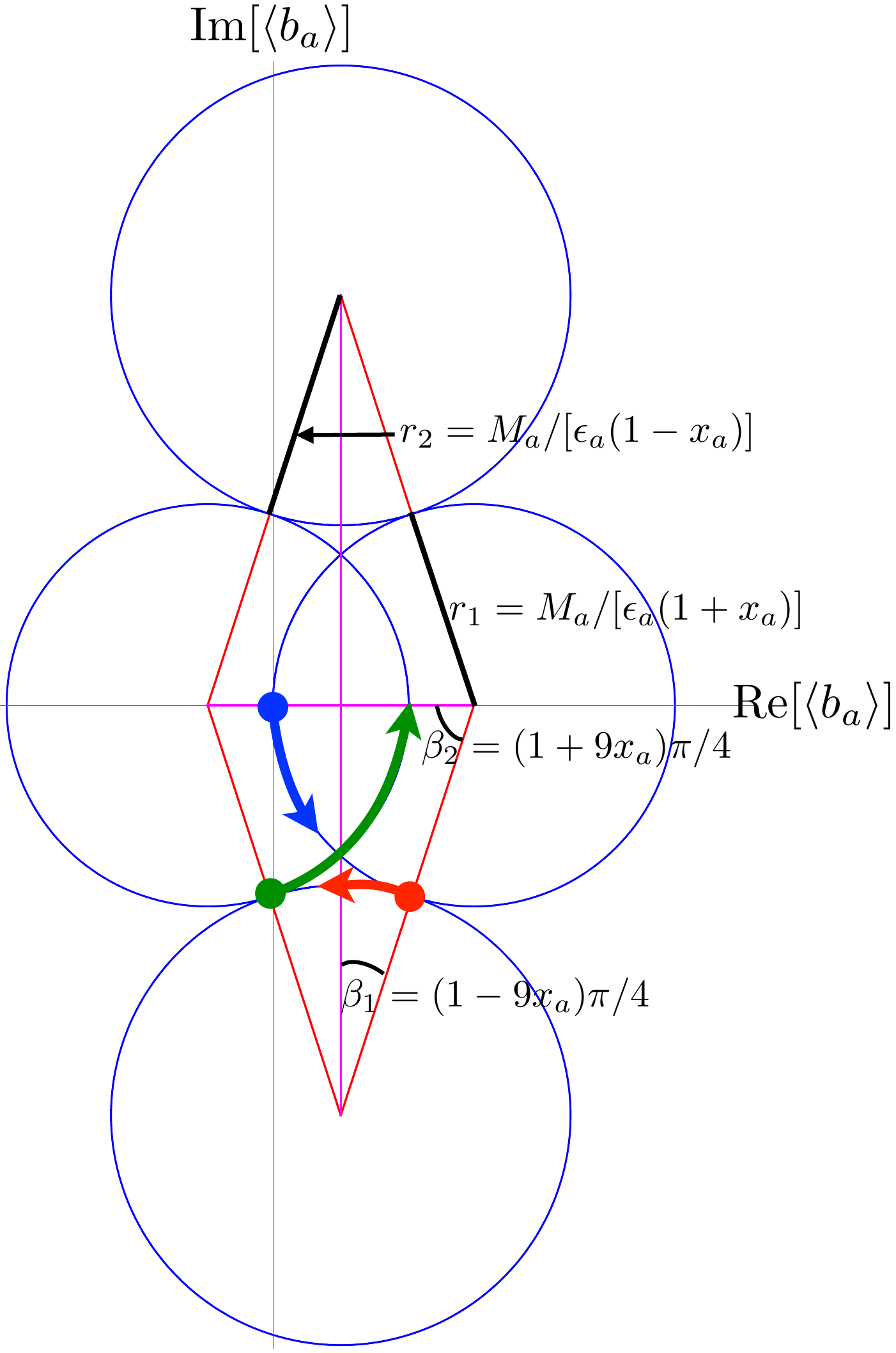} 
\caption{Same as Fig.\ \ref{fig:m0123}(b) (i.e., $m=1$), but now with  $x_a = 1/15$.  
Starting at the origin, $\langle b_a \rangle$ evolves along the blue arrow, making a full circle (of radius  {$r_1 = M_a/[\epsilon_a (1+x_a)]$}) plus $\beta_2 = (1+ 9 x_a) \pi/4$, then along the red arrow, making two full circles (of radius  {$r_2 = M_a/[\epsilon_a (1-x_a)]$}) plus $2 \beta_1 = (1- 9 x_a) \pi/2$, then along the green arrow, making two full circles (or radius $r_1$) plus $2 \beta_2$, etc\dots, until it comes back to the origin. \label{fig:m0imperfect}
}
\end{figure}
i.e., as in the single-mode flowers discussed in Sec.\ \ref{sec:flowers}, we have circles of two radii   {$r_1 = M_a/[\epsilon_a (1 + x_a)]$} and  {$r_2 = M_a/[\epsilon_a(1-x_a)]$}. And, as in single-mode flowers, the smaller circles are over-rotating, while the larger circles are under-rotating. The case of nonzero $x_a$ works similarly for other values of $m$. 

To calculate the total conditional phase, we need to compute the areas of both the symmetric-mode flowers (for the values of $M_s$ and $x_s$ corresponding to the four assignments of $Z_1$ and $Z_2$)  
and the anti-symmetric-mode flowers (for  values of $M_a$ and $x_a$ corresponding to the four assignments of $Z_1$ and $Z_2$) and appropriately combine them, taking into account the fact that in reality $\epsilon_a < 0$ (while the above figures assumed  positive $\epsilon_a$).  
Once the formula for the total conditional phase is obtained, it should, as in Eq.\ (\ref{eq:phipi}), be set to $\pi$ in order to get a gate equivalent (up to single-qubit rotations) to a CZ gate.

{We would like to note that the results of this section are modified only slightly if the two oscillators have slightly different frequencies. In particular, the modes $b_s$ and $b_a$ are now not perfectly symmetric and antisymmetric: $b_s = \alpha b_1 + \beta b_2$, $b_a = \beta b_1 - \alpha b_2$ for some $\alpha, \beta > 0$ with $\alpha^2 + \beta^2 = 1$. The formulas for $\epsilon_s$ and $\epsilon_a$ are also  modified accordingly. As a result, in Eqs.\ (\ref{eq:Msformula},\ref{eq:Msformula2}), $g_1/\sqrt{2} \rightarrow \alpha g_1$ and $g_2/\sqrt{2} \rightarrow \beta g_2$, while, in Eqs.\ (\ref{eq:Maformula},\ref{eq:Maformula2}), $g_1/\sqrt{2} \rightarrow \beta g_1$ and $g_2/\sqrt{2} \rightarrow \alpha g_2$. Similarly, in Eqs.\ (\ref{eq:esa},\ref{eq:tauformula},\ref{eq:epsilonsformula}), $g_1 g_2 \rightarrow 2 \alpha \beta g_1 g_2$. These modifications do not make it any harder to satisfy Eqs.\ (\ref{eq:est2pi},\ref{eq:epsilonaformula2},\ref{eq:esa}). These modifications also do not interfere with our proof that the $P=4$ flower applied to the symmetric mode works simultaneously on the antisymmetric mode provided $n_a$ is odd.}

\section{Canceling coupling to adjacent cavities \label{sec:adjacent}}

In a real architecture,  
we will be implementing a cavity-mediated cross-cross-resonance gate between a pair of qubits 1 and 2, which are also coupled, via additional cavities, to other qubits.  
These additional adjacent oscillators will be coupled during the gate to qubits 1 and 2 via the desired ``good'' interaction and via the ``bad'' dispersive coupling.  
If an adjacent oscillator starts in the vacuum state and is far in frequency from the oscillator mediating the gate, the adjacent oscillator will have negligible effect.
However, if the adjacent oscillator doesn't start in the vacuum state or is close in frequency to the oscillator mediating the gate,  we may need to cancel both the good and the bad coupling to such an adjacent cavity. In this section, we show how the flower scheme can be used to cancel such couplings.

\begin{figure}[t!]
\centering
\includegraphics[width=\linewidth]{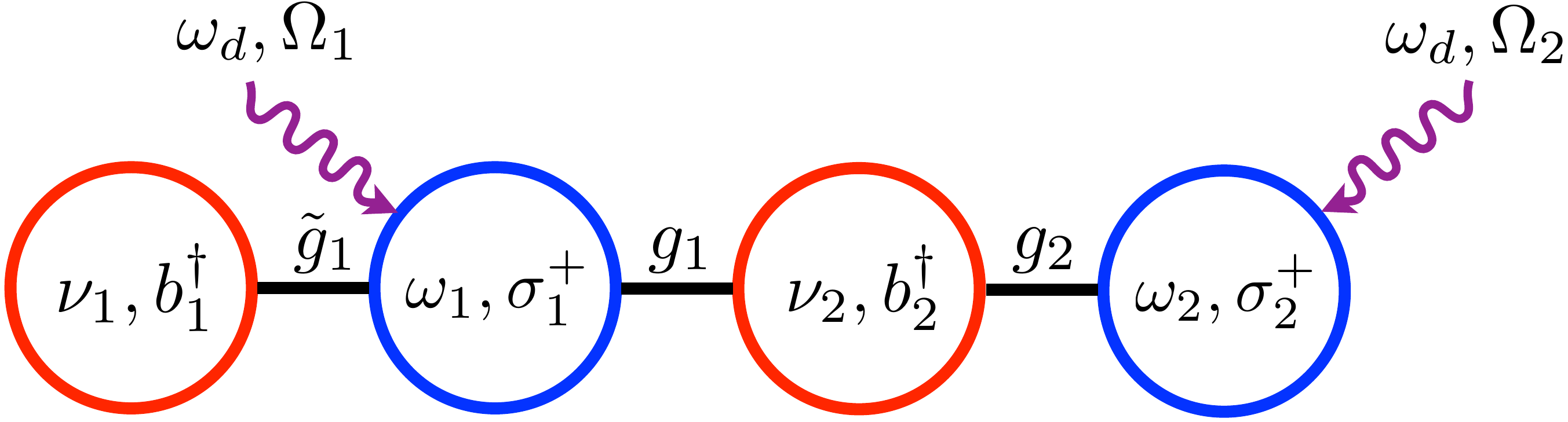} 
\caption{Cavity-mediated cross-cross-resonance gate in the presence of an adjacent cavity. Two qubits with frequencies $\omega_1$ (qubit 1) and $\omega_2$ (qubit 2) and raising operators $\sigma_1^\dagger$ and $\sigma_2^\dagger$ are coupled, with coupling strengths $g_1$ and $g_2$, to a cavity with frequency $\nu_2$ and creation operator $b_2^\dagger$, which mediates a cross-cross-resonance gate between them. The qubits are driven with Rabi frequencies $\Omega_1$ and $\Omega_2$ at the same frequency $\omega_d$. Qubit 1 is also coupled with coupling strength $\tilde g_1$ to an additional cavity with frequency $\nu_1$ and creation operator $b_1^\dagger$. \label{fig:adjacent}
}
\end{figure}
As shown in Fig.\ \ref{fig:adjacent}, suppose qubit 1 
is coupled to qubit 2  
via a cross-cross-resonance gate mediated by an oscillator with frequency $\nu_2$ and  creation operator $b_2^\dagger$. Furthermore, suppose that qubit 1 is also coupled to another oscillator described by creation operator $b_1^\dagger$. The Hamiltonian is then very similar to the one in Eq.\ (\ref{eq:Hsa}) describing two qubits coupled via two modes. The only difference is in coupling coefficients and in the fact that qubit 2 is not coupled to $b_1$. The Hamiltonian is
\ba
H = H_2 + H_1,
\ea
where
\ba
H_2 &=& M_2 (b_2 e^{- i \epsilon_2 (1+ x_2)  t} + h.c.),\\
H_1 &=& M_1 (b_1 e^{- i \epsilon_1 (1 + x_1) t} + h.c.),
\ea
where 
\ba
M_2 &=& \frac{1}{2}(g_1 \sin \theta_1 Z_1 + g_2 \sin \theta_2 Z_2),\\
M_1 &=& \frac{1}{2}\tilde g_1 \sin \theta_1 Z_1,\\
\epsilon_2 x_2 &=& \sum_i f_i Z_i,\\
\epsilon_1 x_1 &=& \tilde f_1 Z_1.
\ea
Here $H_2$ mediates the cross-cross-resonance gate, $H_1$ describes the parasitic coupling to mode $b_1$, $\epsilon_i = \nu_i - \omega_d$, $\sin \theta_i \approx \Omega_i/(\omega_i - \omega_d)$, $f_i \sim g_i^2/(\omega_i - \nu_2)$,  and $\tilde f_1 \sim \tilde g_1^2/(\omega_1 - \nu_1)$. 
So the problem we need to solve is very similar to the problem we solved for the case of the cross-cross-resonance gate mediated by two modes in Sec.\ \ref{sec:2oscillators}. We use $H_2$ to implement a cross-cross-resonance gate via a $P=4$ flower with $\epsilon_2 \tau = 2 \pi$. And then, exactly as in Sec.\ \ref{sec:2oscillators}, we need to make sure that $\epsilon_1 \tau = 2 \pi n_1$ for an odd integer $n_1$. This will ensure that the oscillator $b_1$ comes back to the initial state, and  
there is no residual effect on qubit $1$  
due to $b_1$ because both qubit states undergo a flower of the same area. 

It is worth pointing out that, if each qubit-qubit bond consists of two oscillators {(an approach proposed in Sec.\ \ref{sec:2oscillators}}), then one would need to cancel a lot of such couplings to adjacent modes (2 couplings per bond). For example, suppose qubits are arranged in a square lattice where each bond is mediated by two oscillators. Furthermore, suppose we are implementing a cross-cross-resonance gate between neighboring qubits 1 and 2. Each of these qubits is then connected to 3 additional bonds, with two modes per bond. 
If all modes involved have different frequencies, we need to bring all 14 modes back to their initial states, which requires substantial tuning. Although the number of conditions can be reduced if some of the modes have the same frequency, one needs to be careful in such cases to avoid unwanted resonances.  
The number of conditions can also be reduced if the adjacent modes start in vacuum and are far away in frequency from the mode mediating the gate since such off-resonant modes have a negligible effect. By utilizing pulse shaping, all the conditions could, in principle, be met as in Ref.~\cite{shapira20}, at the cost of additional complexity. Detailed analysis would also be necessary to study robustness to imperfections, such as static deviations of various parameters from their targeted values or shot-to-shot fluctuations of these parameters.

\section{Canceling always-on ZZ interactions {in an architecture based on cavity-mediated cross-cross-resonance gates}   
\label{sec:ZZcancel}}

In this section, we sketch how the cavity-mediated cross-cross-resonance gate might be used in a large array of transmons. In particular, we show how to eliminate always-on ZZ interactions, {i.e.\ qubit-qubit cross-Kerr interactions.} 

\begin{figure}[t!]
\centering
\includegraphics[width=\linewidth]{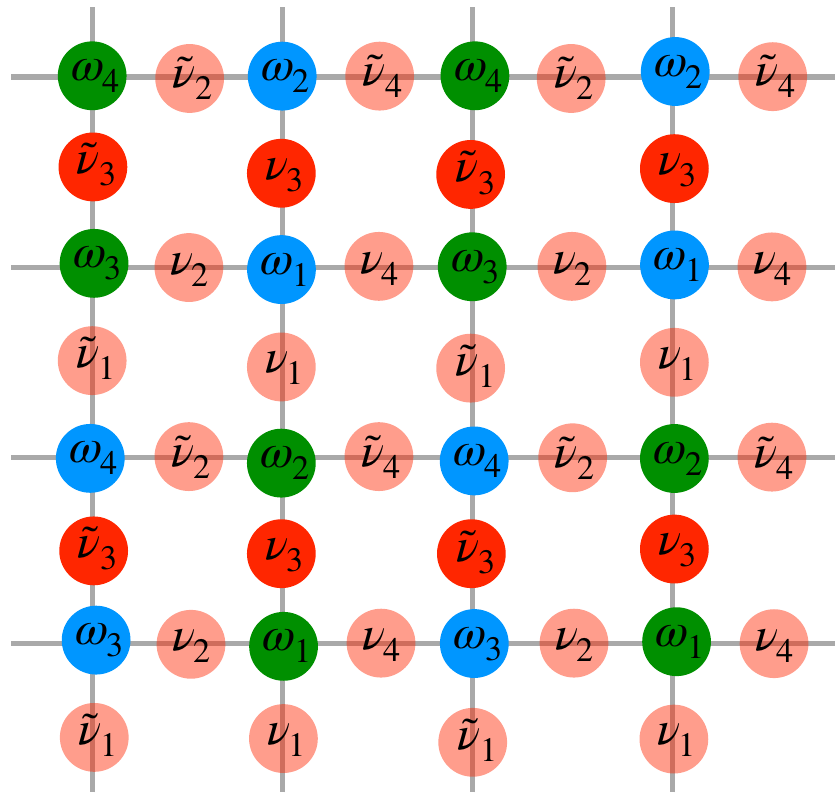} 
\caption{A square lattice of transmons (blue and green) coupled via oscillators (dark red and light red). The frequencies of the transmons take values $\omega_1$, $\omega_2$, $\omega_3$, and $\omega_4$ and are assigned in a periodic fashion as shown. (The degeneracy between qubit frequencies can be broken down further to avoid next-nearest-neighbor cross talk.) The frequencies of the oscillators take values $\nu_1$, $\tilde \nu_1$, $\nu_2$, $\tilde \nu_2$, $\nu_3$, $\tilde \nu_3$, $\nu_4$, and $\tilde \nu_4$ and are assigned in a periodic fashion as shown. 
\label{fig:grid}}
\end{figure}

Consider a 2D grid of transmons and oscillators  
shown in Fig.\ \ref{fig:grid}.
In blue and green, we show the frequencies of transmons. In dark red and light red, we show the frequencies of the oscillators that are used to mediate cross-cross-resonance gates. We assume that all transmons have the same anharmonicity $\eta$. In the next two subsections, we consider two approaches for eliminating always-on ZZ interactions between neighboring transmons. 

Before proceeding, it is worth mentioning that a straightforward way to reduce the always-on ZZ interaction between two transmons is to couple them via two resonators in series (instead of one resonator), as discussed in Sec.\ \ref{sec:2oscillators}, Appendix \ref{sec:ZZ2}, and Ref.\ \cite{kumph24a}. We expect that the approaches for canceling the always-on ZZ interaction discussed in this section can be combined with the two-resonator approach.

\subsection{Setting always-on ZZ interactions to zero \label{sec:straddle}}

In this subsection, we discuss the first approach for canceling always-on ZZ interactions: simply choosing our static system parameters in such a way that always-on ZZ interactions vanish. 

As shown in Appendix \ref{sec:ZZ1}, to cancel always-on ZZ interactions between a pair of neighboring transmons with frequencies $\omega_i$ and $\omega_j$ coupled via an oscillator of frequency $\nu$, we need to solve the quadratic equation  
\ba
(\omega_i - \nu)^2 + (\omega_j - \nu)^2 + (\omega_i + \omega_j - 2 \nu) \eta = 0 \label{eq:zeroZZ}
\ea
for $\nu$.  
This quadratic equation has two roots for $\nu$ whenever $\eta > |\omega_1 - \omega_2|$, {the so-called straddling regime}. While we can avoid resonant flip-flop interactions between neighboring transmons with only two different transmon frequencies $\omega_1 = \omega_4$ and $\omega_2 = \omega_3$ (see Fig.\ \ref{fig:grid}), this would give only two options for oscillator frequencies, the two roots of Eq.\ (\ref{eq:zeroZZ}) for $i = 1$ and $j = 2$.  
At the same time, when we couple two transmons using the cross-cross-resonance gate, to make sure that the other 6 oscillators that are adjacent to the two transmons don't interfere with the gate, it is most convenient to have the frequencies of all of these adjacent oscillators be different from the frequency of the oscillator mediating the gate.  
We therefore need sufficiently many different cavity frequencies, which in turn requires, due to Eq.\ (\ref{eq:zeroZZ}), sufficiently many different transmon frequencies. Choosing four different transmons frequencies arranged periodically as shown in Fig.\ \ref{fig:grid} is sufficient to have all 6 adjacent oscillators always out of resonance with the gate-mediating oscillator.  
Specifically, for $(i,j)=(1,2)$, we call the two roots of Eq.\ (\ref{eq:zeroZZ}) $\nu_1$, $\nu_3$. For $(i,j)=(3,4)$, we call the two roots $\tilde \nu_1$, $\tilde \nu_3$. For $(i,j)=(1,3)$, we call the two roots $\nu_2$, $\nu_4$. For $(i,j)=(2,4)$, we call the two roots $\tilde \nu_2$, $\tilde \nu_4$. We then arrange these 8 oscillator frequencies as shown in Fig.\ \ref{fig:grid}.

Using the three remaining tuning knobs, $\omega_2 - \omega_1$, $\omega_3 - \omega_1$, and $\omega_4 - \omega_1$, one can try to maximize various frequency differences such as those between transmons and adjacent oscillators, between transmons connected by an edge, and between oscillators that sit on edges sharing a vertex.

A quantum computation in this architecture can be carried out by interspersing single-qubit gates with the following four repeating layers of 2-qubit gates: (1) applying simultaneously 2-qubit gates along all or some $\nu_1$ and $\tilde \nu_1$ bonds, (2) applying simultaneously 2-qubit gates along all or some $\nu_2$ and $\tilde \nu_2$ bonds, (3) applying simultaneously 2-qubit gates along all or some $\nu_3$ and $\tilde \nu_3$ bonds, and (4) applying simultaneously 2-qubit gates along all or some $\nu_4$ and $\tilde \nu_4$ bonds. 

If driving $\omega_1$ and $\omega_2$ qubits to implement the $\nu_1$ 2-qubit gate also results in undesired coupling along bond $\nu_3$ (e.g.\ if $|\nu_3 - \nu_1|$ is not large enough), layer (1) of two-qubit gates can be broken down into two steps, so that every other $\nu_1$ gate is implemented in one step and then the remainder in the second step.  
Similarly, when $|\nu_1 - \tilde \nu_1|$ is too small, then simultaneously driving $\nu_1$ and $\tilde \nu_1$ bonds may induce unwanted couplings along $\nu_2$, $\tilde \nu_2$, $\nu_4$, $\tilde \nu_4$ bonds if one or more of the latter four frequencies is also too close to $\nu_1$ and $\tilde \nu_1$. In that case, we can again break down the implementation of the layer (1) into two or more steps. Similar considerations apply to layers (2), (3), and (4).

{The approach for nullifying always-on ZZ interactions explored in this subsection requires engineering each pair of neighboring qubits to be in the straddling regime. This is a significant technical constraint because  flux-tunable transmons have shorter coherence times, while fixed-frequency transmons are challenging to precisely tune \cite{zhang22i}. This motivates the exploration, in the next subsection, of the flowers approach for canceling the always-on ZZ interactions, an approach that does not rely on the straddling regime.}

\subsection{Canceling always-on ZZ interactions with the flowers approach}

In this subsection, we consider the second approach for canceling always-on ZZ interactions: applying $\pi$ pulses as part of the flowers approach.

Suppose we are not using Eq.\ (\ref{eq:zeroZZ}) to cancel the always-on ZZ interactions. Nevertheless, let us assume the frequencies of transmons and oscillators are still chosen according to the two-dimensional pattern in Fig.\ \ref{fig:grid} [although the values of the frequencies themselves are not chosen according to Eq.\ (\ref{eq:zeroZZ})]. Furthermore, let us suppose we want to simultaneously implement 2-qubit gates  along all $\nu_3$ and $\tilde \nu_3$ bonds.  
In Fig.\ \ref{fig:grid}, these bonds are shown in dark red, while the remaining bonds are shown in light red.

\begin{figure}[t]
\centering
\includegraphics[width=0.7 \linewidth]{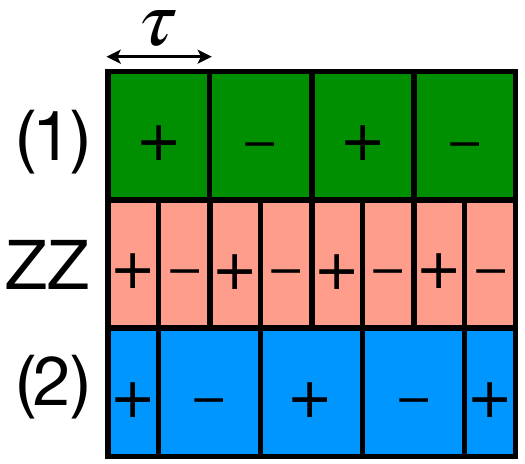} 
\caption{The figure showing cancellation of always-on ZZ interactions. The horizontal axis is time going from left to right. The first row (green) shows the 4 time intervals of duration $\tau$ corresponding to the pulses from sequence (1) applied on green transmons. The sign is the sign of the effective Z. The third row (blue) shows the time intervals corresponding to the pulses from the CPMG-like sequence (2) applied on blue transmons. Finally, the middle row (light red) shows the resulting effective sign of the ZZ interaction between a green transmon and a blue transmon. We see that the ZZ interaction is echoed away. \label{fig:cancelation}
}
\end{figure}

Let us assume that we are implementing the cross-cross-resonance gate using the $P=4$ flowers scheme. As explained in Sec.\ \ref{sec:flowers}, 
we can implement this gate in at least the following two ways: (1) the original $P=4$ flowers approach from Ref.~\cite{manovitz17} waits for time $\tau = 3 \pi/(2 \epsilon)$ before applying the first pair of $\pi$ pulses, then applies the rest of the pulses at the same time interval $\tau$; (2) a CPMG-like modification of this pulse sequence is to wait for time $\tau/2 = 3 \pi/(4 \epsilon)$ before applying the first pair of $\pi$ pulses, and then applying the rest of the pulses still at the time interval $\tau$. We then use pulse sequence (1) on the green qubits in Fig.\ \ref{fig:grid} and pulse sequence (2) on the blue qubits. The key observation is that all the bonds that we do not want to couple along during this gate (i.e.~light-red bonds) couple a green site and a blue site. Because the two pulse sequences are $\tau/2$ out of sinc with each other, the sign of the effective always-on ZZ interaction flips every $\tau/2$, which results in the cancellation of always-on ZZ interactions along all light-red bonds. This is illustrated in Fig.\ \ref{fig:cancelation}.

In the above scheme, if we want to implement a gate not on all dark-red bonds, but only on some, we can still keep applying the pulses on all transmons. This will cancel always-on ZZ interactions on light-red bonds, but will not cancel the ZZ interactions on the inactive dark-red bonds that we keep turned off. One possible approach to canceling the always-on ZZ interactions on the inactive dark-red bonds might be to always implement gates on all dark-red bonds, but tune the total nonlinear phase on some of these bonds to be zero. 

A potential issue with the above 2D scheme is that, when we drive two $\nu_3$ bonds that are in the same column and separated by one $\nu_1$ bond, the two qubits sharing the $\nu_1$ bond may also end up coupled via an unintentional $\nu_1$-mediated cross-cross-resonance gate. One way to resolve this problem is to define two difference frequencies $\nu_3$ and alternate them along a column, which we are free to do since we are not constrained in this subsection by Eq.\ (\ref{eq:zeroZZ}).

\section{Simultaneous cross-cross-resonance gates using a metamaterial \label{sec:meta}}

In this section, we show how our gate works in the case of $N$ qubits (or other anharmonic systems such as transmons) interacting with each other via an array of coupled oscillators, i.e.\ a metamaterial. In particular, for a chain of $N$ qubits connected by $\sim N$ metamaterial cavities, our gate allows for the simultaneous application of the two-qubit cross-cross-resonance gate on up to $N/2$ arbitrary---potentially distant---disjoint pairs of qubits.

The Hamiltonian is 
\ba
H &=& \sum_k \nu_k b^\dagger_k b_k + \sum_{i = 1}^N \Big[\frac{\omega_i}{2} Z_i + \Omega_i \cos(\omega_{d,i} t) X_i \nonumber \\
&& + \sum_k g_{i,k} (\sigma^+_i b_k + h.c.)\Big] \label{eq:Nqubitmeta},
\ea 
where $b_k$ are annihilation operators for the eigenmodes (of energy $\nu_k$) of the metamaterial, and qubit $i$ is driven with frequency $\omega_{d,i}$ and Rabi frequency $\Omega_i$. The qubit-metamaterial couplings are $g_{i,k} = g_i \eta_{i,k}$, where $g_i$ is the coupling of the qubit to the corresponding cavity (with annihilation operator $b_i$) and $\eta_{i,k}$ is the matrix that diagonalizes the metamaterial: $b_i = \sum_k \eta_{i,k} b_k$. 

Although approaches where each two-qubit gate is mediated by multiple modes may be possible (by analogy with trapped-ion approaches \cite{grzesiak20,shapira20}) and may even result in faster gates, the simplest approach to implement $N/2$ simultaneous two-qubit gates on arbitrary disjoint pairs of qubits is to choose, for each gate, a distinct mode $k$ and drive the corresponding pair of qubits with frequency near-resonant with that mode. Assuming all other modes can be ignored, all $N/2$ (or fewer, if desired) simultaneous two-qubit gates can be implemented in parallel exactly as described in Sec.\ \ref{sec:gate}, and the corresponding dispersive coupling can be canceled exactly as described in Secs.\ \ref{sec:integers} and \ref{sec:flowers}. 

It remains to find the condition for neglecting the other modes. Assuming that the nearest-neighbor coupling in the metamaterial is $J$, we have $\sim N$ modes in a band of width $\sim J$, corresponding to mode spacing $\sim J/N$. The ``good'' cross-resonance coupling of qubit $i$ to the desired mode $k$ has strength $g_{i,k} \Omega_i/\Delta_i$, where $\Delta_i$ is the detuning between the qubit and the mode (and can be equivalently thought of as the detuning between the qubit and the drive, since the drive is near-resonant with the mode). The condition for neglecting the other modes is that this strength is much smaller than the detuning $\sim J/N$ to the nearest off-resonant mode:
\ba
g_{i,k} \Omega_i/\Delta_i \ll J/N.\label{eq:ignore}
\ea
Since the gate time $\tau$ is $\tau \sim (g_{i,k} \Omega_i/\Delta_i)^{-1}$, this condition can also be expressed as
\ba
\tau \gg N/J.
\ea
Notice that, 
if we populate all the modes during the gate as analyzed in Ref.~\cite{shapira20}, the corresponding gates can operate at the speed limit given by $\tau \sim N/J$.

\subsection{Dispersive coupling}

In this subsection, we derive the dispersive coupling for $N$ qubits coupled via a metamaterial.

In the limit of weak dressing, the dispersive coupling (i.e.\ the ``bad'' interaction) can be derived in the absence of the cross-resonance drive. To second order in $g_i$, the two-qubit single-mode perturbation-theory derivation of Sec.\ \ref{sec:quantumstark} can be extended trivially to the $N$-qubit multi-mode situation of Eq.\ (\ref{eq:Nqubitmeta}) to give the dispersive coupling Hamiltonian
\ba
H_\textrm{disp} = \sum_i \sum_k f_{i,k} Z_i b^\dagger_k b_k, \label{eq:metastark}
\ea
where
\ba
f_{i,k} \sim \frac{g^2_{i,k}}{\Delta_{i,k}},
\ea
where $\Delta_i = \omega_i - \nu_k$. As in the two-qubit single-mode case of Sec.\ \ref{sec:quantumstark}, the dispersive coupling Hamiltonian has this simple diagonal form only in the absence of near-resonances. In particular, for any pair of qubits $i$ and $j$ and any pair of modes $p$ and $k$ (such that either $i \neq j$, or $p \neq k$, or both), we assume $|\omega_i + \nu_p - \omega_j - \nu_k| \gg g_{i,k} g_{j,p}/|\omega_i - \nu_k|, g_{i,k} g_{j,p}/|\omega_j - \nu_p|$. If this condition were not satisfied, then off-diagonal processes of the form $\sigma^+_i b_k \sigma^-_j b_p^\dagger$ would have been allowed.

\subsection{Detailed analysis in the case of 2 driven qubits \label{sec:metatwo}}

The condition in Eq.~(\ref{eq:ignore})  enables the metamaterial to mediate up to $N/2$ simultaneous cross-cross-resonance two-qubit gates, provided that the dispersive coupling is negligible. If we are not deep enough in the limit of Eq.~(\ref{eq:ignore}) and/or the dispersive coupling is not negligible, there will be associated errors.  
In this subsection, we study these errors for the case where only two qubits are driven.

Starting with Eq.\ (\ref{eq:Nqubitmeta}), we suppose that the qubits we drive are qubits 1 and 2 and that they are driven with frequency $\omega_d$. Furthermore, suppose the mode we use is the $k=0$ mode with frequency $\nu_0$, so $\omega_d \approx \nu_0$. Doing the rotating-wave approximation and moving into the interaction picture with respect to 
\ba
H 
&=& \omega_d \sum_k b^\dagger_k b_k + \frac{\omega_d}{2} \sum_i Z_i
\ea
gives the Hamiltonian 
\ba
H &=& \sum_k \epsilon_k b^\dagger_k b_k + \sum_i \left[\frac{\Delta_i}{2} Z_i + \sum_k g_{i,k} (\sigma^+_i b_k + h.c.)\right] \nonumber \\
&& + \frac{\Omega_1}{2} X_1 + \frac{\Omega_2}{2} X_2,
\ea
where $\Delta_i = \omega_i - \omega_d$ and $\epsilon_k = \nu_k - \omega_d$. 
As usual, we now change the basis of the two driven qubits to arrive at  
\ba
H &=& \sum_k \epsilon_k b^\dagger_k b_k + \sum_{i = 1}^2 \frac{\sqrt{\Delta_i^2 + \Omega_i^2}}{2} Z_i + \sum_{i = 3}^N \frac{\Delta_i}{2} Z_i  \nonumber \\
&& + \sum_{i = 1}^2 \sum_k \frac{g_{i,k}}{2} \left[(X_i \cos \theta_i + Z_i \sin \theta_i + i Y_i) b_k + h.c.\right] \nonumber \\
&&+ \sum_{i = 3}^N \sum_k g_{i,k} (\sigma^+_i b_k + h.c.),
\ea
where, as usual, we assume that the change-of-basis angles $\theta_i$ are small, i.e.\ $\theta_i \approx \Omega_i/\Delta_i \ll 1$.

Taking into account the dispersive coupling given in Eq.\ (\ref{eq:metastark}), the dominant interactions are then given by 
\ba
H &=& \sum_{k} \left[\epsilon_k (1 + x_k) b^\dagger_k b_k + M_k (b_k + b^\dagger_k)\right], \label{eq:HNk1}
\ea
where
\ba
x_k &=& \frac{1}{\epsilon_k} \sum_{i = 1}^N f_{i,k} Z_i,\\
M_k &=& \frac{1}{2} \sum^2_{i = 1} g_{i,k} Z_i \sin \theta_i. \label{eq:HNk3}
\ea

To solve for the time evolution unitary $U(t)$ under $H$, we can treat $x_k$ and $M_k$ as numbers, rather on operators, and find $U(t)$ for the values of $x_k$ and $M_k$ corresponding to all possible assignments of $Z_i$. Once $x_k$ and $M_k$ are treated as numbers, $H$ is a sum of independent terms describing evolution of each mode $b_k$. The full evolution can therefore be found exactly as in Sec.\ \ref{sec:quantumstark}. In particular, in the interaction picture with respect to  {$\sum_k  \epsilon_k (1 + x_k) b^\dagger_k b_k$}, and defining $\epsilon'_k = \epsilon_k (1 + x_k)$, the evolution unitary is 
\ba
U(t) &=& \prod_k e^{\frac{1- e^{i \epsilon'_k t}}{\epsilon'_k} M_k b_k^\dagger - \textrm{h.c.}} e^{i \frac{M_k^2}{(\epsilon'_k)^2} (\epsilon'_k t - \sin(\epsilon'_k t))}.
\ea
So the lab-frame evolution unitary is 
\begin{align}
V(t) U(t), 
\label{eq:metalab}
\end{align}
where
\ba
V(t) = \prod_k e^{- i \epsilon'_k t b^\dagger_k b_k}.
\ea

{In Sec.\ \ref{sec:2oscillators}, we study a special case of Eqs.\ (\ref{eq:HNk1}-\ref{eq:HNk3}) where 2 qubits are coupled via 2 oscillators. In the following Sec.\ \ref{sec:N1}, we will consider in detail the case of $N$ qubits and 1 photonic mode. Then, in Sec.\ \ref{sec:2N}, we will consider the case of 2 qubits and $\sim N$ photonic modes, followed by the general case of $N$ qubits and $\sim N$ photonic modes in Sec.\ \ref{sec:NN}.}

\subsubsection{$N$ qubits and 1 photonic mode \label{sec:N1}}

To start with a simpler case, let us assume that all modes except for the $k=0$ mode are far off-resonance and can be neglected. We will therefore study the effect of other spins, which participate in the dynamics due to the dispersive coupling.

The Hamiltonian in Eqs.\ (\ref{eq:HNk1}-\ref{eq:HNk3}) then simplifies to (taking $g_{1,0} = g_{2,0} = g$ and $\sin \theta_1 = \sin \theta_2 = \sin \theta$) 
\ba
H &=& \epsilon' b^\dagger_0 b_0 + M (b_0 + b^\dagger_0),\label{eq:hamfull2}\\
\epsilon' &=& \epsilon_0 + \sum_{i=1}^N f_{i,0} Z_i,\label{eq:epsprime2}\\ 
M &=& \frac{g \sin \theta}{2} (Z_1 + Z_2).
\ea

\textbf{{Errors induced by} the dispersive coupling.---}Let us first estimate the error from the dispersive coupling if we don’t attempt correcting for it. The evolution unitary in the lab frame in Eq.\ (\ref{eq:metalab}) simplifies to  
\begin{align}
V(t) U(t) = e^{- i \epsilon' t b^\dagger_0 b_0} e^{\frac{1- e^{i \epsilon' t}}{\epsilon'} M b_0^\dagger - \textrm{h.c.}} e^{i \frac{M^2}{\epsilon'^2} (\epsilon' t - \sin(\epsilon' t))}.
\end{align}

Assuming we do only one circle in phase space, the gate takes time $\tau \sim 1/(g \sin \theta)$, and we have $\epsilon_0 = 2 \pi/\tau \sim g \sin \theta$. 

While $\epsilon_0 \tau = 2 \pi$ is chosen such that 
the photons come back to their initial state in phase space in the absence of the dispersive coupling, due to the second factor in $V(\tau) U(\tau)$, they now end up a distance $\sim (M/\epsilon') ((\epsilon'-\epsilon_0)/\epsilon_0)$ away from where they started, thus producing entanglement between the photons on the one hand and spins 1 and 2 (whose state determines the value of $M$) on the other. This term also produces entanglement with the other spins because the value of $\epsilon'$ depends on the state of all the spins. 

The factor $V(\tau)$ in $V(\tau) U(\tau)$, the one we use for taking the system back to the lab frame, also produces entanglement between the photons and all the spins because $\epsilon'$ depends on the state of all the spins. The amount of entanglement depends on how populated the photonic mode is: 
the photonic state gets rotated around the origin by a small angle $\tau \sum_i f_{i,0} Z_i$, which needs to be multiplied by the distance from the origin to get the displacement in phase space. 

Finally, again because of the dispersive coupling, the third factor in $V(\tau) U(\tau)$ results in undesired entanglement between the target qubits $1, 2$ on the one hand and the other qubits $i > 2$ on the other.

We will discuss below approaches for canceling the dispersive coupling, but let us start by computing the infidelity for the case where the dispersive coupling is not canceled. For this analysis, we will assume that the photonic mode starts in vacuum. 
In this case, since we will be computing fidelity with respect to the desired final state (which must also have our photonic mode in vacuum), 
$V(\tau) = e^{- i \epsilon' \tau b^\dagger_0 b_0}$ doesn't contribute to the infidelity. 

Turning to the second factor in $V(\tau) U(\tau)$, $e^{\frac{1- e^{i \epsilon' \tau}}{\epsilon'} M b_0^\dagger - \textrm{h.c.}}$, the overlap between the initial vacuum state and the vacuum state displaced by  
$\sim (M/\epsilon') ((\epsilon'-\epsilon_0)/\epsilon_0)$) is given by 1 minus roughly the square of this displacement. 
Assuming the worst case that the dispersive coupling terms to all $\sim N$ spins constructively interfere, the resulting infidelity is
\ba
\sim (M/\epsilon')^2 ((\epsilon'-\epsilon_0)/\epsilon_0)^2 &\sim& (M/\epsilon_0)^2 (N f_{i,0}/\epsilon_0)^2 \nonumber \\
&\sim& \left(\frac{N g^2}{\Delta \epsilon_0}\right)^2,
\ea
where we assumed that, for all $i$, $f_{i,0} \sim g^2/\Delta$ for some characteristic detuning $\Delta$. So the resulting infidelity is negligible provided
\ba
\frac{N g^2}{\Delta \epsilon_0} \ll 1.
\ea
Since $\epsilon_0 \sim g \Omega/\Delta$, this condition can also be written as
\ba
\frac{N g}{\Omega} \ll 1,
\ea
which, in turn, is equivalent to the intuitive condition that the worst-case dispersive coupling $N f \sim N g^2/\Delta$ in Eq.\ (\ref{eq:epsprime2}) is negligible compared to the strength $M \sim g \Omega/\Delta$ of the cross-resonance interaction in Eq.\ (\ref{eq:hamfull2}).

Finally, turning to the third factor in $V(\tau) U(\tau)$, $e^{i \frac{M^2}{\epsilon'^2} (\epsilon' \tau - \sin(\epsilon' \tau))}$, we write $\epsilon' = \epsilon_0 (1 + x)$ and expand the exponent to lowest nonzero order in $x$. Using the fact that $\sin(\epsilon_0 \tau) = 0$, we have $\epsilon' \tau - \sin(\epsilon' \tau) = \epsilon_0 \tau + O(x^3)$. So the dominant error comes from $\epsilon'$ in the denominator, resulting in
\ba
e^{i \frac{M^2}{\epsilon_0^2} (1 - 2 x) \epsilon_0 \tau}.
\ea
To calculate the resulting infidelity, we do essentially the same calculation as before to find that the infidelity is
\ba
\sim x^2 \sim \left(\frac{N g^2}{\Delta \epsilon_0}\right)^2,
\ea
where we again assumed the worst case of constructive interference. So again, we find that the infidelity is negligible provided
\ba
\frac{N g^2}{\Delta \epsilon_0} \ll 1.
\ea

It is worth pointing out that calculating the worst-case error (i.e.~assuming $x \sim N$) could be too pessimistic and that the error that eventually matters could be an average error, which can be estimated using a random-walk argument where $x$ scales as $x \sim \sqrt{N}$. In this case, the infidelity can be estimated as 
\ba
\sim x^2 \sim \left(\frac{\sqrt{N} g^2}{\Delta \epsilon_0}\right)^2,
\ea
so the condition for neglecting this error is 
\ba
\frac{\sqrt{N} g^2}{\Delta \epsilon_0} \ll 1,
\ea
which, using $\epsilon_0 \sim g \Omega/\Delta$, can be rewritten as
\ba
\frac{\sqrt{N} g}{\Omega} \ll 1.
\ea

\textbf{Integers approach for canceling the dispersive coupling.---}As we have shown, not canceling the dispersive coupling produces entanglement between the photons and all the qubits. In addition to resulting in gate infidelity, this leads to additional complications when the same photonic mode is reused for future gates. 
We will therefore now discuss the two approaches 
for canceling the dispersive coupling. 

Let us first consider the ``integers'' approach. 
To implement the integers approach, it is sufficient to achieve 
\ba
\epsilon_0 t &=& 2 \pi n,\\
f_{i,0} t &=& 2 \pi n_i,
\ea
where $n$ and $n_i$ (for $i = 1, 2, \dots N$) are integers. So what used to be a 3-integers approach for the case of one mode and two qubits, has become an $(N+1)$-integers approach for the case of one mode and $N$ qubits.  
Finding enough tunability to obtain $N+1$ integers seems very challenging.
Furthermore, even if we satisfy all $N+1$ integer conditions, the resulting gate is 
\ba
U(\tau) = e^{i \frac{M^2}{\epsilon'} \tau}.
\ea
Since $\epsilon'$ depends on the state of all $N$ qubits, the gate is actually a multi-qubit gate that entangles all $N$ qubits. 

Let us evaluate the corresponding gate fidelity. To compare actual evolution to the desired evolution, we evaluate state fidelity 
\ba
F = \left|\langle \psi|e^{-i \frac{M^2}{\epsilon_0} \tau} e^{i \frac{M^2}{\epsilon'} \tau} |\psi\rangle\right|^2,
\ea
where $\epsilon' = \epsilon_0 + \sum_i f_{i,0} Z_i $ and $M \sim g \sin \theta \sum_{i= 1,2} Z_i$. The two exponents commute, so we can add them. To lowest order in $f_{i,0}$, we get 
\ba
F = \left|\langle \psi| e^{-i \tau \frac{M^2}{\epsilon_0^2} \sum_i f_{i,0} Z_i}  |\psi\rangle\right|^2.
\ea
To find the worst-case infidelity, we take $|\psi\rangle$ to be an equal superposition of two states for which $\sum_i f_{i,0} Z_i = \pm \sum_i |f_{i,0}|$. The resulting fidelity is then 
\ba
F = \cos^2\left(\tau \frac{M^2}{\epsilon_0^2} \sum_i |f_{i,0}|\right),
\ea
so the error is 
\ba
1-F &\sim& \left[\tau \frac{M^2}{\epsilon_0^2} \sum_i |f_{i,0}| \right]^2 \sim \left[\tau \frac{g^2 \sin^2 \theta}{\epsilon_0^2} N \frac{g^2}{\Delta} \right]^2 \nonumber \\
& \sim & \left( \frac{N g^2}{\Delta \epsilon_0} \right)^2. 
\ea 
Notice that this error has exactly the same scaling as the error that arises when we don’t try to cancel the dispersive coupling (with the same caveat that the average-case error here would also have $N$ instead of the worst-case $N^2$). Thus, the entire effort of integer-based dispersive coupling cancellation does not change the scaling of the infidelity; it only makes the error somewhat cleaner, i.e.~the error is now confined to the spin degrees of freedom, as opposed to being also spread over the photons.

\textbf{Flowers approach for canceling the dispersive coupling.---}Now let us consider the flowers approach. Recalling the definition
\ba
x = \frac{\sum_{i = 1}^N f_{i,0} Z_i}{\epsilon_0},
\ea
we write
\ba
H = \epsilon_0 (1 + x) b^\dagger_0 b_0 + M (b_0 + b^\dagger_0).
\ea
If we conjugate this Hamiltonian by a global $\pi$ pulse on all $N$ qubits, we get 
\ba
H = \epsilon_0 (1 - x) b^\dagger_0 b_0 - M (b_0 + b^\dagger_0),
\ea
so the flowers approach works in exactly the same way as for two qubits, a case we discussed in Sec.\ \ref{sec:flowers}. However, the area of the flower (and therefore the accumulated phase) depends on $x$, which in turn depends on the state of all $N$ qubits, which means we again get a multi-qubit gate that entangles all $N$ qubits. 

Let us again try to estimate the infidelity due to the fact that we get a multi-qubit gate instead of the desired two-qubit gate. As in Sec.\ \ref{sec:flowers}, the area in phase space has two components: one component is composed of parts of a circle, while the other is a polygon. Both components have area that is invariant under $x \rightarrow - x$. This means that the correction to the area scales as $x^2$. The state fidelity is thus 
\begin{align}
F &= \left| \langle \psi| e^{-i 2 A} e^{i 2 A (1 + c x^2)} |\psi\rangle\right|^2 = \left| \langle \psi| e^{i 2 A c x^2} |\psi\rangle\right|^2,
\end{align}
where $A \sim 1$ is the ideal area (in the absence of the dispersive coupling) and $c$ is some constant. Let us assume that $N$ is large, so that $N-2 \approx N$.   
Depending on the states of the $N$ qubits, 
$|x|$ ranges from $0$ to $N |f_{i,0}|/\epsilon_0 \sim N g^2/(\Delta \epsilon_0)$. Preparing $|\psi\rangle$ as an equal superposition of two states, one with $x = 0$ and another with $|x| \sim N g^2/(\Delta \epsilon_0)$, we find 
\ba
1 - F \sim A^2 x^4 \sim \left(\frac{N g^2}{\Delta \epsilon_0} \right)^4 \sim \left(\frac{N g}{\Omega} \right)^4,
\ea
which is the square of the infidelity we get with the integers approach, which is in turn the same as if we don’t cancel the dispersive coupling at all.  
In other words, the flowers approach results in much better performance in this case  than the integers approach. Combined with the fact that the integers approach is also much more challenging (it requires tuning $\sim N$ parameters), the flowers approach is preferred in this case. As before, if we consider the average case instead of the worst case, we can replace $N$ with $\sqrt{N}$. 

It is also worth pointing out that the  
highly non-local multi-qubit-gate error discussed above is a type of error that might be challenging to incorporate into a fault-tolerant error correcting scheme. Indeed, a typical error model is that, when we apply a two-qubit gate, we get errors only on the two qubits that undergo the gate. On the other hand, we have a situation here where all qubits that couple to the metamaterial experience a correlated error while we are trying to apply a two-qubit gate only on two of them. {Fault-tolerant error correcting schemes with such non-local correlated errors are still possible, but require special care.} One could worry that trapped-ion computers based on long chains also have this non-local correlated error problem, but the trapped-ion situation is not quite the same: there, if one shines light only on two of the qubits, the other qubits do not participate in the gates. This is an important difference between trapped ions and our metamaterial, unless we assume we have a tunable coupler connecting each qubit to the metamaterial. {Nevertheless, we think our approach is promising even in the absence of tunable couplers because it allows one to realize $\approx N/2$ simultaneous two-qubit gates on arbitrary disjoint pairs of $N$ qubits connected by the same metamaterial. Since most qubits in this case are already participating in a two-qubit gate, the above concern of causing errors on numerous idle qubits is not an issue.}

\textbf{Applying dynamical decoupling and spin locking on spectator qubits.---}We found above that, with more than two qubits coupled to the same mode, we cannot completely get rid of the dispersive coupling using the  flowers and integers approaches (the spectator qubits end up participating in an entangling gate anyway), although the flowers approach does make the error higher order in the strength of the dispersive coupling.

How can we completely decouple the spectator qubits? One solution is dynamical decoupling: apply fast $\pi$ pulses on the spectator qubits that quickly turn $Z_i \rightarrow - Z_i$, so that the corresponding dispersive coupling averages away leaving us with just two spins coupled to the oscillator. The rate of the applied $\pi$ pulses has to be faster than $f_{i,0} \sim g^2/\Delta$. 
Notice that we cannot use such a simple approach of fast pulses to eliminate the dispersive coupling corresponding to the two spins participating in the gate because this would get rid of the gate, as well. It is worth pointing out that we can replace the dynamical decoupling of the $N-2$ spectator qubits with spin locking, i.e.\ driving them continuously on resonance with a Rabi frequency larger than $f_{i,0} \sim g^2/\Delta$.

Another solution would be to engineer flowers in such a way that the other spins are disentangled at the end. One way to do this is to repeat each flower $2^{N-2}$ times, once for each computational basis state of the $N-2$ spectator qubits: i.e., we repeat the flower sandwiched between all possible $2^{N-2}$ combinations of $\pi$ pulses on the $N-2$ spectator qubits. This way, independently of what computational basis state these spectator qubits have started in, we will have gone through all of them. Of course, such an exponential solution is practical only for small $N$.  
If there are special relationships between the different $f_{i,0}$ (such as, e.g., $f_{i,0} = f_0$), schemes that scale better with $N$ may be possible.

\subsubsection{2 qubits and $\sim N$ photonic modes \label{sec:2N}}

Having discussed the simple case of $N$ qubits and 1 photonic mode, let us describe another simple case of 2 qubits and $\sim N$ photonic modes.

The Hamiltonian in Eqs.\ (\ref{eq:HNk1}-\ref{eq:HNk3}) then simplifies to
\ba
H &=& \sum_{k} \left[\epsilon_k (1 + x_k) b^\dagger_k b_k + M_k (b_k + b^\dagger_k)\right],\\
x_k &=& \frac{1}{\epsilon_k} \sum_{i = 1}^2 f_{i,k} Z_i,\\
M_k &=& \frac{1}{2} \sum^2_{i = 1} g_{i,k}  Z_i \sin \theta_i, 
\ea
i.e., the dispersive coupling coefficient $x_k$ now also  
depends only on $Z_1$ and $Z_2$.

 We suppose that we want to work primarily via mode $k = 0$, and we want to understand the effect of the other modes. Even if there is no dispersive coupling (i.e. $x_k = 0$ for all $k$), inability to neglect other modes leads to significant challenges. Indeed, if we cannot neglect the other modes, we need each of them to come back to its initial state, i.e.~we need $\epsilon_k \tau = 2 \pi n_k$ with integer $n_k$ for each $k$, which is very challenging for a large number of modes (which we assume scales as $\sim N$). 
And then if, on top of this, we want to cancel the dispersive coupling  
using the integers approach, we would need to require $f_{i,k} t = 2 \pi n_{i,k} \tau$ with integer $n_{i,k}$ for each $i$ and $k$. This triples the number of integers, resulting in a 3-integers-per-mode approach. Although it is very challenging to tune the system to a regime where we have 3 integers per mode,  
if this regime is achieved, one can obtain a perfect two-qubit gate  
by making sure the four phases associated with the four computational basis states of the two qubits yield the desired non-linear phase of $\pi$.

We now turn to the flowers approach.  
We suppose that mode $k$ does a flower with $P_k$ petals (where $P_k \geq 4$ is even). Allowing for 
additional $m_k$ loops around each petal, the time interval between $\pi$ pulses is
\ba
\tau = \frac{\pi}{\epsilon_k} + \frac{2 \pi}{P_k \epsilon_k} + \frac{2 \pi m_k}{\epsilon_k}.
\ea
Since the time interval $\tau$ between $\pi$ pulses is the same for all modes, 
the goal is to compensate for the variation in $\epsilon_k$ from one mode to another by choosing $P_k$ and $m_k$. Furthermore, to make sure all flowers come back to the origin, the number of time intervals of duration $\tau$ that the gate takes must be a common multiple  (which we can choose to be the least common multiple) of all $P_k$.  
We leave to future work to find whether such compensation is possible for large numbers of modes. We also emphasize that, as in the case of trapped ions \cite{shapira20}, more complicated control sequences, involving, for example, time-dependent amplitudes and detunings of the cross-resonance drives, may help reduce the infidelity of the cross-cross-resonance gate in the presence of multiple photonic modes and the dispersive coupling.

\subsubsection{$N$ qubits and $\sim N$ photonic modes \label{sec:NN}}

Having considered in detail the case of $N$ qubits and 1 photonic mode (Sec.\ \ref{sec:N1}) and the case of $2$ qubits and $\sim N$ photonic modes (Sec.\ \ref{sec:2N}), we now come back to the general case of $N$ qubits and $\sim N$ photonic modes.

The Hamiltonian is given in Eqs.\ (\ref{eq:HNk1}-\ref{eq:HNk3}). If no additional pulses are applied during the gate, then evolution unitary is given in Eq.\ (\ref{eq:metalab}). As in the case of $2$ qubits and $N$ photonic modes (Sec.\ \ref{sec:2N}), even 
without the dispersive coupling (i.e.\ assuming $x_k = 0$ for all $k$, i.e.~$\epsilon'_k = \epsilon_k$ for all $k$), if we want all modes to come back to the initial state (and not leave qubits entangled with photons), for each mode $k$, we need $\epsilon_k \tau  = 2 \pi n_k$ for some integer $n_k$, which, for large a large number of modes is very challenging to achieve. If we want all modes to come back to the initial state in the presence of the dispersive coupling, then we need another $\sim N^2$ integers $n_{i,k}$ so that $f_{i,k} = 2 \pi n_{i,k} \tau$ for all qubits $i$ and all modes $k$. And even when all of these  challenging conditions are satisfied, we are left with the gate
\ba
U(t) = \prod_k e^{i \frac{M_k^2}{\epsilon_k (1 + x_k)}},
\ea
which, as in the case of $N$ qubits and one mode (Sec.\ \ref{sec:N1}), will entangle all $N$ qubits because, unlike $M_k$ which depend only on $Z_1$ and $Z_2$, $x_k$ depend on all $N$ qubits.  
As for whether the flowers approach works, this depends on the same considerations that we discussed for the case of 2 qubits and $N$ modes in Sec.\ \ref{sec:2N}. As in that section,  gate fidelity could also be improved by pulse shaping \cite{shapira20}.

\section{{Gate parameters and decoherence} \label{sec:imperfections}}

{In this section, we discuss gate parameters and the expected effects of decoherence. In the next section (Sec.\ \ref{sec:simul}), we present numerical simulations to validate the performance of our gate.}

{We start by discussing gate infidelity coming from 
$T_1^{\mathrm{cav}}$, the $T_1$ time of the cavity mediating the cross-cross-resonance gate. The resulting infidelity can be estimated as $\sim \langle a^\dagger a \rangle T /T_1^{\mathrm{cav}}$, where $T$ is the gate duration and $\langle a^\dagger a \rangle$ is the typical cavity occupation during the gate. The resulting error can manifest as residual entanglement between the qubits and the cavity. In the case of the simplest cavity-mediated cross-cross-resonance gate between two-level systems (Sec.\ \ref{sec:qubits}), we have $T \sim \Delta/(g \Omega)$ and $\langle a^\dagger a \rangle \sim 1$ (as the numerical results in the next section show, an effective value $\langle a^\dagger a\rangle\sim 0.1$ is more appropriate for estimating the cavity-loss infidelity $\langle a^\dagger a\rangle T/T_1^{\mathrm{cav}}$). Since the flowers approach (Sec.\ \ref{sec:flowers}) doesn't significantly affect gate duration and the scale of the phase space excursion, these estimates also apply to the flowers approach. In the case of the integers approach (Sec.\ \ref{sec:integers}), we have $\epsilon T \sim n_\epsilon$ and $M^2 T/\epsilon \sim 1$ with $M \sim g \Omega/\Delta$, which gives $T \sim \sqrt{n_\epsilon}/M$, $\langle a^\dagger a \rangle \sim (M/\epsilon)^2 \sim 1/n_\epsilon$, leading to a reduction in infidelity by a factor of $1/\sqrt{n_\epsilon}$ compared to the basic gate of Sec.~\ref{sec:qubits}. When $\sim N$ pairs of cavity-mediated cross-cross-resonance gates are operated simultaneously via $\sim N$ modes, the resulting infidelity will be multiplied by $\sim N$. In the case of transmons (Sec.\ \ref{sec:transmons}), assuming the $M_0$ term in Eq.\ (\ref{eq:Hcc}) is canceled by driving the cavity, the conclusions are the same except $\Omega/\Delta$ is replaced with $\Omega \eta/(\Delta (\Delta - \eta))$. It is also worth pointing out that the evolution under the Hamiltonian in Eq. (\ref{eq:Hcc}), which includes the cross-resonance interaction and the dispersive coupling,  can be solved analytically in the presence of cavity decay because the resulting Lindblad master equation is at most quadratic in cavity creation and annihilation operators (see e.g.\ Ref.\ \cite{jandura24}).}

{We now turn to the discussion of gate infidelity coming from qubit decoherence. Since the cavity-mediated cross-cross-resonance gate is designed for use with fixed-frequency qubits, we assume that the $T_2$ and $T_1$ times of the qubits are of the same order. 
For $N$ qubits simultaneously participating in one or more cross-cross-resonance gates, the contribution to infidelity from {$T_1^\mathrm{q}$, the} $T_1$ {time of the qubits}, will be $\sim N T/${$T_1^\mathrm{q}$}, where $T$ is the gate duration discussed above.}

{As we will discuss in more detail in the next section, we take gate parameters for transmons to be  anharmonicity $\eta/2\pi = 300\:\text{MHz}$, qubit-drive detuning $\Delta/2\pi \approx 1\:\text{GHz}$, drive Rabi frequency $\Omega/2\pi=100\:\text{MHz}$, {and} transmon-resonator coupling $g/2 \pi=40\:\text{MHz}$. {The} duration of the  {$P = 4$ flowers protocol}  is thus {[see Eqs.\ (\ref{eq:Ai},\ref{eq:ttotP4})]} $T \approx 1.45 \pi \Delta (\Delta - \eta)/(g \Omega \eta)\sim 400\:\text{ns}$. Therefore, the infidelity {$\sim 0.1 T/T_1^{\mathrm{cav}}$} associated with cavity decay---assuming the $M_0$ term in Eq.\ (\ref{eq:Hcc}) is canceled by driving the cavity---and  the infidelity {$\sim T/T_1^{\mathrm{q}}$} associated with transmon decoherence  {are both $\lesssim 10^{-3}$ provided  $T_1^{\mathrm{cav}} \gtrsim 40$ $\mu$s and $T_1^{\mathrm{q}} \gtrsim 400$ $\mu$s}.}

{Moderate fabrication imperfections do not affect the performance of our cavity-mediated cross-cross-resonance gate provided that the parameters of the fabricated system can be precisely measured. {We expect that  drive-induced unwanted state transitions \cite{dai26} are avoidable for our parameters and can be mitigated even further by reducing the drive Rabi frequency.}}

\section{{Numerical Simulations} \label{sec:simul}}

{In this section, we present numerical simulations that validate the performance of our cavity-mediated cross-cross-resonance gate. We simulated three versions of the gate: one with two qubits (two-level systems) coupled via a cavity, as in Eq.~\eqref{eq:ham1}, one with four qubits (two-level systems) coupled via two cavities, based on Eq.~\eqref{eq:Nqubitmeta}, and one with two transmons coupled via a cavity, as in Eq.~\eqref{eq:CR1}. All simulations were done using QuantumToolbox.jl~\cite{QuantumToolbox.jl2025}.}

\begin{figure}[t!]
\centering
\includegraphics[width=1\linewidth]{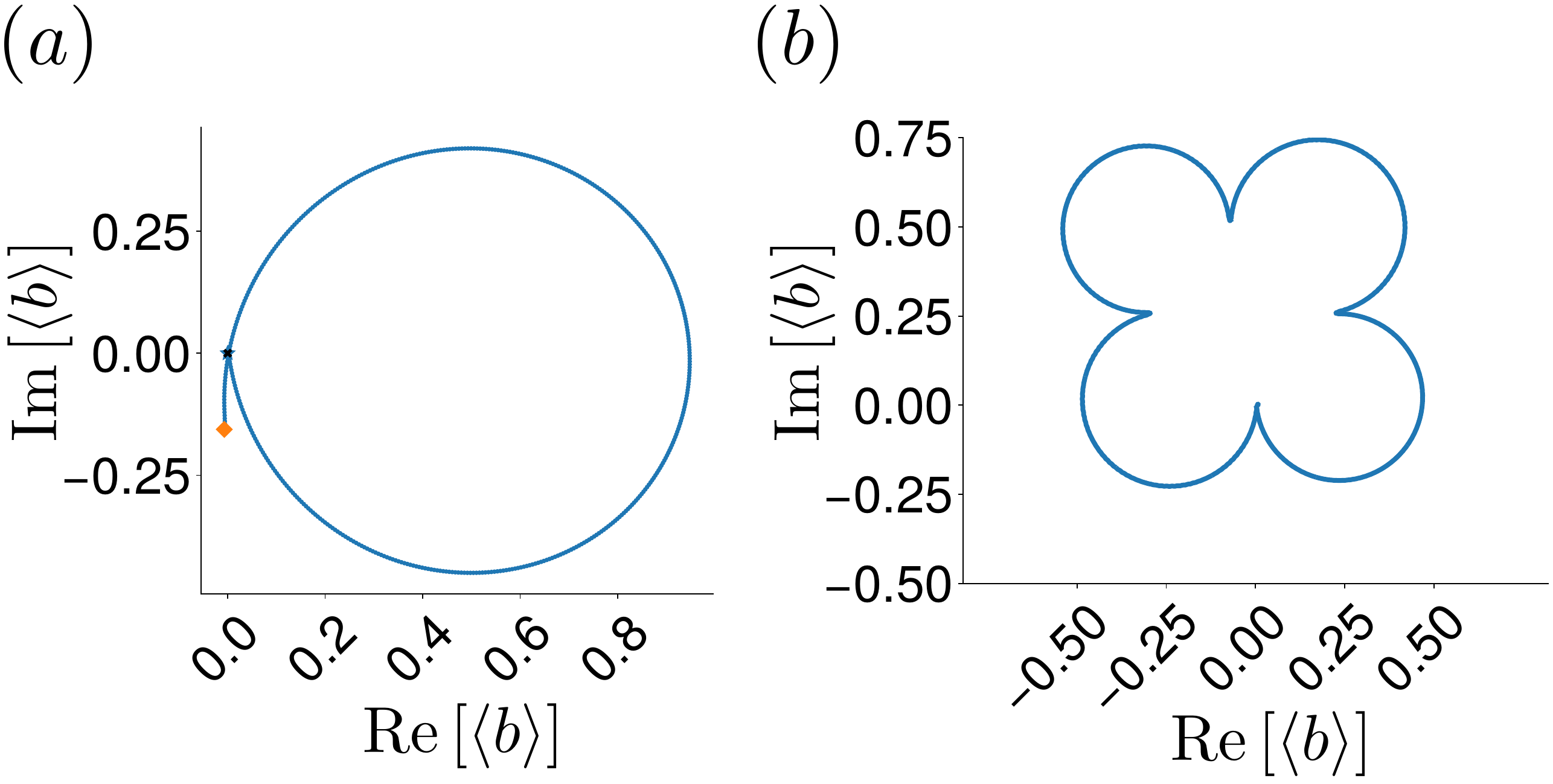} 
\caption{{Cross-cross-resonance gate between two qubits (two-level systems). We present  the cavity phase space trajectory for the $|00\rangle$ state. (a) The basic gate in the presence of  dispersive coupling. The starting point is denoted with a blue circle, while the ending point is denoted with an orange diamond. We see that the trajectory does not return to the origin, leading to residual entanglement with the cavity. 
(b) The $P=4$ flowers protocol. In this case, the trajectory does return to the origin, despite the dispersive coupling.} \label{fig:qubit_flowers_sim}
}
\end{figure}
{We start with the simulation of the cavity-mediated cross-cross-resonance gate between two qubits (two-level systems). 
We take $\Delta_1/2\pi = 920\:\text{MHz}$, $\Delta_2/2\pi = 1000\:\text{MHz}$, $g_1/2\pi=10\:\text{MHz}$, $g_2/2\pi=12\:\text{MHz}$, and we drive both qubits with an amplitude $\Omega/2\pi=208\:\text{MHz}$, with a flat-top pulse with a Nuttall window~\footnote{The Nuttall window is defined as $w(n) = a_0 - a_1 \cos(2\pi n/N_s) +a_2 \cos (4\pi n/N_s) - a_3 \cos (6\pi n/N_s)$ with $a_0 = 0.355768$, $a_1=0.487396$, $a_2=0.144232$, and $a_3=0.012604$. $N_s$ is the number of time steps that the window is broken into.} for the ramp-up (rise) and ramp-down (fall) parts of the pulse (rise/fall time fraction~$3\%$ of flat duration).
To calibrate a bare cross-resonance gate, we do sequential one-dimensional sweeps over the cavity detuning~$\epsilon$
around the analytic estimate,
gate time~$T$, and drive
amplitude~$\Omega$. In the case of the $P=4$ flowers protocol, this optimization is followed by a sweep over
$(\epsilon,\,T_{\mathrm{seg}})$ to globally minimize the average gate infidelity. Here $T_{\mathrm{seg}}$ is the duration of each of the 4 segments, so that $T=4T_{\mathrm{seg}}$. Open-system performance is assessed with Monte Carlo wavefunction simulations that include qubit relaxation and cavity photon loss. 

For the closed system, the resulting trajectory of the cavity in phase space is shown in  Fig.~\ref{fig:qubit_flowers_sim} for the basic gate in (a) and for the $P=4$ flowers protocol in (b). The basic gate reaches an average gate infidelity of $1.3\times10^{-2}$, dominated by residual entanglement with the cavity due to the dispersive coupling. The resulting detuning is $\epsilon/2\pi = 4.7\:\mathrm{MHz}$, and the optimal gate time is $T = 223\:\mathrm{ns}$. The flowers protocol gets to an infidelity of $9.6\times10^{-4}$, still dominated by the residual entanglement. The resulting detuning is then $\epsilon/2\pi=9.7\:\mathrm{MHz}$, and the gate duration is $T=4T_{\mathrm{seg}}= 310\:\mathrm{ns}$. When adding $T_1$ to both qubits {(with equal $T_1^{\mathrm{q}}$)} and the cavity {($T_1^{\mathrm{cav}}$)}, we have an {additional decay-induced} infidelity of   {$0.85 \, T/T_1^{\mathrm{q}} + 0.18 \, T/T_1^{\mathrm{cav}}$} for the basic gate and {$0.84 \, T/T_1^{\mathrm{q}} + 0.12 \, T/T_1^{\mathrm{cav}}$} for the flowers protocol.  {This means that $T_1^\mathrm{q} = 200 (250)$ $\mu$s and $T_1^\mathrm{cav} = 40 (35)$ $\mu$s each contribute only $\approx 10^{-3}$ to the infidelity for the basic gate (flowers protocol). 
The reduction (from $0.18$ to $0.12$) in sensitivity to cavity loss when using the flowers protocol is consistent with the fact that the time-averaged occupation of the cavity for the ideal flowers protocol is lower than for the ideal basic gate.}

\begin{figure}[t!]
\centering
\includegraphics[width=1\linewidth]{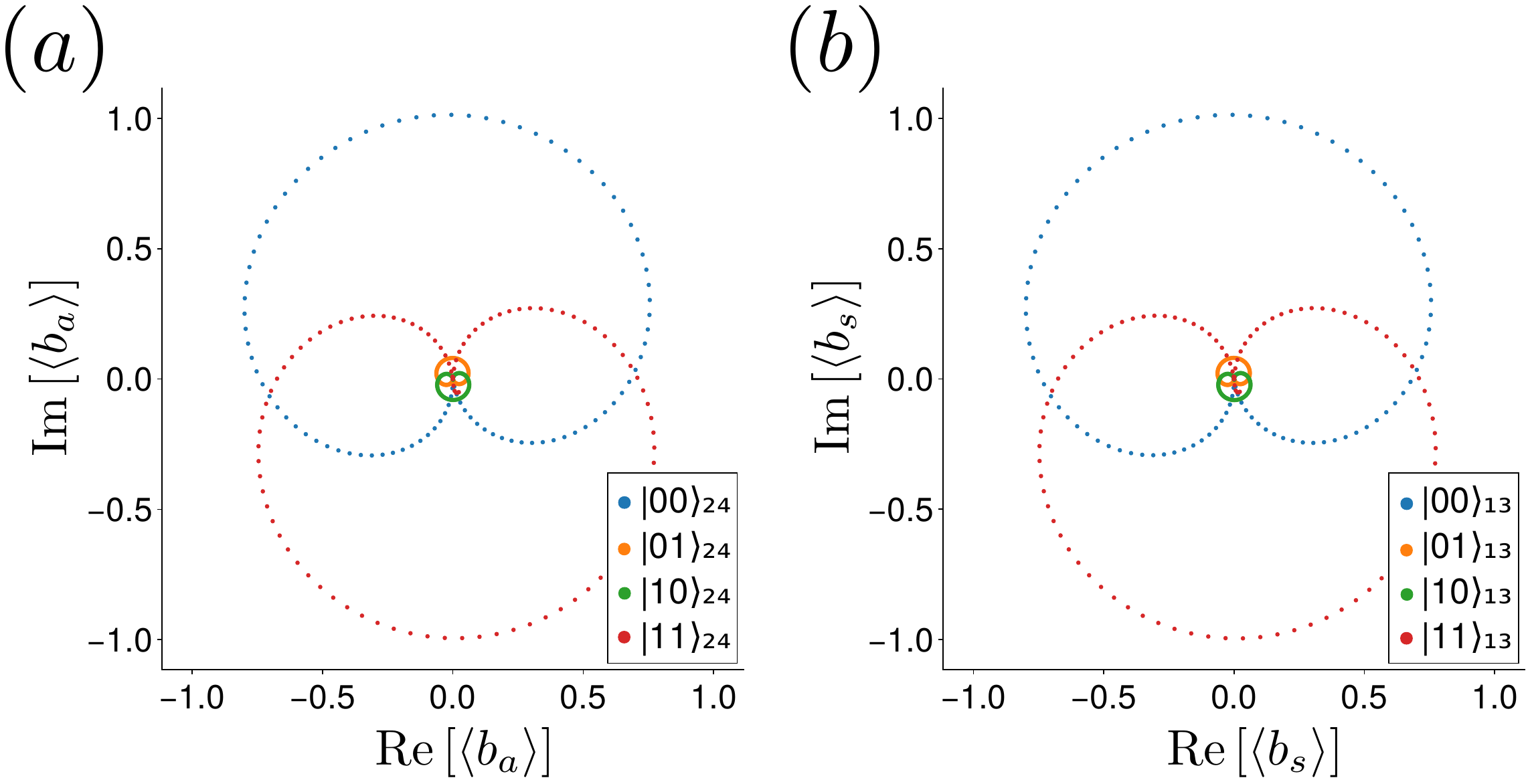} 
\caption{{Parallel cross-cross-resonance gates between two pairs of qubits mediated by two cavities. We present  the phase space trajectory for all relevant two-qubit states for each mode. (a) Gate between Q1 and Q3, which is happening simultaneously with 
(b) the gate between Q2 and Q4. For the plot in (a), Q1 and Q3 start in $\ket{00}$. Similarly, for the plot in (b), Q2 and Q4 start in $\ket{00}$.}
\label{fig:parallel_gates}
}

\end{figure}
{For the four-qubit simulation in Fig.~\ref{fig:parallel_gates}, we consider four qubits and a pair of coupled cavities, where qubits Q1 and Q2 are coupled to one cavity, while qubits Q3 and Q4 are coupled to the other cavity. We define symmetric cavity mode $b_s$ at energy $-J$ and anti-symmetric cavity mode $b_a$ at energy $+J$, with $J/2\pi=65\:\text{MHz}$. These two normal modes enable two simultaneous entangling gates: Q1--Q3 via $b_s$ and Q2--Q4 via $b_a$.  The qubit detunings relative to the bare cavity frequency at $J=0$ are $\Delta_1/2\pi=920\:\text{MHz}$, $\Delta_2/2\pi=1000\:\text{MHz}$, $\Delta_3/2\pi=1100\:\text{MHz}$, $\Delta_4/2\pi=1170\:\text{MHz}$, with equal qubit--cavity couplings $g/2\pi=2\:\text{MHz}$, a base drive amplitude $\Omega/2\pi=190\:\text{MHz}$, and a flat-top Nuttall envelope (rise/fall time fraction~$5\%$ of flat duration).
Gate parameters are calibrated by first sweeping the drive frequencies to locate the cavity resonances, then performing sequential one-dimensional sweeps of the gate time and drive amplitudes for each gate in isolation---where negligible cross-talk to the other cavity mode is verified---and finally re-optimizing the gate time and both drive amplitudes with both drives active simultaneously to account for any residual inter-mode coupling. Each of the two qubit gates, when running in parallel, achieves a closed-system infidelity of $1.5\times10^{-3}$, when averaged over the possible states of the other pair. The gate time is $T=1389\:\mathrm{ns}$, and the detuning is $\epsilon/2\pi=0.76\:\mathrm{MHz}$ for both gates.}

\begin{figure}[t!]
\centering
\includegraphics[width=1\linewidth]{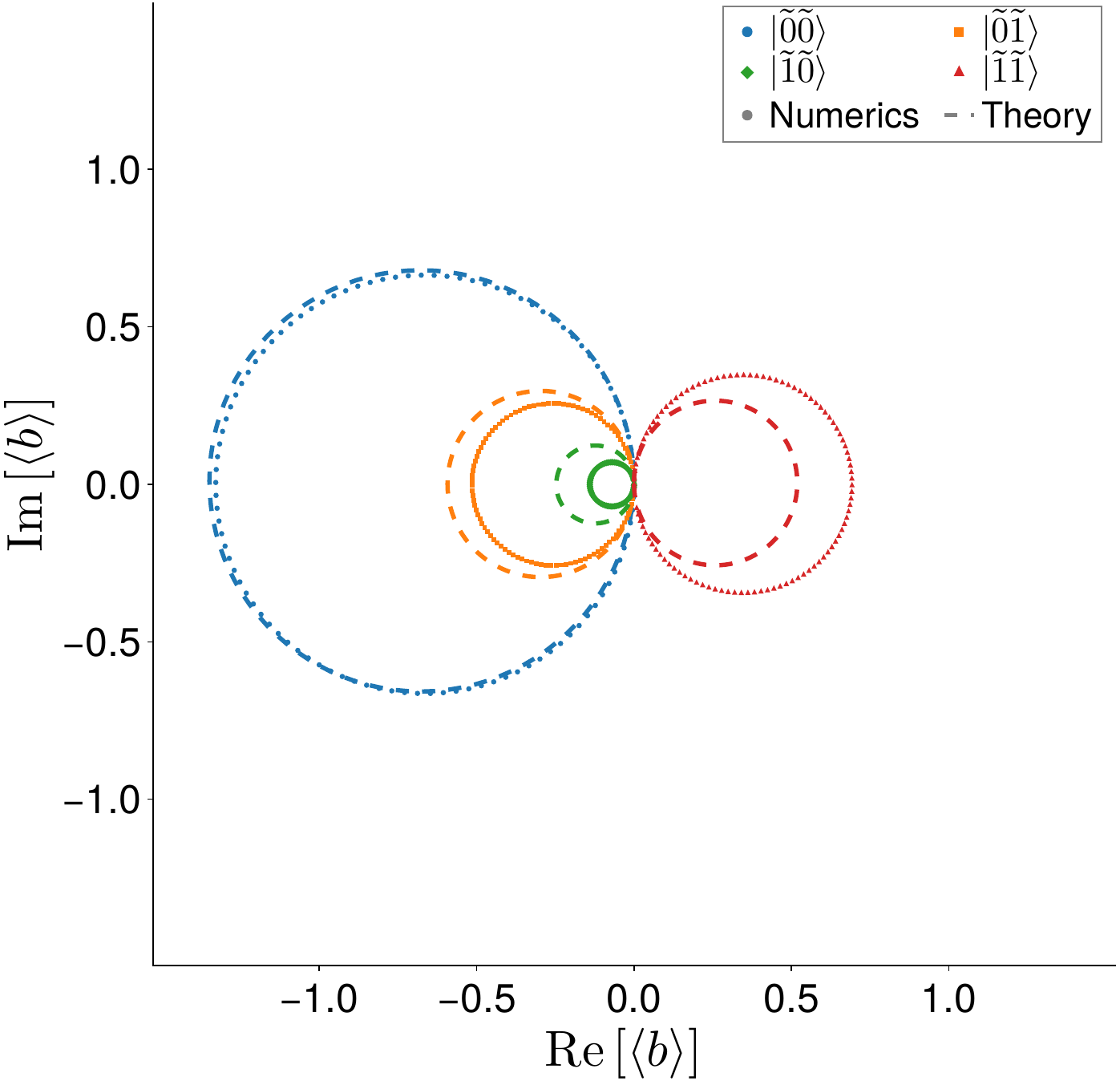} 
\caption{{Cross-cross-resonance gate between two transmons mediated by a cavity, with an added cancellation drive on the cavity. We present the cavity phase space trajectories for all four two-qubit states, compared to the theoretically estimated curves.} \label{fig:transmon_ccr}
}
\end{figure}
{Finally, in Fig.~\ref{fig:transmon_ccr}, we go beyond the two-level approximation for the qubits. We model each transmon as a four-level system with anharmonicities $\eta_1/2\pi=328\:\text{MHz}$, $\eta_2/2\pi=300\:\text{MHz}$, detunings $\Delta_1/2\pi=920\:\text{MHz}$,
$\Delta_2/2\pi=1000\:\text{MHz}$, equal qubit--cavity couplings $g/2\pi=2\:\text{MHz}$, and a
drive amplitude $\Omega/2\pi=110\:\text{MHz}$.  The two-transmon system is first
diagonalized in the driven (dressed) frame to identify the four logical states $\{|\tilde 0 \tilde 0\rangle,|\tilde 0 \tilde 1\rangle,|\tilde 1 \tilde 0\rangle,|\tilde 1 \tilde 1\rangle\}$; these are then tensored with the cavity vacuum [we have to keep as many as $25$ Fock states due to the $M_0$ term in Eq.\ (\ref{eq:Hcc})] to form the computational basis of the full qubit--cavity--qubit Hilbert space.  Since we work directly in the dressed basis, we don't need to ramp $\Omega$ up and down in this simulation. The bare cross-cross-resonance gate, driven at the analytically estimated cavity detuning~$\epsilon$, reaches an infidelity of $5.1\times10^{-2}$, dominated by residual qubit--cavity entanglement arising from dispersive coupling, which is increased in effect due to the $M_0$ term in Eq.\ (\ref{eq:Hcc}), making the cavity population much higher. To suppress this error, we add a direct cavity tone whose amplitude cancels the $M_0$ term in Eq.\ (\ref{eq:Hcc}), leaving only the desired entanglement-generating $M$ term.  The cancellation amplitude $\Omega_{\rm cancel}$ and detuning are jointly optimized via a two-dimensional sweep over $(\epsilon,\,\Omega_{\rm cancel})$, reducing the infidelity to $1.3\times10^{-2}$, still limited by the dispersive coupling, but to a lower level. Increasing the drive $\Omega/2\pi$ to $220\:\text{MHz}$ reduces the infidelity to $3\times10^{-3}$. The gate time in this case is $T = 1939\:\mathrm{ns}$, and the detuning is $\epsilon/2\pi = 0.516\:\mathrm{MHz}$.}

{To make the gate between two transmons less sensitive to decoherence, we make the gate faster by increasing $g/2\pi$ from the above value of $2$ MHz to $40$ MHz. We then use the $P = 4$ flowers protocol to cancel the resulting much larger dispersive coupling. Using $\Omega/2\pi = 100$ MHz and repeating the two-dimensional sweep over $(\epsilon,\Omega_{\rm cancel})$, we find $T = 469$ ns, $\epsilon/2\pi = 17.52\:\mathrm{MHz}$, and an infidelity of $1.5 \times 10^{-3}$. When adding $T_1$ to both transmons and the cavity, we get an additional decay-induced infidelity of $0.816 \, T/T_1^{\mathrm{q}} + 0.064 \, T/T_1^{\mathrm{cav}}$. This means that $T_1^\mathrm{q} = 400$ $\mu$s and $T_1^\mathrm{cav} = 30$ $\mu$s each contribute only $\approx 10^{-3}$ to the infidelity.}

\section{Conclusions and Outlook \label{sec:conclusions}}

We proposed a cavity-mediated cross-cross-resonance gate between two qubits or other nonlinear elements. We showed that this cross-cross-resonance gate allows one to realize simultaneous gates between multiple pairs of qubits coupled via the same metamaterial composed of an array of coupled cavities.

{While we focused in this paper on using an array of coupled oscillators, i.e.\ a metamaterial, to mediate interactions between qubits, our formalism applies just as well to the use of coplanar waveguide resonators to mediate the coupling. Indeed, coplanar waveguide resonators offer large frequency spacing between modes \cite{goppl08a} and may therefore enable faster cross-cross-resonance gates compared to what a metamaterial can offer.}

 While we separately considered the integers and flowers approaches for canceling the dispersive coupling, one can consider combining them. In Ref.\ \cite{golan25}, we also show that the strength of  the (``bad'') dispersive coupling can be suppressed relative to the desired (``good'') interaction by spin-locking the qubits or by working with dual-rail qubits. {While we discuss imperfections (Sec.\ \ref{sec:imperfections}) and do numerical simulations (Sec.\ \ref{sec:simul}) with a focus on systems involving at most four qubits, it would be important to analyze in detail the operation of entire architectures, such as the one shown in Fig.\ \ref{fig:grid}.}

Many exciting research directions remain for the case of qubits coupled via a metamaterial, a scenario that would also benefit from detailed numerical simulations  in the presence of dominant imperfections {and beyond the simplest example analyzed in Fig.\ \ref{fig:parallel_gates}}. To make sure that multiple modes of a metamaterial come back to their initial states at the end of a gate, we can consider borrowing ingenious approaches developed by the trapped-ion community (see e.g.\ Refs.~\cite{grzesiak20,shapira20}). 
However, it remains to be seen whether these approaches can be made to work in combination with dispersive coupling cancellation techniques. It would be useful to extend the detailed analysis of Sec.\ \ref{sec:metatwo}, where we assumed that only one pair of qubits is driven, to the case where multiple pairs of qubits are driven simultaneously. While we assumed for most of Sec.\ \ref{sec:meta} that the metamaterial is made of nearest-neighbor-coupled cavities, it would be interesting to explore the use of long-range interactions between the cavities to engineer dispersions that have modes separated from their neighbors by large gaps, allowing for faster cross-cross-resonance gates. It would also be interesting to explore to what degree the ideas in this manuscript can be extended to the case where the metamaterial is constructed out of qubits \cite{bose07,yao11}, a scenario that does work at least in the minimal example of two qubits coupled via another qubit, as we showed at the end of Sec.\ \ref{sec:qubits}.

While we focused in this paper on 2-qubit gates, it is natural to use a single cavity or a metamaterial to implement multi-qubit gates, by analogy with trapped ions (see e.g.~Ref.\ \cite{lu19}). Indeed, if $N > 2$ qubits are coupled to the same harmonic oscillator mode and are all driven by a cross-resonance drive of the same frequency, we get the same effective Hamiltonian as in Eq.\ (\ref{eq:noquantumstark}), except the sum over $j$ now runs over $N$ qubits, and the resulting gate is still given by Eq.\ (\ref{eq:phase}), except the index $j$ in the definition of $M$ runs over $N$ qubits. We can then choose the gate parameters to implement the  unitary $U(\tau) = \exp(-i (\pi/4) \sum_{i < j} Z_i Z_j)$. Up to single-qubit rotations, this unitary is equal to $\frac{1}{\sqrt{2}} (1 + i^{N + E} Z^{\otimes N})$, where $E = 1 (2)$ for even (odd) $N$ \cite{leibfried18}. Together with single-qubit gates, a single application of this unitary can imprint on one of the $N$ qubits any desired stabilizer-code syndrome corresponding to the remaining $N-1$ qubits \cite{leibfried18}. It is also worth mentioning that the same oscillator mode can simultaneously mediate two independent multi-qubit gates by driving some of the qubits $\pi/2$ out of phase relative to the other qubits, as discussed around Eq.\ (\ref{eq:hz1z2cosxi}). 

It would be interesting to extend the gates proposed in this work in such a way that they can take advantage of collective enhancement enabled by highly populated metamaterial modes and therefore operate much faster, by analogy with the corresponding fast trapped-ion gates \cite{burd21}. Preparing the metamaterial modes in non-vacuum initial states may also help suppress errors associated with cavity decay, by analogy with resonator-induced phase gates \cite{puri16}. 

\begin{acknowledgments}

We thank the staff from across the AWS Center for
Quantum Computing that enabled this project. We also
thank Simone Severini, Bill
Vass, James Hamilton, Nafea Bshara, and Peter DeSantis at AWS for their involvement and support of the research
activities at the AWS Center for Quantum Computing. {During the preparation of this manuscript, the authors used Claude (Anthropic) to assist with debugging and extending the numerical simulation code. All AI-assisted code changes and numerical results were verified by the authors, who take full responsibility for the content of this publication.}

\end{acknowledgments}

\onecolumngrid
\appendix
 
\section{{Detailed derivation of the $P=4$ flower} \label{sec:P4derivation}}

{In this section, we derive the evolution of $b(t)$, shown in Eqs.\ (\ref{eq:tau1}-\ref{eq:tau4}), for the case of the $P = 4$ flower.}

The  rotating-frame-to-lab-frame unitaries {used to arrive at Eq.\ (\ref{eq:Hprf}) and Eq.\ (\ref{eq:Hmrf})} are
\ba
V_+(t) &=& e^{- i \epsilon_+ t b^\dagger b},\\
V_-(t) &=& e^{- i \epsilon_- t b^\dagger b},
\ea
respectively. Defining the evolution under $H$ in Eq.~(\ref{eq:Hprf}) for time $\tau$  as $U_+(\tau)$ and the evolution under $H_-$ in Eq.~(\ref{eq:Hmrf}) for time $\tau$ as $U_-(\tau)$, the full lab-frame evolution after time $4 \tau$ is
\ba
U = V_- U_-(\tau) V_+ U_+(\tau) V_- U_-(\tau) V_+ U_+(\tau),
\ea
where we introduced the shorthand notation $V_- = V_-(\tau)$ and $V_+ = V_+(\tau)$. We now rewrite this evolution as follows:
\ba
U &=& V_- V_+ V_- V_+ [(V_+ V_- V_+)^\dagger U_-(\tau) V_+ V_- V_+] \nonumber\\
&& \times [(V_- V_+)^\dagger U_+(\tau) V_- V_+] [V_+^\dagger U_-(\tau) V_+] U_+(\tau),
\ea
where we inserted three different resolutions of identity. We first notice that 
\ba
V_- V_+ V_- V_+ &=& e^{- i (2 \epsilon_- + 2 \epsilon_+) \tau b^\dagger b} = e^{- i 4 \epsilon \tau b^\dagger b} \nonumber \\
&=& e^{- i 6 \pi b^\dagger b} = 1.
\ea
The unitary $V_+^\dagger U_-(\tau) V_+$ corresponds to evolution from time $t=0$ to time $t=\tau$ under
\ba
V_+^\dagger H_- V_+ = - M b e^{- i \epsilon_- t} e^{- i \epsilon_+ \tau} + h.c.,
\ea
which is equivalent to evolution from $t = \tau$ to $t = 2 \tau$ under
\ba
H^{(2)} = - M b e^{- i \epsilon_- t} e^{- i (\epsilon_+-\epsilon_-) \tau} + h.c.. \label{eq:H(2)}
\ea
The unitary $(V_- V_+)^\dagger U_+(\tau) V_- V_+$ corresponds to evolution from time $t=0$ to time $t=\tau$ under
\ba
(V_- V_+)^\dagger H V_- V_+ = M b e^{- i \epsilon_+ t} e^{- i (\epsilon_+ + \epsilon_-) \tau} + h.c.,
\ea
which is equivalent to evolution from $t = 2 \tau$ to $t = 3 \tau$ under
\ba
H^{(3)} = M b e^{- i \epsilon_+ t} e^{- i (\epsilon_- - \epsilon_+) \tau} + h.c..\label{eq:H(3)}
\ea
Finally, the unitary $[(V_+ V_- V_+)^\dagger U_-(\tau) V_+ V_- V_+]$ corresponds to evolution from time $t=0$ to time $t=\tau$ under
\begin{align}
(V_+ V_- V_+)^\dagger H_- V_+ V_- V_+ = -M b e^{- i \epsilon_- t} e^{- i (2 \epsilon_+ + \epsilon_-) \tau} + h.c.,
\end{align}
which is equivalent to evolution from $t = 3 \tau$ to $t = 4 \tau$ under
\begin{align}
H^{(4)} &= -M b e^{- i \epsilon_- t} e^{- i (2 \epsilon_+ - 2 \epsilon_-) \tau} + h.c..\label{eq:H(4)}
\end{align}

To summarize, the entire lab-frame evolution unitary $U$ corresponds to evolution from $t = 0$ to $t = 4 \tau$, where, from $t = 0$ to $t = \tau$ we evolve under $H$ in Eq.~(\ref{eq:Hprf}), from $t = \tau$ to $t = 2 \tau$ under $H^{(2)}$ in Eq.~(\ref{eq:H(2)}), from $t = 2 \tau$ to $t = 3 \tau$ under $H^{(3)}$ in Eq.~(\ref{eq:H(3)}), and from $t = 3 \tau$ to $t = 4 \tau$ under $H^{(4)}$ in Eq.~(\ref{eq:H(4)}). The full Hamiltonian from $t = 0$ to $t = 4 \tau$ is therefore of the form in Eq.\ (\ref{eq:generalSM}), where $f(t)$ and $g(t)$ change discontinuously at $t = \tau$, $2 \tau$, and $3 \tau$.  {To show that the $P=4$ flower works,} we therefore need to show that the trajectory in phase space closes at $t = 4 \tau$, in which case the phase picked up would be twice the area enclosed. 

Heisenberg evolution under
\ba
H = M e^{-i \phi} e^{-i \epsilon t} b + h.c.
\ea
is
\ba
b(t) = b(0) + \frac{M e^{i \phi}}{\epsilon} \left(1 - e^{i \epsilon t}\right)
\ea
and corresponds to counterclockwise rotation around a circle of radius $M/\epsilon$ whose center is shifted from the initial position by $M e^{i \phi}/\epsilon$ in the complex plane. {Using this, we can immediately express the evolution we derived in this section using Eqs.\ (\ref{eq:tau1}-\ref{eq:tau4}).}

\section{{Flowers with $P \neq 4$}  \label{sec:Pneq4}}

{In this section, we first show that all flowers with even $P \geq 4$ successfully cancel the dispersive coupling during a cavity-mediated cross-cross-resonance gate. We then slightly generalize the flowers approach to allow the time interval before the first pair of $\pi$ pulses to be shorter than the rest of the time intervals between pairs of $\pi$ pulses. Finally, we show that flowers with $P=2$ and with odd $P$ do not work.}

We  begin by showing that all flowers with even $P \geq 4$ work. Exactly as in Ref.~\cite{manovitz17}, we apply $P$ pairs of $\pi$ pulses at time intervals $\tau = \pi/\epsilon + 2 \pi/(P \epsilon)$. In Fig.\ \ref{fig:N6x0}, we show that, for the case $x = 0$ considered in Ref.\ \cite{manovitz17}, the $P = 6$ evolution takes the photonic state to where it started. 
\begin{figure}[t!]
\centering
\includegraphics[width= 0.5 \linewidth]{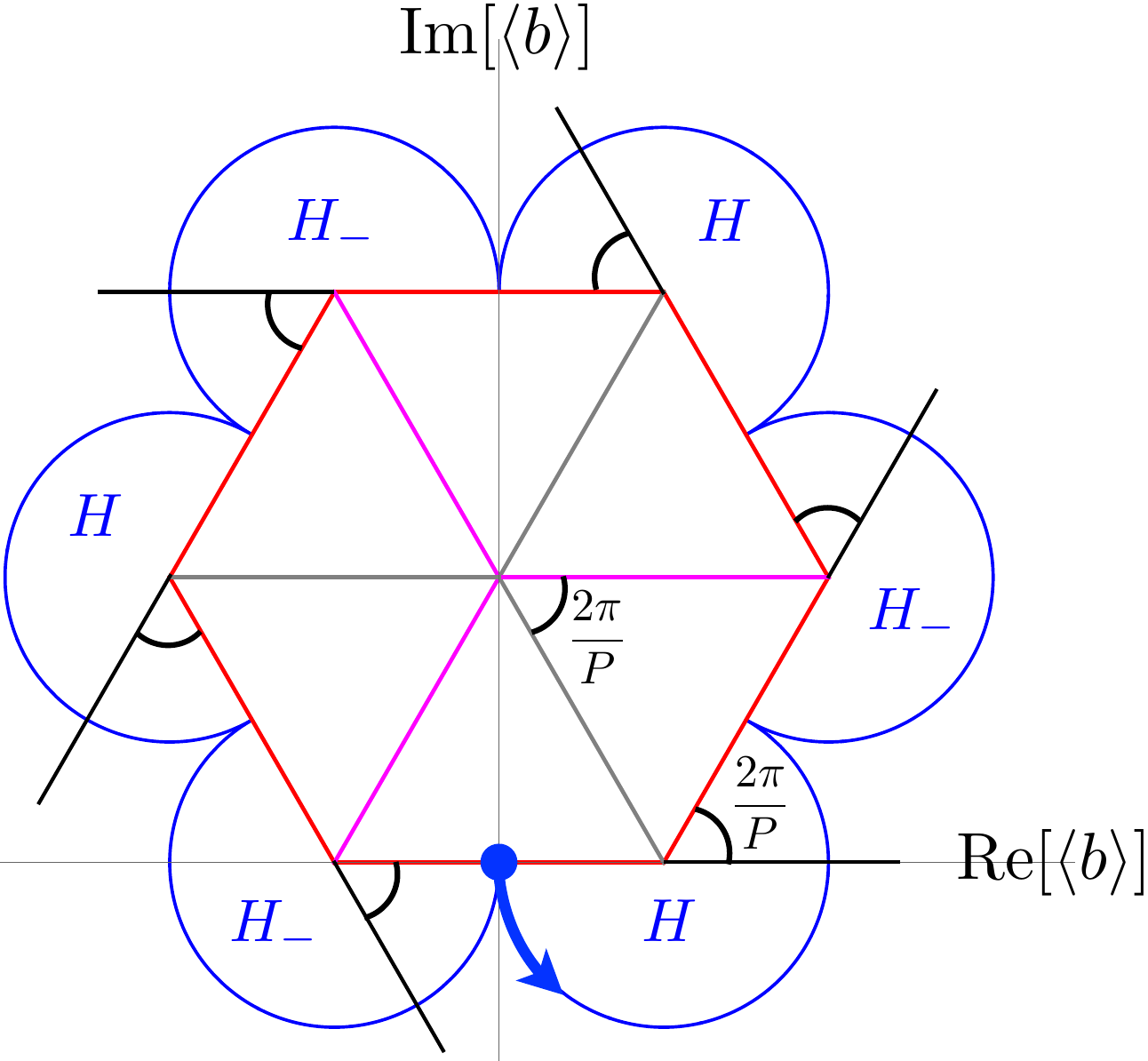} 
\caption{The $P = 6$ flower for the case $x = 0$. Starting at the origin in phase space, $\langle b \rangle$  evolves counterclockwise for angle $\pi + 2 \pi/P$ around $P$ circles. \label{fig:N6x0}
}
\end{figure}
Figure \ref{fig:N6} shows the $P = 6$ flower for $x \neq 0$.
Now $\beta_1 = (2 \pi - \epsilon (1+x) \tau)/2 $ and $\beta_2 = (2 \pi - \epsilon (1-x) \tau)/2$. The key is that $\beta_1+\beta_2$ is independent of $x$ (i.e.\ over-rotation and under-rotation by $\epsilon x \tau$ cancel), so, like in the original $x = 0$ protocol, the third angle in the triangle is 
\ba
\pi - (\beta_1 + \beta_2) &=& \pi - (2 \pi - \epsilon \tau) \\
&= & \pi - (2 \pi - (\pi + 2 \pi/P)) \\
&=& 2 \pi/P.
\ea
So we have $P$ congruent triangles making up a $P$-sided polygon. 

\begin{figure}[t!]
\centering
\includegraphics[width= 0.5 \linewidth]{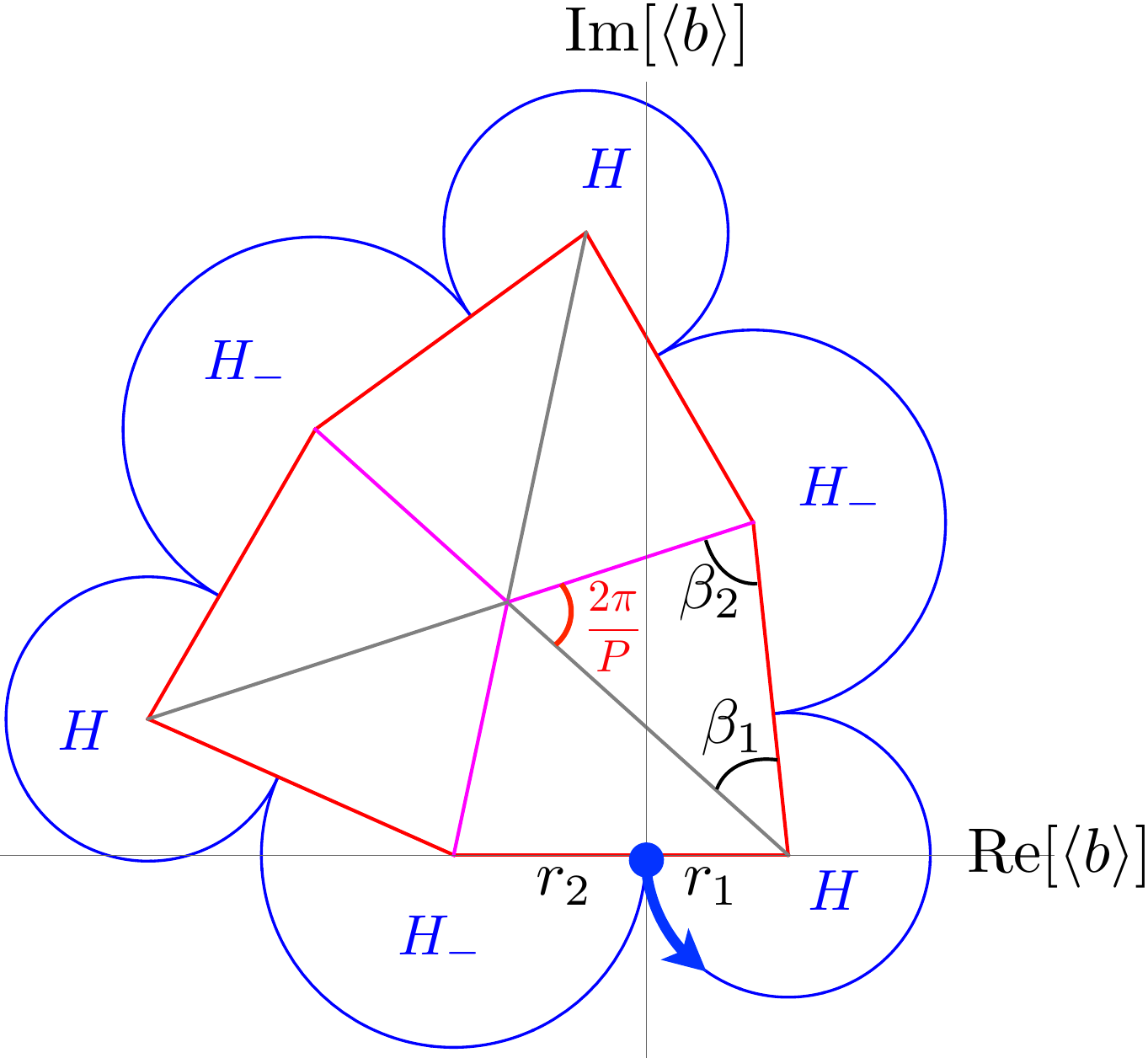} 
\caption{The $P = 6$ flower for the case $x = 0.15$. Starting at the origin in phase space, $\langle b \rangle$  evolves counterclockwise around $P$ circles: for angle $(1+x) (\pi + 3 \pi/P)$ around the small circles and for angle $(1-x) (\pi + 3 \pi/P)$ around the  large circles. The phase-space trajectory comes back to the origin. \label{fig:N6}
}
\end{figure}

The phase of the resulting (up to single-qubit rotations) controlled-phase gate can be computed from the areas of the flowers. In particular, the phase accumulated for a particular initial state of $Z_i$ is twice the area of the corresponding flower. Exactly as in Ref.~\cite{manovitz17}, the area of the flower can be computed by calculating separately the area of the red $P$-sided polygon and the areas of the parts of blue circles outside the polygon. The area of the polygon is
\ba
A_p = P \frac{r^2}{2} \frac{\sin \beta_1 \sin \beta_2}{\sin (2 \pi/P)},\label{eq:Ap}
\ea
where $r = r_1 + r_2$. The combined area of the relevant parts of the circles is 
\ba
A_c &=& \frac{P}{2} \pi r_1^2  \frac{\epsilon (1+ x) \tau}{2 \pi} +\frac{P}{2} \pi r_2^2  \frac{\epsilon (1- x) \tau}{2 \pi} \\
&=& \frac{P+2}{2} \pi \left(\frac{M}{\epsilon} \right)^2\frac{1}{1-x^2}.\label{eq:Ac}
\ea
{Since the plots are made in the $\{\langle x\rangle,\langle p \rangle\}/\sqrt{2} = \{\textrm{Re}[\langle b \rangle], \textrm{Im}[\langle b \rangle]\}$ plane,} the accumulated phase is $2 (A_p + A_c)$. The phase is the same for $(Z_1,Z_2) = (1,1)$ and $(Z_1,Z_2) = (-1,-1)$. Similarly, the phase is the same for $(Z_1,Z_2) = (1,-1)$ and $(Z_1,Z_2) = (-1,1)$. Between the two groups, the values of $x$ and $M$ are different. We will denote these values by $x_{++} = (f_1 +f_2)/\epsilon$, $x_{+-} = (f_1 - f_2)/\epsilon$, $M_{++} = (g_1 \sin \theta_1 + g_2 \sin \theta_2)/2$, $M_{+-} = (g_1 \sin \theta_1 - g_2 \sin \theta_2)/2$. To obtain a gate equivalent (up to single-qubit rotations) to a CZ gate, we need the following condition
\ba
|\phi(1,1) + \phi(-1,-1) - \phi(1,-1) - \phi(-1,1)| = \pi, \label{eq:phipi}
\ea
where $\phi(Z_1,Z_2)$ is the phase accumulated for qubit state ($Z_1$,$Z_2$). This translates to
\ba
|4 (A_p(x_{++}, M_{++}) + A_c(x_{++}, M_{++})) \nonumber \\
- 4 (A_p(x_{+-}, M_{+-}) + A_c(x_{+-}, M_{+-}))| = \pi.\label{eq:piflower}
\ea
Let us do a sanity check. Suppose $\sin \theta_1 = \sin \theta_2 = 1$, $g_1 = g_2 = g$, and $f_1 = f_2 = 0$. So $x_{++} = x_{+-} = M_{+-} = 0$ and $M_{++} = g$. The condition becomes 
\ba
4 \left|P \frac{2 g^2}{\epsilon^2} \frac{\cot(\pi/P)}{2} + \frac{P + 2}{2} \pi \frac{g^2}{\epsilon^2}\right| = \pi \label{eq:sanity}.
\ea
To match the definitions in Ref.~\cite{manovitz17}, we take $g = 2 \tilde \Omega$ and obtain a condition equivalent to Eq.~(3) in Ref.~\cite{manovitz17}, as desired. 

Equation (\ref{eq:piflower}) for the total phase $\pi$ can be solved for $\epsilon$, which can then be plugged into the formula for the total duration of the gate: $T_\textrm{tot} = P \tau = \pi (P+2)/\epsilon$. {If instead of using Eq.\ (\ref{eq:piflower}), we use Eq.\ (\ref{eq:sanity}), we find 
\ba
T_\textrm{tot} = \frac{\pi}{g} \frac{P + 2}{\sqrt{2 (P + 2) + 4 P \cot(\pi/P)/\pi}}, \label{eq:ttot}
\ea
in agreement with Ref.~\cite{manovitz17}. This gate duration grows monotonically with $P$. In the limit $P \rightarrow \infty$, it approaches $T_\textrm{tot} = \pi^2/(2 g)$, which is slower than the ideal gate (without dispersive coupling) [see Eq.\ (\ref{eq:fastest})] by a factor of only $\pi/2 \approx 1.57$, which, in turn, is not appreciably different from the corresponding factor of $\sqrt{3}/\sqrt{3 + 4/\pi} \approx 1.45$ for $P = 4$.}

Assuming that $x$ is small, the formula for $T_\textrm{tot}$ will be approximately the same as for $x=0$.  In our case, we also need to include $\sin \theta_1$ and $\sin \theta_2$, which we set to 1 when doing the sanity  {checks in Eqs.\ (\ref{eq:sanity},\ref{eq:ttot}).} However, for $x=0$, the effect of $\sin \theta_1$ and $\sin \theta_2$ is simply to multiply the total time by $1/\sqrt{\sin \theta_1 \sin \theta_2}$. Since the smallest $P$ for which the cancellation works is $P=4$, we should just use $P=4$ to get the fastest gate. The time this gate will take (for $x \approx 0$) is then 
\ba
T_\textrm{tot} &=& \frac{\pi}{\sqrt{g_1 g_2 \sin \theta_1 \sin \theta_2}} \frac{3}{\sqrt{3 + \frac{4}{\pi}}} \nonumber \\
&\approx& 1.45 \frac{\pi}{\sqrt{g_1 g_2 \sin \theta_1 \sin \theta_2}}.\label{eq:ttotP4}
\ea
If $\pi$ pulses can be applied well, this {``flowers''} approach is likely more promising than the ``integers'' approach, which will be much slower, unless $n_\epsilon$ can be made small, which is not possible if $f_i \ll \epsilon$, as discussed in Sec.\ \ref{sec:integers}. 

While $P=4$ gives the fastest gate, we may want to use $P>4$ if we are also interested in canceling $Z$ noise on the qubits, as originally demonstrated in Ref.~\cite{manovitz17} {and as we have alluded to below Eq.\ (\ref{eq:beta2})}. $Z$ noise is described by the Hamiltonian $H_Z = \sum_i h_i(t) Z_i$, and we can cancel it if $h_i(t)$ varies sufficiently slowly that we can apply $\pi$ pulses on a faster time scale. As {Eq.\ (\ref{eq:ttot})}  shows for $x=0$ (and as will be the case for small nonzero $x$), although the gate duration increases monotonically with $P$, the gate doesn’t  get appreciably slower at higher $P$.  {It is worth noting that successful application of flowers with larger $P$ requires faster and higher-fidelity $\pi$ pulses.}

We can slightly generalize the flowers approach as follows. Recall that, in the  flowers approach described so far, we apply $P$ pairs of $\pi$ pulses at equal time intervals $\tau = \pi/\epsilon + 2 \pi/(P \epsilon)$. This means that the last pair of pulses is applied at the very end of the time evolution, i.e.~no more time evolution takes place after this last pair of pulses, which are just used to return the system to the original rotating frame.  
We can generalize the flowers approach by making the first time interval (i.e.~wait time before the first pair of pulses) shorter, say $\tilde \tau < \tau$ instead of $\tau$. Then we apply the remaining pulses at the usual time intervals $\tau$. And then, after the last pair of pulses, we evolve for time $\tau - \tilde \tau$. What this modification does is that it simply rotates the entire (rotating-frame) flower in phase space. Figure \ref{fig:modifiedN4} shows the $P=4$ example for $\tilde \tau = \tau/2$, 
while Fig.\ \ref{fig:modifiedN6} shows the $P=6$ example for $\tilde \tau = \tau/2$. 
\begin{figure}[t!]
\centering
\includegraphics[width=0.5 \linewidth]{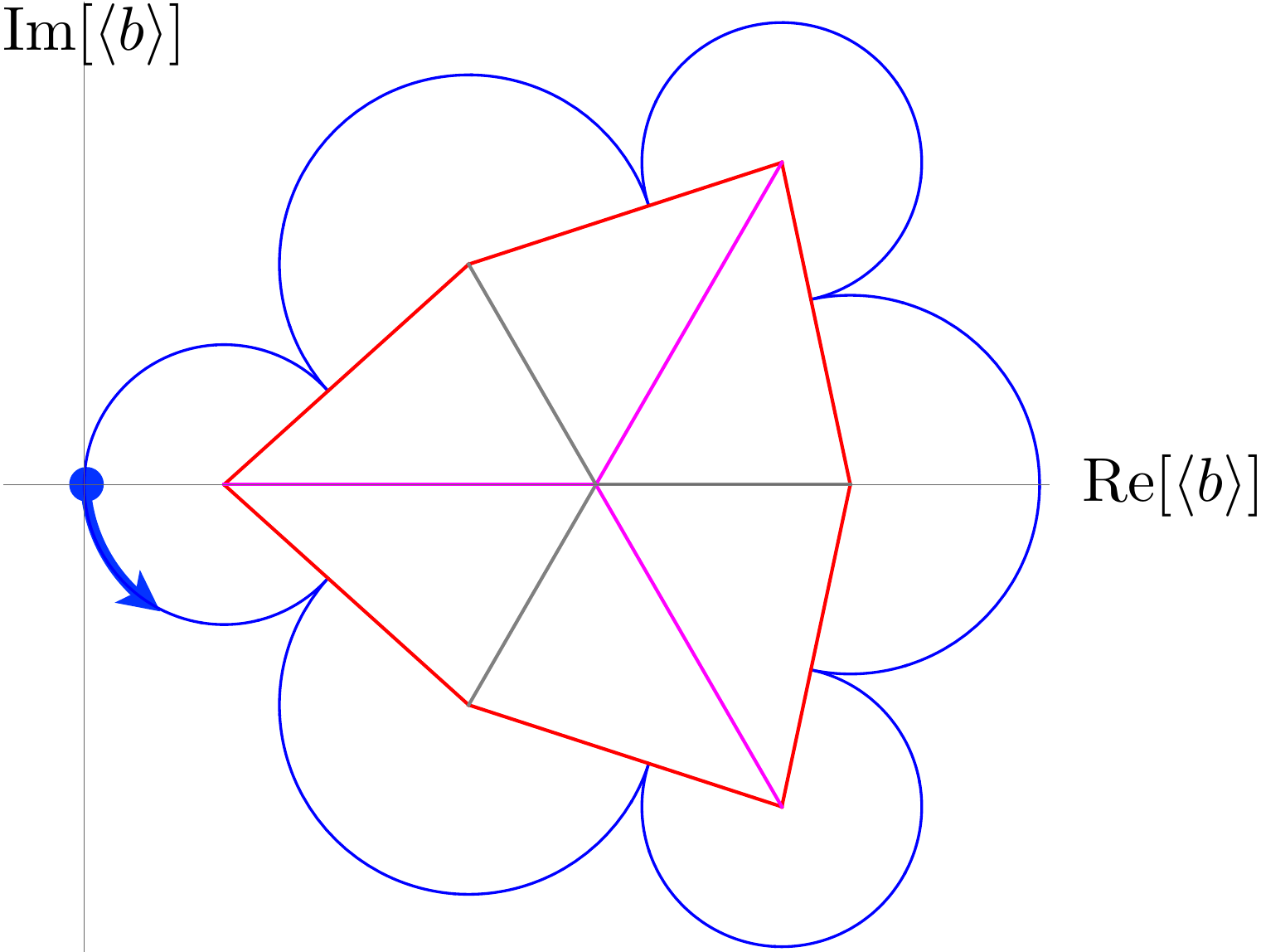} 
\caption{A modification of the $P=6$ flower from Fig.\ \ref{fig:N6}: the first time interval is $\tilde \tau = \tau/2$ [where $\tau = \pi/\epsilon + 2 \pi/(P \epsilon)$], while the last time interval is $\tau - \tilde \tau$. This modification simply rotates the original flower in phase space. \label{fig:modifiedN6}
}
\end{figure}

Setting $\tilde \tau = \tau/2$ transforms our sequence of $\pi$ pulses into a Carr-Purcell-Meiboom-Gill-like (CMPG-like) \cite{carr54,meiboom58} 
pulse sequence for canceling phase ($Z$) noise on the qubits. One potential small disadvantage of using a CPMG-like sequence (or any other sequence with $\tilde \tau < \tau$) over the original pulse sequence is that we need to apply all $P$ pairs of pulses during the gate. In contrast, in the original pulse sequence, the last pair of pulses occurs at the very end, so it simply brings us to the original rotating frame and can be absorbed into subsequent gates. Thus, the original pulse sequence can effectively be implemented with $P-1$ pairs of pulses instead of $P$.

To mitigate imperfections in the implementation of the $\pi$ pulses themselves, we can alternate $\pi$ pulses around $X$ axis and $Y$ axis. Since we only care about flipping $Z$ to $-Z$, this doesn’t affect the ideal flowers. In the case of the $P=4$ CPMG-like flower (shown in Fig.\ \ref{fig:modifiedN4}), alternating between $X$ and $Y$ $\pi$ pulses results in the XY4(s) pulse sequence, while in the case of the original $P=4$ flower (shown in Fig.\ \ref{fig:N4}), this results in the XY4(a) pulse sequence (see, for example, Fig.~2 in Ref.~\cite{ali-ahmed13}).
  
{We now note} that the $P=2$ flower (i.e.\ a flower with 2 petals) {and flowers with odd $P$ (i.e.\ flowers with an odd number of petals)}   {work} in the context of Ref.\ \cite{manovitz17} for  {suppressing} slowly varying $Z$ noise, but  {do} not work for canceling the dispersive coupling. As shown in Fig.\ \ref{fig:N2x0}, 
\begin{figure}[t!]
\centering
\includegraphics[width=0.4\linewidth]{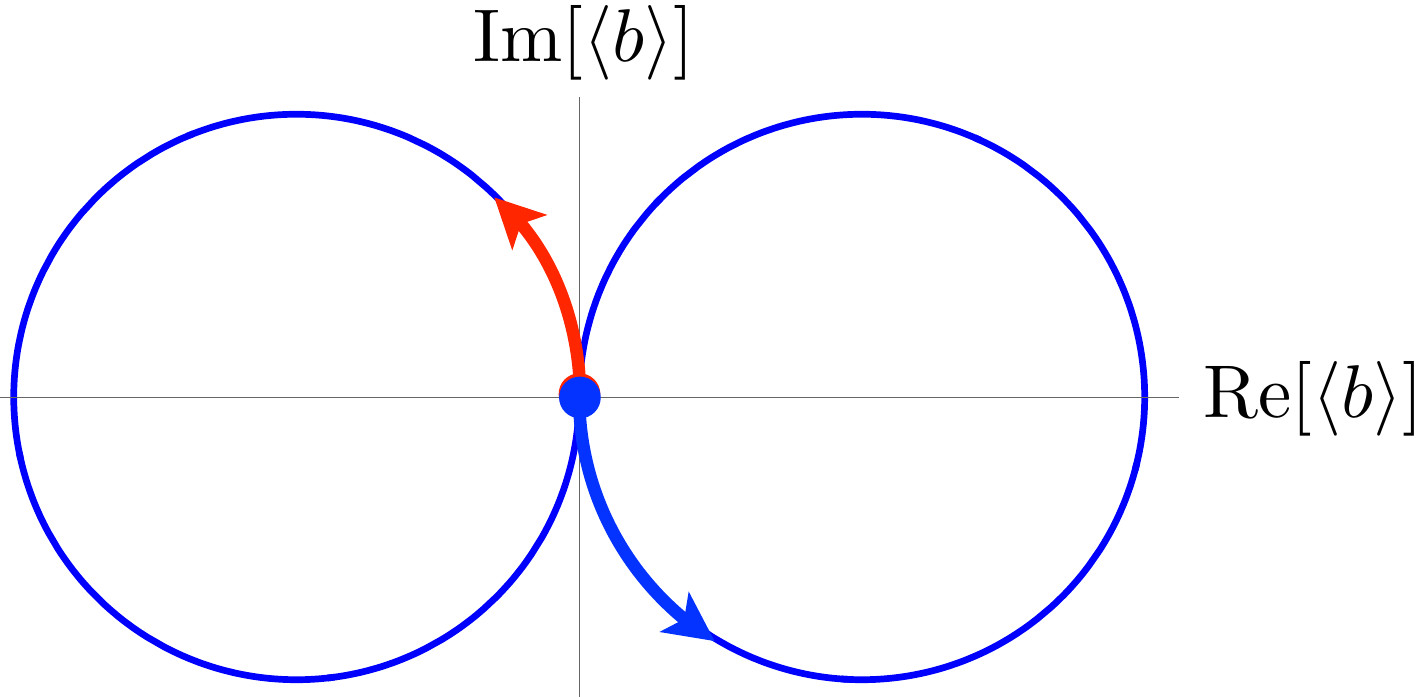} 
\caption{$P=2$ flower for the case $x= 0$. Starting at the origin in phase space, $\langle b \rangle$ evolves counterclockwise along the blue arrow and completes a full circle returning to the origin. Then $\langle b \rangle$ evolves counterclockwise along the red arrow and completes another full circle, again returning to the origin.  \label{fig:N2x0}
}
\end{figure}
for $x = 0$, the $P=2$ case corresponds to a pair of pulses applied after a full $\epsilon \tau = 2 \pi$ rotation, followed by the second pair of pulses applied after another full $2 \pi$ rotation. On the other hand, as shown in Fig.\ \ref{fig:N2}, 
\begin{figure}[t!]
\centering
\includegraphics[width=0.4\linewidth]{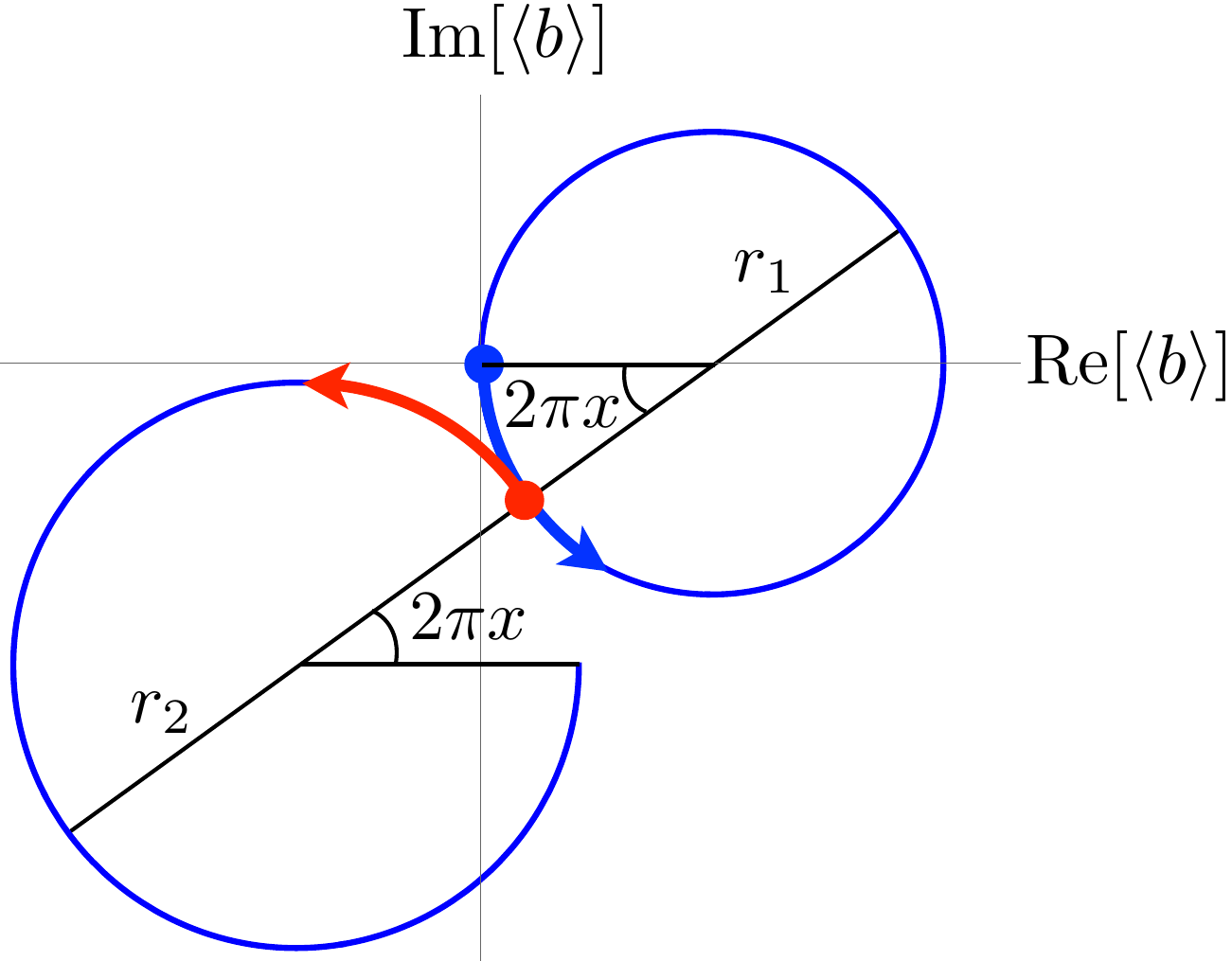} 
\caption{$P=2$ flower does not work for nonzero $x$, as shown here for the case $x= 0.1$. Starting at the origin in phase space, $\langle b \rangle$ evolves counterclockwise along the small circle of radius $r_1$ (blue arrow) and over-rotates by $2\pi x$. Then $\langle b \rangle$ evolves counterclockwise along the large circle of radius $r_2$ (red arrow) and under-rotates by $2 \pi x$. The system does not return to the origin.  \label{fig:N2}
}
\end{figure}
for $x = 0.1$, the first (small) circle over-rotates by $2 \pi x$. The second (large) circle under-rotates by $2 \pi x$. However, the large circle is tangent to the small circle, i.e.\ it shares only one point with the small circle, so it cannot possibly return the system back to the origin. {As for odd $P$, starting at the origin in phase space, odd values of $P$ do not bring our system back to the origin because we have more applications of $H$ than of $H_{-}$, so that over-rotations and under-rotations cannot compensate each other.}  

\section{{The ``flowers'' approach for transmons} \label{sec:flowerstransmons}}

{In this section, we show that the flowers approach works for transmons despite the presence of the $M_0$ term in Eq.\ (\ref{eq:Hcc}) in addition to the desired $M$ term.}

{First, the $M_0$ term can be easily canceled by driving the cavity. Second, the flowers approach works even in the presence of the $M_0$ term, as we will now show.}

{In the presence of the $M_0$ term, Eqs.\ (\ref{eq:tau1}-\ref{eq:tau4}) describing the $P = 4$ flower get modified as follows: 
\begin{align}
b(t) &= b(0) + \frac{(M+M_0)}{\epsilon_+} \left(1 - e^{i \epsilon_+ t}\right), \label{eq:tau1b}\\ 
b(\tau + t) &= b(\tau) + \frac{(-M+M_0) e^{i \epsilon_+ \tau}}{\epsilon_-} \left(1 - e^{i \epsilon_- t}\right), \label{eq:tau2b}\\ 
b(2 \tau + t) &= b(2 \tau) + \frac{(M+M_0) e^{i (\epsilon_++\epsilon_-) \tau}}{\epsilon_+} \left(1 - e^{i \epsilon_+ t}\right),\label{eq:tau3b}\\ 
b(3 \tau + t) &= b(3 \tau) + \frac{(-M+M_0) e^{i (2 \epsilon_++\epsilon_-) \tau}}{\epsilon_-} \left(1 - e^{i \epsilon_- t}\right).\label{eq:tau4b}
\end{align}}

{One can then check analytically that $b(3 \tau + \tau)$ in Eq.\ (\ref{eq:tau4b}) is exactly equal to $b(0)$. Let us also explain intuitively why this happens. Since the displacements are linear in $M$ and $M_0$, and since the $M$ displacements alone take $b(t)$ back to $b(0)$, it remains to show that the $M_0$ displacements alone take $b(t)$ back to $b(0)$:
\begin{align}
b(t) &= b(0) + \frac{M_0}{\epsilon_+} \left(1 - e^{i \epsilon_+ t}\right), \label{eq:tau1c}\\ 
b(\tau + t) &= b(\tau) + \frac{M_0 e^{i \epsilon_+ \tau}}{\epsilon_-} \left(1 - e^{i \epsilon_- t}\right), \label{eq:tau2c}\\ 
b(2 \tau + t) &= b(2 \tau) + \frac{M_0 e^{i (\epsilon_++\epsilon_-) \tau}}{\epsilon_+} \left(1 - e^{i \epsilon_+ t}\right),\label{eq:tau3c}\\ 
b(3 \tau + t) &= b(3 \tau) + \frac{M_0 e^{i (2 \epsilon_++\epsilon_-) \tau}}{\epsilon_-} \left(1 - e^{i \epsilon_- t}\right).\label{eq:tau4c}
\end{align}}

\begin{figure}[t!]
\centering
\includegraphics[width= 0.5 \linewidth]{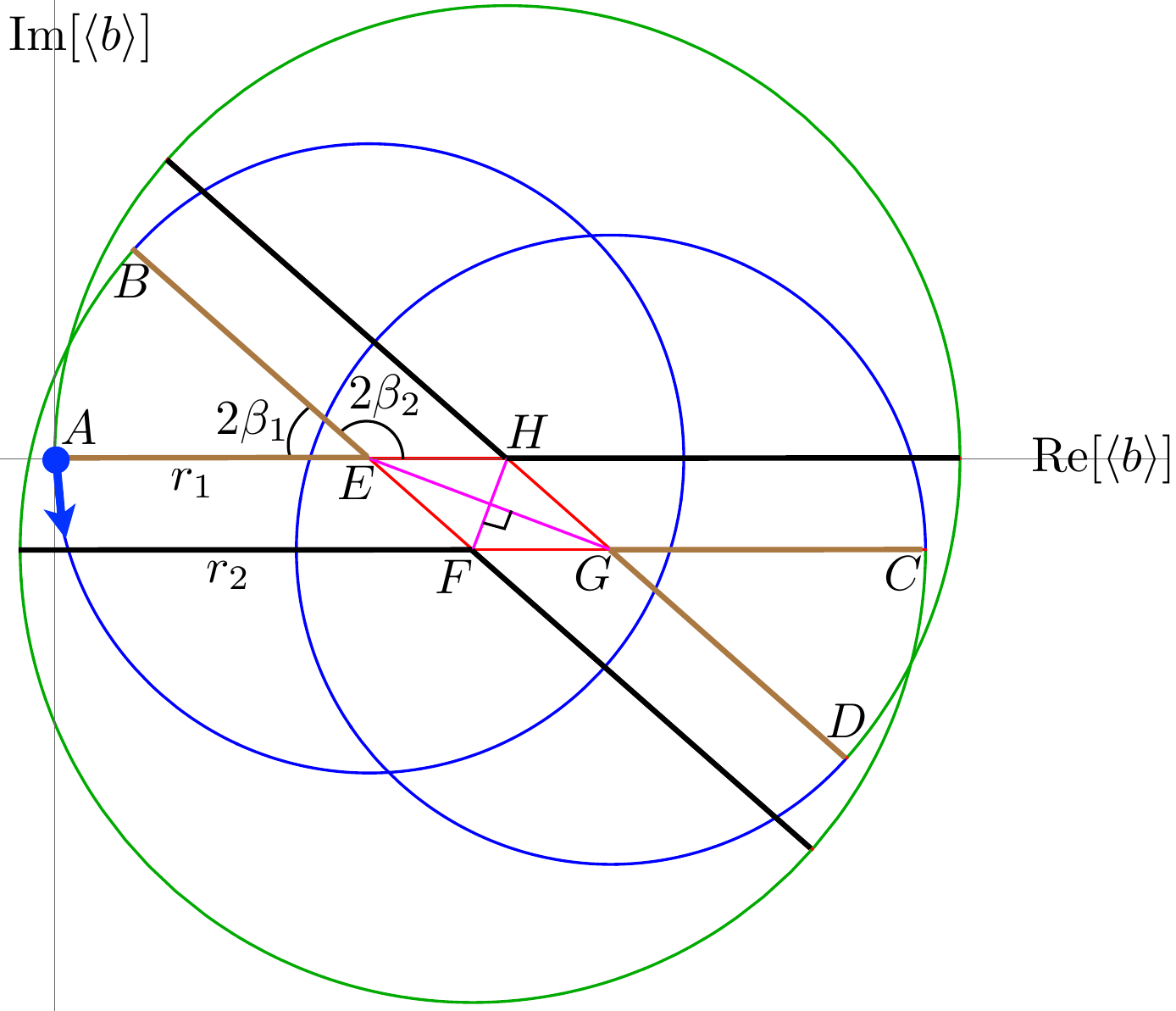} 
\caption{{The modification of the $P = 4$ flower from Fig.\ \ref{fig:N4} for the case $x = 0.18$ (reduced from $x=0.2$ to avoid accidental crossings) where only the sign of $x$ changes after each $\tau$ but the sign of $M$ (which is now called $M_0$) does not change [see Eqs.\ (\ref{eq:tau1c}-\ref{eq:tau4c})]. Starting at the origin $A$ in phase space, $\langle b \rangle$  evolves counterclockwise around four circles: for angle $(1+x) 3 \pi/2 = 2 \pi - 2 \beta_1$ around the small blue circle of radius $r_1$ centered at $E$ to arrive at $B$ [see Eq.\ (\ref{eq:tau1c})], for angle $(1-x) 3 \pi/2 = 2 \pi - 2 \beta_2$ around the large green circle of radius $r_2$ centered at $F$ to arrive at $C$ [see Eq.\ (\ref{eq:tau2c})], for angle $2 \pi - 2 \beta_1$ around the small blue circle of radius $r_1$ centered at $G$ to arrive at $D$ [see Eq.\ (\ref{eq:tau3c})], and for angle $2 \pi - 2 \beta_2$ around the large green circle of radius $r_2$ centered at $H$ to arrive back at the origin $A$ [see Eq.\ (\ref{eq:tau4c})]. Each brown segment has length $r_1$, while each black segment has length $r_2$. The rhombus $EFGH$ has sides of length $r_2-r_1$, in contrast to Fig.\ \ref{fig:N4}, where the rhombus has sides of length $r_2 + r_1$.  \label{fig:N4transmons}}
}
\end{figure}
{It is easy to see analytically that $b(3 \tau + \tau)$ in Eq.\ (\ref{eq:tau4c}) is exactly equal to $b(0)$. Indeed, since $e^{i (\epsilon_+ + \epsilon_-) \tau} = -1$, the second term on the right-hand side of Eq.\ (\ref{eq:tau3c}) evaluated at $\tau$ is the negative of the second term on the right-hand side of Eq.\ (\ref{eq:tau1c}) evaluated at $\tau$. For the same reason, the second term on the right-hand side of Eq.\ (\ref{eq:tau4c}) evaluated at $\tau$ is the negative of the second term on the right-hand side of Eq.\ (\ref{eq:tau2c}) evaluated at $\tau$. It is also instructive to  plot}  the evolution under Eqs.\ (\ref{eq:tau1c}-\ref{eq:tau4c}), {which we do in Fig.\ \ref{fig:N4transmons}.} {This figure is the modification of Fig.\ \ref{fig:N4} to the case where the sign of $x$ still changes after each $\tau$ but the sign of $M$ (which is now called $M_0$) does not. We refer the reader to the caption of Fig.\ \ref{fig:N4transmons} for a detailed explanation of the trajectory that the photonic state takes in phase space to arrive back at the origin. As in the case of Fig.\ \ref{fig:N4}, the over-rotation around small circles or radius $r_1$ (i.e.\ deviation of $2 \beta_1$ from $\pi/2$) cancels with the under-rotation around large circles of radius $r_2$ (i.e.\ deviation of $2 \beta_2$ from $\pi/2$) since $2 \beta_1 + 2 \beta_2 = \pi$ independently of the value of $x$. The main difference between Fig.\ \ref{fig:N4} and Fig.\ \ref{fig:N4transmons} is that the red rhombus in the former has sides of length $r_2 + r_1$, while it has sides of length $r_2 - r_1$ in the latter.}

\begin{figure}[t!]
\centering
\includegraphics[width= 0.5 \linewidth]{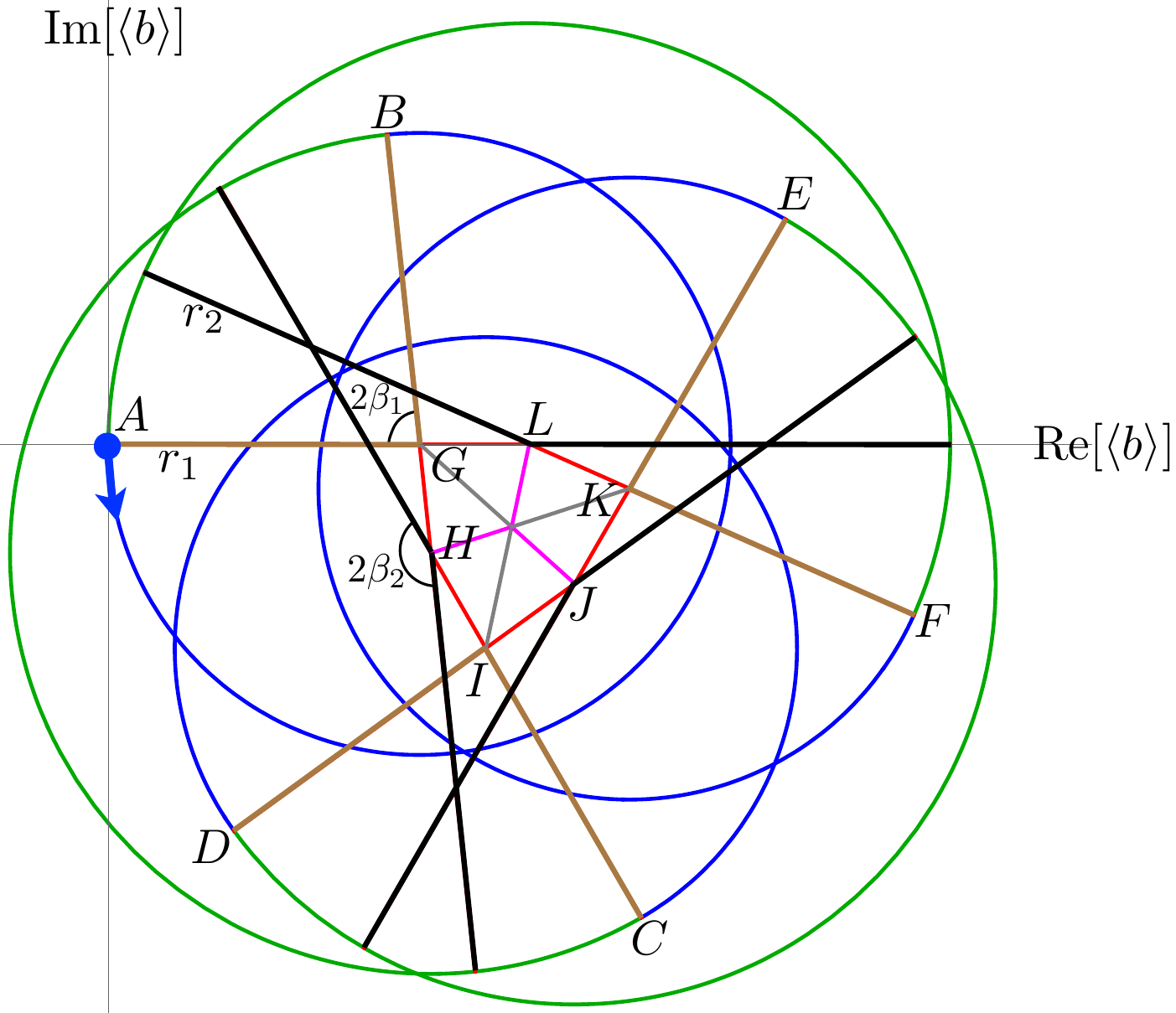} 
\caption{{The modification of the $P = 6$ flower from Fig.\ \ref{fig:N6} with $x = 0.15$, where only the sign of $x$ changes after each $\tau$ but the sign of $M$ (which is now called $M_0$) does not change. Starting at the origin $A$ in phase space, $\langle b \rangle$  evolves counterclockwise around six circles: for angle $2 \pi - 2 \beta_1$ around the small blue circle of radius $r_1$ centered at $G$ to arrive at $B$, for angle $2 \pi - 2 \beta_2$ around the large green circle of radius $r_2$ centered at $H$ to arrive at $C$, for angle $2 \pi - 2 \beta_1$ around the small blue circle of radius $r_1$ centered at $I$ to arrive at $D$, for angle $2 \pi - 2 \beta_2$ around the large green circle of radius $r_2$ centered at $J$ to arrive at $E$, for angle $2 \pi - 2 \beta_1$ around the small blue circle of radius $r_1$ centered at $K$ to arrive at $F$, and for angle $2 \pi - 2 \beta_2$ around the large green circle of radius $r_2$ centered at $L$ to arrive back at the origin $A$. Each brown segment has length $r_1$, while each black segment has length $r_2$. The red $P$-sided polygon $GHIJKL$ is congruent to the corresponding polygon in Fig.\ \ref{fig:N6} but has sides of length $r_2-r_1$ instead of $r_2 + r_1$. \label{fig:N6transmons}}
}
\end{figure}
{In fact, all flowers with even $P \geq 4$ work for transmons despite the presence of the $M_0$ term in Eq.\ (\ref{eq:Hcc}). As in the special case of $P = 4$ just described, it is sufficient to show that the $M_0$ displacements alone (without the $M$ displacements) take $b(t)$ back to $b(0)$. As shown in Fig.\ \ref{fig:N6transmons} for the case $P = 6$, the $M_0$ flower (where $M_0$ does not change sign every $\tau$) works similarly to the $M$ flower in Fig.\ \ref{fig:N6}: the centers of the $P$ circles form the vertices of a $P$-sided polygon made up of $P$ congruent triangles with angles $\beta_1$, $\beta_2$, and $2 \pi/P$, except the sides of this polygon now have length $r_2 - r_1$ instead of $r_2 + r_1$.}

{Having shown that all flowers with even $P \geq 4$ take $b(t)$ back to the origin despite the presence of the $M_0$ term in Eq.\ (\ref{eq:Hcc}), we will now show that, in the limit of sufficiently weak dispersive coupling, the resulting two-qubit gate is only slightly affected by the $M_0$ term. In the frame where the $b^\dagger b$ term is rotated away, the evolution in phase space is described by a Hamiltonian of the form given in Eq.\ (\ref{eq:generalSM}), with
\begin{align}
f(t) &= f_1(t) + f_2(t), \\
g(t) &= g_1(t) + g_2(t),
\end{align}
where $f_1(t)$ and $g_1(t)$ are proportional to $M$, while $f_2(t)$ and $g_2(t)$ are proportional to $M_0$. $f(t)$ and $g(t)$ oscillate at frequency $\epsilon_+ = \epsilon (1+x)$ during half of the petals and at frequency $\epsilon_- = \epsilon (1-x)$ during the other half, where $x = \sum_i f_i Z_i/\epsilon$.}

{At the end of the evolution, when $b(t)$ is back at the origin, the unitary is given by Eq.\ (\ref{eq:eiA}), where $A(t)$ is given in Eq.\ (\ref{eq:Aoft}). Writing $F(t) = F_1(t) + F_2(t)$, where $F_1(t)$ is proportional to $M$ and $F_2$ is proportional to $M_0$, we have
\begin{align}
A(t) = - \int_0^t (F_1(t') + F_2(t')) (g_1(t') + g_2(t')) dt'.
\end{align}}

{Consider first the case of no dispersive coupling, i.e.\ when $x = 0$, so that $\epsilon_+ = \epsilon_- = \epsilon$. In this case, $A(t)$ depends on $Z_1$ and $Z_2$ only through $M = \sum_{i=1,2} \frac{g_i}{2} Z_i A_i$ in $F_1$ and $g_1$. Therefore, the terms $F_1(t') g_2(t')$ and $F_2(t') g_1(t')$ contribute only to single-qubit $Z$ rotations while $F_2(t') g_2(t')$ gives a constant phase. In other words, only $F_1(t') g_1(t')$ contributes to a two-qubit gate, meaning that the two-qubit gate is unaffected by the $M_0$ term up to single-qubit $Z$ rotations.   
}

{When the dispersive coupling is not zero, there is an additional dependence in $A(t)$ on $Z_1$ and $Z_2$ via $\epsilon_+$ and $\epsilon_-$. To obtain a gate equivalent (up to single-qubit rotations) to a CZ gate, we need the condition
\ba
|A(1,1) + A(-1,-1) - A(1,-1) - A(-1,1)| = \pi,
\ea
where $A(Z_1,Z_2)$ is $A(t)$ evaluated at the qubit state $(Z_1,Z_2)$. Since we are computing here the area using Eq.\ (\ref{eq:Aoft}) and not using the plots with rescaled phase space, there is no factor of 2 multiplying each $A$. 
Since the condition for a CZ gate is not affected by $M_0$ for $x=0$, if the dispersive coupling (i.e.~$x$) is sufficiently weak, the condition for a CZ gate is affected by $M_0$ only slightly.}

\section{Uhrig flowers \label{sec:uhrig}}

In this section, we show that, while Ref.\ \cite{manovitz17} can be generalized to the Uhrig \cite{uhrig07} pulse sequence, this pulse sequence cannot be used to cancel the dispersive coupling.

The Uhrig pulse sequence \cite{uhrig07} is the optimized version of the CPMG sequence. To define the Uhrig pulse sequence, we take any integer $p \geq 1$ and define 
\ba
t_i = \tau_c \sin^2\left[\frac{ \pi i}{2 (p+1)}\right]
\ea
for $i = 0, 1, \dots , p+1$. We then apply $\pi$ pulses at times $t_1$ through $t_{p+1} = \tau_c$, where, as usual, the $\pi$ pulses at the final time $\tau_c$ can be absorbed into subsequent gates. Notice that $t_0 = 0$. The $p=1$ Uhrig pulse sequence is just the Hahn echo (i.e.~a $\pi$ pulse at $t_1 = \tau_c/2$). The $p=2$ Uhrig pulse sequence is the simplest CPMG sequence with $t_1 = \tau_c/4$ and $t_2 = 3 \tau_c/4$.

We will now show how  Ref.~\cite{manovitz17}---which is equivalent to Sec.\ \ref{sec:flowers} without the dispersive coupling, i.e.\ with $x = 0$---generalizes from CPMG to Uhrig, resulting in Uhrig flowers. The goal is to find the smallest $\tau_c$ that, using the notation and language of Sec.\ \ref{sec:flowers}, brings us back to the origin in phase space (assuming for simplicity that we started at the origin). For the $p = 1$ Uhrig sequence (i.e.~Hahn echo), the smallest $\tau_c$ is $\tau_c = 4 \pi/\epsilon$, so that $t_1 = 2 \pi/\epsilon$, which is the trivial solution: we wait until we are back at the origin before applying the pulse. 

The $p=2$ Uhrig sequence is just the simplest CPMG sequence, and the smallest $\tau_c$ is $\tau_c = 4 \pi/\epsilon$, which recovers the $P=2$ CPMG flower from  Ref.~\cite{manovitz17}. We show this flower in Fig.\ \ref{fig:uhrig}(a).
\begin{figure}[t!]
\centering
\includegraphics[width=0.99 \linewidth]{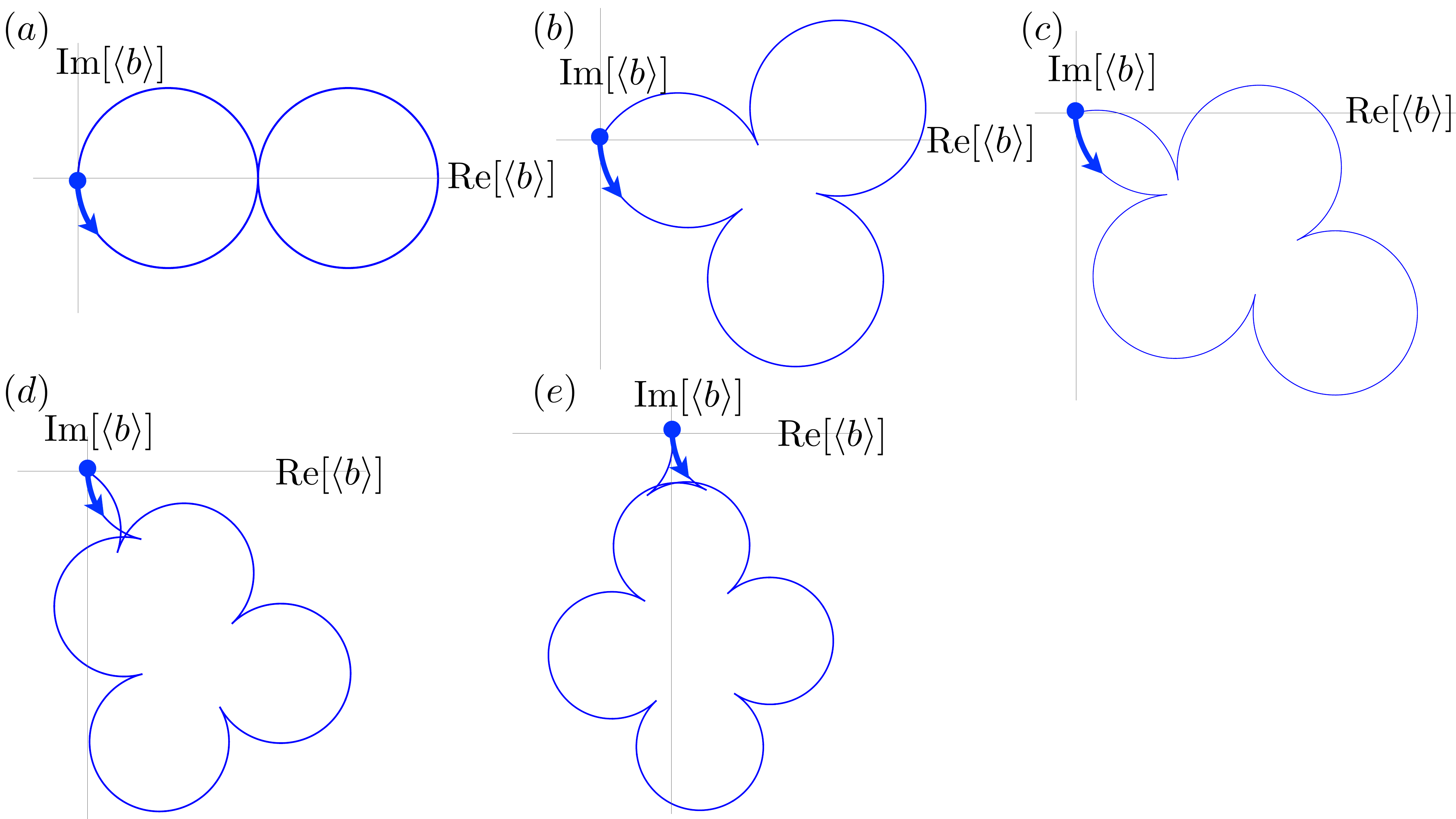} 
\caption{Time evolution of the real and imaginary parts of $\langle b \rangle$ (where $b$ is a bosonic annihilation operator) under the (a) $p=2$ Uhrig sequence with $\tau_c = 4 \pi/\epsilon$, (b) $p=3$ Uhrig sequence with $\tau_c = 2.42259... \times 2 \pi/\epsilon$, (c) $p=4$ Uhrig sequence with $\tau_c = 2.79084... \times 2 \pi/\epsilon$, (d) $p=5$ Uhrig sequence with $\tau_c = 3.1629... \times 2 \pi/\epsilon$, and (e) $p=6$ Uhrig sequence with $\tau_c = 3.52887... \times 2 \pi/\epsilon$. \label{fig:uhrig}
}
\end{figure}

The $p=3$ Uhrig sequence gives the first nontrivial solution. The smallest $\tau_c$ that brings us back to the origin is $\tau_c = 2.42259... \times 2 \pi/\epsilon$, and the corresponding Uhrig flower is shown in Fig.\ \ref{fig:uhrig}(b).

For the $p=4$ Uhrig sequence, the smallest $\tau_c$ is $\tau_c = 2.79084... \times 2 \pi/\epsilon$, and the corresponding Uhrig flower is shown in Fig.\ \ref{fig:uhrig}(c). 
For the $p=5$ Uhrig sequence, the smallest $\tau_c$ is $\tau_c = 3.1629... \times 2 \pi/\epsilon$, and the corresponding Uhrig flower is shown in Fig.\ \ref{fig:uhrig}(d). 
For the $p=6$ Uhrig sequence, the smallest $\tau_c$ is $\tau_c = 3.52887... \times 2 \pi/\epsilon$, and the corresponding Uhrig flower is shown in Fig.\ \ref{fig:uhrig}(e).

We leave it to future work to find analytic expressions for the values of $\tau_c$ and to study what the Uhrig flowers look like for larger values of $p$.

While we have shown in this section that Ref.\ \cite{manovitz17} generalizes to the Uhrig pulse sequence, the Uhrig flowers---unlike the flowers studied in Sec.\ \ref{sec:flowers}---do not work for canceling the dispersive coupling: a nonzero dispersive coupling results in an open trajectory in phase space. Nevertheless, if dispersive coupling were not a problem (e.g.~we canceled it by other means or it is negligible), but phase ($Z$) noise on the qubits were a problem, then one could consider doing Uhrig flowers instead of the original flowers from Ref.\  \cite{manovitz17}.

\section{Always-on ZZ interaction for two transmons coupled via an oscillator \label{sec:ZZ1}}
In this section, we compute the always-on ZZ interaction between two transmons coupled via an oscillator.

Consider two transmons with annihilation operators $c_i$ ($i = 1,2$) coupled to each other via an oscillator with annihilation operator $b$. The Hamiltonian is 
\ba
H = \nu b^\dagger b + \sum_{i=1,2} \left(\omega_i c^\dagger_i c_i - \frac{\eta_i}{2} c^\dagger_i c^\dagger_i c_i  c_i\right) + \sum_i g_i (c_i^\dagger b + h.c.), \label{eq:Hrwa}
\ea
where $\nu$ is the frequency of the oscillator, $\omega_i$ is the frequency of transmon $i$, $\eta_i$ is the anharmonicity of transmon $i$, and $g_i$ is the coupling strength of transmon $i$ to the oscillator. {We note that, in Eq.\ (\ref{eq:Hrwa}), we invoked the rotating-wave approximation by dropping counter-rotating terms $\propto c^\dagger_i b^\dagger$ and $\propto c_i b$, which is a good approximation provided that the couplings $g_i$  and the transmon-oscillator detunings $|\omega_i - \nu|$ are much smaller than the oscillator frequency $\nu$ \cite{blais21}. We refer the reader to Ref.\ \cite{solgun22} for a more general method for calculating always-on ZZ interactions.}

We assume that the oscillator starts in the vacuum state, while the transmons start in the $\ket{0}$-$\ket{1}$ manifold. We write states in the occupation basis as $|\rm{transmon\;1}, \rm{transmon\;2}, \rm{oscillator}\rangle$.  State $|0,0,0\rang$ is an eigenstate of the Hamiltonian with eigenvalue 0. More generally, the total number of excitations is conserved. 

We then assume that the states $|1,0,0\rang$ and $|0,1,0\rang$ are not resonant with each other or with $|0,0,1\rang$ and that $g_i$ are small enough that $|1,0,0\rang$ and $|0,1,0\rang$ get only weakly dressed. We can then calculate the shift  $E_1$ of $|1,0,0\rang$ and the shift $E_2$ of $|0,1,0\rang$ to fourth order in $g_i$.  To calculate $E_1$,  we apply the Schrieffer-Wolff transformation (see Appendix \ref{sec:ZZ2} for a review of the Schrieffer-Wolff formalism \cite{bravyi11b}) to the following Hamiltonian, written in the basis $\{|1,0,0\rang,|0,1,0\rang,|0,0,1\rang\}$: 
\ba
H_1 =\left(\begin{array}{ccc} 0 & 0 & g_1 \\ 0 & \omega_2-\omega_1 & g_2 \\ g_1 & g_2 & \nu - \omega_1 \end{array}\right).
\ea
We find 
\ba
E_1 = \frac{g_1^2}{\Delta_1} - \frac{g_1^4}{\Delta_1^3} + \frac{g_1^2 g_2^2}{(\Delta_1 - \Delta_2) \Delta_1^2},
\ea
where $\Delta_i = \omega_i - \nu$. Similarly, 
\ba
E_2 = \frac{g_2^2}{\Delta_2} - \frac{g_2^4}{\Delta_2^3} - \frac{g_1^2 g_2^2}{(\Delta_1 - \Delta_2) \Delta_2^2}.
\ea

Similarly, we assume that $|1,1,0\rang$ gets only weakly dressed due to its coupling to the other five 2-excitation states. To calculate the shift $E_{12}$ of $|1,1,0\rang$ to fourth order in $g_i$, we apply the Schrieffer-Wolff transformation to the following Hamiltonian, written in the basis $\{|1,1,0\rang,|1,0,1\rang,|0,1,1\rang,|0,0,2\rang,|2,0,0\rang,|0,2,0\rangle\}$: 
\ba
H_{12} = \left(\begin{array}{cccccc} 0 & g_2 & g_1 & 0 & 0 & 0 \\ g_2 & \nu - \omega_2 & 0 & \sqrt{2} g_1 & \sqrt{2} g_1 & 0 \\ g_1 & 0 & \nu - \omega_1 & \sqrt{2} g_2 & 0 & \sqrt{2} g_2 \\ 0 & \sqrt{2} g_1 & \sqrt{2} g_2 & 2 \nu - \omega_1 - \omega_2 & 0 & 0 \\ 0 & \sqrt{2} g_1 & 0 & 0 & \omega_1 - \omega_2 - \eta_1 & 0 \\ 0 & 0 & \sqrt{2} g_2 & 0 & 0 & \omega_2 - \omega_1 - \eta_2 \end{array} \right).\label{eq:h12app}
\ea
The always-on ZZ interaction between the two transmons can then be written as $H = J_{ZZ} Z_1 Z_2$, where 
\ba 
J_{ZZ} = \frac{1}{4}(E_{12} - E_1 - E_2) = \frac{(\Delta_1^2 \eta_1 + \Delta_2^2 \eta_2 + (\Delta_1 + \Delta_2) \eta_1 \eta_2) g_1^2 g_2^2}{2 \Delta_1^2 \Delta_2^2 (\eta_1 + \Delta_2 - \Delta_1)(\eta_2 + \Delta_1 - \Delta_2)}.\label{eq:jzz}
\ea 
If we can set this to zero, the always-on ZZ interaction would be zero to fourth order in $g$. 
The numerator, and hence $J_{ZZ}$, can be set to zero by solving for $\Delta_2$ the quadratic equation $\Delta_1^2 \eta_1 +  \Delta_2^2 \eta_2 + (\Delta_1 + \Delta_2) \eta_1 \eta_2 = 0$. This has a solution provided $\eta_1 \eta_2 \geq 4 \Delta_1(\eta_2 + \Delta_1)$, which we can in turn ensure with an appropriate choice of $\Delta_1$. Since we are trying to set a single number to zero and since each two-qubit interaction will have its own cavity with its own frequency $\nu$ associated with it (see e.g.~Fig.~\ref{fig:grid}), we can judicially choose the frequencies of the transmons in a way that allows us to select the cavity frequency on each bond to set the ZZ on each bond to zero.  

In the qubit limit (i.e.~$\eta_1, \eta_2 \rightarrow \infty$), Eq.\ (\ref{eq:jzz}) simplifies to 
\ba
J_{ZZ} = \frac{  (\Delta_1 + \Delta_2) g_1^2 g_2^2}{2 \Delta_1^2 \Delta_2^2 }, \label{eq:jzzetainf}
\ea
which can, in principle, be set to zero when $\Delta_2 = - \Delta_1$. However, the regime $\Delta_2 = - \Delta_1$ does not work since, in this regime, two excited transmons can resonantly transfer their excitations onto the cavity, violating the assumption that $\ket{1,1,0}$ is only weakly dressed. 

Let us also calculate $J_{ZZ}$ for the case where the oscillator is anharmonic. The only change in this case is that $2 \nu$ in the fourth entry on the diagonal of $H_{12}$ in Eq.\ (\ref{eq:h12app}) is replaced with $2 \nu - \tilde \eta$, where $\tilde \eta$ is the anharmonicity of the oscillator. We then find 
\ba 
J_{ZZ} &=& \frac{1}{4}(E_{12} - E_1 - E_2) \nonumber \\
&=& \frac{[(\Delta_1 + \Delta_2)(\Delta_1^2 \eta_1 + \Delta_2^2 \eta_2 + (\Delta_1 + \Delta_2) \eta_1 \eta_2) + ((\Delta_1 - \Delta_2)^2 (\Delta_1 + \Delta_2) + \Delta_2^2 \eta_1 + \Delta_1^2 \eta_2) \tilde \eta] g_1^2 g_2^2}{2 \Delta_1^2 \Delta_2^2 (\eta_1 + \Delta_2 - \Delta_1)(\eta_2 + \Delta_1 - \Delta_2)(\tilde \eta + \Delta_1 + \Delta_2)}, \label{eq:zzetatil}
\ea
which, as expected, reduces to Eq.\ (\ref{eq:jzz}) when $\tilde \eta = 0$. {It is easy to check that Eq.\ (\ref{eq:zzetatil}) is 1/4 of the right-hand side of Eq.\ (C2) in Ref.~\cite{mundada19} evaluated at $g_{1+} = g_1$, $g_{2+} = g_2$, $\Delta_{1+} = \Delta_1$, $\Delta_{2+} = \Delta_2$, $\alpha_+ = \tilde \eta$, $\Delta_{12} = \Delta_1 - \Delta_2$, $\Delta_{21} = \Delta_2 - \Delta_1$, $\alpha_1 = \eta_1$, $\alpha_2 = \eta_2$, and $g_{1-} = g_{2-} = 0$. When $g_{1-}$ and $g_{2-}$ are nonzero, Eq.\ (C2) in Ref.~\cite{mundada19} generalizes Eq.\ (\ref{eq:zzetatil}) to the case where the two transmons are coupled not via one anharmonic oscillator but via two anharmonic oscillators in parallel.} 

\subsection{Alternative derivation of Eq.\ (\ref{eq:zzetatil})\label{sec:alt}}

In this subsection, we present an alternative derivation of Eq.\ (\ref{eq:zzetatil}). 

We begin by rewriting our Hamiltonian as a sum of the quadratic part $H_2$ and the quartic part $H_4$: 
\ba
H &=& H_2 + H_4, \\
H_2 &=& \nu b^\dagger b + \sum_i \omega_i c^\dagger_i c_i + \sum_i g_i (c_i^\dagger b + h.c.),\\
H_4 &=& -\frac{\tilde \eta}{2} b^\dagger b^\dagger b b - \sum_i\frac{\eta_i}{2} c^\dagger_i c^\dagger_i c_i c_i.
\ea
We now exactly diagonalize $H_2$:
\ba
H_2 =  \tilde \nu \tilde b^\dagger \tilde b + \sum_i \tilde \omega_i \tilde  c^\dagger_i \tilde c_i,
\ea
where the unitary basis change is
\ba
\begin{pmatrix}c_{1}\\ c_{2}\\ b \end{pmatrix}= U \begin{pmatrix}\tilde c_{1}\\ \tilde c_{2}\\ \tilde b \end{pmatrix} =\begin{pmatrix}u_{11} & u_{12} & u_{1b}\\ u_{21} & u_{22} & u_{2b}\\ \gamma_{1} & \gamma_{2} & \gamma_{b} \end{pmatrix}\begin{pmatrix}\tilde{c}_{1}\\ \tilde{c}_{2}\\ \tilde{b} \end{pmatrix}.
\ea
Because the matrix we are diagonalizing is a real symmetric matrix, we can choose the unitary basis change matrix $U$ to be real, which we will assume from now on. Let us compute this matrix to second order in $g$. To do this, we will first derive $U^{-1}$ and then invert it, which is equivalent to taking a transpose since $U$ is unitary and real. Using the Schrieffer-Wolff transformation, we find  $U^{-1}$: 
\ba
U^{-1} = \begin{pmatrix}1 + O(g^2) & \frac{g_1 g_2}{\Delta_1 (\Delta_1 - \Delta_2)} & \frac{g_1}{\Delta_1} \\ \frac{g_1 g_2}{\Delta_2 (\Delta_2 - \Delta_1)} & 1 + O(g^2) & \frac{g_2}{\Delta_2}\\ - \frac{g_1}{\Delta_1} & - \frac{g_2}{\Delta_2}  & 1+O(g^2) \end{pmatrix} + O(g^3).
\ea
We can now invert this to find 
\ba
U = (U^{-1})^T = \begin{pmatrix}u_{11} & u_{12} & u_{1b}\\ u_{21} & u_{22} & u_{2b}\\ \gamma_{1} & \gamma_{2} & \gamma_{b} \end{pmatrix} = \begin{pmatrix}1 + O(g^2) &  \frac{g_1 g_2}{\Delta_2 (\Delta_2 - \Delta_1)} & -\frac{g_1}{\Delta_1} \\ \frac{g_1 g_2}{\Delta_1 (\Delta_1 - \Delta_2)} & 1 + O(g^2) & -\frac{g_2}{\Delta_2}\\ \frac{g_1}{\Delta_1} & \frac{g_2}{\Delta_2}  & 1 + O(g^2) \end{pmatrix} + O(g^3).
\ea
We see that $u_{11}$, $u_{22}$, and $\gamma_b$ are equal to 1 with $O(g^2)$ corrections. We also see that $\gamma_1$, $\gamma_2$, $u_{1b}$, and $u_{2b}$ are of order $g$, while $u_{12}, u_{21}$ are of order $g^2$. 

We are now ready to compute the ZZ interaction. Recall that we are interested in computing, to fourth order in $g$, the energy difference $J_{ZZ} = (E_{12} - E_1 - E_2)/4$, where $E_1$ is the energy of the state adiabatically connected (as we turn on $g$) to $c^\dagger_1 |vac\rang$, $E_2$ is the energy of the state adiabatically connected to $c^\dagger_2 |vac\rang$, while $E_{12}$ is the energy of the state adiabatically connected to $c^\dagger_1 c^\dagger_2|vac\rang$. Here $\ket{vac} = \ket{0,0,0}$ is the vacuum state with no excitations. 

To compute $J_{ZZ}$, we re-express $H_4$ in terms of $\tilde c_i$ and $\tilde b$:
\ba
H_4 = -\frac{\tilde \eta}{2} (\gamma_1 \tilde c_1^\dagger + \gamma_2 \tilde c_2^\dagger + \gamma_b \tilde b^\dagger)^2 (\gamma_1 \tilde c_1 + \gamma_2 \tilde c_2 + \gamma_b \tilde b)^2 - \sum_i\frac{\eta_i}{2} (u_{i1} \tilde c_1^\dagger + u_{i2} \tilde c_2^\dagger + u_{ib} \tilde b^\dagger)^2 (u_{i1} \tilde c_1 + u_{i2} \tilde c_2 + u_{ib} \tilde b)^2.\label{eq:hh4}
\ea
One contribution to $J_{ZZ}$ then simply comes from taking the expectation value of $H_4$ in state $\tilde c^\dagger_1 \tilde c^\dagger_2|vac\rang$ and dividing by 4, which gives
\ba
J_{ZZ,1} = - \frac{\tilde \eta}{2}  \gamma_1^2 \gamma_2^2 - \sum_i \frac{\eta_i}{2}  u_{i1}^2 u_{i2}^2. \label{eq:zz1}
\ea
If the entries of $U$ are computed exactly, this is the value of ZZ interaction to first order in $\eta$ and to all orders in $g$. This is  complementary to Eq.\ (\ref{eq:zzetatil}), which is correct to 4th order in $g$ and to all orders in $\eta$. We refer the reader to Sec.\ \ref{sec:etasquared}, where we extend the derivation of Eq.\ (\ref{eq:zz1}) to second order in $\eta$.

Let us now continue with the derivation in this subsection to arrive at Eq.\ (\ref{eq:zzetatil}). The first contribution to $J_{ZZ}$ is then $J_{ZZ,1}$ in Eq.\ (\ref{eq:zz1}). To order $g^4$, we can set $u_{11} = u_{22} = 1$, so that 
\ba
J_{ZZ,1} &=& - \frac{\tilde \eta}{2}  \gamma_1^2 \gamma_2^2 - \frac{1}{2}(\eta_1 u_{12}^2 + \eta_2 u_{21}^2)\\
&=& - \tilde \eta \frac{g_1^2 g_2^2}{2\Delta_1^2 \Delta_2^2} -  \eta_1 \frac{g_1^2 g_2^2}{2 \Delta_2^2 (\Delta_2 - \Delta_1)^2}-  \eta_2 \frac{g_1^2 g_2^2}{2 \Delta_1^2 (\Delta_2 - \Delta_1)^2} \\
&=& - \frac{(\eta_1 \Delta_1^2 + \eta_2 \Delta_2^2 + \tilde \eta (\Delta_1 - \Delta_2)^2)g_1^2 g_2^2}{2 \Delta_1^2 \Delta_2^2 (\Delta_2 - \Delta_1)^2},\label{eq:jzz1expanded}
\ea
where we plugged in expressions for matrix elements of $U$ that are correct to lowest nonzero order in $g$. This is the correct answer for $J_{ZZ}$ to order $g^4$ after it has also been expanded to first order in $\eta$, as one can check by expanding Eq.\ (\ref{eq:zzetatil}) to first order in $\eta$. However, Eq.\ (\ref{eq:zzetatil}) is correct to all orders in $\eta$ (not just to first order). In order to derive the missing terms, we need to examine which states $H_4$ can take us to virtually from the initial state $\tilde c^\dagger_1 \tilde c^\dagger_2|vac\rang$. Since we will now be doing second-order perturbation theory, and since we are interested in contributions that are fourth-order in $g$, we will only care about matrix elements (from $\tilde c^\dagger_1 \tilde c^\dagger_2|vac\rang$ to other states) that are of second or lower order in $g$. The only such matrix elements and intermediate states are state $\frac{1}{\sqrt{2}}(\tilde b^\dagger)^2|vac\rang$ with matrix element $-\sqrt{2} \tilde \eta \gamma_b^2 \gamma_1 \gamma_2$, state $\frac{1}{\sqrt{2}}(\tilde c_1^\dagger)^2|vac\rang$ with matrix element $-\sqrt{2} \eta_1 u_{11}^3 u_{12}$, and state $\frac{1}{\sqrt{2}}(\tilde c_2^\dagger)^2|vac\rang$ with matrix element $-\sqrt{2} \eta_2 u_{22}^3 u_{21}$. Since all three of these matrix elements are of order $g^2$, we can replace $\gamma_b \rightarrow 1, u_{11} \rightarrow 1, u_{22} \rightarrow 1$. Furthermore, we will be doing the simplest possible second-order perturbation theory where these three virtual processes are going to Stark shift the initial state $\tilde c^\dagger_1 \tilde c^\dagger_2|vac\rang $ quadratically in the corresponding matrix element. Furthermore, since we will already have a term of order $g^4$ in the numerators, it is sufficient to evaluate the denominators to zeroth order in $g$. The resulting contribution to $J_{ZZ}$ is 
\ba
J_{ZZ,2} &=& - \frac{\tilde \eta^2 \gamma_1^2 \gamma_2^2}{2(-\Delta_1 - \Delta_2 - \tilde \eta)} - \frac{ \eta_1^2 u_{12}^2}{2 (\Delta_1 - \eta_1 - \Delta_2)} - \frac{ \eta_2^2 u_{21}^2}{2 (\Delta_2 - \eta_2 - \Delta_1)} \\
 &=& \frac{\tilde \eta^2 g_1^2 g_2^2}{2 \Delta_1^2 \Delta_2^2(\Delta_1 + \Delta_2 + \tilde \eta)} +\frac{ \eta_1^2 g_1^2 g_2^2}{2 \Delta_2^2 (\Delta_2 - \Delta_1)^2(\eta_1 + \Delta_2 - \Delta_1)} + \frac{ \eta_2^2 g_1^2 g_2^2}{2 \Delta_1^2 (\Delta_1 - \Delta_2)^2(\eta_2+ \Delta_1-\Delta_2)},\label{eq:jzz2expanded}
\ea
where we plugged in matrix elements of $U$ that are correct to lowest nonzero order in $g$. One can then easily check that adding $J_{ZZ,1}$ from Eq.\ (\ref{eq:jzz1expanded}) and $J_{ZZ,2}$ from Eq.\ (\ref{eq:jzz2expanded}) recovers $J_{ZZ}$ from Eq.\ (\ref{eq:zzetatil}), which is correct to fourth order in $g$ and to all orders in $\eta$.

\subsection{Including direct interaction between data qubits}

In this subsection, we analyze how ZZ interactions derived in Eq.\ (\ref{eq:zzetatil}) are affected by the presence of direct interaction between data qubits. The Hamiltonian is now, {again within the rotating-wave approximation,} 
\ba
H = \nu b^\dagger b - \frac{\tilde \eta}{2} b^\dagger b^\dagger b b + \sum_i \left(\omega_i c^\dagger_i c_i - \frac{\eta_i}{2} c^\dagger_i  c^\dagger_i c_i c_i \right) + \sum_i g_i (c_i^\dagger b + h.c.) + \tilde g(c^\dagger_1 c_2 + h.c.), \label{eq:Hwithdirectcoupling}
\ea
where $\tilde g$ is the strength of the direct interaction.

We assume again that the computational basis states are perturbed only slightly, so that we can use perturbation theory. And we would like to again derive the ZZ interaction to fourth order in $g$, where $g_1$, $g_2$, and $\tilde g$ are assumed to all be proportional to the small parameter $g$. The vacuum state $\ket{0,0,0}$ is again not perturbed. In the basis $\{|1,0,0\rang,|0,1,0\rang,|0,0,1\rang\}$, the single-excitation Hamiltonian is 
\ba
H_1 =\left(\begin{array}{ccc} 0 & \tilde g & g_1 \\ \tilde g & \omega_2-\omega_1 & g_2 \\ g_1 & g_2 & \nu - \omega_1 \end{array}\right),
\ea
and we find, again using the Schrieffer-Wolff transformation, 
\begin{align}
E_1 &= \frac{g_1^2}{\Delta_1} + \frac{\tilde g^2}{\Delta_1 - \Delta_2} + \frac{2 g_1 g_2 \tilde g}{\Delta_1 (\Delta_1 - \Delta_2)} - \frac{((\Delta_1 - \Delta_2) g_1^2 + \Delta_1 \tilde g^2) (\Delta_2^2 g_1^2 + \Delta_1 \Delta_2 (g_2^2-2 g_1^2) +      \Delta_1^2 (g_1^2 - g_2^2 + \tilde g^2))}{\Delta_1^3 (\Delta_1 - \Delta_2)^3},
\end{align}
where $\Delta_i = \omega_i - \nu$. Notice that, when $\tilde g \neq 0$, we have a term $\sim g^3$. The expression for $E_2$ is the same except the subscripts 1 and 2 are swapped.

To calculate the shift $E_{12}$ of $|1,1,0\rang$ to fourth order in $g$, we apply the Schrieffer-Wolff on the following Hamiltonian, written in the basis $\{|1,1,0\rang,|1,0,1\rang,|0,1,1\rang,|0,0,2\rang,|2,0,0\rang,|0,2,0\rangle\}$: 
\ba
H_{12} = \left(\begin{array}{cccccc} 0 & g_2 & g_1 & 0 & \sqrt{2} \tilde g & \sqrt{2} \tilde g \\ g_2 & \nu - \omega_2 & \tilde g & \sqrt{2} g_1 & \sqrt{2} g_1 & 0 \\ g_1 & \tilde g & \nu - \omega_1 & \sqrt{2} g_2 & 0 & \sqrt{2} g_2 \\ 0 & \sqrt{2} g_1 & \sqrt{2} g_2 & 2 \nu - \omega_1 - \omega_2 - \tilde \eta & 0 & 0 \\ \sqrt{2} \tilde g & \sqrt{2} g_1 & 0 & 0 & \omega_1 - \omega_2 - \eta_1 & 0 \\ \sqrt{2} \tilde g & 0 & \sqrt{2} g_2 & 0 & 0 & \omega_2 - \omega_1 - \eta_2 \end{array} \right)
\ea
and find that 
\ba
\frac{1}{4} (E_{12} - E_1 - E_2) = J^{(0)}_{ZZ} + J_{ZZ}^{(1)} + J_{ZZ}^{(2)} + J_{ZZ}^{(4)},
\ea
where we expanded the answer in powers of $\tilde g$ (the term cubic in $\tilde g$ is zero). 

In particular,
\ba
J_{ZZ}^{(0)} = \frac{[(\Delta_1 + \Delta_2)(\Delta_1^2 \eta_1 + \Delta_2^2 \eta_2 + (\Delta_1 + \Delta_2) \eta_1 \eta_2) + ((\Delta_1 - \Delta_2)^2 (\Delta_1 + \Delta_2) + \Delta_2^2 \eta_1 + \Delta_1^2 \eta_2) \tilde \eta] g_1^2 g_2^2}{2 \Delta_1^2 \Delta_2^2 (\eta_1 + \Delta_2 - \Delta_1)(\eta_2 + \Delta_1 - \Delta_2)(\tilde \eta + \Delta_1 + \Delta_2)} \label{eq:jzzsup0}
\ea
is the $\tilde g = 0$ answer in Eq.\ (\ref{eq:zzetatil}). The intuition for this term is that an excitation hops from $c_1$ to $c_2$ via $g_1$ and $g_2$ and then back. 

The term linear in $\tilde g$,
\ba
J_{ZZ}^{(1)} = -\frac{(\Delta_1 \eta_1 + \Delta_2 \eta_2 + \eta_1 \eta_2) g_1 g_2 \tilde g}{\Delta_1 \Delta_2 (\Delta_1 - \Delta_2 - \eta_1) (\Delta_1 - \Delta_2 + \eta_2)},\label{eq:jzzsup1} 
\ea
is third order in our perturbation theory. The intuition for this term is that an excitation hops from $c_1$ to $c_2$ via $\tilde g$ and then back via $g_1$ and $g_2$ (or vice versa). Depending on the size of $\tilde g$ relative to $g_{1}^2/\Delta$ (where we assume $g_1 \sim g_2$ and $\Delta$ is a typical energy scale in the denominators), $J_{ZZ}^{(1)}$ may or may not dominate over the original fourth-order term $J_{ZZ}^{(0)}$. 

The term quadratic in $\tilde g$ consists of a second-order term and a fourth-order term: 
\ba
J_{ZZ}^{(2)} &=& \frac{\tilde g^2}{2 (\Delta_2 - \Delta_1 + \eta_1)} + \frac{\tilde g^2}{2 (\Delta_1 - \Delta_2 + \eta_2)} + \nonumber \\
&& + \frac{\tilde g^2 g_1^2}{2 \Delta_1^2} \left(\frac{(2 \Delta_1 - \Delta_2) (\Delta_2 + \eta_1)}{\Delta_2 (\Delta_2-\Delta_1 + \eta_1)^2} -     \frac{\Delta_1}{(\Delta_1 - \Delta_2 + \eta_2)^2} - \frac{1}{\Delta_1 - \Delta_2 + \eta_2}\right)+  \nonumber \\
&& +\frac{\tilde g^2 g_2^2}{2 \Delta_2^2} \left(\frac{(2 \Delta_2 - \Delta_1) (\Delta_1 + \eta_2)}{\Delta_1 (\Delta_1-\Delta_2 + \eta_2)^2} -     \frac{\Delta_2}{(\Delta_2 - \Delta_1 + \eta_1)^2} - \frac{1}{\Delta_2 - \Delta_1 + \eta_1}\right).\label{eq:jzzsup2} 
\ea
The intuition for the quadratic term is that an excitation hops from $c_1$ to $c_2$ via $\tilde g$ and back via $\tilde g$. The intuition for the quartic term is that, in addition to the process in the previous sentence, we also hop from one of the data qubits to the coupler and back to that data qubit. Unless we tune the quadratic term ($\propto \tilde g^2$) to cancellation, it will dominate over the quartic terms ($\propto \tilde g^2 g_1^2$ and $\propto \tilde g^2 g_2^2$). 

Finally, the term quartic in $\tilde g$ is 
\ba
J_{ZZ}^{(4)} = \frac{(\eta_1 + \eta_2) (2 \Delta_1^2 + 2 \Delta_2^2 + \eta_1^2+ \eta_2^2 + 2 (\Delta_2-\Delta_1) (\eta_1 - \eta_2) -     4 \Delta_1 \Delta_2) \tilde g^4}{(\Delta_1 - \Delta_2 - \eta_1)^3 (\Delta_1 - \Delta_2 +     \eta_2)^3}.
\ea
Unless the quadratic term $\propto \tilde g^2$ in $J_{ZZ}^{(2)}$ is tuned to cancellation, that quadratic term will dominate over $J_{ZZ}^{(4)}$.  

Putting these together: Assuming that we don’t tune various terms to cancellation, assuming $\tilde g \ll g_1 \sim g_2$, and assuming we are not close to resonances, there are generically two distinct regimes. Taking $\Delta$ to be a typical energy scale in the denominators, these two regimes are
\begin{enumerate}
\item If $\tilde g \ll g_1^2/\Delta$, then the ZZ interaction is dominated by the $\tilde g = 0$ answer $J_{ZZ}^{(0)}$.
\item If $\tilde g \gg g_1^2/\Delta$, then the ZZ interaction is dominated by $\frac{ \tilde g^2}{2 (\Delta_2 - \Delta_1 + \eta_1)} + \frac{\tilde g^2}{2 (\Delta_1 - \Delta_2 + \eta_2)}$ from $J_{ZZ}^{(2)}$.
\end{enumerate}

We note that our expressions for $J_{ZZ}^{(0)}$, $J_{ZZ}^{(1)}$, and $J_{ZZ}^{(2)}$ are in agreement with Appendix D of Ref.\ \cite{heunisch23a}. 
In particular, using the dictionary
\ba
U_1 = -\eta_1, U_2 = -\eta_2, U_c = -\tilde \eta, g_{12} = \tilde g, g_{1c} = g_1, g_{2c} = g_2, \Delta_{12} = -\Delta_{21} = \Delta_1-\Delta_2, \Delta_{1c} = \Delta_1, \Delta_{2c} = \Delta_2,
\ea
$\zeta^{(2)}$ from Ref.\ \cite{heunisch23a} is equal to  four times the first line of $J_{ZZ}^{(2)}$ in Eq.\ (\ref{eq:jzzsup2}),  $\zeta^{(3)}$ is  equal to four times $J_{ZZ}^{(1)}$ in Eq.\ (\ref{eq:jzzsup1}), and  $\zeta^{(4)}$ is equal to four times $J_{ZZ}^{(0)}$ in Eq.\ (\ref{eq:jzzsup0}). It is worth mentioning that, if $\tilde g \gg g_1^2/\Delta$, then, for the purposes of computing $J_{ZZ}$, we cannot ignore $J_{ZZ}^{(4)}$ and the last two lines of $J_{ZZ}^{(2)}$.

\subsection{Derivation to second order in $\eta$ and to all orders in $g$ \label{sec:etasquared}}

In this subsection, we derive $J_{ZZ}$ to second order in $\eta$ and to all orders in $g$.

As in Sec.\ \ref{sec:alt}, we write our Hamiltonian as a sum of the quadratic part $H_2$ and the quartic part $H_4$: 
\ba
H &=& H_2 + H_4,\\
H_2 &=& \nu b^\dagger b + \sum_i \omega_i c^\dagger_i c_i + \sum_i g_i (c_i^\dagger b + h.c.) + \tilde g (c_1^\dagger c_2 + h.c.),\\
H_4 &=& -\frac{\tilde \eta}{2} b^\dagger b^\dagger b b - \sum_i\frac{\eta_i}{2} c^\dagger_i c^\dagger_i c_i c_i,
\ea
where, in contrast to Sec.\ \ref{sec:alt}, we also included the direct coupling $\tilde g$. We assume that the same small parameter $g$ controls $g_1$, $g_2$, and $\tilde g$ and that the same small parameter $\eta$ controls $\eta_1$, $\eta_2$, $\tilde \eta$.

As in Sec.\ \ref{sec:alt}, we start by exactly diagonalizing $H_2$:
\ba
H_2 =  \tilde \nu \tilde b^\dagger \tilde b + \sum_i \tilde \omega_i \tilde  c^\dagger_i \tilde c_i,
\ea
where the unitary basis change is
\ba
\begin{pmatrix}c_{1}\\ c_{2}\\ b \end{pmatrix}= U \begin{pmatrix}\tilde c_{1}\\ \tilde c_{2}\\ \tilde b \end{pmatrix} =\begin{pmatrix}u_{11} & u_{12} & u_{1b}\\ u_{21} & u_{22} & u_{2b}\\ \gamma_{1} & \gamma_{2} & \gamma_{b} \end{pmatrix}\begin{pmatrix}\tilde{c}_{1}\\ \tilde{c}_{2}\\ \tilde{b} \end{pmatrix}.
\ea
A convenient feature of this derivation is that adding direct qubit-qubit coupling $\tilde g$ here is easy---we just need to rediagonalize the quadratic part without even changing the notation. In fact, we could even diagonalize the quadratic Hamiltonian across an entire array of qubits.

Since we want to be working now to all orders in $g$, we are not going to compute approximations to the matrix elements of $U$, but assume instead that they have been exactly calculated numerically. As in Sec.\ \ref{sec:alt}, we can assume that $U$ is real.

We are interested in computing, to second order in $\eta$, the energy difference $J_{ZZ} = (E_{12} - E_1 - E_2)/4$, where $E_1$ is the energy of the state adiabatically connected (as we turn on $\eta$) to $\tilde c^\dagger_1 |vac\rang$, $E_2$ is the energy of the state adiabatically connected to $\tilde c^\dagger_2 |vac\rang$, while $E_{12}$ is the energy of the state adiabatically connected to $\tilde c^\dagger_1 \tilde c^\dagger_2|vac\rang$. Here $|vac\rang$ is the vacuum state with no excitations. 

We first note that, by taking the expectation values of $H_2$ in states $\tilde c^\dagger_1 |vac\rang, \tilde c^\dagger_2 |vac\rang, \tilde c^\dagger_1 \tilde c^\dagger_2|vac\rang$, we will not get a contribution to $J_{ZZ}$. This is the expected result that, in the absence of anharmonicity, there will be no ZZ interaction. Furthermore, the vacuum $|vac\rang$ and the single-excitation states $\tilde c^\dagger_1 |vac\rang, \tilde c^\dagger_2 |vac\rang$ are unaffected by $H_4$ to all orders in $H_4$ since $H_4$ annihilates all states with fewer than 2 excitations. We can thus restrict our analysis to the effect of $H_4$ on $\tilde c^\dagger_1 \tilde c^\dagger_2|vac\rang$. 

Now, exactly as in Eq.\ (\ref{eq:hh4}), let us re-express $H_4$ in terms of $\tilde c_i$ and $\tilde b$:
\ba
H_4 = -\frac{\tilde \eta}{2} (\gamma_1 \tilde c_1^\dagger + \gamma_2 \tilde c_2^\dagger + \gamma_b \tilde b^\dagger)^2 (\gamma_1 \tilde c_1 + \gamma_2 \tilde c_2 + \gamma_b \tilde b)^2 - \sum_i\frac{\eta_i}{2} (u_{i1} \tilde c_1^\dagger + u_{i2} \tilde c_2^\dagger + u_{ib} \tilde b^\dagger)^2 (u_{i1} \tilde c_1 + u_{i2} \tilde c_2 + u_{ib} \tilde b)^2.
\ea
Exactly as in Eq.\ (\ref{eq:zz1}), the first-order in $\eta$ contribution to $J_{ZZ}$ is just one quarter of the expectation value of $H_4$ in state $\tilde c^\dagger_1 \tilde c^\dagger_2|vac\rang$, which is
\ba
J_{ZZ,1} = - \frac{\tilde \eta}{2} \gamma_1^2 \gamma_2^2 - \sum_i \frac{\eta_i}{2} u_{i1}^2 u_{i2}^2.\label{eq:jzz12}
\ea

The contribution to $J_{ZZ}$ at second order in $\eta$ comes from the simplest second-order perturbation theory where we virtually go to each one of the five possible intermediate states ($\tilde c^\dagger_1 \tilde b^\dagger|vac\rang$, $\tilde c^\dagger_2 \tilde b^\dagger |vac\rang$, $\frac{1}{\sqrt{2}} (\tilde c^\dagger_1)^2|vac\rang$, $\frac{1}{\sqrt{2}} (\tilde c^\dagger_2)^2|vac\rang$, $\frac{1}{\sqrt{2}} (\tilde b^\dagger)^2|vac\rang$)  and back. We therefore get 
\ba
J_{ZZ,2} &=& - \frac{(\tilde \eta  \gamma_1^2 \gamma_2 \gamma_b + \sum_i  \eta_i u_{i1}^2 u_{i2} u_{ib})^2}{\tilde \nu - \tilde \omega_2} - \frac{( \tilde \eta  \gamma_1 \gamma^2_2 \gamma_b + \sum_i  \eta_i u_{i1} u_{i2}^2 u_{ib})^2}{\tilde \nu - \tilde \omega_1}- \nonumber \\
&&- \frac{(\tilde \eta  \gamma_1^3 \gamma_2 + \sum_i \eta_i u_{i1}^3 u_{i2})^2}{2 (\tilde \omega_1 - \tilde \omega_2)}- \frac{(\tilde \eta  \gamma_1 \gamma^3_2 + \sum_i \eta_i u_{i1} u_{i2}^3)^2}{2 (\tilde \omega_2 - \tilde \omega_1)}-  \nonumber \\
&&- \frac{(\tilde \eta  \gamma_1 \gamma_2 \gamma_b^2+ \sum_i \eta_i u_{i1} u_{i2} u_{ib}^2)^2}{2 (2 \tilde \nu - \tilde \omega_1 - \tilde \omega_2)}.\label{eq:jzz22}
\ea
Notice that, because we are doing the calculation only to second order in $\eta$, we are calculating the denominators only to zeroth order in $\eta$. The resulting total $J_{ZZ,1} + J_{ZZ,2}$ is correct to second order in $\eta$ and to all orders in $g$.

As a sanity check, when we expand $J_{ZZ,1} + J_{ZZ,2}$ (computed from Eqs.\ (\ref{eq:jzz12},\ref{eq:jzz22})) to fourth order in $g$, we recover Eq.\ (\ref{eq:zzetatil}) expanded to second order $\eta$. 

If we have many transmons in an array, then both the first-order (in $\eta$) and the second-order contributions become more complicated. In particular, the first-order contribution will now have a sum over all anharmonicities (not just of the two transmons that are populated and of the coupler connecting them). Similarly, the second-order contribution will now have a term corresponding to each intermediate state, and the number of such intermediate states will be quadratic in the number of elements (transmons and couplers) in the array (since we need to consider all possible ways of distributing 2 excitations across the elements). This contrasts with the derivation that uses perturbation theory in $g$---that derivation is spatially local in the sense that, for any constant order in $g$ (e.g.~$g^4$ or $g^6$), neighbors that are sufficiently far away don’t contribute.

\section{Always-on ZZ interaction for two transmons coupled via two oscillators \label{sec:ZZ2}}

In this section, we compute the always-on ZZ interaction for two transmons coupled via two oscillators.

Consider two transmons with annihilation operators $c_i$ coupled each to its own oscillator $b_i$, and assume that the two oscillators are coupled to each other. {Working, as in Appendix \ref{sec:ZZ1}, within the rotating-wave approximation,} the Hamiltonian is 
\ba
H = \nu \sum_i b_i^\dagger b_i + \sum_i \left(\omega_i c^\dagger_i c_i - \frac{\eta_i}{2} c^\dagger_i c^\dagger_i c_i c_i \right) + \sum_i g_i (c_i^\dagger b_i + h.c.) + \tilde g (b_1^\dagger b_2 + h.c.). \label{eq:H2oscillators}
\ea

If we treat $g_1$, $g_2$, and $\tilde g$ all three as small parameters proportional to some small $g$, we find that the ZZ interaction vanishes to fourth order in $g$ independently of the choices of the other parameters. This is obvious because an excitation from transmon 1 needs three hops to get to transmon 2, so it needs 6 hops to go there and back. This is why the ZZ interaction, which we will derive below, appears only at sixth order.

Before deriving the ZZ interaction to sixth order, let us study a simpler case where we treat only $g_1$ and $g_2$ as small parameters proportional to some small $g$ (while assuming that $\tilde g$ is not small). In this case, the ZZ interaction appears at fourth in $g$. 

To derive this fourth-order ZZ interaction, we use the same approach as in Appendix \ref{sec:ZZ1}. We will use the basis notation $|\rm{transmon1},\rm{transmon2},\rm{oscillator1},\rm{oscillator2}\rang$. To derive the shift $E_1$ of state $|1,0,0,0\rang$, we write the single-excitation Hamiltonian in the basis $\{|1,0,0,0\rang,|0,1,0,0\rangle,|0,0,1,0\rangle,|0,0,0,1\rang\}$:
\ba
H_1 = \left(\begin{array}{cccc} 0 & 0 & g_1 & 0 \\ 0 & \Delta_2 - \Delta_1 & 0 & g_2 \\ g_1 & 0 &  - \Delta_1 & \tilde g \\ 0 & g_2 & \tilde g & - \Delta_1 \end{array}\right),
\ea
where $\Delta_i = \omega_i - \nu$.
To do the Schrieffer-Wolff transformation [which we review below, starting with Eq.\ (\ref{eq:schr})] for the case where $\tilde g$ is \textit{not} treated as a small parameter, we first rotate the basis to diagonalize $\tilde g$ coupling, i.e.~we replace the last two states with their symmetric and antisymmetric linear combinations. A similar procedure is used to derive the shift $E_2$ of state $|0,1,0,0\rang$.

To derive the shift $E_{12}$ of state $|1100\rang$, we write the two-excitation Hamiltonian in the basis $\{|1100\rang,|1010\rang,|1001\rang,|0110\rang,|0101\rang,|0011\rang,|0020\rang,|0002\rang,|2000\rang,|0200\rang\}$:
\begin{align}
H_{12} = \left(\begin{array}{cccccccccc} 0 & 0 & g_2 & g_1 & 0 & 0 & 0 & 0 & 0 & 0 \\ 0 & -\Delta_2 & \tilde g & 0 & 0 & 0 & \sqrt{2} g_1 & 0 & \sqrt{2} g_1 & 0 \\ g_2 & \tilde g & - \Delta_2 & 0 & 0 & g_1 & 0 & 0 & 0 & 0 \\ g_1 & 0 & 0 & -\Delta_1 & \tilde g & g_2 & 0 & 0 & 0 & 0 \\ 0 & 0 & 0 & \tilde g & -\Delta_1 & 0 & 0 & \sqrt{2} g_2 & 0 & \sqrt{2} g_2 \\ 0 & 0 & g_1 & g_2 & 0 & - (\Delta_1 + \Delta_2) & \sqrt{2} \tilde g & \sqrt{2} \tilde g & 0 & 0 \\ 0 & \sqrt{2} g_1 & 0 & 0 & 0 & \sqrt{2} \tilde g & -(\Delta_1 + \Delta_2) & 0 & 0 & 0 \\ 0 & 0 & 0 & 0 & \sqrt{2} g_2 & \sqrt{2} \tilde g & 0 & -(\Delta_1 + \Delta_2) & 0 & 0 \\ 0 & \sqrt{2} g_1 & 0 & 0 & 0 & 0 & 0 & 0 & \Delta_1 - \Delta_2 - \eta_1 & 0 \\ 0 & 0 & 0 & 0 & \sqrt{2} g_2 & 0 & 0 & 0 & 0 & \Delta_2 - \Delta_1 - \eta_2 \end{array}\right).
\end{align}
To do the Schrieffer-Wolff transformation for the case where $\tilde g$ is \textit{not} treated as a small parameter, we first rotate the basis to diagonalize $\tilde g$ coupling, which can easily be done analytically.

We then find
\begin{align}
J_{ZZ} &= \frac{1}{4}(E_{12} - E_1 - E_2) \nonumber \\
&= \frac{g_1^2 g_2^2 \tilde g^2 (-\Delta_2^4 \eta_2 - \eta_1 \Delta_1^4 - \eta_1 \eta_2 (\Delta_1 + \Delta_2) (\Delta_1^2 + \Delta_2^2) +     2 (\Delta_1^2 \eta_1 + \Delta_2^2 \eta_2 +  (\Delta_1 + \Delta_2) \eta_1 \eta_2) \tilde g^2 - (\eta_1 +        \eta_2) \tilde g^4)}{2(\Delta_1 - \Delta_2 - \eta_1) (\Delta_1 - \Delta_2 + \eta_2) (\Delta_1^2 - \tilde g^2)^2 (\Delta_2^2 -     \tilde g^2)^2 }.\label{eq:jzz2cavities1}
\end{align}
{It is easy to check that Eq.\ (\ref{eq:jzz2cavities1}) is 1/4 of the right-hand side of Eq.\ (C2) in Ref.~\cite{mundada19} evaluated at $g_{1+} = g_{1-} = g_1/\sqrt{2}$, $g_{2+} = -g_{2-} = g_2/\sqrt{2}$, $\Delta_{1+} = \Delta_1-\tilde g$, $\Delta_{1-} = \Delta_1+\tilde g$, $\Delta_{2+} = \Delta_2-\tilde g$, $\Delta_{2-} = \Delta_2+\tilde g$, $\Delta_{12} = \Delta_1 - \Delta_2$, $\Delta_{21} = \Delta_2 - \Delta_1$, $\alpha_1 = \eta_1$, $\alpha_2 = \eta_2$, and $\alpha_+ = \alpha_- = 0$.}

If $\eta_1 = \eta_2 = \eta $,  {Eq.\ (\ref{eq:jzz2cavities1})} simplifies to 
\ba
J_{ZZ} = \frac{\eta g_1^2 g_2^2 \tilde g^2 (\Delta_2^4 + \Delta_1^4 + \eta (\Delta_1 + \Delta_2) (\Delta_1^2 + \Delta_2^2) - 2 (\Delta_1^2 + \Delta_2^2 +  (\Delta_1 + \Delta_2) \eta) \tilde g^2 + 2 \tilde g^4)}{2 (\eta^2-(\Delta_1 - \Delta_2)^2) (\Delta_1^2 - \tilde g^2)^2 (\Delta_2^2 -     \tilde g^2)^2 }.
\ea
As in the case of a single cavity mediating the coupling (see Appendix \ref{sec:ZZ1}), parameters can be chosen such that the numerator vanishes. 

In the qubit limit (i.e.~$\eta \rightarrow \infty$), $J_{ZZ}$ simplifies to 
\ba
J_{ZZ} = \frac{g_1^2 g_2^2 \tilde g^2 (\Delta_1 + \Delta_2) (\Delta_1^2 + \Delta_2^2 - 2 \tilde g^2)}{2 (\Delta_1^2 - \tilde g^2)^2(\Delta_2^2 - \tilde g^2)^2}.
\ea
Unlike in the case of one cavity mediating the coupling [Eq.\ (\ref{eq:jzzetainf})], here we can get a cancellation when $\Delta_1$ and $\Delta_2$ are of the same sign (by setting $\Delta_1^2 + \Delta_2^2 - 2 \tilde g^2 = 0$).

Let us now compute the 6th order always-on ZZ interactions for the case where $\tilde g$, $g_1$, and $g_2$ are all assumed to be proportional to the same small parameter $g$. We first review the general Schrieffer-Wolff formalism \cite{bravyi11b} for going to 6th order. 
We start by writing 
\ba
H = H_0 + H_x + H_y,\label{eq:schr}
\ea
where 
\ba
H_0 &=& \left(\begin{array}{cc} 0 & 0 \\ 0 & W \end{array}\right), \\
H_x &=& \left(\begin{array}{cc} 0 & X \\ X^\dagger & 0 \end{array}\right), \\
H_y &=& \left(\begin{array}{cc} 0 & 0 \\ 0 & Y \end{array}\right), \\
\tilde H_0 &=&  \left(\begin{array}{cc} 0 & 0 \\ 0 & W^{-1} \end{array}\right),
\ea
where each entry stands for a block and where the basis consists of $M_l$ basis states for the low-energy subspace (first block) followed by $M_h$ basis states for the high-energy subspace (second block). The $M_h \times M_h$ matrix $W$ is assumed to be diagonal with all entries (energies) on the diagonal being nonzero---these energies separate the high-energy subspace from the low-energy subspace (but these energies can be both positive and negative, so the words ``high energy'' really mean energy away from 0, but could be both above and below 0). We also assume that the matrices $X$ and $Y$ are proportional to a small parameter $t$. The goal is to find an $M_l \times M_l$ effective Hamiltonian whose eigenvalues are equal to order $t^6$ to the $M_l$ eigenvalues of $H$ closest to zero. These will be the eigenvalues that the $M_l$ zero eigenvalues of $H$ flow to as we adiabatically turn on $t$. 

We will do three consecutive Schrieffer-Wolff transformations. To do the first one of them, we define a matrix of order $t$, 
\ba
S = \left[\tilde H_0, H_x\right],
\ea
which satisfies
\ba
[S,H_0] = - H_x.
\ea
It is easy to check that $S^\dagger = - S$, so that $e^S$ is unitary, and so that we can define a slightly rotated Hamiltonian
\ba
H_2 = e^S H e^{-S},
\ea
which has the same eigenvalues as $H$. Expanding in the small parameter $t$ to 6th order, we find
\ba
H_2 &=& H + [S,H] + \frac{1}{2!} [S,[S,H]] + \frac{1}{3!} [S,[S,[S,H]] + \frac{1}{4!} [S,[S,[S,[S,H]]]] + \frac{1}{5!} [S,[S,[S,[S,[S,H]]]]] \nonumber \\
&& + \frac{1}{6!} [S,[S,[S,[S,[S,[S,H_0]]]]]] + O(t^7) .
\ea
Plugging in $H = H_0 + H_x + H_y$ and using $[S,H_0] = - H_x$, we find 
\ba
H_2 = H_0 + H_{2x} + H_{2y} + O(t^7),
\ea
where
\begin{align}
H_{2x} &= [S,H_y] + \left(\frac{1}{2!} - \frac{1}{3!}\right)[S,[S,H_x]] +\frac{1}{3!} [S,[S,[S,H_y]]] + \left(\frac{1}{4!} - \frac{1}{5!}\right)[S,[S,[S,[S,H_x]]]] + \frac{1}{5!}[S,[S,[S,[S,[S,H_y]]]]],\\
H_{2y} &= H_{y} +  \left(1 - \frac{1}{2!}\right) [S,H_x] + \frac{1}{2!} [S,[S,H_y]] +\left(\frac{1}{3!} - \frac{1}{4!}\right)[S,[S,[S,H_x]]] + \frac{1}{4!} [S,[S,[S,[S,H_y]]]] \nonumber \\
& + \left(\frac{1}{5!} - \frac{1}{6!}\right)[S,[S,[S,[S,[S,H_x]]]]].
\end{align}
The terms are grouped in such a way that we have arrived at a problem that has the same form as the one we started with, i.e.~$H_{2x}$ has the same block-off-diagonal structure as $H_x$, and $H_{2y}$ has the same block-diagonal structure as $H_y$ (except now it is also potentially nonzero on the first diagonal block), except now $H_{2x}$ is of order $t^2$ instead of $t$. In other words, we decoupled the low-energy and the high-energy subspaces to order $t$. 

We now repeat this procedure and define 
\ba
S_2 = [\tilde H_0, H_{2x}],
\ea
which is of order $t^2$ and satisfies 
\ba
[S_2,H_0] = - H_{2x}.
\ea
We now do  a small unitary rotation of the Hamiltonian by defining 
\ba
H_3 = e^{S_2} H_2 e^{- S_2},
\ea
which we can write as
\ba
H_3 = H_2 + [S_2,H_2] + \frac{1}{2!} [S_2,[S_2,H_2]] + \frac{1}{3!} [S_2,[S_2,[S_2,H_0]]] + O(t^7).
\ea
As above, we plug in our expressions for $H_2$ and $S_2$ and find 
\ba
H_3 = H_0 + H_{3x} + H_{3y} + O(t^7),
\ea
where 
\ba
H_{3x} &=& [S_2,H_{2y}] + \left(\frac{1}{2!} - \frac{1}{3!}\right)[S_2,[S_2,H_{2x}]], \\
H_{3y} &=& H_{2y} +  \left(1 - \frac{1}{2!}\right) [S_2,H_{2x}] + \frac{1}{2!} [S_2,[S_2,H_{2y}]].
\ea 
Again $H_{2y}$ is block-diagonal, while $H_{3x}$ is block off-diagonal and is now of order $t^3$. So we repeat the Schrieffer-Wolff transformation for the last (third) time. We define 
\ba
S_3 = [\tilde H_0, H_{3x}],
\ea
which is of order $t^3$ and satisfies 
\ba
[S_3,H_0] = - H_{3x}.
\ea
We now do a small unitary rotation of the Hamiltonian by defining 
\ba
H_4 = e^{S_3} H_3 e^{- S_3},
\ea
which we can write as
\ba
H_4 = H_3 + [S_3,H_3] + \frac{1}{2!} [S_3,[S_3,H_0]] + O(t^7).
\ea
As above, we plug in our expressions for $H_3$ and $S_3$ and find 
\ba
H_4 = H_0 + H_{4x} + H_{4y} + O(t^7),
\ea
where 
\ba
H_{4x} &=& [S_3,H_{3y}],\\
H_{4y} &=& H_{3y} +  \left(1 - \frac{1}{2!}\right) [S_3,H_{3x}].
\ea
Again $H_{4y}$ is block-diagonal, while $H_{4x}$ is block off-diagonal and is now of order $t^4$. Since we are interested in the low-energy eigenvalues to order $t^3$, $H_{4x}$ can be dropped (it contributes at most at order $t^8$ since it needs to take us to the high-energy subspace and back). On the other hand, $H_{4y}$ evaluated within the low-energy subspace is our answer.

We now apply this general formalism to our problem of computing $J_{ZZ}$ to 6th order in $g$. We will do this procedure three times: (1) for the single excitation that starts on transmon 1, (2) for the single excitation that starts on transmon 2, and (3) for two excitations, where both transmons start in the excited state. In all of these cases, the low-energy subspace is one-dimensional, so we are just computing an energy shift.

For the case where the single excitation starts on transmon 1, we have 
\ba
H_0 &=& \left(\begin{array}{cccc} 0 & 0 & 0& 0 \\ 0 & \Delta_2 - \Delta_1 & 0 & 0 \\ 0 & 0 &  - \Delta_1 & 0 \\ 0 & 0 & 0 & - \Delta_1 \end{array}\right),\\
H_x &=& \left(\begin{array}{cccc} 0 & 0 & g_1 & 0 \\ 0 & 0 & 0 & 0 \\ g_1 & 0 &  0 & 0 \\ 0 & 0& 0 & 0 \end{array}\right),\\
H_y &=& \left(\begin{array}{cccc} 0 & 0 & 0 & 0 \\ 0 & 0 & 0 & g_2 \\ 0& 0 &  0 & \tilde g \\ 0 & g_2 & \tilde g & 0 \end{array}\right),\\
\tilde H_0 &=& \left(\begin{array}{cccc} 0 & 0 & 0& 0 \\ 0 & \frac{1}{\Delta_2 - \Delta_1} & 0 & 0 \\ 0 & 0 &  - \frac{1}{\Delta_1} & 0 \\ 0 & 0 & 0 & - \frac{1}{\Delta_1} \end{array}\right).
\ea
Then taking the first entry of the matrix $H_{4y}$ and expanding it to 6th order in the small parameter $t$ (assuming $g_1$, $g_2$, and $\tilde g$ are all proportional to $t$), we find the energy shift of state $\ket{1,0,0,0}$:
\ba
E_1 = \frac{g_1^2}{\Delta_1} - \frac{g_1^2 (g_1^2 - \tilde g^2)}{\Delta_1^3} + \frac{ 2 g_1^6  - 4 g_1^4 \tilde g^2 + g_1^2  \tilde g^4}{\Delta_1^5} - \frac{g_1^2  g_2^2 \tilde g^2}{\Delta_1^4 (\Delta_2-\Delta_1)}. 
\ea
Repeating the same calculation for the case where the single excitation is on transmon 2 (corresponding to state $\ket{0,1,0,0}$), we get the same answer except $g_1$ is replaced with $g_2$ and $\Delta_1$ is replaced with $\Delta_2$: 
\ba
E_2 = \frac{g_2^2}{\Delta_2} - \frac{g_2^2 (g_2^2 - \tilde g^2)}{\Delta_2^3} + \frac{ 2 g_2^6  - 4 g_2^4 \tilde g^2 + g_2^2  \tilde g^4}{\Delta_2^5} - \frac{g_1^2  g_2^2 \tilde g^2}{\Delta_2^4 (\Delta_1-\Delta_2)}.
\ea
It remains to derive the energy of the two-excitation state to 6th order in the small parameter. We have
\begin{align}
H_{0} &= \left(\begin{array}{cccccccccc} 0 & 0 & 0 & 0 & 0 & 0 & 0 & 0 & 0 & 0 \\ 0 & -\Delta_2 & 0 & 0 & 0 & 0 & 0 & 0 & 0 & 0 \\ 0 & 0 & - \Delta_2 & 0 & 0 & 0 & 0 & 0 & 0 & 0 \\ 0 & 0 & 0 & -\Delta_1 & 0 & 0 & 0 & 0 & 0 & 0 \\ 0 & 0 & 0 & 0 & -\Delta_1 & 0 & 0 & 0 & 0 & 0 \\ 0 & 0 & 0 & 0 & 0 & - (\Delta_1 + \Delta_2) & 0 & 0 & 0 & 0 \\ 0 & 0 & 0 & 0 & 0 & 0 & -(\Delta_1 + \Delta_2) & 0 & 0 & 0 \\ 0 & 0 & 0 & 0 & 0 & 0 & 0 & -(\Delta_1 + \Delta_2) & 0 & 0 \\ 0 & 0 & 0 & 0 & 0 & 0 & 0 & 0 & \Delta_1 - \Delta_2 - \eta_1 & 0 \\ 0 & 0 & 0 & 0 & 0 & 0 & 0 & 0 & 0 & \Delta_2 - \Delta_1 - \eta_2 \end{array}\right),\\
H_x &= \left(\begin{array}{cccccccccc} 0 & 0 & g_2 & g_1 & 0 & 0 & 0 & 0 & 0 & 0 \\ 0 & 0 & 0 & 0 & 0 & 0 & 0 & 0 & 0 & 0 \\ g_2 & 0 & 0 & 0 & 0 & 0 & 0 & 0 & 0 & 0  \\ g_1 & 0 & 0 & 0 & 0 & 0 & 0 & 0 & 0 & 0  \\ 0 & 0 & 0 & 0 & 0 & 0 & 0 & 0 & 0 & 0 \\ 0 & 0 & 0 & 0 & 0 & 0 & 0 & 0 & 0 & 0 \\ 0 & 0 & 0 & 0 & 0 & 0 & 0 & 0 & 0 & 0 \\ 0 & 0 & 0 & 0 & 0 & 0 & 0 & 0 & 0 & 0 \\ 0 & 0 & 0 & 0 & 0 & 0 & 0 & 0 & 0 & 0 \\ 0 & 0 & 0 & 0 & 0 & 0 & 0 & 0 & 0 & 0 \end{array}\right),\\
H_y &= \left(\begin{array}{cccccccccc} 0 & 0 & 0 & 0& 0 & 0 & 0 & 0 & 0 & 0 \\ 0 & 0 & \tilde g & 0 & 0 & 0 & \sqrt{2} g_1 & 0 & \sqrt{2} g_1 & 0 \\ 0 & \tilde g & 0 & 0 & 0 & g_1 & 0 & 0 & 0 & 0 \\ 0 & 0 & 0 & 0 & \tilde g & g_2 & 0 & 0 & 0 & 0 \\ 0 & 0 & 0 & \tilde g & 0 & 0 & 0 & \sqrt{2} g_2 & 0 & \sqrt{2} g_2 \\ 0 & 0 & g_1 & g_2 & 0 & 0 & \sqrt{2} \tilde g & \sqrt{2} \tilde g & 0 & 0 \\ 0 & \sqrt{2} g_1 & 0 & 0 & 0 & \sqrt{2} \tilde g & 0 & 0 & 0 & 0 \\ 0 & 0 & 0 & 0 & \sqrt{2} g_2 & \sqrt{2} \tilde g & 0 & 0 & 0 & 0 \\ 0 & \sqrt{2} g_1 & 0 & 0 & 0 & 0 & 0 & 0 & 0 & 0 \\ 0 & 0 & 0 & 0 & \sqrt{2} g_2 & 0 & 0 & 0 & 0 & 0 \end{array}\right),\\
\tilde H_{0} &= \left(\begin{array}{cccccccccc} 0 & 0 & 0 & 0 & 0 & 0 & 0 & 0 & 0 & 0 \\ 0 & -\frac{1}{\Delta_2} & 0 & 0 & 0 & 0 & 0 & 0 & 0 & 0 \\ 0 & 0 & - \frac{1}{\Delta_2} & 0 & 0 & 0 & 0 & 0 & 0 & 0 \\ 0 & 0 & 0 & -\frac{1}{\Delta_1} & 0 & 0 & 0 & 0 & 0 & 0 \\ 0 & 0 & 0 & 0 & -\frac{1}{\Delta_1} & 0 & 0 & 0 & 0 & 0 \\ 0 & 0 & 0 & 0 & 0 & - \frac{1}{\Delta_1 + \Delta_2} & 0 & 0 & 0 & 0 \\ 0 & 0 & 0 & 0 & 0 & 0 & -\frac{1}{\Delta_1 + \Delta_2} & 0 & 0 & 0 \\ 0 & 0 & 0 & 0 & 0 & 0 & 0 & -\frac{1}{\Delta_1 + \Delta_2} & 0 & 0 \\ 0 & 0 & 0 & 0 & 0 & 0 & 0 & 0 & \frac{1}{\Delta_1 - \Delta_2 - \eta_1} & 0 \\ 0 & 0 & 0 & 0 & 0 & 0 & 0 & 0 & 0 & \frac{1}{\Delta_2 - \Delta_1 - \eta_2} \end{array}\right).
\end{align}
Taking the first entry of the matrix $H_{4x}$ and expanding it to 6th order in the small parameter $t$ (assuming $g_1$, $g_2$, and $\tilde g$ are all proportional to $t$), we find $E_{12}$. Instead of giving the expression for $E_{12}$ itself, we give the expression for the strength of the resulting always-on ZZ interaction: 
\ba
J_{ZZ} = \frac{1}{4} (E_{12} - E_1 - E_2) =  -\frac{ (\Delta_1^4 \eta_1 +     \Delta_2^4 \eta_2 + (\Delta_1 + \Delta_2) (\Delta_1^2 +  \Delta_2^2) \eta_1 \eta_2) g_1^2 g_2^2 \tilde g^2}{2 \Delta_1^4 \Delta_2^4 (\Delta_1 - \Delta_2 - \eta_1) (\Delta_1 - \Delta_2 + \eta_2)}. \label{eq:zz6thorder}
\ea
As expected, the $g^2$ and $g^4$ terms vanished, and the lowest-order interaction is of order $g^6$, where the excitation needs to hop over from transmon 1 via the cavities to transmon 2 and back, which gives $g_1^2 g_2^2 \tilde g^2$. In the limit of infinite anharmonicity (i.e. the qubit limit), the interaction simplifies to 
\ba
J_{ZZ} = \frac{(\Delta_1 + \Delta_2) (\Delta_1^2 + \Delta_2^2) g_1^2 g_2^2 \tilde g^2}{2 \Delta_1^4 \Delta_2^4}.
\ea

{It is worth pointing out that Eq.\ (\ref{eq:zz6thorder}) is simply the expansion of Eq.\ (\ref{eq:jzz2cavities1}) to second order in $\tilde g$. In other words, the 6th-order perturbation theory was unnecessary for deriving Eq.\ (\ref{eq:zz6thorder}), since we were able to derive a 4th-order (in $g_1$ and $g_2$) result in Eq.\ (\ref{eq:jzz2cavities1}) correct to all orders in $\tilde g$. However, we chose to keep the 6th-order perturbation theory formalism since it is necessary for more complicated situations where such 4th-order shortcuts are not available.}

\baselineskip=12pt
\normalem
\bibliographystyle{apsrev4-2}

\begin{thebibliography}{120}%
\makeatletter
\providecommand \@ifxundefined [1]{%
 \@ifx{#1\undefined}
}%
\providecommand \@ifnum [1]{%
 \ifnum #1\expandafter \@firstoftwo
 \else \expandafter \@secondoftwo
 \fi
}%
\providecommand \@ifx [1]{%
 \ifx #1\expandafter \@firstoftwo
 \else \expandafter \@secondoftwo
 \fi
}%
\providecommand \natexlab [1]{#1}%
\providecommand \enquote  [1]{``#1''}%
\providecommand \bibnamefont  [1]{#1}%
\providecommand \bibfnamefont [1]{#1}%
\providecommand \citenamefont [1]{#1}%
\providecommand \href@noop [0]{\@secondoftwo}%
\providecommand \href [0]{\begingroup \@sanitize@url \@href}%
\providecommand \@href[1]{\@@startlink{#1}\@@href}%
\providecommand \@@href[1]{\endgroup#1\@@endlink}%
\providecommand \@sanitize@url [0]{\catcode `\\12\catcode `\$12\catcode
  `\&12\catcode `\#12\catcode `\^12\catcode `\_12\catcode `\%12\relax}%
\providecommand \@@startlink[1]{}%
\providecommand \@@endlink[0]{}%
\providecommand \url  [0]{\begingroup\@sanitize@url \@url }%
\providecommand \@url [1]{\endgroup\@href {#1}{\urlprefix }}%
\providecommand \urlprefix  [0]{URL }%
\providecommand \Eprint [0]{\href }%
\providecommand \doibase [0]{https://doi.org/}%
\providecommand \selectlanguage [0]{\@gobble}%
\providecommand \bibinfo  [0]{\@secondoftwo}%
\providecommand \bibfield  [0]{\@secondoftwo}%
\providecommand \translation [1]{[#1]}%
\providecommand \BibitemOpen [0]{}%
\providecommand \bibitemStop [0]{}%
\providecommand \bibitemNoStop [0]{.\EOS\space}%
\providecommand \EOS [0]{\spacefactor3000\relax}%
\providecommand \BibitemShut  [1]{\csname bibitem#1\endcsname}%
\let\auto@bib@innerbib\@empty
\bibitem [{\citenamefont {Krantz}\ \emph {et~al.}(2019)\citenamefont {Krantz},
  \citenamefont {Kjaergaard}, \citenamefont {Yan}, \citenamefont {Orlando},
  \citenamefont {Gustavsson},\ and\ \citenamefont {Oliver}}]{krantz19}%
  \BibitemOpen
  \bibfield  {author} {\bibinfo {author} {\bibfnamefont {P.}~\bibnamefont
  {Krantz}}, \bibinfo {author} {\bibfnamefont {M.}~\bibnamefont {Kjaergaard}},
  \bibinfo {author} {\bibfnamefont {F.}~\bibnamefont {Yan}}, \bibinfo {author}
  {\bibfnamefont {T.~P.}\ \bibnamefont {Orlando}}, \bibinfo {author}
  {\bibfnamefont {S.}~\bibnamefont {Gustavsson}},\ and\ \bibinfo {author}
  {\bibfnamefont {W.~D.}\ \bibnamefont {Oliver}},\ }\href@noop {} {\bibfield
  {journal} {\bibinfo  {journal} {Appl. Phys. Rev.}\ }\textbf {\bibinfo
  {volume} {6}},\ \bibinfo {pages} {021318} (\bibinfo {year}
  {2019})}\BibitemShut {NoStop}%
\bibitem [{\citenamefont {Kjaergaard}\ \emph {et~al.}(2020)\citenamefont
  {Kjaergaard}, \citenamefont {Schwartz}, \citenamefont {Braum{\"u}ller},
  \citenamefont {Krantz}, \citenamefont {Wang}, \citenamefont {Gustavsson},\
  and\ \citenamefont {Oliver}}]{kjaergaard20}%
  \BibitemOpen
  \bibfield  {author} {\bibinfo {author} {\bibfnamefont {M.}~\bibnamefont
  {Kjaergaard}}, \bibinfo {author} {\bibfnamefont {M.~E.}\ \bibnamefont
  {Schwartz}}, \bibinfo {author} {\bibfnamefont {J.}~\bibnamefont
  {Braum{\"u}ller}}, \bibinfo {author} {\bibfnamefont {P.}~\bibnamefont
  {Krantz}}, \bibinfo {author} {\bibfnamefont {J.~I.~J.}\ \bibnamefont {Wang}},
  \bibinfo {author} {\bibfnamefont {S.}~\bibnamefont {Gustavsson}},\ and\
  \bibinfo {author} {\bibfnamefont {W.~D.}\ \bibnamefont {Oliver}},\
  }\href@noop {} {\bibfield  {journal} {\bibinfo  {journal} {Annu. Rev.
  Condens. Matter Phys.}\ }\textbf {\bibinfo {volume} {11}},\ \bibinfo {pages}
  {369} (\bibinfo {year} {2020})}\BibitemShut {NoStop}%
\bibitem [{\citenamefont {Blais}\ \emph {et~al.}(2021)\citenamefont {Blais},
  \citenamefont {Grimsmo}, \citenamefont {Girvin},\ and\ \citenamefont
  {Wallraff}}]{blais21}%
  \BibitemOpen
  \bibfield  {author} {\bibinfo {author} {\bibfnamefont {A.}~\bibnamefont
  {Blais}}, \bibinfo {author} {\bibfnamefont {A.~L.}\ \bibnamefont {Grimsmo}},
  \bibinfo {author} {\bibfnamefont {S.~M.}\ \bibnamefont {Girvin}},\ and\
  \bibinfo {author} {\bibfnamefont {A.}~\bibnamefont {Wallraff}},\ }\href@noop
  {} {\bibfield  {journal} {\bibinfo  {journal} {Rev. Mod. Phys.}\ }\textbf
  {\bibinfo {volume} {93}},\ \bibinfo {pages} {025005} (\bibinfo {year}
  {2021})}\BibitemShut {NoStop}%
\bibitem [{\citenamefont {Blais}\ \emph {et~al.}(2003)\citenamefont {Blais},
  \citenamefont {Van Den~Brink},\ and\ \citenamefont {Zagoskin}}]{blais03}%
  \BibitemOpen
  \bibfield  {author} {\bibinfo {author} {\bibfnamefont {A.}~\bibnamefont
  {Blais}}, \bibinfo {author} {\bibfnamefont {A.~M.}\ \bibnamefont {Van
  Den~Brink}},\ and\ \bibinfo {author} {\bibfnamefont {A.~M.}\ \bibnamefont
  {Zagoskin}},\ }\href@noop {} {\bibfield  {journal} {\bibinfo  {journal}
  {Phys. Rev. Lett.}\ }\textbf {\bibinfo {volume} {90}},\ \bibinfo {pages}
  {127901} (\bibinfo {year} {2003})}\BibitemShut {NoStop}%
\bibitem [{\citenamefont {Bialczak}\ \emph {et~al.}(2010)\citenamefont
  {Bialczak}, \citenamefont {Ansmann}, \citenamefont {Hofheinz}, \citenamefont
  {Lucero}, \citenamefont {Neeley}, \citenamefont {O'Connell}, \citenamefont
  {Sank}, \citenamefont {Wang}, \citenamefont {Wenner}, \citenamefont
  {Steffen}, \citenamefont {Cleland},\ and\ \citenamefont
  {Martinis}}]{bialczak10}%
  \BibitemOpen
  \bibfield  {author} {\bibinfo {author} {\bibfnamefont {R.~C.}\ \bibnamefont
  {Bialczak}}, \bibinfo {author} {\bibfnamefont {M.}~\bibnamefont {Ansmann}},
  \bibinfo {author} {\bibfnamefont {M.}~\bibnamefont {Hofheinz}}, \bibinfo
  {author} {\bibfnamefont {E.}~\bibnamefont {Lucero}}, \bibinfo {author}
  {\bibfnamefont {M.}~\bibnamefont {Neeley}}, \bibinfo {author} {\bibfnamefont
  {A.~D.}\ \bibnamefont {O'Connell}}, \bibinfo {author} {\bibfnamefont
  {D.}~\bibnamefont {Sank}}, \bibinfo {author} {\bibfnamefont {H.}~\bibnamefont
  {Wang}}, \bibinfo {author} {\bibfnamefont {J.}~\bibnamefont {Wenner}},
  \bibinfo {author} {\bibfnamefont {M.}~\bibnamefont {Steffen}}, \bibinfo
  {author} {\bibfnamefont {A.~N.}\ \bibnamefont {Cleland}},\ and\ \bibinfo
  {author} {\bibfnamefont {J.~M.}\ \bibnamefont {Martinis}},\ }\href@noop {}
  {\bibfield  {journal} {\bibinfo  {journal} {Nat. Phys.}\ }\textbf {\bibinfo
  {volume} {6}},\ \bibinfo {pages} {409} (\bibinfo {year} {2010})}\BibitemShut
  {NoStop}%
\bibitem [{\citenamefont {Dewes}\ \emph {et~al.}(2012)\citenamefont {Dewes},
  \citenamefont {Ong}, \citenamefont {Schmitt}, \citenamefont {Lauro},
  \citenamefont {Boulant}, \citenamefont {Bertet}, \citenamefont {Vion},\ and\
  \citenamefont {Esteve}}]{dewes12}%
  \BibitemOpen
  \bibfield  {author} {\bibinfo {author} {\bibfnamefont {A.}~\bibnamefont
  {Dewes}}, \bibinfo {author} {\bibfnamefont {F.~R.}\ \bibnamefont {Ong}},
  \bibinfo {author} {\bibfnamefont {V.}~\bibnamefont {Schmitt}}, \bibinfo
  {author} {\bibfnamefont {R.}~\bibnamefont {Lauro}}, \bibinfo {author}
  {\bibfnamefont {N.}~\bibnamefont {Boulant}}, \bibinfo {author} {\bibfnamefont
  {P.}~\bibnamefont {Bertet}}, \bibinfo {author} {\bibfnamefont
  {D.}~\bibnamefont {Vion}},\ and\ \bibinfo {author} {\bibfnamefont
  {D.}~\bibnamefont {Esteve}},\ }\href@noop {} {\bibfield  {journal} {\bibinfo
  {journal} {Phys. Rev. Lett.}\ }\textbf {\bibinfo {volume} {108}},\ \bibinfo
  {pages} {057002} (\bibinfo {year} {2012})}\BibitemShut {NoStop}%
\bibitem [{\citenamefont {Barends}\ \emph {et~al.}(2014)\citenamefont
  {Barends}, \citenamefont {Kelly}, \citenamefont {Megrant}, \citenamefont
  {Veitia}, \citenamefont {Sank}, \citenamefont {Jeffrey}, \citenamefont
  {White}, \citenamefont {Mutus}, \citenamefont {Fowler}, \citenamefont
  {Campbell}, \citenamefont {Chen}, \citenamefont {Chen}, \citenamefont
  {Chiaro}, \citenamefont {Dunsworth}, \citenamefont {Neill}, \citenamefont
  {O'Malley}, \citenamefont {Roushan}, \citenamefont {Vainsencher},
  \citenamefont {Wenner}, \citenamefont {Korotkov}, \citenamefont {Cleland},\
  and\ \citenamefont {Martinis}}]{barends14}%
  \BibitemOpen
  \bibfield  {author} {\bibinfo {author} {\bibfnamefont {R.}~\bibnamefont
  {Barends}}, \bibinfo {author} {\bibfnamefont {J.}~\bibnamefont {Kelly}},
  \bibinfo {author} {\bibfnamefont {A.}~\bibnamefont {Megrant}}, \bibinfo
  {author} {\bibfnamefont {A.}~\bibnamefont {Veitia}}, \bibinfo {author}
  {\bibfnamefont {D.}~\bibnamefont {Sank}}, \bibinfo {author} {\bibfnamefont
  {E.}~\bibnamefont {Jeffrey}}, \bibinfo {author} {\bibfnamefont {T.~C.}\
  \bibnamefont {White}}, \bibinfo {author} {\bibfnamefont {J.}~\bibnamefont
  {Mutus}}, \bibinfo {author} {\bibfnamefont {A.~G.}\ \bibnamefont {Fowler}},
  \bibinfo {author} {\bibfnamefont {B.}~\bibnamefont {Campbell}}, \bibinfo
  {author} {\bibfnamefont {Y.}~\bibnamefont {Chen}}, \bibinfo {author}
  {\bibfnamefont {Z.}~\bibnamefont {Chen}}, \bibinfo {author} {\bibfnamefont
  {B.}~\bibnamefont {Chiaro}}, \bibinfo {author} {\bibfnamefont
  {A.}~\bibnamefont {Dunsworth}}, \bibinfo {author} {\bibfnamefont
  {C.}~\bibnamefont {Neill}}, \bibinfo {author} {\bibfnamefont
  {P.}~\bibnamefont {O'Malley}}, \bibinfo {author} {\bibfnamefont
  {P.}~\bibnamefont {Roushan}}, \bibinfo {author} {\bibfnamefont
  {A.}~\bibnamefont {Vainsencher}}, \bibinfo {author} {\bibfnamefont
  {J.}~\bibnamefont {Wenner}}, \bibinfo {author} {\bibfnamefont {A.~N.}\
  \bibnamefont {Korotkov}}, \bibinfo {author} {\bibfnamefont {A.~N.}\
  \bibnamefont {Cleland}},\ and\ \bibinfo {author} {\bibfnamefont {J.~M.}\
  \bibnamefont {Martinis}},\ }\href@noop {} {\bibfield  {journal} {\bibinfo
  {journal} {Nature (London)}\ }\textbf {\bibinfo {volume} {508}},\ \bibinfo
  {pages} {500} (\bibinfo {year} {2014})}\BibitemShut {NoStop}%
\bibitem [{\citenamefont {Chen}\ \emph {et~al.}(2014)\citenamefont {Chen},
  \citenamefont {Neill}, \citenamefont {Roushan}, \citenamefont {Leung},
  \citenamefont {Fang}, \citenamefont {Barends}, \citenamefont {Kelly},
  \citenamefont {Campbell}, \citenamefont {Chen}, \citenamefont {Chiaro},
  \citenamefont {Dunsworth}, \citenamefont {Jeffrey}, \citenamefont {Megrant},
  \citenamefont {Mutus}, \citenamefont {O'Malley}, \citenamefont {Quintana},
  \citenamefont {Sank}, \citenamefont {Vainsencher}, \citenamefont {Wenner},
  \citenamefont {White}, \citenamefont {Geller}, \citenamefont {Cleland},\ and\
  \citenamefont {Martinis}}]{chen14}%
  \BibitemOpen
  \bibfield  {author} {\bibinfo {author} {\bibfnamefont {Y.}~\bibnamefont
  {Chen}}, \bibinfo {author} {\bibfnamefont {C.}~\bibnamefont {Neill}},
  \bibinfo {author} {\bibfnamefont {P.}~\bibnamefont {Roushan}}, \bibinfo
  {author} {\bibfnamefont {N.}~\bibnamefont {Leung}}, \bibinfo {author}
  {\bibfnamefont {M.}~\bibnamefont {Fang}}, \bibinfo {author} {\bibfnamefont
  {R.}~\bibnamefont {Barends}}, \bibinfo {author} {\bibfnamefont
  {J.}~\bibnamefont {Kelly}}, \bibinfo {author} {\bibfnamefont
  {B.}~\bibnamefont {Campbell}}, \bibinfo {author} {\bibfnamefont
  {Z.}~\bibnamefont {Chen}}, \bibinfo {author} {\bibfnamefont {B.}~\bibnamefont
  {Chiaro}}, \bibinfo {author} {\bibfnamefont {A.}~\bibnamefont {Dunsworth}},
  \bibinfo {author} {\bibfnamefont {E.}~\bibnamefont {Jeffrey}}, \bibinfo
  {author} {\bibfnamefont {A.}~\bibnamefont {Megrant}}, \bibinfo {author}
  {\bibfnamefont {J.~Y.}\ \bibnamefont {Mutus}}, \bibinfo {author}
  {\bibfnamefont {P.~J.~J.}\ \bibnamefont {O'Malley}}, \bibinfo {author}
  {\bibfnamefont {C.~M.}\ \bibnamefont {Quintana}}, \bibinfo {author}
  {\bibfnamefont {D.}~\bibnamefont {Sank}}, \bibinfo {author} {\bibfnamefont
  {A.}~\bibnamefont {Vainsencher}}, \bibinfo {author} {\bibfnamefont
  {J.}~\bibnamefont {Wenner}}, \bibinfo {author} {\bibfnamefont {T.~C.}\
  \bibnamefont {White}}, \bibinfo {author} {\bibfnamefont {M.~R.}\ \bibnamefont
  {Geller}}, \bibinfo {author} {\bibfnamefont {A.~N.}\ \bibnamefont
  {Cleland}},\ and\ \bibinfo {author} {\bibfnamefont {J.~M.}\ \bibnamefont
  {Martinis}},\ }\href@noop {} {\bibfield  {journal} {\bibinfo  {journal}
  {Phys. Rev. Lett.}\ }\textbf {\bibinfo {volume} {113}},\ \bibinfo {pages}
  {220502} (\bibinfo {year} {2014})}\BibitemShut {NoStop}%
\bibitem [{\citenamefont {Yan}\ \emph {et~al.}(2018)\citenamefont {Yan},
  \citenamefont {Krantz}, \citenamefont {Sung}, \citenamefont {Kjaergaard},
  \citenamefont {Campbell}, \citenamefont {Wang}, \citenamefont {Orlando},
  \citenamefont {Gustavsson},\ and\ \citenamefont {Oliver}}]{yan18}%
  \BibitemOpen
  \bibfield  {author} {\bibinfo {author} {\bibfnamefont {F.}~\bibnamefont
  {Yan}}, \bibinfo {author} {\bibfnamefont {P.}~\bibnamefont {Krantz}},
  \bibinfo {author} {\bibfnamefont {Y.}~\bibnamefont {Sung}}, \bibinfo {author}
  {\bibfnamefont {M.}~\bibnamefont {Kjaergaard}}, \bibinfo {author}
  {\bibfnamefont {D.}~\bibnamefont {Campbell}}, \bibinfo {author}
  {\bibfnamefont {J.~I.~J.}\ \bibnamefont {Wang}}, \bibinfo {author}
  {\bibfnamefont {T.~P.}\ \bibnamefont {Orlando}}, \bibinfo {author}
  {\bibfnamefont {S.}~\bibnamefont {Gustavsson}},\ and\ \bibinfo {author}
  {\bibfnamefont {W.~D.}\ \bibnamefont {Oliver}},\ }\href@noop {} {\bibfield
  {journal} {\bibinfo  {journal} {Phys. Rev. Appl.}\ }\textbf {\bibinfo
  {volume} {10}},\ \bibinfo {pages} {054062} (\bibinfo {year}
  {2018})}\BibitemShut {NoStop}%
\bibitem [{\citenamefont {Arute}\ \emph {et~al.}(2019)\citenamefont {Arute},
  \citenamefont {Arya}, \citenamefont {Babbush}, \citenamefont {Bacon},
  \citenamefont {Bardin}, \citenamefont {Barends}, \citenamefont {Biswas},
  \citenamefont {Boixo}, \citenamefont {Brandao}, \citenamefont {Buell},
  \citenamefont {Burkett}, \citenamefont {Chen}, \citenamefont {Chen},
  \citenamefont {Chiaro}, \citenamefont {Collins}, \citenamefont {Courtney},
  \citenamefont {Dunsworth}, \citenamefont {Farhi}, \citenamefont {Foxen},
  \citenamefont {Fowler}, \citenamefont {Gidney}, \citenamefont {Giustina},
  \citenamefont {Graff}, \citenamefont {Guerin}, \citenamefont {Habegger},
  \citenamefont {Harrigan}, \citenamefont {Hartmann}, \citenamefont {Ho},
  \citenamefont {Hoffmann}, \citenamefont {Huang}, \citenamefont {Humble},
  \citenamefont {Isakov}, \citenamefont {Jeffrey}, \citenamefont {Jiang},
  \citenamefont {Kafri}, \citenamefont {Kechedzhi}, \citenamefont {Kelly},
  \citenamefont {Klimov}, \citenamefont {Knysh}, \citenamefont {Korotkov},
  \citenamefont {Kostritsa}, \citenamefont {Landhuis}, \citenamefont
  {Lindmark}, \citenamefont {Lucero}, \citenamefont {Lyakh}, \citenamefont
  {Mandr{\`a}}, \citenamefont {McClean}, \citenamefont {McEwen}, \citenamefont
  {Megrant}, \citenamefont {Mi}, \citenamefont {Michielsen}, \citenamefont
  {Mohseni}, \citenamefont {Mutus}, \citenamefont {Naaman}, \citenamefont
  {Neeley}, \citenamefont {Neill}, \citenamefont {Niu}, \citenamefont {Ostby},
  \citenamefont {Petukhov}, \citenamefont {Platt}, \citenamefont {Quintana},
  \citenamefont {Rieffel}, \citenamefont {Roushan}, \citenamefont {Rubin},
  \citenamefont {Sank}, \citenamefont {Satzinger}, \citenamefont {Smelyanskiy},
  \citenamefont {Sung}, \citenamefont {Trevithick}, \citenamefont
  {Vainsencher}, \citenamefont {Villalonga}, \citenamefont {White},
  \citenamefont {Yao}, \citenamefont {Yeh}, \citenamefont {Zalcman},
  \citenamefont {Neven},\ and\ \citenamefont {Martinis}}]{arute19}%
  \BibitemOpen
  \bibfield  {author} {\bibinfo {author} {\bibfnamefont {F.}~\bibnamefont
  {Arute}}, \bibinfo {author} {\bibfnamefont {K.}~\bibnamefont {Arya}},
  \bibinfo {author} {\bibfnamefont {R.}~\bibnamefont {Babbush}}, \bibinfo
  {author} {\bibfnamefont {D.}~\bibnamefont {Bacon}}, \bibinfo {author}
  {\bibfnamefont {J.~C.}\ \bibnamefont {Bardin}}, \bibinfo {author}
  {\bibfnamefont {R.}~\bibnamefont {Barends}}, \bibinfo {author} {\bibfnamefont
  {R.}~\bibnamefont {Biswas}}, \bibinfo {author} {\bibfnamefont
  {S.}~\bibnamefont {Boixo}}, \bibinfo {author} {\bibfnamefont {F.~G. S.~L.}\
  \bibnamefont {Brandao}}, \bibinfo {author} {\bibfnamefont {D.~A.}\
  \bibnamefont {Buell}}, \bibinfo {author} {\bibfnamefont {B.}~\bibnamefont
  {Burkett}}, \bibinfo {author} {\bibfnamefont {Y.}~\bibnamefont {Chen}},
  \bibinfo {author} {\bibfnamefont {Z.}~\bibnamefont {Chen}}, \bibinfo {author}
  {\bibfnamefont {B.}~\bibnamefont {Chiaro}}, \bibinfo {author} {\bibfnamefont
  {R.}~\bibnamefont {Collins}}, \bibinfo {author} {\bibfnamefont
  {W.}~\bibnamefont {Courtney}}, \bibinfo {author} {\bibfnamefont
  {A.}~\bibnamefont {Dunsworth}}, \bibinfo {author} {\bibfnamefont
  {E.}~\bibnamefont {Farhi}}, \bibinfo {author} {\bibfnamefont
  {B.}~\bibnamefont {Foxen}}, \bibinfo {author} {\bibfnamefont
  {A.}~\bibnamefont {Fowler}}, \bibinfo {author} {\bibfnamefont
  {C.}~\bibnamefont {Gidney}}, \bibinfo {author} {\bibfnamefont
  {M.}~\bibnamefont {Giustina}}, \bibinfo {author} {\bibfnamefont
  {R.}~\bibnamefont {Graff}}, \bibinfo {author} {\bibfnamefont
  {K.}~\bibnamefont {Guerin}}, \bibinfo {author} {\bibfnamefont
  {S.}~\bibnamefont {Habegger}}, \bibinfo {author} {\bibfnamefont {M.~P.}\
  \bibnamefont {Harrigan}}, \bibinfo {author} {\bibfnamefont {M.~J.}\
  \bibnamefont {Hartmann}}, \bibinfo {author} {\bibfnamefont {A.}~\bibnamefont
  {Ho}}, \bibinfo {author} {\bibfnamefont {M.}~\bibnamefont {Hoffmann}},
  \bibinfo {author} {\bibfnamefont {T.}~\bibnamefont {Huang}}, \bibinfo
  {author} {\bibfnamefont {T.~S.}\ \bibnamefont {Humble}}, \bibinfo {author}
  {\bibfnamefont {S.~V.}\ \bibnamefont {Isakov}}, \bibinfo {author}
  {\bibfnamefont {E.}~\bibnamefont {Jeffrey}}, \bibinfo {author} {\bibfnamefont
  {Z.}~\bibnamefont {Jiang}}, \bibinfo {author} {\bibfnamefont
  {D.}~\bibnamefont {Kafri}}, \bibinfo {author} {\bibfnamefont
  {K.}~\bibnamefont {Kechedzhi}}, \bibinfo {author} {\bibfnamefont
  {J.}~\bibnamefont {Kelly}}, \bibinfo {author} {\bibfnamefont {P.~V.}\
  \bibnamefont {Klimov}}, \bibinfo {author} {\bibfnamefont {S.}~\bibnamefont
  {Knysh}}, \bibinfo {author} {\bibfnamefont {A.}~\bibnamefont {Korotkov}},
  \bibinfo {author} {\bibfnamefont {F.}~\bibnamefont {Kostritsa}}, \bibinfo
  {author} {\bibfnamefont {D.}~\bibnamefont {Landhuis}}, \bibinfo {author}
  {\bibfnamefont {M.}~\bibnamefont {Lindmark}}, \bibinfo {author}
  {\bibfnamefont {E.}~\bibnamefont {Lucero}}, \bibinfo {author} {\bibfnamefont
  {D.}~\bibnamefont {Lyakh}}, \bibinfo {author} {\bibfnamefont
  {S.}~\bibnamefont {Mandr{\`a}}}, \bibinfo {author} {\bibfnamefont {J.~R.}\
  \bibnamefont {McClean}}, \bibinfo {author} {\bibfnamefont {M.}~\bibnamefont
  {McEwen}}, \bibinfo {author} {\bibfnamefont {A.}~\bibnamefont {Megrant}},
  \bibinfo {author} {\bibfnamefont {X.}~\bibnamefont {Mi}}, \bibinfo {author}
  {\bibfnamefont {K.}~\bibnamefont {Michielsen}}, \bibinfo {author}
  {\bibfnamefont {M.}~\bibnamefont {Mohseni}}, \bibinfo {author} {\bibfnamefont
  {J.}~\bibnamefont {Mutus}}, \bibinfo {author} {\bibfnamefont
  {O.}~\bibnamefont {Naaman}}, \bibinfo {author} {\bibfnamefont
  {M.}~\bibnamefont {Neeley}}, \bibinfo {author} {\bibfnamefont
  {C.}~\bibnamefont {Neill}}, \bibinfo {author} {\bibfnamefont {M.~Y.}\
  \bibnamefont {Niu}}, \bibinfo {author} {\bibfnamefont {E.}~\bibnamefont
  {Ostby}}, \bibinfo {author} {\bibfnamefont {A.}~\bibnamefont {Petukhov}},
  \bibinfo {author} {\bibfnamefont {J.~C.}\ \bibnamefont {Platt}}, \bibinfo
  {author} {\bibfnamefont {C.}~\bibnamefont {Quintana}}, \bibinfo {author}
  {\bibfnamefont {E.~G.}\ \bibnamefont {Rieffel}}, \bibinfo {author}
  {\bibfnamefont {P.}~\bibnamefont {Roushan}}, \bibinfo {author} {\bibfnamefont
  {N.~C.}\ \bibnamefont {Rubin}}, \bibinfo {author} {\bibfnamefont
  {D.}~\bibnamefont {Sank}}, \bibinfo {author} {\bibfnamefont {K.~J.}\
  \bibnamefont {Satzinger}}, \bibinfo {author} {\bibfnamefont {V.}~\bibnamefont
  {Smelyanskiy}}, \bibinfo {author} {\bibfnamefont {K.~J.}\ \bibnamefont
  {Sung}}, \bibinfo {author} {\bibfnamefont {M.~D.}\ \bibnamefont
  {Trevithick}}, \bibinfo {author} {\bibfnamefont {A.}~\bibnamefont
  {Vainsencher}}, \bibinfo {author} {\bibfnamefont {B.}~\bibnamefont
  {Villalonga}}, \bibinfo {author} {\bibfnamefont {T.}~\bibnamefont {White}},
  \bibinfo {author} {\bibfnamefont {Z.~J.}\ \bibnamefont {Yao}}, \bibinfo
  {author} {\bibfnamefont {P.}~\bibnamefont {Yeh}}, \bibinfo {author}
  {\bibfnamefont {A.}~\bibnamefont {Zalcman}}, \bibinfo {author} {\bibfnamefont
  {H.}~\bibnamefont {Neven}},\ and\ \bibinfo {author} {\bibfnamefont {J.~M.}\
  \bibnamefont {Martinis}},\ }\href@noop {} {\bibfield  {journal} {\bibinfo
  {journal} {Nature (London)}\ }\textbf {\bibinfo {volume} {574}},\ \bibinfo
  {pages} {505} (\bibinfo {year} {2019})}\BibitemShut {NoStop}%
\bibitem [{\citenamefont {Majer}\ \emph {et~al.}(2007)\citenamefont {Majer},
  \citenamefont {Chow}, \citenamefont {Gambetta}, \citenamefont {Koch},
  \citenamefont {Johnson}, \citenamefont {Schreier}, \citenamefont {Frunzio},
  \citenamefont {Schuster}, \citenamefont {Houck}, \citenamefont {Wallraff},
  \citenamefont {Blais}, \citenamefont {Devoret}, \citenamefont {Girvin},\ and\
  \citenamefont {Schoelkopf}}]{majer07}%
  \BibitemOpen
  \bibfield  {author} {\bibinfo {author} {\bibfnamefont {J.}~\bibnamefont
  {Majer}}, \bibinfo {author} {\bibfnamefont {J.~M.}\ \bibnamefont {Chow}},
  \bibinfo {author} {\bibfnamefont {J.~M.}\ \bibnamefont {Gambetta}}, \bibinfo
  {author} {\bibfnamefont {J.}~\bibnamefont {Koch}}, \bibinfo {author}
  {\bibfnamefont {B.~R.}\ \bibnamefont {Johnson}}, \bibinfo {author}
  {\bibfnamefont {J.~A.}\ \bibnamefont {Schreier}}, \bibinfo {author}
  {\bibfnamefont {L.}~\bibnamefont {Frunzio}}, \bibinfo {author} {\bibfnamefont
  {D.~I.}\ \bibnamefont {Schuster}}, \bibinfo {author} {\bibfnamefont {A.~A.}\
  \bibnamefont {Houck}}, \bibinfo {author} {\bibfnamefont {A.}~\bibnamefont
  {Wallraff}}, \bibinfo {author} {\bibfnamefont {A.}~\bibnamefont {Blais}},
  \bibinfo {author} {\bibfnamefont {M.~H.}\ \bibnamefont {Devoret}}, \bibinfo
  {author} {\bibfnamefont {S.~M.}\ \bibnamefont {Girvin}},\ and\ \bibinfo
  {author} {\bibfnamefont {R.~J.}\ \bibnamefont {Schoelkopf}},\ }\href@noop {}
  {\bibfield  {journal} {\bibinfo  {journal} {Nature (London)}\ }\textbf
  {\bibinfo {volume} {449}},\ \bibinfo {pages} {443} (\bibinfo {year}
  {2007})}\BibitemShut {NoStop}%
\bibitem [{\citenamefont {Gambetta}\ \emph {et~al.}(2011)\citenamefont
  {Gambetta}, \citenamefont {Houck},\ and\ \citenamefont {Blais}}]{gambetta11}%
  \BibitemOpen
  \bibfield  {author} {\bibinfo {author} {\bibfnamefont {J.~M.}\ \bibnamefont
  {Gambetta}}, \bibinfo {author} {\bibfnamefont {A.~A.}\ \bibnamefont
  {Houck}},\ and\ \bibinfo {author} {\bibfnamefont {A.}~\bibnamefont {Blais}},\
  }\href@noop {} {\bibfield  {journal} {\bibinfo  {journal} {Phys. Rev. Lett.}\
  }\textbf {\bibinfo {volume} {106}},\ \bibinfo {pages} {030502} (\bibinfo
  {year} {2011})}\BibitemShut {NoStop}%
\bibitem [{\citenamefont {Srinivasan}\ \emph {et~al.}(2011)\citenamefont
  {Srinivasan}, \citenamefont {Hoffman}, \citenamefont {Gambetta},\ and\
  \citenamefont {Houck}}]{srinivasan11}%
  \BibitemOpen
  \bibfield  {author} {\bibinfo {author} {\bibfnamefont {S.~J.}\ \bibnamefont
  {Srinivasan}}, \bibinfo {author} {\bibfnamefont {A.~J.}\ \bibnamefont
  {Hoffman}}, \bibinfo {author} {\bibfnamefont {J.~M.}\ \bibnamefont
  {Gambetta}},\ and\ \bibinfo {author} {\bibfnamefont {A.~A.}\ \bibnamefont
  {Houck}},\ }\href@noop {} {\bibfield  {journal} {\bibinfo  {journal} {Phys.
  Rev. Lett.}\ }\textbf {\bibinfo {volume} {106}},\ \bibinfo {pages} {083601}
  (\bibinfo {year} {2011})}\BibitemShut {NoStop}%
\bibitem [{\citenamefont {DiCarlo}\ \emph {et~al.}(2009)\citenamefont
  {DiCarlo}, \citenamefont {Chow}, \citenamefont {Gambetta}, \citenamefont
  {Bishop}, \citenamefont {Johnson}, \citenamefont {Schuster}, \citenamefont
  {Majer}, \citenamefont {Blais}, \citenamefont {Frunzio}, \citenamefont
  {Girvin},\ and\ \citenamefont {Schoelkopf}}]{dicarlo09}%
  \BibitemOpen
  \bibfield  {author} {\bibinfo {author} {\bibfnamefont {L.}~\bibnamefont
  {DiCarlo}}, \bibinfo {author} {\bibfnamefont {J.~M.}\ \bibnamefont {Chow}},
  \bibinfo {author} {\bibfnamefont {J.~M.}\ \bibnamefont {Gambetta}}, \bibinfo
  {author} {\bibfnamefont {L.~S.}\ \bibnamefont {Bishop}}, \bibinfo {author}
  {\bibfnamefont {B.~R.}\ \bibnamefont {Johnson}}, \bibinfo {author}
  {\bibfnamefont {D.~I.}\ \bibnamefont {Schuster}}, \bibinfo {author}
  {\bibfnamefont {J.}~\bibnamefont {Majer}}, \bibinfo {author} {\bibfnamefont
  {A.}~\bibnamefont {Blais}}, \bibinfo {author} {\bibfnamefont
  {L.}~\bibnamefont {Frunzio}}, \bibinfo {author} {\bibfnamefont {S.~M.}\
  \bibnamefont {Girvin}},\ and\ \bibinfo {author} {\bibfnamefont {R.~J.}\
  \bibnamefont {Schoelkopf}},\ }\href@noop {} {\bibfield  {journal} {\bibinfo
  {journal} {Nature (London)}\ }\textbf {\bibinfo {volume} {460}},\ \bibinfo
  {pages} {240} (\bibinfo {year} {2009})}\BibitemShut {NoStop}%
\bibitem [{\citenamefont {Strauch}\ \emph {et~al.}(2003)\citenamefont
  {Strauch}, \citenamefont {Johnson}, \citenamefont {Dragt}, \citenamefont
  {Lobb}, \citenamefont {Anderson},\ and\ \citenamefont
  {Wellstood}}]{strauch03}%
  \BibitemOpen
  \bibfield  {author} {\bibinfo {author} {\bibfnamefont {F.~W.}\ \bibnamefont
  {Strauch}}, \bibinfo {author} {\bibfnamefont {P.~R.}\ \bibnamefont
  {Johnson}}, \bibinfo {author} {\bibfnamefont {A.~J.}\ \bibnamefont {Dragt}},
  \bibinfo {author} {\bibfnamefont {C.~J.}\ \bibnamefont {Lobb}}, \bibinfo
  {author} {\bibfnamefont {J.~R.}\ \bibnamefont {Anderson}},\ and\ \bibinfo
  {author} {\bibfnamefont {F.~C.}\ \bibnamefont {Wellstood}},\ }\href@noop {}
  {\bibfield  {journal} {\bibinfo  {journal} {Phys. Rev. Lett.}\ }\textbf
  {\bibinfo {volume} {91}},\ \bibinfo {pages} {167005} (\bibinfo {year}
  {2003})}\BibitemShut {NoStop}%
\bibitem [{\citenamefont {DiCarlo}\ \emph {et~al.}(2010)\citenamefont
  {DiCarlo}, \citenamefont {Reed}, \citenamefont {Sun}, \citenamefont
  {Johnson}, \citenamefont {Chow}, \citenamefont {Gambetta}, \citenamefont
  {Frunzio}, \citenamefont {Girvin}, \citenamefont {Devoret},\ and\
  \citenamefont {Schoelkopf}}]{dicarlo10}%
  \BibitemOpen
  \bibfield  {author} {\bibinfo {author} {\bibfnamefont {L.}~\bibnamefont
  {DiCarlo}}, \bibinfo {author} {\bibfnamefont {M.~D.}\ \bibnamefont {Reed}},
  \bibinfo {author} {\bibfnamefont {L.}~\bibnamefont {Sun}}, \bibinfo {author}
  {\bibfnamefont {B.~R.}\ \bibnamefont {Johnson}}, \bibinfo {author}
  {\bibfnamefont {J.~M.}\ \bibnamefont {Chow}}, \bibinfo {author}
  {\bibfnamefont {J.~M.}\ \bibnamefont {Gambetta}}, \bibinfo {author}
  {\bibfnamefont {L.}~\bibnamefont {Frunzio}}, \bibinfo {author} {\bibfnamefont
  {S.~M.}\ \bibnamefont {Girvin}}, \bibinfo {author} {\bibfnamefont {M.~H.}\
  \bibnamefont {Devoret}},\ and\ \bibinfo {author} {\bibfnamefont {R.~J.}\
  \bibnamefont {Schoelkopf}},\ }\href@noop {} {\bibfield  {journal} {\bibinfo
  {journal} {Nature (London)}\ }\textbf {\bibinfo {volume} {467}},\ \bibinfo
  {pages} {574} (\bibinfo {year} {2010})}\BibitemShut {NoStop}%
\bibitem [{\citenamefont {Yamamoto}\ \emph {et~al.}(2010)\citenamefont
  {Yamamoto}, \citenamefont {Neeley}, \citenamefont {Lucero}, \citenamefont
  {Bialczak}, \citenamefont {Kelly}, \citenamefont {Lenander}, \citenamefont
  {Mariantoni}, \citenamefont {O'Connell}, \citenamefont {Sank}, \citenamefont
  {Wang}, \citenamefont {Weides}, \citenamefont {Wenner}, \citenamefont {Yin},
  \citenamefont {Cleland},\ and\ \citenamefont {Martinis}}]{yamamoto10}%
  \BibitemOpen
  \bibfield  {author} {\bibinfo {author} {\bibfnamefont {T.}~\bibnamefont
  {Yamamoto}}, \bibinfo {author} {\bibfnamefont {M.}~\bibnamefont {Neeley}},
  \bibinfo {author} {\bibfnamefont {E.}~\bibnamefont {Lucero}}, \bibinfo
  {author} {\bibfnamefont {R.~C.}\ \bibnamefont {Bialczak}}, \bibinfo {author}
  {\bibfnamefont {J.}~\bibnamefont {Kelly}}, \bibinfo {author} {\bibfnamefont
  {M.}~\bibnamefont {Lenander}}, \bibinfo {author} {\bibfnamefont
  {M.}~\bibnamefont {Mariantoni}}, \bibinfo {author} {\bibfnamefont {A.~D.}\
  \bibnamefont {O'Connell}}, \bibinfo {author} {\bibfnamefont {D.}~\bibnamefont
  {Sank}}, \bibinfo {author} {\bibfnamefont {H.}~\bibnamefont {Wang}}, \bibinfo
  {author} {\bibfnamefont {M.}~\bibnamefont {Weides}}, \bibinfo {author}
  {\bibfnamefont {J.}~\bibnamefont {Wenner}}, \bibinfo {author} {\bibfnamefont
  {Y.}~\bibnamefont {Yin}}, \bibinfo {author} {\bibfnamefont {A.~N.}\
  \bibnamefont {Cleland}},\ and\ \bibinfo {author} {\bibfnamefont {J.~M.}\
  \bibnamefont {Martinis}},\ }\href@noop {} {\bibfield  {journal} {\bibinfo
  {journal} {Phys. Rev. B}\ }\textbf {\bibinfo {volume} {82}},\ \bibinfo
  {pages} {184515} (\bibinfo {year} {2010})}\BibitemShut {NoStop}%
\bibitem [{\citenamefont {Neg{\^i}rneac}\ \emph {et~al.}(2021)\citenamefont
  {Neg{\^i}rneac}, \citenamefont {Ali}, \citenamefont {Muthusubramanian},
  \citenamefont {Battistel}, \citenamefont {Sagastizabal}, \citenamefont
  {Moreira}, \citenamefont {Marques}, \citenamefont {Vlothuizen}, \citenamefont
  {Beekman}, \citenamefont {Zachariadis}, \citenamefont {Haider}, \citenamefont
  {Bruno},\ and\ \citenamefont {DiCarlo}}]{negirneac21}%
  \BibitemOpen
  \bibfield  {author} {\bibinfo {author} {\bibfnamefont {V.}~\bibnamefont
  {Neg{\^i}rneac}}, \bibinfo {author} {\bibfnamefont {H.}~\bibnamefont {Ali}},
  \bibinfo {author} {\bibfnamefont {N.}~\bibnamefont {Muthusubramanian}},
  \bibinfo {author} {\bibfnamefont {F.}~\bibnamefont {Battistel}}, \bibinfo
  {author} {\bibfnamefont {R.}~\bibnamefont {Sagastizabal}}, \bibinfo {author}
  {\bibfnamefont {M.~S.}\ \bibnamefont {Moreira}}, \bibinfo {author}
  {\bibfnamefont {J.~F.}\ \bibnamefont {Marques}}, \bibinfo {author}
  {\bibfnamefont {W.~J.}\ \bibnamefont {Vlothuizen}}, \bibinfo {author}
  {\bibfnamefont {M.}~\bibnamefont {Beekman}}, \bibinfo {author} {\bibfnamefont
  {C.}~\bibnamefont {Zachariadis}}, \bibinfo {author} {\bibfnamefont
  {N.}~\bibnamefont {Haider}}, \bibinfo {author} {\bibfnamefont
  {A.}~\bibnamefont {Bruno}},\ and\ \bibinfo {author} {\bibfnamefont
  {L.}~\bibnamefont {DiCarlo}},\ }\href@noop {} {\bibfield  {journal} {\bibinfo
   {journal} {Phys. Rev. Lett.}\ }\textbf {\bibinfo {volume} {126}},\ \bibinfo
  {pages} {220502} (\bibinfo {year} {2021})}\BibitemShut {NoStop}%
\bibitem [{\citenamefont {Paraoanu}(2006)}]{paraoanu06}%
  \BibitemOpen
  \bibfield  {author} {\bibinfo {author} {\bibfnamefont {G.~S.}\ \bibnamefont
  {Paraoanu}},\ }\href@noop {} {\bibfield  {journal} {\bibinfo  {journal}
  {Phys. Rev. B}\ }\textbf {\bibinfo {volume} {74}},\ \bibinfo {pages} {140504}
  (\bibinfo {year} {2006})}\BibitemShut {NoStop}%
\bibitem [{\citenamefont {Rigetti}\ and\ \citenamefont
  {Devoret}(2010)}]{rigetti10}%
  \BibitemOpen
  \bibfield  {author} {\bibinfo {author} {\bibfnamefont {C.}~\bibnamefont
  {Rigetti}}\ and\ \bibinfo {author} {\bibfnamefont {M.}~\bibnamefont
  {Devoret}},\ }\href@noop {} {\bibfield  {journal} {\bibinfo  {journal} {Phys.
  Rev. B}\ }\textbf {\bibinfo {volume} {81}},\ \bibinfo {pages} {134507}
  (\bibinfo {year} {2010})}\BibitemShut {NoStop}%
\bibitem [{\citenamefont {Cross}\ and\ \citenamefont
  {Gambetta}(2015)}]{cross15}%
  \BibitemOpen
  \bibfield  {author} {\bibinfo {author} {\bibfnamefont {A.~W.}\ \bibnamefont
  {Cross}}\ and\ \bibinfo {author} {\bibfnamefont {J.~M.}\ \bibnamefont
  {Gambetta}},\ }\href@noop {} {\bibfield  {journal} {\bibinfo  {journal}
  {Phys. Rev. A}\ }\textbf {\bibinfo {volume} {91}},\ \bibinfo {pages} {032325}
  (\bibinfo {year} {2015})}\BibitemShut {NoStop}%
\bibitem [{\citenamefont {Puri}\ and\ \citenamefont {Blais}(2016)}]{puri16}%
  \BibitemOpen
  \bibfield  {author} {\bibinfo {author} {\bibfnamefont {S.}~\bibnamefont
  {Puri}}\ and\ \bibinfo {author} {\bibfnamefont {A.}~\bibnamefont {Blais}},\
  }\href@noop {} {\bibfield  {journal} {\bibinfo  {journal} {Phys. Rev. Lett.}\
  }\textbf {\bibinfo {volume} {116}},\ \bibinfo {pages} {180501} (\bibinfo
  {year} {2016})}\BibitemShut {NoStop}%
\bibitem [{\citenamefont {Paik}\ \emph {et~al.}(2016)\citenamefont {Paik},
  \citenamefont {Mezzacapo}, \citenamefont {Sandberg}, \citenamefont {McClure},
  \citenamefont {Abdo}, \citenamefont {C{\'o}rcoles}, \citenamefont {Dial},
  \citenamefont {Bogorin}, \citenamefont {Plourde}, \citenamefont {Steffen},
  \citenamefont {Cross}, \citenamefont {Gambetta},\ and\ \citenamefont
  {Chow}}]{paik16}%
  \BibitemOpen
  \bibfield  {author} {\bibinfo {author} {\bibfnamefont {H.}~\bibnamefont
  {Paik}}, \bibinfo {author} {\bibfnamefont {A.}~\bibnamefont {Mezzacapo}},
  \bibinfo {author} {\bibfnamefont {M.}~\bibnamefont {Sandberg}}, \bibinfo
  {author} {\bibfnamefont {D.~T.}\ \bibnamefont {McClure}}, \bibinfo {author}
  {\bibfnamefont {B.}~\bibnamefont {Abdo}}, \bibinfo {author} {\bibfnamefont
  {A.~D.}\ \bibnamefont {C{\'o}rcoles}}, \bibinfo {author} {\bibfnamefont
  {O.}~\bibnamefont {Dial}}, \bibinfo {author} {\bibfnamefont {D.~F.}\
  \bibnamefont {Bogorin}}, \bibinfo {author} {\bibfnamefont {B.~L.~T.}\
  \bibnamefont {Plourde}}, \bibinfo {author} {\bibfnamefont {M.}~\bibnamefont
  {Steffen}}, \bibinfo {author} {\bibfnamefont {A.~W.}\ \bibnamefont {Cross}},
  \bibinfo {author} {\bibfnamefont {J.~M.}\ \bibnamefont {Gambetta}},\ and\
  \bibinfo {author} {\bibfnamefont {J.~M.}\ \bibnamefont {Chow}},\ }\href@noop
  {} {\bibfield  {journal} {\bibinfo  {journal} {Phys. Rev. Lett.}\ }\textbf
  {\bibinfo {volume} {117}},\ \bibinfo {pages} {250502} (\bibinfo {year}
  {2016})}\BibitemShut {NoStop}%
\bibitem [{\citenamefont {Malekakhlagh}\ \emph {et~al.}(2022)\citenamefont
  {Malekakhlagh}, \citenamefont {Shanks},\ and\ \citenamefont
  {Paik}}]{malekakhlagh22a}%
  \BibitemOpen
  \bibfield  {author} {\bibinfo {author} {\bibfnamefont {M.}~\bibnamefont
  {Malekakhlagh}}, \bibinfo {author} {\bibfnamefont {W.}~\bibnamefont
  {Shanks}},\ and\ \bibinfo {author} {\bibfnamefont {H.}~\bibnamefont {Paik}},\
  }\href@noop {} {\bibfield  {journal} {\bibinfo  {journal} {Phys. Rev. A}\
  }\textbf {\bibinfo {volume} {105}},\ \bibinfo {pages} {022607} (\bibinfo
  {year} {2022})}\BibitemShut {NoStop}%
\bibitem [{\citenamefont {Kumph}\ \emph {et~al.}(2024)\citenamefont {Kumph},
  \citenamefont {Raftery}, \citenamefont {Finck}, \citenamefont {Blair},
  \citenamefont {Carniol}, \citenamefont {Carnevale}, \citenamefont {Keefe},
  \citenamefont {Arena}, \citenamefont {Hall}, \citenamefont {McKay},\ and\
  \citenamefont {Stehlik}}]{kumph24a}%
  \BibitemOpen
  \bibfield  {author} {\bibinfo {author} {\bibfnamefont {M.}~\bibnamefont
  {Kumph}}, \bibinfo {author} {\bibfnamefont {J.}~\bibnamefont {Raftery}},
  \bibinfo {author} {\bibfnamefont {A.}~\bibnamefont {Finck}}, \bibinfo
  {author} {\bibfnamefont {J.}~\bibnamefont {Blair}}, \bibinfo {author}
  {\bibfnamefont {A.}~\bibnamefont {Carniol}}, \bibinfo {author} {\bibfnamefont
  {S.}~\bibnamefont {Carnevale}}, \bibinfo {author} {\bibfnamefont {G.~A.}\
  \bibnamefont {Keefe}}, \bibinfo {author} {\bibfnamefont {V.}~\bibnamefont
  {Arena}}, \bibinfo {author} {\bibfnamefont {S.}~\bibnamefont {Hall}},
  \bibinfo {author} {\bibfnamefont {D.}~\bibnamefont {McKay}},\ and\ \bibinfo
  {author} {\bibfnamefont {G.}~\bibnamefont {Stehlik}},\ }\href@noop {}
  {\bibfield  {journal} {\bibinfo  {journal} {arXiv:2406.11770}\ } (\bibinfo
  {year} {2024})}\BibitemShut {NoStop}%
\bibitem [{\citenamefont {Huang}\ \emph {et~al.}(2024)\citenamefont {Huang},
  \citenamefont {Kim}, \citenamefont {Roy}, \citenamefont {Lu}, \citenamefont
  {Romanenko}, \citenamefont {Zhu},\ and\ \citenamefont
  {Grassellino}}]{huang2024fast}%
  \BibitemOpen
  \bibfield  {author} {\bibinfo {author} {\bibfnamefont {Z.}~\bibnamefont
  {Huang}}, \bibinfo {author} {\bibfnamefont {T.}~\bibnamefont {Kim}}, \bibinfo
  {author} {\bibfnamefont {T.}~\bibnamefont {Roy}}, \bibinfo {author}
  {\bibfnamefont {Y.}~\bibnamefont {Lu}}, \bibinfo {author} {\bibfnamefont
  {A.}~\bibnamefont {Romanenko}}, \bibinfo {author} {\bibfnamefont
  {S.}~\bibnamefont {Zhu}},\ and\ \bibinfo {author} {\bibfnamefont
  {A.}~\bibnamefont {Grassellino}},\ }\href@noop {} {\bibfield  {journal}
  {\bibinfo  {journal} {Phys. Rev. Appl.}\ }\textbf {\bibinfo {volume} {22}},\
  \bibinfo {pages} {034007} (\bibinfo {year} {2024})}\BibitemShut {NoStop}%
\bibitem [{\citenamefont {Leek}\ \emph {et~al.}(2009)\citenamefont {Leek},
  \citenamefont {Filipp}, \citenamefont {Maurer}, \citenamefont {Baur},
  \citenamefont {Bianchetti}, \citenamefont {Fink}, \citenamefont {G{\"o}ppl},
  \citenamefont {Steffen},\ and\ \citenamefont {Wallraff}}]{leek09}%
  \BibitemOpen
  \bibfield  {author} {\bibinfo {author} {\bibfnamefont {P.~J.}\ \bibnamefont
  {Leek}}, \bibinfo {author} {\bibfnamefont {S.}~\bibnamefont {Filipp}},
  \bibinfo {author} {\bibfnamefont {P.}~\bibnamefont {Maurer}}, \bibinfo
  {author} {\bibfnamefont {M.}~\bibnamefont {Baur}}, \bibinfo {author}
  {\bibfnamefont {R.}~\bibnamefont {Bianchetti}}, \bibinfo {author}
  {\bibfnamefont {J.~M.}\ \bibnamefont {Fink}}, \bibinfo {author}
  {\bibfnamefont {M.}~\bibnamefont {G{\"o}ppl}}, \bibinfo {author}
  {\bibfnamefont {L.}~\bibnamefont {Steffen}},\ and\ \bibinfo {author}
  {\bibfnamefont {A.}~\bibnamefont {Wallraff}},\ }\href@noop {} {\bibfield
  {journal} {\bibinfo  {journal} {Phys. Rev. B}\ }\textbf {\bibinfo {volume}
  {79}},\ \bibinfo {pages} {180511} (\bibinfo {year} {2009})}\BibitemShut
  {NoStop}%
\bibitem [{\citenamefont {Poletto}\ \emph {et~al.}(2012)\citenamefont
  {Poletto}, \citenamefont {Gambetta}, \citenamefont {Merkel}, \citenamefont
  {Smolin}, \citenamefont {Chow}, \citenamefont {C{\'o}rcoles}, \citenamefont
  {Keefe}, \citenamefont {Rothwell}, \citenamefont {Rozen}, \citenamefont
  {Abraham}, \citenamefont {Rigetti},\ and\ \citenamefont
  {Steffen}}]{poletto12}%
  \BibitemOpen
  \bibfield  {author} {\bibinfo {author} {\bibfnamefont {S.}~\bibnamefont
  {Poletto}}, \bibinfo {author} {\bibfnamefont {J.~M.}\ \bibnamefont
  {Gambetta}}, \bibinfo {author} {\bibfnamefont {S.~T.}\ \bibnamefont
  {Merkel}}, \bibinfo {author} {\bibfnamefont {J.~A.}\ \bibnamefont {Smolin}},
  \bibinfo {author} {\bibfnamefont {J.~M.}\ \bibnamefont {Chow}}, \bibinfo
  {author} {\bibfnamefont {A.~D.}\ \bibnamefont {C{\'o}rcoles}}, \bibinfo
  {author} {\bibfnamefont {G.~A.}\ \bibnamefont {Keefe}}, \bibinfo {author}
  {\bibfnamefont {M.~B.}\ \bibnamefont {Rothwell}}, \bibinfo {author}
  {\bibfnamefont {J.~R.}\ \bibnamefont {Rozen}}, \bibinfo {author}
  {\bibfnamefont {D.~W.}\ \bibnamefont {Abraham}}, \bibinfo {author}
  {\bibfnamefont {C.}~\bibnamefont {Rigetti}},\ and\ \bibinfo {author}
  {\bibfnamefont {M.}~\bibnamefont {Steffen}},\ }\href@noop {} {\bibfield
  {journal} {\bibinfo  {journal} {Phys. Rev. Lett.}\ }\textbf {\bibinfo
  {volume} {109}},\ \bibinfo {pages} {240505} (\bibinfo {year}
  {2012})}\BibitemShut {NoStop}%
\bibitem [{\citenamefont {Chow}\ \emph {et~al.}(2013)\citenamefont {Chow},
  \citenamefont {Gambetta}, \citenamefont {Cross}, \citenamefont {Merkel},
  \citenamefont {Rigetti},\ and\ \citenamefont {Steffen}}]{chow13}%
  \BibitemOpen
  \bibfield  {author} {\bibinfo {author} {\bibfnamefont {J.~M.}\ \bibnamefont
  {Chow}}, \bibinfo {author} {\bibfnamefont {J.~M.}\ \bibnamefont {Gambetta}},
  \bibinfo {author} {\bibfnamefont {A.~W.}\ \bibnamefont {Cross}}, \bibinfo
  {author} {\bibfnamefont {S.~T.}\ \bibnamefont {Merkel}}, \bibinfo {author}
  {\bibfnamefont {C.}~\bibnamefont {Rigetti}},\ and\ \bibinfo {author}
  {\bibfnamefont {M.}~\bibnamefont {Steffen}},\ }\href@noop {} {\bibfield
  {journal} {\bibinfo  {journal} {New J. Phys.}\ }\textbf {\bibinfo {volume}
  {15}},\ \bibinfo {pages} {115012} (\bibinfo {year} {2013})}\BibitemShut
  {NoStop}%
\bibitem [{\citenamefont {Zeytino{\u g}lu}\ \emph {et~al.}(2015)\citenamefont
  {Zeytino{\u g}lu}, \citenamefont {Pechal}, \citenamefont {Berger},
  \citenamefont {Abdumalikov}, \citenamefont {Wallraff},\ and\ \citenamefont
  {Filipp}}]{zeytinoglu15}%
  \BibitemOpen
  \bibfield  {author} {\bibinfo {author} {\bibfnamefont {S.}~\bibnamefont
  {Zeytino{\u g}lu}}, \bibinfo {author} {\bibfnamefont {M.}~\bibnamefont
  {Pechal}}, \bibinfo {author} {\bibfnamefont {S.}~\bibnamefont {Berger}},
  \bibinfo {author} {\bibfnamefont {A.~A.}\ \bibnamefont {Abdumalikov}},
  \bibinfo {author} {\bibfnamefont {A.}~\bibnamefont {Wallraff}},\ and\
  \bibinfo {author} {\bibfnamefont {S.}~\bibnamefont {Filipp}},\ }\href@noop {}
  {\bibfield  {journal} {\bibinfo  {journal} {Phys. Rev. A}\ }\textbf {\bibinfo
  {volume} {91}},\ \bibinfo {pages} {043846} (\bibinfo {year}
  {2015})}\BibitemShut {NoStop}%
\bibitem [{\citenamefont {Egger}\ \emph {et~al.}(2019)\citenamefont {Egger},
  \citenamefont {Ganzhorn}, \citenamefont {Salis}, \citenamefont {Fuhrer},
  \citenamefont {M{\"u}ller}, \citenamefont {Barkoutsos}, \citenamefont {Moll},
  \citenamefont {Tavernelli},\ and\ \citenamefont {Filipp}}]{egger19}%
  \BibitemOpen
  \bibfield  {author} {\bibinfo {author} {\bibfnamefont {D.}~\bibnamefont
  {Egger}}, \bibinfo {author} {\bibfnamefont {M.}~\bibnamefont {Ganzhorn}},
  \bibinfo {author} {\bibfnamefont {G.}~\bibnamefont {Salis}}, \bibinfo
  {author} {\bibfnamefont {A.}~\bibnamefont {Fuhrer}}, \bibinfo {author}
  {\bibfnamefont {P.}~\bibnamefont {M{\"u}ller}}, \bibinfo {author}
  {\bibfnamefont {{\relax P.Kl}.}~\bibnamefont {Barkoutsos}}, \bibinfo {author}
  {\bibfnamefont {N.}~\bibnamefont {Moll}}, \bibinfo {author} {\bibfnamefont
  {I.}~\bibnamefont {Tavernelli}},\ and\ \bibinfo {author} {\bibfnamefont
  {S.}~\bibnamefont {Filipp}},\ }\href@noop {} {\bibfield  {journal} {\bibinfo
  {journal} {Phys. Rev. Appl.}\ }\textbf {\bibinfo {volume} {11}},\ \bibinfo
  {pages} {014017} (\bibinfo {year} {2019})}\BibitemShut {NoStop}%
\bibitem [{\citenamefont {Bertet}\ \emph {et~al.}(2006)\citenamefont {Bertet},
  \citenamefont {Harmans},\ and\ \citenamefont {Mooij}}]{bertet06}%
  \BibitemOpen
  \bibfield  {author} {\bibinfo {author} {\bibfnamefont {P.}~\bibnamefont
  {Bertet}}, \bibinfo {author} {\bibfnamefont {C.~J. P.~M.}\ \bibnamefont
  {Harmans}},\ and\ \bibinfo {author} {\bibfnamefont {J.~E.}\ \bibnamefont
  {Mooij}},\ }\href@noop {} {\bibfield  {journal} {\bibinfo  {journal} {Phys.
  Rev. B}\ }\textbf {\bibinfo {volume} {73}},\ \bibinfo {pages} {064512}
  (\bibinfo {year} {2006})}\BibitemShut {NoStop}%
\bibitem [{\citenamefont {Niskanen}\ \emph {et~al.}(2006)\citenamefont
  {Niskanen}, \citenamefont {Nakamura},\ and\ \citenamefont
  {Tsai}}]{niskanen06}%
  \BibitemOpen
  \bibfield  {author} {\bibinfo {author} {\bibfnamefont {A.~O.}\ \bibnamefont
  {Niskanen}}, \bibinfo {author} {\bibfnamefont {Y.}~\bibnamefont {Nakamura}},\
  and\ \bibinfo {author} {\bibfnamefont {J.-S.}\ \bibnamefont {Tsai}},\
  }\href@noop {} {\bibfield  {journal} {\bibinfo  {journal} {Phys. Rev. B}\
  }\textbf {\bibinfo {volume} {73}},\ \bibinfo {pages} {094506} (\bibinfo
  {year} {2006})}\BibitemShut {NoStop}%
\bibitem [{\citenamefont {Liu}\ \emph {et~al.}(2007)\citenamefont {Liu},
  \citenamefont {Wei}, \citenamefont {Johansson}, \citenamefont {Tsai},\ and\
  \citenamefont {Nori}}]{liu07a}%
  \BibitemOpen
  \bibfield  {author} {\bibinfo {author} {\bibfnamefont {Y.-X.}\ \bibnamefont
  {Liu}}, \bibinfo {author} {\bibfnamefont {L.~F.}\ \bibnamefont {Wei}},
  \bibinfo {author} {\bibfnamefont {J.~R.}\ \bibnamefont {Johansson}}, \bibinfo
  {author} {\bibfnamefont {J.~S.}\ \bibnamefont {Tsai}},\ and\ \bibinfo
  {author} {\bibfnamefont {F.}~\bibnamefont {Nori}},\ }\href@noop {} {\bibfield
   {journal} {\bibinfo  {journal} {Phys. Rev. B}\ }\textbf {\bibinfo {volume}
  {76}},\ \bibinfo {pages} {144518} (\bibinfo {year} {2007})}\BibitemShut
  {NoStop}%
\bibitem [{\citenamefont {Niskanen}\ \emph {et~al.}(2007)\citenamefont
  {Niskanen}, \citenamefont {Harrabi}, \citenamefont {Yoshihara}, \citenamefont
  {Nakamura}, \citenamefont {Lloyd},\ and\ \citenamefont {Tsai}}]{niskanen07}%
  \BibitemOpen
  \bibfield  {author} {\bibinfo {author} {\bibfnamefont {A.~O.}\ \bibnamefont
  {Niskanen}}, \bibinfo {author} {\bibfnamefont {K.}~\bibnamefont {Harrabi}},
  \bibinfo {author} {\bibfnamefont {F.}~\bibnamefont {Yoshihara}}, \bibinfo
  {author} {\bibfnamefont {Y.}~\bibnamefont {Nakamura}}, \bibinfo {author}
  {\bibfnamefont {S.}~\bibnamefont {Lloyd}},\ and\ \bibinfo {author}
  {\bibfnamefont {J.~S.}\ \bibnamefont {Tsai}},\ }\href@noop {} {\bibfield
  {journal} {\bibinfo  {journal} {Science}\ }\textbf {\bibinfo {volume}
  {316}},\ \bibinfo {pages} {723} (\bibinfo {year} {2007})}\BibitemShut
  {NoStop}%
\bibitem [{\citenamefont {Beaudoin}\ \emph {et~al.}(2012)\citenamefont
  {Beaudoin}, \citenamefont {Da~Silva}, \citenamefont {Dutton},\ and\
  \citenamefont {Blais}}]{beaudoin12}%
  \BibitemOpen
  \bibfield  {author} {\bibinfo {author} {\bibfnamefont {F.}~\bibnamefont
  {Beaudoin}}, \bibinfo {author} {\bibfnamefont {M.~P.}\ \bibnamefont
  {Da~Silva}}, \bibinfo {author} {\bibfnamefont {Z.}~\bibnamefont {Dutton}},\
  and\ \bibinfo {author} {\bibfnamefont {A.}~\bibnamefont {Blais}},\
  }\href@noop {} {\bibfield  {journal} {\bibinfo  {journal} {Phys. Rev. A}\
  }\textbf {\bibinfo {volume} {86}},\ \bibinfo {pages} {022305} (\bibinfo
  {year} {2012})}\BibitemShut {NoStop}%
\bibitem [{\citenamefont {Strand}\ \emph {et~al.}(2013)\citenamefont {Strand},
  \citenamefont {Ware}, \citenamefont {Beaudoin}, \citenamefont {Ohki},
  \citenamefont {Johnson}, \citenamefont {Blais},\ and\ \citenamefont
  {Plourde}}]{strand13}%
  \BibitemOpen
  \bibfield  {author} {\bibinfo {author} {\bibfnamefont {J.~D.}\ \bibnamefont
  {Strand}}, \bibinfo {author} {\bibfnamefont {M.}~\bibnamefont {Ware}},
  \bibinfo {author} {\bibfnamefont {F.}~\bibnamefont {Beaudoin}}, \bibinfo
  {author} {\bibfnamefont {T.~A.}\ \bibnamefont {Ohki}}, \bibinfo {author}
  {\bibfnamefont {B.~R.}\ \bibnamefont {Johnson}}, \bibinfo {author}
  {\bibfnamefont {A.}~\bibnamefont {Blais}},\ and\ \bibinfo {author}
  {\bibfnamefont {B.~L.~T.}\ \bibnamefont {Plourde}},\ }\href@noop {}
  {\bibfield  {journal} {\bibinfo  {journal} {Phys. Rev. B}\ }\textbf {\bibinfo
  {volume} {87}},\ \bibinfo {pages} {220505} (\bibinfo {year}
  {2013})}\BibitemShut {NoStop}%
\bibitem [{\citenamefont {Kapit}(2015)}]{kapit15}%
  \BibitemOpen
  \bibfield  {author} {\bibinfo {author} {\bibfnamefont {E.}~\bibnamefont
  {Kapit}},\ }\href@noop {} {\bibfield  {journal} {\bibinfo  {journal} {Phys.
  Rev. A}\ }\textbf {\bibinfo {volume} {92}},\ \bibinfo {pages} {012302}
  (\bibinfo {year} {2015})}\BibitemShut {NoStop}%
\bibitem [{\citenamefont {Sirois}\ \emph {et~al.}(2015)\citenamefont {Sirois},
  \citenamefont {{Castellanos-Beltran}}, \citenamefont {DeFeo}, \citenamefont
  {Ranzani}, \citenamefont {Lecocq}, \citenamefont {Simmonds}, \citenamefont
  {Teufel},\ and\ \citenamefont {Aumentado}}]{sirois15}%
  \BibitemOpen
  \bibfield  {author} {\bibinfo {author} {\bibfnamefont {A.~J.}\ \bibnamefont
  {Sirois}}, \bibinfo {author} {\bibfnamefont {M.~A.}\ \bibnamefont
  {{Castellanos-Beltran}}}, \bibinfo {author} {\bibfnamefont {M.~P.}\
  \bibnamefont {DeFeo}}, \bibinfo {author} {\bibfnamefont {L.}~\bibnamefont
  {Ranzani}}, \bibinfo {author} {\bibfnamefont {F.}~\bibnamefont {Lecocq}},
  \bibinfo {author} {\bibfnamefont {R.~W.}\ \bibnamefont {Simmonds}}, \bibinfo
  {author} {\bibfnamefont {J.~D.}\ \bibnamefont {Teufel}},\ and\ \bibinfo
  {author} {\bibfnamefont {J.}~\bibnamefont {Aumentado}},\ }\href@noop {}
  {\bibfield  {journal} {\bibinfo  {journal} {Appl. Phys. Lett.}\ }\textbf
  {\bibinfo {volume} {106}},\ \bibinfo {pages} {172603} (\bibinfo {year}
  {2015})}\BibitemShut {NoStop}%
\bibitem [{\citenamefont {McKay}\ \emph {et~al.}(2016)\citenamefont {McKay},
  \citenamefont {Filipp}, \citenamefont {Mezzacapo}, \citenamefont {Magesan},
  \citenamefont {Chow},\ and\ \citenamefont {Gambetta}}]{mckay16}%
  \BibitemOpen
  \bibfield  {author} {\bibinfo {author} {\bibfnamefont {D.~C.}\ \bibnamefont
  {McKay}}, \bibinfo {author} {\bibfnamefont {S.}~\bibnamefont {Filipp}},
  \bibinfo {author} {\bibfnamefont {A.}~\bibnamefont {Mezzacapo}}, \bibinfo
  {author} {\bibfnamefont {E.}~\bibnamefont {Magesan}}, \bibinfo {author}
  {\bibfnamefont {J.~M.}\ \bibnamefont {Chow}},\ and\ \bibinfo {author}
  {\bibfnamefont {J.~M.}\ \bibnamefont {Gambetta}},\ }\href@noop {} {\bibfield
  {journal} {\bibinfo  {journal} {Phys. Rev. Appl.}\ }\textbf {\bibinfo
  {volume} {6}},\ \bibinfo {pages} {064007} (\bibinfo {year}
  {2016})}\BibitemShut {NoStop}%
\bibitem [{\citenamefont {Naik}\ \emph {et~al.}(2017)\citenamefont {Naik},
  \citenamefont {Leung}, \citenamefont {Chakram}, \citenamefont {Groszkowski},
  \citenamefont {Lu}, \citenamefont {Earnest}, \citenamefont {McKay},
  \citenamefont {Koch},\ and\ \citenamefont {Schuster}}]{naik17}%
  \BibitemOpen
  \bibfield  {author} {\bibinfo {author} {\bibfnamefont {R.~K.}\ \bibnamefont
  {Naik}}, \bibinfo {author} {\bibfnamefont {N.}~\bibnamefont {Leung}},
  \bibinfo {author} {\bibfnamefont {S.}~\bibnamefont {Chakram}}, \bibinfo
  {author} {\bibfnamefont {P.}~\bibnamefont {Groszkowski}}, \bibinfo {author}
  {\bibfnamefont {Y.}~\bibnamefont {Lu}}, \bibinfo {author} {\bibfnamefont
  {N.}~\bibnamefont {Earnest}}, \bibinfo {author} {\bibfnamefont {D.~C.}\
  \bibnamefont {McKay}}, \bibinfo {author} {\bibfnamefont {J.}~\bibnamefont
  {Koch}},\ and\ \bibinfo {author} {\bibfnamefont {D.~I.}\ \bibnamefont
  {Schuster}},\ }\href@noop {} {\bibfield  {journal} {\bibinfo  {journal} {Nat.
  Commun.}\ }\textbf {\bibinfo {volume} {8}},\ \bibinfo {pages} {1904}
  (\bibinfo {year} {2017})}\BibitemShut {NoStop}%
\bibitem [{\citenamefont {Caldwell}\ \emph {et~al.}(2018)\citenamefont
  {Caldwell}, \citenamefont {Didier}, \citenamefont {Ryan}, \citenamefont
  {Sete}, \citenamefont {Hudson}, \citenamefont {Karalekas}, \citenamefont
  {Manenti}, \citenamefont {Da~Silva}, \citenamefont {Sinclair}, \citenamefont
  {Acala}, \citenamefont {Alidoust}, \citenamefont {Angeles}, \citenamefont
  {Bestwick}, \citenamefont {Block}, \citenamefont {Bloom}, \citenamefont
  {Bradley}, \citenamefont {Bui}, \citenamefont {Capelluto}, \citenamefont
  {Chilcott}, \citenamefont {Cordova}, \citenamefont {Crossman}, \citenamefont
  {Curtis}, \citenamefont {Deshpande}, \citenamefont {Bouayadi}, \citenamefont
  {Girshovich}, \citenamefont {Hong}, \citenamefont {Kuang}, \citenamefont
  {Lenihan}, \citenamefont {Manning}, \citenamefont {Marchenkov}, \citenamefont
  {Marshall}, \citenamefont {Maydra}, \citenamefont {Mohan}, \citenamefont
  {O'Brien}, \citenamefont {Osborn}, \citenamefont {Otterbach}, \citenamefont
  {Papageorge}, \citenamefont {Paquette}, \citenamefont {Pelstring},
  \citenamefont {Polloreno}, \citenamefont {Prawiroatmodjo}, \citenamefont
  {Rawat}, \citenamefont {Reagor}, \citenamefont {Renzas}, \citenamefont
  {Rubin}, \citenamefont {Russell}, \citenamefont {Rust}, \citenamefont
  {Scarabelli}, \citenamefont {Scheer}, \citenamefont {Selvanayagam},
  \citenamefont {Smith}, \citenamefont {Staley}, \citenamefont {Suska},
  \citenamefont {Tezak}, \citenamefont {Thompson}, \citenamefont {To},
  \citenamefont {Vahidpour}, \citenamefont {Vodrahalli}, \citenamefont
  {Whyland}, \citenamefont {Yadav}, \citenamefont {Zeng},\ and\ \citenamefont
  {Rigetti}}]{caldwell18}%
  \BibitemOpen
  \bibfield  {author} {\bibinfo {author} {\bibfnamefont {S.~A.}\ \bibnamefont
  {Caldwell}}, \bibinfo {author} {\bibfnamefont {N.}~\bibnamefont {Didier}},
  \bibinfo {author} {\bibfnamefont {C.~A.}\ \bibnamefont {Ryan}}, \bibinfo
  {author} {\bibfnamefont {E.~A.}\ \bibnamefont {Sete}}, \bibinfo {author}
  {\bibfnamefont {A.}~\bibnamefont {Hudson}}, \bibinfo {author} {\bibfnamefont
  {P.}~\bibnamefont {Karalekas}}, \bibinfo {author} {\bibfnamefont
  {R.}~\bibnamefont {Manenti}}, \bibinfo {author} {\bibfnamefont {M.~P.}\
  \bibnamefont {Da~Silva}}, \bibinfo {author} {\bibfnamefont {R.}~\bibnamefont
  {Sinclair}}, \bibinfo {author} {\bibfnamefont {E.}~\bibnamefont {Acala}},
  \bibinfo {author} {\bibfnamefont {N.}~\bibnamefont {Alidoust}}, \bibinfo
  {author} {\bibfnamefont {J.}~\bibnamefont {Angeles}}, \bibinfo {author}
  {\bibfnamefont {A.}~\bibnamefont {Bestwick}}, \bibinfo {author}
  {\bibfnamefont {M.}~\bibnamefont {Block}}, \bibinfo {author} {\bibfnamefont
  {B.}~\bibnamefont {Bloom}}, \bibinfo {author} {\bibfnamefont
  {A.}~\bibnamefont {Bradley}}, \bibinfo {author} {\bibfnamefont
  {C.}~\bibnamefont {Bui}}, \bibinfo {author} {\bibfnamefont {L.}~\bibnamefont
  {Capelluto}}, \bibinfo {author} {\bibfnamefont {R.}~\bibnamefont {Chilcott}},
  \bibinfo {author} {\bibfnamefont {J.}~\bibnamefont {Cordova}}, \bibinfo
  {author} {\bibfnamefont {G.}~\bibnamefont {Crossman}}, \bibinfo {author}
  {\bibfnamefont {M.}~\bibnamefont {Curtis}}, \bibinfo {author} {\bibfnamefont
  {S.}~\bibnamefont {Deshpande}}, \bibinfo {author} {\bibfnamefont {T.~E.}\
  \bibnamefont {Bouayadi}}, \bibinfo {author} {\bibfnamefont {D.}~\bibnamefont
  {Girshovich}}, \bibinfo {author} {\bibfnamefont {S.}~\bibnamefont {Hong}},
  \bibinfo {author} {\bibfnamefont {K.}~\bibnamefont {Kuang}}, \bibinfo
  {author} {\bibfnamefont {M.}~\bibnamefont {Lenihan}}, \bibinfo {author}
  {\bibfnamefont {T.}~\bibnamefont {Manning}}, \bibinfo {author} {\bibfnamefont
  {A.}~\bibnamefont {Marchenkov}}, \bibinfo {author} {\bibfnamefont
  {J.}~\bibnamefont {Marshall}}, \bibinfo {author} {\bibfnamefont
  {R.}~\bibnamefont {Maydra}}, \bibinfo {author} {\bibfnamefont
  {Y.}~\bibnamefont {Mohan}}, \bibinfo {author} {\bibfnamefont
  {W.}~\bibnamefont {O'Brien}}, \bibinfo {author} {\bibfnamefont
  {C.}~\bibnamefont {Osborn}}, \bibinfo {author} {\bibfnamefont
  {J.}~\bibnamefont {Otterbach}}, \bibinfo {author} {\bibfnamefont
  {A.}~\bibnamefont {Papageorge}}, \bibinfo {author} {\bibfnamefont {J.-P.}\
  \bibnamefont {Paquette}}, \bibinfo {author} {\bibfnamefont {M.}~\bibnamefont
  {Pelstring}}, \bibinfo {author} {\bibfnamefont {A.}~\bibnamefont
  {Polloreno}}, \bibinfo {author} {\bibfnamefont {G.}~\bibnamefont
  {Prawiroatmodjo}}, \bibinfo {author} {\bibfnamefont {V.}~\bibnamefont
  {Rawat}}, \bibinfo {author} {\bibfnamefont {M.}~\bibnamefont {Reagor}},
  \bibinfo {author} {\bibfnamefont {R.}~\bibnamefont {Renzas}}, \bibinfo
  {author} {\bibfnamefont {N.}~\bibnamefont {Rubin}}, \bibinfo {author}
  {\bibfnamefont {D.}~\bibnamefont {Russell}}, \bibinfo {author} {\bibfnamefont
  {M.}~\bibnamefont {Rust}}, \bibinfo {author} {\bibfnamefont {D.}~\bibnamefont
  {Scarabelli}}, \bibinfo {author} {\bibfnamefont {M.}~\bibnamefont {Scheer}},
  \bibinfo {author} {\bibfnamefont {M.}~\bibnamefont {Selvanayagam}}, \bibinfo
  {author} {\bibfnamefont {R.}~\bibnamefont {Smith}}, \bibinfo {author}
  {\bibfnamefont {A.}~\bibnamefont {Staley}}, \bibinfo {author} {\bibfnamefont
  {M.}~\bibnamefont {Suska}}, \bibinfo {author} {\bibfnamefont
  {N.}~\bibnamefont {Tezak}}, \bibinfo {author} {\bibfnamefont {D.~C.}\
  \bibnamefont {Thompson}}, \bibinfo {author} {\bibfnamefont {T.-W.}\
  \bibnamefont {To}}, \bibinfo {author} {\bibfnamefont {M.}~\bibnamefont
  {Vahidpour}}, \bibinfo {author} {\bibfnamefont {N.}~\bibnamefont
  {Vodrahalli}}, \bibinfo {author} {\bibfnamefont {T.}~\bibnamefont {Whyland}},
  \bibinfo {author} {\bibfnamefont {K.}~\bibnamefont {Yadav}}, \bibinfo
  {author} {\bibfnamefont {W.}~\bibnamefont {Zeng}},\ and\ \bibinfo {author}
  {\bibfnamefont {C.}~\bibnamefont {Rigetti}},\ }\href@noop {} {\bibfield
  {journal} {\bibinfo  {journal} {Phys. Rev. Appl.}\ }\textbf {\bibinfo
  {volume} {10}},\ \bibinfo {pages} {034050} (\bibinfo {year}
  {2018})}\BibitemShut {NoStop}%
\bibitem [{\citenamefont {Didier}\ \emph {et~al.}(2018)\citenamefont {Didier},
  \citenamefont {Sete}, \citenamefont {{da Silva}},\ and\ \citenamefont
  {Rigetti}}]{didier18}%
  \BibitemOpen
  \bibfield  {author} {\bibinfo {author} {\bibfnamefont {N.}~\bibnamefont
  {Didier}}, \bibinfo {author} {\bibfnamefont {E.~A.}\ \bibnamefont {Sete}},
  \bibinfo {author} {\bibfnamefont {M.~P.}\ \bibnamefont {{da Silva}}},\ and\
  \bibinfo {author} {\bibfnamefont {C.}~\bibnamefont {Rigetti}},\ }\href@noop
  {} {\bibfield  {journal} {\bibinfo  {journal} {Phys. Rev. A}\ }\textbf
  {\bibinfo {volume} {97}},\ \bibinfo {pages} {022330} (\bibinfo {year}
  {2018})}\BibitemShut {NoStop}%
\bibitem [{\citenamefont {Reagor}\ \emph {et~al.}(2018)\citenamefont {Reagor},
  \citenamefont {Osborn}, \citenamefont {Tezak}, \citenamefont {Staley},
  \citenamefont {Prawiroatmodjo}, \citenamefont {Scheer}, \citenamefont
  {Alidoust}, \citenamefont {Sete}, \citenamefont {Didier}, \citenamefont
  {Da~Silva}, \citenamefont {Acala}, \citenamefont {Angeles}, \citenamefont
  {Bestwick}, \citenamefont {Block}, \citenamefont {Bloom}, \citenamefont
  {Bradley}, \citenamefont {Bui}, \citenamefont {Caldwell}, \citenamefont
  {Capelluto}, \citenamefont {Chilcott}, \citenamefont {Cordova}, \citenamefont
  {Crossman}, \citenamefont {Curtis}, \citenamefont {Deshpande}, \citenamefont
  {El~Bouayadi}, \citenamefont {Girshovich}, \citenamefont {Hong},
  \citenamefont {Hudson}, \citenamefont {Karalekas}, \citenamefont {Kuang},
  \citenamefont {Lenihan}, \citenamefont {Manenti}, \citenamefont {Manning},
  \citenamefont {Marshall}, \citenamefont {Mohan}, \citenamefont {O'Brien},
  \citenamefont {Otterbach}, \citenamefont {Papageorge}, \citenamefont
  {Paquette}, \citenamefont {Pelstring}, \citenamefont {Polloreno},
  \citenamefont {Rawat}, \citenamefont {Ryan}, \citenamefont {Renzas},
  \citenamefont {Rubin}, \citenamefont {Russel}, \citenamefont {Rust},
  \citenamefont {Scarabelli}, \citenamefont {Selvanayagam}, \citenamefont
  {Sinclair}, \citenamefont {Smith}, \citenamefont {Suska}, \citenamefont {To},
  \citenamefont {Vahidpour}, \citenamefont {Vodrahalli}, \citenamefont
  {Whyland}, \citenamefont {Yadav}, \citenamefont {Zeng},\ and\ \citenamefont
  {Rigetti}}]{reagor18}%
  \BibitemOpen
  \bibfield  {author} {\bibinfo {author} {\bibfnamefont {M.}~\bibnamefont
  {Reagor}}, \bibinfo {author} {\bibfnamefont {C.~B.}\ \bibnamefont {Osborn}},
  \bibinfo {author} {\bibfnamefont {N.}~\bibnamefont {Tezak}}, \bibinfo
  {author} {\bibfnamefont {A.}~\bibnamefont {Staley}}, \bibinfo {author}
  {\bibfnamefont {G.}~\bibnamefont {Prawiroatmodjo}}, \bibinfo {author}
  {\bibfnamefont {M.}~\bibnamefont {Scheer}}, \bibinfo {author} {\bibfnamefont
  {N.}~\bibnamefont {Alidoust}}, \bibinfo {author} {\bibfnamefont {E.~A.}\
  \bibnamefont {Sete}}, \bibinfo {author} {\bibfnamefont {N.}~\bibnamefont
  {Didier}}, \bibinfo {author} {\bibfnamefont {M.~P.}\ \bibnamefont
  {Da~Silva}}, \bibinfo {author} {\bibfnamefont {E.}~\bibnamefont {Acala}},
  \bibinfo {author} {\bibfnamefont {J.}~\bibnamefont {Angeles}}, \bibinfo
  {author} {\bibfnamefont {A.}~\bibnamefont {Bestwick}}, \bibinfo {author}
  {\bibfnamefont {M.}~\bibnamefont {Block}}, \bibinfo {author} {\bibfnamefont
  {B.}~\bibnamefont {Bloom}}, \bibinfo {author} {\bibfnamefont
  {A.}~\bibnamefont {Bradley}}, \bibinfo {author} {\bibfnamefont
  {C.}~\bibnamefont {Bui}}, \bibinfo {author} {\bibfnamefont {S.}~\bibnamefont
  {Caldwell}}, \bibinfo {author} {\bibfnamefont {L.}~\bibnamefont {Capelluto}},
  \bibinfo {author} {\bibfnamefont {R.}~\bibnamefont {Chilcott}}, \bibinfo
  {author} {\bibfnamefont {J.}~\bibnamefont {Cordova}}, \bibinfo {author}
  {\bibfnamefont {G.}~\bibnamefont {Crossman}}, \bibinfo {author}
  {\bibfnamefont {M.}~\bibnamefont {Curtis}}, \bibinfo {author} {\bibfnamefont
  {S.}~\bibnamefont {Deshpande}}, \bibinfo {author} {\bibfnamefont
  {T.}~\bibnamefont {El~Bouayadi}}, \bibinfo {author} {\bibfnamefont
  {D.}~\bibnamefont {Girshovich}}, \bibinfo {author} {\bibfnamefont
  {S.}~\bibnamefont {Hong}}, \bibinfo {author} {\bibfnamefont {A.}~\bibnamefont
  {Hudson}}, \bibinfo {author} {\bibfnamefont {P.}~\bibnamefont {Karalekas}},
  \bibinfo {author} {\bibfnamefont {K.}~\bibnamefont {Kuang}}, \bibinfo
  {author} {\bibfnamefont {M.}~\bibnamefont {Lenihan}}, \bibinfo {author}
  {\bibfnamefont {R.}~\bibnamefont {Manenti}}, \bibinfo {author} {\bibfnamefont
  {T.}~\bibnamefont {Manning}}, \bibinfo {author} {\bibfnamefont
  {J.}~\bibnamefont {Marshall}}, \bibinfo {author} {\bibfnamefont
  {Y.}~\bibnamefont {Mohan}}, \bibinfo {author} {\bibfnamefont
  {W.}~\bibnamefont {O'Brien}}, \bibinfo {author} {\bibfnamefont
  {J.}~\bibnamefont {Otterbach}}, \bibinfo {author} {\bibfnamefont
  {A.}~\bibnamefont {Papageorge}}, \bibinfo {author} {\bibfnamefont {J.-P.}\
  \bibnamefont {Paquette}}, \bibinfo {author} {\bibfnamefont {M.}~\bibnamefont
  {Pelstring}}, \bibinfo {author} {\bibfnamefont {A.}~\bibnamefont
  {Polloreno}}, \bibinfo {author} {\bibfnamefont {V.}~\bibnamefont {Rawat}},
  \bibinfo {author} {\bibfnamefont {C.~A.}\ \bibnamefont {Ryan}}, \bibinfo
  {author} {\bibfnamefont {R.}~\bibnamefont {Renzas}}, \bibinfo {author}
  {\bibfnamefont {N.}~\bibnamefont {Rubin}}, \bibinfo {author} {\bibfnamefont
  {D.}~\bibnamefont {Russel}}, \bibinfo {author} {\bibfnamefont
  {M.}~\bibnamefont {Rust}}, \bibinfo {author} {\bibfnamefont {D.}~\bibnamefont
  {Scarabelli}}, \bibinfo {author} {\bibfnamefont {M.}~\bibnamefont
  {Selvanayagam}}, \bibinfo {author} {\bibfnamefont {R.}~\bibnamefont
  {Sinclair}}, \bibinfo {author} {\bibfnamefont {R.}~\bibnamefont {Smith}},
  \bibinfo {author} {\bibfnamefont {M.}~\bibnamefont {Suska}}, \bibinfo
  {author} {\bibfnamefont {T.-W.}\ \bibnamefont {To}}, \bibinfo {author}
  {\bibfnamefont {M.}~\bibnamefont {Vahidpour}}, \bibinfo {author}
  {\bibfnamefont {N.}~\bibnamefont {Vodrahalli}}, \bibinfo {author}
  {\bibfnamefont {T.}~\bibnamefont {Whyland}}, \bibinfo {author} {\bibfnamefont
  {K.}~\bibnamefont {Yadav}}, \bibinfo {author} {\bibfnamefont
  {W.}~\bibnamefont {Zeng}},\ and\ \bibinfo {author} {\bibfnamefont {C.~T.}\
  \bibnamefont {Rigetti}},\ }\href@noop {} {\bibfield  {journal} {\bibinfo
  {journal} {Sci. Adv.}\ }\textbf {\bibinfo {volume} {4}},\ \bibinfo {pages}
  {eaao3603} (\bibinfo {year} {2018})}\BibitemShut {NoStop}%
\bibitem [{\citenamefont {Sung}\ \emph {et~al.}(2021)\citenamefont {Sung},
  \citenamefont {Ding}, \citenamefont {Braum{\"u}ller}, \citenamefont
  {Veps{\"a}l{\"a}inen}, \citenamefont {Kannan}, \citenamefont {Kjaergaard},
  \citenamefont {Greene}, \citenamefont {Samach}, \citenamefont {McNally},
  \citenamefont {Kim} \emph {et~al.}}]{sung2021realization}%
  \BibitemOpen
  \bibfield  {author} {\bibinfo {author} {\bibfnamefont {Y.}~\bibnamefont
  {Sung}}, \bibinfo {author} {\bibfnamefont {L.}~\bibnamefont {Ding}}, \bibinfo
  {author} {\bibfnamefont {J.}~\bibnamefont {Braum{\"u}ller}}, \bibinfo
  {author} {\bibfnamefont {A.}~\bibnamefont {Veps{\"a}l{\"a}inen}}, \bibinfo
  {author} {\bibfnamefont {B.}~\bibnamefont {Kannan}}, \bibinfo {author}
  {\bibfnamefont {M.}~\bibnamefont {Kjaergaard}}, \bibinfo {author}
  {\bibfnamefont {A.}~\bibnamefont {Greene}}, \bibinfo {author} {\bibfnamefont
  {G.~O.}\ \bibnamefont {Samach}}, \bibinfo {author} {\bibfnamefont
  {C.}~\bibnamefont {McNally}}, \bibinfo {author} {\bibfnamefont
  {D.}~\bibnamefont {Kim}}, \emph {et~al.},\ }\href@noop {} {\bibfield
  {journal} {\bibinfo  {journal} {Phys. Rev. X}\ }\textbf {\bibinfo {volume}
  {11}},\ \bibinfo {pages} {021058} (\bibinfo {year} {2021})}\BibitemShut
  {NoStop}%
\bibitem [{\citenamefont {Stehlik}\ \emph {et~al.}(2021)\citenamefont
  {Stehlik}, \citenamefont {Zajac}, \citenamefont {Underwood}, \citenamefont
  {Phung}, \citenamefont {Blair}, \citenamefont {Carnevale}, \citenamefont
  {Klaus}, \citenamefont {Keefe}, \citenamefont {Carniol}, \citenamefont
  {Kumph} \emph {et~al.}}]{stehlik2021tunable}%
  \BibitemOpen
  \bibfield  {author} {\bibinfo {author} {\bibfnamefont {J.}~\bibnamefont
  {Stehlik}}, \bibinfo {author} {\bibfnamefont {D.}~\bibnamefont {Zajac}},
  \bibinfo {author} {\bibfnamefont {D.}~\bibnamefont {Underwood}}, \bibinfo
  {author} {\bibfnamefont {T.}~\bibnamefont {Phung}}, \bibinfo {author}
  {\bibfnamefont {J.}~\bibnamefont {Blair}}, \bibinfo {author} {\bibfnamefont
  {S.}~\bibnamefont {Carnevale}}, \bibinfo {author} {\bibfnamefont
  {D.}~\bibnamefont {Klaus}}, \bibinfo {author} {\bibfnamefont
  {G.}~\bibnamefont {Keefe}}, \bibinfo {author} {\bibfnamefont
  {A.}~\bibnamefont {Carniol}}, \bibinfo {author} {\bibfnamefont
  {M.}~\bibnamefont {Kumph}}, \emph {et~al.},\ }\href@noop {} {\bibfield
  {journal} {\bibinfo  {journal} {Phys. Rev. Lett.}\ }\textbf {\bibinfo
  {volume} {127}},\ \bibinfo {pages} {080505} (\bibinfo {year}
  {2021})}\BibitemShut {NoStop}%
\bibitem [{\citenamefont {de~Groot}\ \emph {et~al.}(2010)\citenamefont
  {de~Groot}, \citenamefont {Lisenfeld}, \citenamefont {Schouten},
  \citenamefont {Ashhab}, \citenamefont {Lupa{\c s}cu}, \citenamefont
  {Harmans},\ and\ \citenamefont {Mooij}}]{degroot10}%
  \BibitemOpen
  \bibfield  {author} {\bibinfo {author} {\bibfnamefont {P.~C.}\ \bibnamefont
  {de~Groot}}, \bibinfo {author} {\bibfnamefont {J.}~\bibnamefont {Lisenfeld}},
  \bibinfo {author} {\bibfnamefont {R.~N.}\ \bibnamefont {Schouten}}, \bibinfo
  {author} {\bibfnamefont {S.}~\bibnamefont {Ashhab}}, \bibinfo {author}
  {\bibfnamefont {A.}~\bibnamefont {Lupa{\c s}cu}}, \bibinfo {author}
  {\bibfnamefont {C.~J. P.~M.}\ \bibnamefont {Harmans}},\ and\ \bibinfo
  {author} {\bibfnamefont {J.~E.}\ \bibnamefont {Mooij}},\ }\href@noop {}
  {\bibfield  {journal} {\bibinfo  {journal} {Nat. Phys.}\ }\textbf {\bibinfo
  {volume} {6}},\ \bibinfo {pages} {763} (\bibinfo {year} {2010})}\BibitemShut
  {NoStop}%
\bibitem [{\citenamefont {Chow}\ \emph {et~al.}(2011)\citenamefont {Chow},
  \citenamefont {C\'orcoles}, \citenamefont {Gambetta}, \citenamefont
  {Rigetti}, \citenamefont {Johnson}, \citenamefont {Smolin}, \citenamefont
  {Rozen}, \citenamefont {Keefe}, \citenamefont {Rothwell}, \citenamefont
  {Ketchen},\ and\ \citenamefont {Steffen}}]{chow11}%
  \BibitemOpen
  \bibfield  {author} {\bibinfo {author} {\bibfnamefont {J.~M.}\ \bibnamefont
  {Chow}}, \bibinfo {author} {\bibfnamefont {A.~D.}\ \bibnamefont
  {C\'orcoles}}, \bibinfo {author} {\bibfnamefont {J.~M.}\ \bibnamefont
  {Gambetta}}, \bibinfo {author} {\bibfnamefont {C.}~\bibnamefont {Rigetti}},
  \bibinfo {author} {\bibfnamefont {B.~R.}\ \bibnamefont {Johnson}}, \bibinfo
  {author} {\bibfnamefont {J.~A.}\ \bibnamefont {Smolin}}, \bibinfo {author}
  {\bibfnamefont {J.~R.}\ \bibnamefont {Rozen}}, \bibinfo {author}
  {\bibfnamefont {G.~A.}\ \bibnamefont {Keefe}}, \bibinfo {author}
  {\bibfnamefont {M.~B.}\ \bibnamefont {Rothwell}}, \bibinfo {author}
  {\bibfnamefont {M.~B.}\ \bibnamefont {Ketchen}},\ and\ \bibinfo {author}
  {\bibfnamefont {M.}~\bibnamefont {Steffen}},\ }\href@noop {} {\bibfield
  {journal} {\bibinfo  {journal} {Phys. Rev. Lett.}\ }\textbf {\bibinfo
  {volume} {107}},\ \bibinfo {pages} {080502} (\bibinfo {year}
  {2011})}\BibitemShut {NoStop}%
\bibitem [{\citenamefont {Kandala}\ \emph {et~al.}(2021)\citenamefont
  {Kandala}, \citenamefont {Wei}, \citenamefont {Srinivasan}, \citenamefont
  {Magesan}, \citenamefont {Carnevale}, \citenamefont {Keefe}, \citenamefont
  {Klaus}, \citenamefont {Dial},\ and\ \citenamefont {McKay}}]{kandala21}%
  \BibitemOpen
  \bibfield  {author} {\bibinfo {author} {\bibfnamefont {A.}~\bibnamefont
  {Kandala}}, \bibinfo {author} {\bibfnamefont {K.~X.}\ \bibnamefont {Wei}},
  \bibinfo {author} {\bibfnamefont {S.}~\bibnamefont {Srinivasan}}, \bibinfo
  {author} {\bibfnamefont {E.}~\bibnamefont {Magesan}}, \bibinfo {author}
  {\bibfnamefont {S.}~\bibnamefont {Carnevale}}, \bibinfo {author}
  {\bibfnamefont {G.~A.}\ \bibnamefont {Keefe}}, \bibinfo {author}
  {\bibfnamefont {D.}~\bibnamefont {Klaus}}, \bibinfo {author} {\bibfnamefont
  {O.}~\bibnamefont {Dial}},\ and\ \bibinfo {author} {\bibfnamefont {D.~C.}\
  \bibnamefont {McKay}},\ }\href@noop {} {\bibfield  {journal} {\bibinfo
  {journal} {Phys. Rev. Lett.}\ }\textbf {\bibinfo {volume} {127}},\ \bibinfo
  {pages} {130501} (\bibinfo {year} {2021})}\BibitemShut {NoStop}%
\bibitem [{\citenamefont {Wei}\ \emph {et~al.}(2022)\citenamefont {Wei},
  \citenamefont {Magesan}, \citenamefont {Lauer}, \citenamefont {Srinivasan},
  \citenamefont {Bogorin}, \citenamefont {Carnevale}, \citenamefont {Keefe},
  \citenamefont {Kim}, \citenamefont {Klaus}, \citenamefont {Landers},
  \citenamefont {Sundaresan}, \citenamefont {Wang}, \citenamefont {Zhang},
  \citenamefont {Steffen}, \citenamefont {Dial}, \citenamefont {McKay},\ and\
  \citenamefont {Kandala}}]{wei21}%
  \BibitemOpen
  \bibfield  {author} {\bibinfo {author} {\bibfnamefont {K.~X.}\ \bibnamefont
  {Wei}}, \bibinfo {author} {\bibfnamefont {E.}~\bibnamefont {Magesan}},
  \bibinfo {author} {\bibfnamefont {I.}~\bibnamefont {Lauer}}, \bibinfo
  {author} {\bibfnamefont {S.}~\bibnamefont {Srinivasan}}, \bibinfo {author}
  {\bibfnamefont {D.~F.}\ \bibnamefont {Bogorin}}, \bibinfo {author}
  {\bibfnamefont {S.}~\bibnamefont {Carnevale}}, \bibinfo {author}
  {\bibfnamefont {G.~A.}\ \bibnamefont {Keefe}}, \bibinfo {author}
  {\bibfnamefont {Y.}~\bibnamefont {Kim}}, \bibinfo {author} {\bibfnamefont
  {D.}~\bibnamefont {Klaus}}, \bibinfo {author} {\bibfnamefont
  {W.}~\bibnamefont {Landers}}, \bibinfo {author} {\bibfnamefont
  {N.}~\bibnamefont {Sundaresan}}, \bibinfo {author} {\bibfnamefont
  {C.}~\bibnamefont {Wang}}, \bibinfo {author} {\bibfnamefont {E.~J.}\
  \bibnamefont {Zhang}}, \bibinfo {author} {\bibfnamefont {M.}~\bibnamefont
  {Steffen}}, \bibinfo {author} {\bibfnamefont {O.~E.}\ \bibnamefont {Dial}},
  \bibinfo {author} {\bibfnamefont {D.~C.}\ \bibnamefont {McKay}},\ and\
  \bibinfo {author} {\bibfnamefont {A.}~\bibnamefont {Kandala}},\ }\href@noop
  {} {\bibfield  {journal} {\bibinfo  {journal} {Phys. Rev. Lett.}\ }\textbf
  {\bibinfo {volume} {129}},\ \bibinfo {pages} {060501} (\bibinfo {year}
  {2022})}\BibitemShut {NoStop}%
\bibitem [{\citenamefont {Chow}\ \emph {et~al.}(2012)\citenamefont {Chow},
  \citenamefont {Gambetta}, \citenamefont {C\'orcoles}, \citenamefont {Merkel},
  \citenamefont {Smolin}, \citenamefont {Rigetti}, \citenamefont {Poletto},
  \citenamefont {Keefe}, \citenamefont {Rothwell}, \citenamefont {Rozen},
  \citenamefont {Ketchen},\ and\ \citenamefont {Steffen}}]{chow12}%
  \BibitemOpen
  \bibfield  {author} {\bibinfo {author} {\bibfnamefont {J.~M.}\ \bibnamefont
  {Chow}}, \bibinfo {author} {\bibfnamefont {J.~M.}\ \bibnamefont {Gambetta}},
  \bibinfo {author} {\bibfnamefont {A.~D.}\ \bibnamefont {C\'orcoles}},
  \bibinfo {author} {\bibfnamefont {S.~T.}\ \bibnamefont {Merkel}}, \bibinfo
  {author} {\bibfnamefont {J.~A.}\ \bibnamefont {Smolin}}, \bibinfo {author}
  {\bibfnamefont {C.}~\bibnamefont {Rigetti}}, \bibinfo {author} {\bibfnamefont
  {S.}~\bibnamefont {Poletto}}, \bibinfo {author} {\bibfnamefont {G.~A.}\
  \bibnamefont {Keefe}}, \bibinfo {author} {\bibfnamefont {M.~B.}\ \bibnamefont
  {Rothwell}}, \bibinfo {author} {\bibfnamefont {J.~R.}\ \bibnamefont {Rozen}},
  \bibinfo {author} {\bibfnamefont {M.~B.}\ \bibnamefont {Ketchen}},\ and\
  \bibinfo {author} {\bibfnamefont {M.}~\bibnamefont {Steffen}},\ }\href@noop
  {} {\bibfield  {journal} {\bibinfo  {journal} {Phys. Rev. Lett.}\ }\textbf
  {\bibinfo {volume} {109}},\ \bibinfo {pages} {060501} (\bibinfo {year}
  {2012})}\BibitemShut {NoStop}%
\bibitem [{\citenamefont {C\'orcoles}\ \emph {et~al.}(2013)\citenamefont
  {C\'orcoles}, \citenamefont {Gambetta}, \citenamefont {Chow}, \citenamefont
  {Smolin}, \citenamefont {Ware}, \citenamefont {Strand}, \citenamefont
  {Plourde},\ and\ \citenamefont {Steffen}}]{corcoles13}%
  \BibitemOpen
  \bibfield  {author} {\bibinfo {author} {\bibfnamefont {A.~D.}\ \bibnamefont
  {C\'orcoles}}, \bibinfo {author} {\bibfnamefont {J.~M.}\ \bibnamefont
  {Gambetta}}, \bibinfo {author} {\bibfnamefont {J.~M.}\ \bibnamefont {Chow}},
  \bibinfo {author} {\bibfnamefont {J.~A.}\ \bibnamefont {Smolin}}, \bibinfo
  {author} {\bibfnamefont {M.}~\bibnamefont {Ware}}, \bibinfo {author}
  {\bibfnamefont {J.}~\bibnamefont {Strand}}, \bibinfo {author} {\bibfnamefont
  {B.~L.~T.}\ \bibnamefont {Plourde}},\ and\ \bibinfo {author} {\bibfnamefont
  {M.}~\bibnamefont {Steffen}},\ }\href@noop {} {\bibfield  {journal} {\bibinfo
   {journal} {Phys. Rev. A}\ }\textbf {\bibinfo {volume} {87}},\ \bibinfo
  {pages} {030301} (\bibinfo {year} {2013})}\BibitemShut {NoStop}%
\bibitem [{\citenamefont {Takita}\ \emph {et~al.}(2016)\citenamefont {Takita},
  \citenamefont {C\'orcoles}, \citenamefont {Magesan}, \citenamefont {Abdo},
  \citenamefont {Brink}, \citenamefont {Cross}, \citenamefont {Chow},\ and\
  \citenamefont {Gambetta}}]{takita16}%
  \BibitemOpen
  \bibfield  {author} {\bibinfo {author} {\bibfnamefont {M.}~\bibnamefont
  {Takita}}, \bibinfo {author} {\bibfnamefont {A.~D.}\ \bibnamefont
  {C\'orcoles}}, \bibinfo {author} {\bibfnamefont {E.}~\bibnamefont {Magesan}},
  \bibinfo {author} {\bibfnamefont {B.}~\bibnamefont {Abdo}}, \bibinfo {author}
  {\bibfnamefont {M.}~\bibnamefont {Brink}}, \bibinfo {author} {\bibfnamefont
  {A.}~\bibnamefont {Cross}}, \bibinfo {author} {\bibfnamefont {J.~M.}\
  \bibnamefont {Chow}},\ and\ \bibinfo {author} {\bibfnamefont {J.~M.}\
  \bibnamefont {Gambetta}},\ }\href@noop {} {\bibfield  {journal} {\bibinfo
  {journal} {Phys. Rev. Lett.}\ }\textbf {\bibinfo {volume} {117}},\ \bibinfo
  {pages} {210505} (\bibinfo {year} {2016})}\BibitemShut {NoStop}%
\bibitem [{\citenamefont {Sheldon}\ \emph {et~al.}(2016)\citenamefont
  {Sheldon}, \citenamefont {Magesan}, \citenamefont {Chow},\ and\ \citenamefont
  {Gambetta}}]{sheldon16}%
  \BibitemOpen
  \bibfield  {author} {\bibinfo {author} {\bibfnamefont {S.}~\bibnamefont
  {Sheldon}}, \bibinfo {author} {\bibfnamefont {E.}~\bibnamefont {Magesan}},
  \bibinfo {author} {\bibfnamefont {J.~M.}\ \bibnamefont {Chow}},\ and\
  \bibinfo {author} {\bibfnamefont {J.~M.}\ \bibnamefont {Gambetta}},\
  }\href@noop {} {\bibfield  {journal} {\bibinfo  {journal} {Phys. Rev. A}\
  }\textbf {\bibinfo {volume} {93}},\ \bibinfo {pages} {060302} (\bibinfo
  {year} {2016})}\BibitemShut {NoStop}%
\bibitem [{\citenamefont {Patterson}\ \emph {et~al.}(2019)\citenamefont
  {Patterson}, \citenamefont {Rahamim}, \citenamefont {Tsunoda}, \citenamefont
  {Spring}, \citenamefont {Jebari}, \citenamefont {Ratter}, \citenamefont
  {Mergenthaler}, \citenamefont {Tancredi}, \citenamefont {Vlastakis},
  \citenamefont {Esposito},\ and\ \citenamefont {Leek}}]{patterson19}%
  \BibitemOpen
  \bibfield  {author} {\bibinfo {author} {\bibfnamefont {A.}~\bibnamefont
  {Patterson}}, \bibinfo {author} {\bibfnamefont {J.}~\bibnamefont {Rahamim}},
  \bibinfo {author} {\bibfnamefont {T.}~\bibnamefont {Tsunoda}}, \bibinfo
  {author} {\bibfnamefont {P.}~\bibnamefont {Spring}}, \bibinfo {author}
  {\bibfnamefont {S.}~\bibnamefont {Jebari}}, \bibinfo {author} {\bibfnamefont
  {K.}~\bibnamefont {Ratter}}, \bibinfo {author} {\bibfnamefont
  {M.}~\bibnamefont {Mergenthaler}}, \bibinfo {author} {\bibfnamefont
  {G.}~\bibnamefont {Tancredi}}, \bibinfo {author} {\bibfnamefont
  {B.}~\bibnamefont {Vlastakis}}, \bibinfo {author} {\bibfnamefont
  {M.}~\bibnamefont {Esposito}},\ and\ \bibinfo {author} {\bibfnamefont
  {P.}~\bibnamefont {Leek}},\ }\href@noop {} {\bibfield  {journal} {\bibinfo
  {journal} {Phys. Rev. Appl.}\ }\textbf {\bibinfo {volume} {12}},\ \bibinfo
  {pages} {064013} (\bibinfo {year} {2019})}\BibitemShut {NoStop}%
\bibitem [{\citenamefont {Hazra}\ \emph {et~al.}(2020)\citenamefont {Hazra},
  \citenamefont {Salunkhe}, \citenamefont {Bhattacharjee}, \citenamefont
  {Bothara}, \citenamefont {Kundu}, \citenamefont {Roy}, \citenamefont
  {Patankar},\ and\ \citenamefont {Vijay}}]{hazra20}%
  \BibitemOpen
  \bibfield  {author} {\bibinfo {author} {\bibfnamefont {S.}~\bibnamefont
  {Hazra}}, \bibinfo {author} {\bibfnamefont {K.~V.}\ \bibnamefont {Salunkhe}},
  \bibinfo {author} {\bibfnamefont {A.}~\bibnamefont {Bhattacharjee}}, \bibinfo
  {author} {\bibfnamefont {G.}~\bibnamefont {Bothara}}, \bibinfo {author}
  {\bibfnamefont {S.}~\bibnamefont {Kundu}}, \bibinfo {author} {\bibfnamefont
  {T.}~\bibnamefont {Roy}}, \bibinfo {author} {\bibfnamefont {M.~P.}\
  \bibnamefont {Patankar}},\ and\ \bibinfo {author} {\bibfnamefont
  {R.}~\bibnamefont {Vijay}},\ }\href@noop {} {\bibfield  {journal} {\bibinfo
  {journal} {Appl. Phys. Lett.}\ }\textbf {\bibinfo {volume} {116}},\ \bibinfo
  {pages} {152601} (\bibinfo {year} {2020})}\BibitemShut {NoStop}%
\bibitem [{\citenamefont {Dogan}\ \emph {et~al.}(2023)\citenamefont {Dogan},
  \citenamefont {Rosenstock}, \citenamefont {Guevel}, \citenamefont {Xiong},
  \citenamefont {Mencia}, \citenamefont {Somoroff}, \citenamefont {Nesterov},
  \citenamefont {Vavilov}, \citenamefont {Manucharyan},\ and\ \citenamefont
  {Wang}}]{dogan22}%
  \BibitemOpen
  \bibfield  {author} {\bibinfo {author} {\bibfnamefont {E.}~\bibnamefont
  {Dogan}}, \bibinfo {author} {\bibfnamefont {D.}~\bibnamefont {Rosenstock}},
  \bibinfo {author} {\bibfnamefont {L.~L.}\ \bibnamefont {Guevel}}, \bibinfo
  {author} {\bibfnamefont {H.}~\bibnamefont {Xiong}}, \bibinfo {author}
  {\bibfnamefont {R.~A.}\ \bibnamefont {Mencia}}, \bibinfo {author}
  {\bibfnamefont {A.}~\bibnamefont {Somoroff}}, \bibinfo {author}
  {\bibfnamefont {K.~N.}\ \bibnamefont {Nesterov}}, \bibinfo {author}
  {\bibfnamefont {M.~G.}\ \bibnamefont {Vavilov}}, \bibinfo {author}
  {\bibfnamefont {V.~E.}\ \bibnamefont {Manucharyan}},\ and\ \bibinfo {author}
  {\bibfnamefont {C.}~\bibnamefont {Wang}},\ }\href@noop {} {\bibfield
  {journal} {\bibinfo  {journal} {Phys. Rev. Appl.}\ }\textbf {\bibinfo
  {volume} {20}} (\bibinfo {year} {2023})}\BibitemShut {NoStop}%
\bibitem [{\citenamefont {de~Groot}\ \emph {et~al.}(2012)\citenamefont
  {de~Groot}, \citenamefont {Ashhab}, \citenamefont {Lupa{\c s}cu},
  \citenamefont {DiCarlo}, \citenamefont {Nori}, \citenamefont {Harmans},\ and\
  \citenamefont {Mooij}}]{degroot12}%
  \BibitemOpen
  \bibfield  {author} {\bibinfo {author} {\bibfnamefont {P.~C.}\ \bibnamefont
  {de~Groot}}, \bibinfo {author} {\bibfnamefont {S.}~\bibnamefont {Ashhab}},
  \bibinfo {author} {\bibfnamefont {A.}~\bibnamefont {Lupa{\c s}cu}}, \bibinfo
  {author} {\bibfnamefont {L.}~\bibnamefont {DiCarlo}}, \bibinfo {author}
  {\bibfnamefont {F.}~\bibnamefont {Nori}}, \bibinfo {author} {\bibfnamefont
  {C.~J. P.~M.}\ \bibnamefont {Harmans}},\ and\ \bibinfo {author}
  {\bibfnamefont {J.~E.}\ \bibnamefont {Mooij}},\ }\href@noop {} {\bibfield
  {journal} {\bibinfo  {journal} {New J. Phys.}\ }\textbf {\bibinfo {volume}
  {14}},\ \bibinfo {pages} {073038} (\bibinfo {year} {2012})}\BibitemShut
  {NoStop}%
\bibitem [{\citenamefont {Kirchhoff}\ \emph {et~al.}(2018)\citenamefont
  {Kirchhoff}, \citenamefont {Ke\ss{}ler}, \citenamefont {Liebermann},
  \citenamefont {Ass\'emat}, \citenamefont {Machnes}, \citenamefont {Motzoi},\
  and\ \citenamefont {Wilhelm}}]{kirchhoff18}%
  \BibitemOpen
  \bibfield  {author} {\bibinfo {author} {\bibfnamefont {S.}~\bibnamefont
  {Kirchhoff}}, \bibinfo {author} {\bibfnamefont {T.}~\bibnamefont
  {Ke\ss{}ler}}, \bibinfo {author} {\bibfnamefont {P.~J.}\ \bibnamefont
  {Liebermann}}, \bibinfo {author} {\bibfnamefont {E.}~\bibnamefont
  {Ass\'emat}}, \bibinfo {author} {\bibfnamefont {S.}~\bibnamefont {Machnes}},
  \bibinfo {author} {\bibfnamefont {F.}~\bibnamefont {Motzoi}},\ and\ \bibinfo
  {author} {\bibfnamefont {F.~K.}\ \bibnamefont {Wilhelm}},\ }\href@noop {}
  {\bibfield  {journal} {\bibinfo  {journal} {Phys. Rev. A}\ }\textbf {\bibinfo
  {volume} {97}},\ \bibinfo {pages} {042348} (\bibinfo {year}
  {2018})}\BibitemShut {NoStop}%
\bibitem [{\citenamefont {Tripathi}\ \emph {et~al.}(2019)\citenamefont
  {Tripathi}, \citenamefont {Khezri},\ and\ \citenamefont
  {Korotkov}}]{tripathi19}%
  \BibitemOpen
  \bibfield  {author} {\bibinfo {author} {\bibfnamefont {V.}~\bibnamefont
  {Tripathi}}, \bibinfo {author} {\bibfnamefont {M.}~\bibnamefont {Khezri}},\
  and\ \bibinfo {author} {\bibfnamefont {A.~N.}\ \bibnamefont {Korotkov}},\
  }\href@noop {} {\bibfield  {journal} {\bibinfo  {journal} {Phys. Rev. A}\
  }\textbf {\bibinfo {volume} {100}},\ \bibinfo {pages} {012301} (\bibinfo
  {year} {2019})}\BibitemShut {NoStop}%
\bibitem [{\citenamefont {Magesan}\ and\ \citenamefont
  {Gambetta}(2020)}]{magesan20}%
  \BibitemOpen
  \bibfield  {author} {\bibinfo {author} {\bibfnamefont {E.}~\bibnamefont
  {Magesan}}\ and\ \bibinfo {author} {\bibfnamefont {J.~M.}\ \bibnamefont
  {Gambetta}},\ }\href@noop {} {\bibfield  {journal} {\bibinfo  {journal}
  {Phys. Rev. A}\ }\textbf {\bibinfo {volume} {101}},\ \bibinfo {pages}
  {052308} (\bibinfo {year} {2020})}\BibitemShut {NoStop}%
\bibitem [{\citenamefont {Sundaresan}\ \emph {et~al.}(2020)\citenamefont
  {Sundaresan}, \citenamefont {Lauer}, \citenamefont {Pritchett}, \citenamefont
  {Magesan}, \citenamefont {Jurcevic},\ and\ \citenamefont
  {Gambetta}}]{sundaresan20}%
  \BibitemOpen
  \bibfield  {author} {\bibinfo {author} {\bibfnamefont {N.}~\bibnamefont
  {Sundaresan}}, \bibinfo {author} {\bibfnamefont {I.}~\bibnamefont {Lauer}},
  \bibinfo {author} {\bibfnamefont {E.}~\bibnamefont {Pritchett}}, \bibinfo
  {author} {\bibfnamefont {E.}~\bibnamefont {Magesan}}, \bibinfo {author}
  {\bibfnamefont {P.}~\bibnamefont {Jurcevic}},\ and\ \bibinfo {author}
  {\bibfnamefont {J.~M.}\ \bibnamefont {Gambetta}},\ }\href@noop {} {\bibfield
  {journal} {\bibinfo  {journal} {PRX Quantum}\ }\textbf {\bibinfo {volume}
  {1}},\ \bibinfo {pages} {020318} (\bibinfo {year} {2020})}\BibitemShut
  {NoStop}%
\bibitem [{\citenamefont {Malekakhlagh}\ \emph {et~al.}(2020)\citenamefont
  {Malekakhlagh}, \citenamefont {Magesan},\ and\ \citenamefont
  {McKay}}]{malekakhlagh20}%
  \BibitemOpen
  \bibfield  {author} {\bibinfo {author} {\bibfnamefont {M.}~\bibnamefont
  {Malekakhlagh}}, \bibinfo {author} {\bibfnamefont {E.}~\bibnamefont
  {Magesan}},\ and\ \bibinfo {author} {\bibfnamefont {D.~C.}\ \bibnamefont
  {McKay}},\ }\href@noop {} {\bibfield  {journal} {\bibinfo  {journal} {Phys.
  Rev. A}\ }\textbf {\bibinfo {volume} {102}},\ \bibinfo {pages} {042605}
  (\bibinfo {year} {2020})}\BibitemShut {NoStop}%
\bibitem [{\citenamefont {Malekakhlagh}\ and\ \citenamefont
  {Magesan}(2022)}]{malekakhlagh22}%
  \BibitemOpen
  \bibfield  {author} {\bibinfo {author} {\bibfnamefont {M.}~\bibnamefont
  {Malekakhlagh}}\ and\ \bibinfo {author} {\bibfnamefont {E.}~\bibnamefont
  {Magesan}},\ }\href@noop {} {\bibfield  {journal} {\bibinfo  {journal} {Phys.
  Rev. A}\ }\textbf {\bibinfo {volume} {105}},\ \bibinfo {pages} {012602}
  (\bibinfo {year} {2022})}\BibitemShut {NoStop}%
\bibitem [{\citenamefont {Nesterov}\ \emph {et~al.}(2022)\citenamefont
  {Nesterov}, \citenamefont {Wang}, \citenamefont {Manucharyan},\ and\
  \citenamefont {Vavilov}}]{nesterov22}%
  \BibitemOpen
  \bibfield  {author} {\bibinfo {author} {\bibfnamefont {K.~N.}\ \bibnamefont
  {Nesterov}}, \bibinfo {author} {\bibfnamefont {C.}~\bibnamefont {Wang}},
  \bibinfo {author} {\bibfnamefont {V.~E.}\ \bibnamefont {Manucharyan}},\ and\
  \bibinfo {author} {\bibfnamefont {M.~G.}\ \bibnamefont {Vavilov}},\
  }\href@noop {} {\bibfield  {journal} {\bibinfo  {journal} {Phys. Rev. Appl.}\
  }\textbf {\bibinfo {volume} {18}},\ \bibinfo {pages} {034063} (\bibinfo
  {year} {2022})}\BibitemShut {NoStop}%
\bibitem [{\citenamefont {Heya}\ and\ \citenamefont {Kanazawa}(2021)}]{heya21}%
  \BibitemOpen
  \bibfield  {author} {\bibinfo {author} {\bibfnamefont {K.}~\bibnamefont
  {Heya}}\ and\ \bibinfo {author} {\bibfnamefont {N.}~\bibnamefont
  {Kanazawa}},\ }\href@noop {} {\bibfield  {journal} {\bibinfo  {journal} {PRX
  Quantum}\ }\textbf {\bibinfo {volume} {2}},\ \bibinfo {pages} {040336}
  (\bibinfo {year} {2021})}\BibitemShut {NoStop}%
\bibitem [{\citenamefont {Petrescu}\ \emph {et~al.}(2023)\citenamefont
  {Petrescu}, \citenamefont {Calonnec}, \citenamefont {Leroux}, \citenamefont
  {Paolo}, \citenamefont {Mundada}, \citenamefont {Sussman}, \citenamefont
  {Vrajitoarea}, \citenamefont {Houck},\ and\ \citenamefont
  {Blais}}]{petrescu21}%
  \BibitemOpen
  \bibfield  {author} {\bibinfo {author} {\bibfnamefont {A.}~\bibnamefont
  {Petrescu}}, \bibinfo {author} {\bibfnamefont {C.~L.}\ \bibnamefont
  {Calonnec}}, \bibinfo {author} {\bibfnamefont {C.}~\bibnamefont {Leroux}},
  \bibinfo {author} {\bibfnamefont {A.~D.}\ \bibnamefont {Paolo}}, \bibinfo
  {author} {\bibfnamefont {P.}~\bibnamefont {Mundada}}, \bibinfo {author}
  {\bibfnamefont {S.}~\bibnamefont {Sussman}}, \bibinfo {author} {\bibfnamefont
  {A.}~\bibnamefont {Vrajitoarea}}, \bibinfo {author} {\bibfnamefont {A.~A.}\
  \bibnamefont {Houck}},\ and\ \bibinfo {author} {\bibfnamefont
  {A.}~\bibnamefont {Blais}},\ }\href@noop {} {\bibfield  {journal} {\bibinfo
  {journal} {Phys. Rev. Appl.}\ }\textbf {\bibinfo {volume} {19}},\ \bibinfo
  {pages} {044003} (\bibinfo {year} {2023})}\BibitemShut {NoStop}%
\bibitem [{\citenamefont {Sorensen}\ and\ \citenamefont
  {Molmer}(2000)}]{sorensen00}%
  \BibitemOpen
  \bibfield  {author} {\bibinfo {author} {\bibfnamefont {A.}~\bibnamefont
  {Sorensen}}\ and\ \bibinfo {author} {\bibfnamefont {K.}~\bibnamefont
  {Molmer}},\ }\href@noop {} {\bibfield  {journal} {\bibinfo  {journal} {Phys.
  Rev. A}\ }\textbf {\bibinfo {volume} {62}},\ \bibinfo {pages} {022311}
  (\bibinfo {year} {2000})}\BibitemShut {NoStop}%
\bibitem [{\citenamefont {Breuckmann}(2021)}]{breuckmann21}%
  \BibitemOpen
  \bibfield  {author} {\bibinfo {author} {\bibfnamefont {N.~P.}\ \bibnamefont
  {Breuckmann}},\ }\href@noop {} {\bibfield  {journal} {\bibinfo  {journal}
  {PRX Quantum}\ }\textbf {\bibinfo {volume} {2}},\ \bibinfo {pages} {040101}
  (\bibinfo {year} {2021})}\BibitemShut {NoStop}%
\bibitem [{\citenamefont {Gottesman}(2014)}]{gottesman14}%
  \BibitemOpen
  \bibfield  {author} {\bibinfo {author} {\bibfnamefont {D.}~\bibnamefont
  {Gottesman}},\ }\href@noop {} {\bibfield  {journal} {\bibinfo  {journal}
  {Quantum Info. Comput.}\ }\textbf {\bibinfo {volume} {14}},\ \bibinfo {pages}
  {1338} (\bibinfo {year} {2014})}\BibitemShut {NoStop}%
\bibitem [{\citenamefont {Tremblay}\ \emph {et~al.}(2022)\citenamefont
  {Tremblay}, \citenamefont {Delfosse},\ and\ \citenamefont
  {Beverland}}]{tremblay22}%
  \BibitemOpen
  \bibfield  {author} {\bibinfo {author} {\bibfnamefont {M.~A.}\ \bibnamefont
  {Tremblay}}, \bibinfo {author} {\bibfnamefont {N.}~\bibnamefont {Delfosse}},\
  and\ \bibinfo {author} {\bibfnamefont {M.~E.}\ \bibnamefont {Beverland}},\
  }\href@noop {} {\bibfield  {journal} {\bibinfo  {journal} {Phys. Rev. Lett.}\
  }\textbf {\bibinfo {volume} {129}},\ \bibinfo {pages} {050504} (\bibinfo
  {year} {2022})}\BibitemShut {NoStop}%
\bibitem [{\citenamefont {Panteleev}\ and\ \citenamefont
  {Kalachev}(2022)}]{panteleev22}%
  \BibitemOpen
  \bibfield  {author} {\bibinfo {author} {\bibfnamefont {P.}~\bibnamefont
  {Panteleev}}\ and\ \bibinfo {author} {\bibfnamefont {G.}~\bibnamefont
  {Kalachev}},\ }in\ \href@noop {} {\emph {\bibinfo {booktitle} {Proceedings of
  the 54th {{Annual ACM SIGACT Symposium}} on {{Theory}} of {{Computing}}}}}\
  (\bibinfo  {publisher} {ACM},\ \bibinfo {address} {Rome Italy},\ \bibinfo
  {year} {2022})\ pp.\ \bibinfo {pages} {375--388}\BibitemShut {NoStop}%
\bibitem [{\citenamefont {Leverrier}\ and\ \citenamefont
  {Zemor}(2022)}]{leverrier22a}%
  \BibitemOpen
  \bibfield  {author} {\bibinfo {author} {\bibfnamefont {A.}~\bibnamefont
  {Leverrier}}\ and\ \bibinfo {author} {\bibfnamefont {G.}~\bibnamefont
  {Zemor}},\ }in\ \href@noop {} {\emph {\bibinfo {booktitle} {2022 {{IEEE}}
  63rd {{Annual Symposium}} on {{Foundations}} of {{Computer Science}}
  ({{FOCS}})}}}\ (\bibinfo  {publisher} {IEEE},\ \bibinfo {address} {Denver,
  CO, USA},\ \bibinfo {year} {2022})\ pp.\ \bibinfo {pages}
  {872--883}\BibitemShut {NoStop}%
\bibitem [{\citenamefont {Bravyi}\ \emph {et~al.}(2024)\citenamefont {Bravyi},
  \citenamefont {Cross}, \citenamefont {Gambetta}, \citenamefont {Maslov},
  \citenamefont {Rall},\ and\ \citenamefont {Yoder}}]{bravyi24b}%
  \BibitemOpen
  \bibfield  {author} {\bibinfo {author} {\bibfnamefont {S.}~\bibnamefont
  {Bravyi}}, \bibinfo {author} {\bibfnamefont {A.~W.}\ \bibnamefont {Cross}},
  \bibinfo {author} {\bibfnamefont {J.~M.}\ \bibnamefont {Gambetta}}, \bibinfo
  {author} {\bibfnamefont {D.}~\bibnamefont {Maslov}}, \bibinfo {author}
  {\bibfnamefont {P.}~\bibnamefont {Rall}},\ and\ \bibinfo {author}
  {\bibfnamefont {T.~J.}\ \bibnamefont {Yoder}},\ }\href@noop {} {\bibfield
  {journal} {\bibinfo  {journal} {Nature (London)}\ }\textbf {\bibinfo {volume}
  {627}},\ \bibinfo {pages} {778} (\bibinfo {year} {2024})}\BibitemShut
  {NoStop}%
\bibitem [{\citenamefont {Cain}\ \emph {et~al.}(2026)\citenamefont {Cain},
  \citenamefont {Xu}, \citenamefont {King}, \citenamefont {Picard},
  \citenamefont {Levine}, \citenamefont {Endres}, \citenamefont {Preskill},
  \citenamefont {Huang},\ and\ \citenamefont {Bluvstein}}]{cain2026shor}%
  \BibitemOpen
  \bibfield  {author} {\bibinfo {author} {\bibfnamefont {M.}~\bibnamefont
  {Cain}}, \bibinfo {author} {\bibfnamefont {Q.}~\bibnamefont {Xu}}, \bibinfo
  {author} {\bibfnamefont {R.}~\bibnamefont {King}}, \bibinfo {author}
  {\bibfnamefont {L.~R.}\ \bibnamefont {Picard}}, \bibinfo {author}
  {\bibfnamefont {H.}~\bibnamefont {Levine}}, \bibinfo {author} {\bibfnamefont
  {M.}~\bibnamefont {Endres}}, \bibinfo {author} {\bibfnamefont
  {J.}~\bibnamefont {Preskill}}, \bibinfo {author} {\bibfnamefont {H.-Y.}\
  \bibnamefont {Huang}},\ and\ \bibinfo {author} {\bibfnamefont
  {D.}~\bibnamefont {Bluvstein}},\ }\href@noop {} {\bibfield  {journal}
  {\bibinfo  {journal} {arXiv:2603.28627}\ } (\bibinfo {year}
  {2026})}\BibitemShut {NoStop}%
\bibitem [{\citenamefont {Zhao}\ \emph {et~al.}(2026)\citenamefont {Zhao},
  \citenamefont {Duckering}, \citenamefont {Gu}, \citenamefont {Maskara},\ and\
  \citenamefont {Zhou}}]{zhao2026towards}%
  \BibitemOpen
  \bibfield  {author} {\bibinfo {author} {\bibfnamefont {C.}~\bibnamefont
  {Zhao}}, \bibinfo {author} {\bibfnamefont {C.}~\bibnamefont {Duckering}},
  \bibinfo {author} {\bibfnamefont {A.}~\bibnamefont {Gu}}, \bibinfo {author}
  {\bibfnamefont {N.}~\bibnamefont {Maskara}},\ and\ \bibinfo {author}
  {\bibfnamefont {H.}~\bibnamefont {Zhou}},\ }\href@noop {} {\bibfield
  {journal} {\bibinfo  {journal} {arXiv:2604.16209}\ } (\bibinfo {year}
  {2026})}\BibitemShut {NoStop}%
\bibitem [{\citenamefont {Kasai}(2026)}]{kasai2026breaking}%
  \BibitemOpen
  \bibfield  {author} {\bibinfo {author} {\bibfnamefont {K.}~\bibnamefont
  {Kasai}},\ }\href@noop {} {\bibfield  {journal} {\bibinfo  {journal}
  {arXiv:2601.08824}\ } (\bibinfo {year} {2026})}\BibitemShut {NoStop}%
\bibitem [{\citenamefont {Wang}\ \emph {et~al.}(2026)\citenamefont {Wang},
  \citenamefont {Lu}, \citenamefont {Zhang}, \citenamefont {Liu}, \citenamefont
  {Chen}, \citenamefont {Wang}, \citenamefont {Wu}, \citenamefont {Xu},
  \citenamefont {Zhu}, \citenamefont {Jin} \emph
  {et~al.}}]{wang2026demonstration}%
  \BibitemOpen
  \bibfield  {author} {\bibinfo {author} {\bibfnamefont {K.}~\bibnamefont
  {Wang}}, \bibinfo {author} {\bibfnamefont {Z.}~\bibnamefont {Lu}}, \bibinfo
  {author} {\bibfnamefont {C.}~\bibnamefont {Zhang}}, \bibinfo {author}
  {\bibfnamefont {G.}~\bibnamefont {Liu}}, \bibinfo {author} {\bibfnamefont
  {J.}~\bibnamefont {Chen}}, \bibinfo {author} {\bibfnamefont {Y.}~\bibnamefont
  {Wang}}, \bibinfo {author} {\bibfnamefont {Y.}~\bibnamefont {Wu}}, \bibinfo
  {author} {\bibfnamefont {S.}~\bibnamefont {Xu}}, \bibinfo {author}
  {\bibfnamefont {X.}~\bibnamefont {Zhu}}, \bibinfo {author} {\bibfnamefont
  {F.}~\bibnamefont {Jin}}, \emph {et~al.},\ }\href@noop {} {\bibfield
  {journal} {\bibinfo  {journal} {Nat. Phys.}\ }\textbf {\bibinfo {volume}
  {22}},\ \bibinfo {pages} {308} (\bibinfo {year} {2026})}\BibitemShut
  {NoStop}%
\bibitem [{\citenamefont {Norris}\ \emph {et~al.}(2025)\citenamefont {Norris},
  \citenamefont {Dalton}, \citenamefont {Zanuz}, \citenamefont {Rommens},
  \citenamefont {Flasby}, \citenamefont {Panah}, \citenamefont {Swiadek},
  \citenamefont {Scarato}, \citenamefont {Hellings}, \citenamefont {Besse}
  \emph {et~al.}}]{norris2025performance}%
  \BibitemOpen
  \bibfield  {author} {\bibinfo {author} {\bibfnamefont {G.~J.}\ \bibnamefont
  {Norris}}, \bibinfo {author} {\bibfnamefont {K.}~\bibnamefont {Dalton}},
  \bibinfo {author} {\bibfnamefont {D.~C.}\ \bibnamefont {Zanuz}}, \bibinfo
  {author} {\bibfnamefont {A.}~\bibnamefont {Rommens}}, \bibinfo {author}
  {\bibfnamefont {A.}~\bibnamefont {Flasby}}, \bibinfo {author} {\bibfnamefont
  {M.~B.}\ \bibnamefont {Panah}}, \bibinfo {author} {\bibfnamefont
  {F.}~\bibnamefont {Swiadek}}, \bibinfo {author} {\bibfnamefont
  {C.}~\bibnamefont {Scarato}}, \bibinfo {author} {\bibfnamefont
  {C.}~\bibnamefont {Hellings}}, \bibinfo {author} {\bibfnamefont {J.-C.}\
  \bibnamefont {Besse}}, \emph {et~al.},\ }\href@noop {} {\bibfield  {journal}
  {\bibinfo  {journal} {arXiv:2503.12603}\ } (\bibinfo {year}
  {2025})}\BibitemShut {NoStop}%
\bibitem [{\citenamefont {Smith}\ \emph {et~al.}(2022)\citenamefont {Smith},
  \citenamefont {Ravi}, \citenamefont {Baker},\ and\ \citenamefont
  {Chong}}]{smith2022scaling}%
  \BibitemOpen
  \bibfield  {author} {\bibinfo {author} {\bibfnamefont {K.~N.}\ \bibnamefont
  {Smith}}, \bibinfo {author} {\bibfnamefont {G.~S.}\ \bibnamefont {Ravi}},
  \bibinfo {author} {\bibfnamefont {J.~M.}\ \bibnamefont {Baker}},\ and\
  \bibinfo {author} {\bibfnamefont {F.~T.}\ \bibnamefont {Chong}},\ }in\
  \href@noop {} {\emph {\bibinfo {booktitle} {2022 55th IEEE/ACM International
  Symposium on Microarchitecture (MICRO)}}}\ (\bibinfo {organization} {IEEE},\
  \bibinfo {year} {2022})\ pp.\ \bibinfo {pages} {1092--1109}\BibitemShut
  {NoStop}%
\bibitem [{\citenamefont {Gold}\ \emph {et~al.}(2021)\citenamefont {Gold},
  \citenamefont {Paquette}, \citenamefont {Stockklauser}, \citenamefont
  {Reagor}, \citenamefont {Alam}, \citenamefont {Bestwick}, \citenamefont
  {Didier}, \citenamefont {Nersisyan}, \citenamefont {Oruc}, \citenamefont
  {Razavi} \emph {et~al.}}]{gold2021entanglement}%
  \BibitemOpen
  \bibfield  {author} {\bibinfo {author} {\bibfnamefont {A.}~\bibnamefont
  {Gold}}, \bibinfo {author} {\bibfnamefont {J.}~\bibnamefont {Paquette}},
  \bibinfo {author} {\bibfnamefont {A.}~\bibnamefont {Stockklauser}}, \bibinfo
  {author} {\bibfnamefont {M.~J.}\ \bibnamefont {Reagor}}, \bibinfo {author}
  {\bibfnamefont {M.~S.}\ \bibnamefont {Alam}}, \bibinfo {author}
  {\bibfnamefont {A.}~\bibnamefont {Bestwick}}, \bibinfo {author}
  {\bibfnamefont {N.}~\bibnamefont {Didier}}, \bibinfo {author} {\bibfnamefont
  {A.}~\bibnamefont {Nersisyan}}, \bibinfo {author} {\bibfnamefont
  {F.}~\bibnamefont {Oruc}}, \bibinfo {author} {\bibfnamefont {A.}~\bibnamefont
  {Razavi}}, \emph {et~al.},\ }\href@noop {} {\bibfield  {journal} {\bibinfo
  {journal} {npj Quantum Inf.}\ }\textbf {\bibinfo {volume} {7}},\ \bibinfo
  {pages} {142} (\bibinfo {year} {2021})}\BibitemShut {NoStop}%
\bibitem [{\citenamefont {Field}\ \emph {et~al.}(2024)\citenamefont {Field},
  \citenamefont {Chen}, \citenamefont {Scharmann}, \citenamefont {Sete},
  \citenamefont {Oruc}, \citenamefont {Vu}, \citenamefont {Kosenko},
  \citenamefont {Mutus}, \citenamefont {Poletto},\ and\ \citenamefont
  {Bestwick}}]{field2024modular}%
  \BibitemOpen
  \bibfield  {author} {\bibinfo {author} {\bibfnamefont {M.}~\bibnamefont
  {Field}}, \bibinfo {author} {\bibfnamefont {A.~Q.}\ \bibnamefont {Chen}},
  \bibinfo {author} {\bibfnamefont {B.}~\bibnamefont {Scharmann}}, \bibinfo
  {author} {\bibfnamefont {E.~A.}\ \bibnamefont {Sete}}, \bibinfo {author}
  {\bibfnamefont {F.}~\bibnamefont {Oruc}}, \bibinfo {author} {\bibfnamefont
  {K.}~\bibnamefont {Vu}}, \bibinfo {author} {\bibfnamefont {V.}~\bibnamefont
  {Kosenko}}, \bibinfo {author} {\bibfnamefont {J.~Y.}\ \bibnamefont {Mutus}},
  \bibinfo {author} {\bibfnamefont {S.}~\bibnamefont {Poletto}},\ and\ \bibinfo
  {author} {\bibfnamefont {A.}~\bibnamefont {Bestwick}},\ }\href@noop {}
  {\bibfield  {journal} {\bibinfo  {journal} {Phys. Rev. Appl.}\ }\textbf
  {\bibinfo {volume} {21}},\ \bibinfo {pages} {054063} (\bibinfo {year}
  {2024})}\BibitemShut {NoStop}%
\bibitem [{\citenamefont {Deng}\ \emph {et~al.}(2025)\citenamefont {Deng},
  \citenamefont {Zheng}, \citenamefont {Liao}, \citenamefont {Zhou},
  \citenamefont {Ge}, \citenamefont {Zhao}, \citenamefont {Lan}, \citenamefont
  {Tan}, \citenamefont {Zhang}, \citenamefont {Li} \emph
  {et~al.}}]{deng2025long}%
  \BibitemOpen
  \bibfield  {author} {\bibinfo {author} {\bibfnamefont {X.}~\bibnamefont
  {Deng}}, \bibinfo {author} {\bibfnamefont {W.}~\bibnamefont {Zheng}},
  \bibinfo {author} {\bibfnamefont {X.}~\bibnamefont {Liao}}, \bibinfo {author}
  {\bibfnamefont {H.}~\bibnamefont {Zhou}}, \bibinfo {author} {\bibfnamefont
  {Y.}~\bibnamefont {Ge}}, \bibinfo {author} {\bibfnamefont {J.}~\bibnamefont
  {Zhao}}, \bibinfo {author} {\bibfnamefont {D.}~\bibnamefont {Lan}}, \bibinfo
  {author} {\bibfnamefont {X.}~\bibnamefont {Tan}}, \bibinfo {author}
  {\bibfnamefont {Y.}~\bibnamefont {Zhang}}, \bibinfo {author} {\bibfnamefont
  {S.}~\bibnamefont {Li}}, \emph {et~al.},\ }\href@noop {} {\bibfield
  {journal} {\bibinfo  {journal} {Phys. Rev. Lett.}\ }\textbf {\bibinfo
  {volume} {134}},\ \bibinfo {pages} {020801} (\bibinfo {year}
  {2025})}\BibitemShut {NoStop}%
\bibitem [{\citenamefont {Wu}\ \emph {et~al.}(2024)\citenamefont {Wu},
  \citenamefont {Yan}, \citenamefont {Andersson}, \citenamefont {Anferov},
  \citenamefont {Chou}, \citenamefont {Conner}, \citenamefont {Grebel},
  \citenamefont {Joshi}, \citenamefont {Li}, \citenamefont {Miller} \emph
  {et~al.}}]{wu2024modular}%
  \BibitemOpen
  \bibfield  {author} {\bibinfo {author} {\bibfnamefont {X.}~\bibnamefont
  {Wu}}, \bibinfo {author} {\bibfnamefont {H.}~\bibnamefont {Yan}}, \bibinfo
  {author} {\bibfnamefont {G.}~\bibnamefont {Andersson}}, \bibinfo {author}
  {\bibfnamefont {A.}~\bibnamefont {Anferov}}, \bibinfo {author} {\bibfnamefont
  {M.-H.}\ \bibnamefont {Chou}}, \bibinfo {author} {\bibfnamefont {C.~R.}\
  \bibnamefont {Conner}}, \bibinfo {author} {\bibfnamefont {J.}~\bibnamefont
  {Grebel}}, \bibinfo {author} {\bibfnamefont {Y.~J.}\ \bibnamefont {Joshi}},
  \bibinfo {author} {\bibfnamefont {S.}~\bibnamefont {Li}}, \bibinfo {author}
  {\bibfnamefont {J.~M.}\ \bibnamefont {Miller}}, \emph {et~al.},\ }\href@noop
  {} {\bibfield  {journal} {\bibinfo  {journal} {Phys. Rev. X}\ }\textbf
  {\bibinfo {volume} {14}},\ \bibinfo {pages} {041030} (\bibinfo {year}
  {2024})}\BibitemShut {NoStop}%
\bibitem [{\citenamefont {Song}\ \emph {et~al.}(2024)\citenamefont {Song},
  \citenamefont {Yang}, \citenamefont {Liu}, \citenamefont {Zhang},
  \citenamefont {Xue}, \citenamefont {Mi}, \citenamefont {Zhang}, \citenamefont
  {Yan}, \citenamefont {Jin},\ and\ \citenamefont {Yu}}]{song2024realization}%
  \BibitemOpen
  \bibfield  {author} {\bibinfo {author} {\bibfnamefont {J.}~\bibnamefont
  {Song}}, \bibinfo {author} {\bibfnamefont {S.}~\bibnamefont {Yang}}, \bibinfo
  {author} {\bibfnamefont {P.}~\bibnamefont {Liu}}, \bibinfo {author}
  {\bibfnamefont {H.-L.}\ \bibnamefont {Zhang}}, \bibinfo {author}
  {\bibfnamefont {G.-M.}\ \bibnamefont {Xue}}, \bibinfo {author} {\bibfnamefont
  {Z.-Y.}\ \bibnamefont {Mi}}, \bibinfo {author} {\bibfnamefont {W.-G.}\
  \bibnamefont {Zhang}}, \bibinfo {author} {\bibfnamefont {F.}~\bibnamefont
  {Yan}}, \bibinfo {author} {\bibfnamefont {Y.-R.}\ \bibnamefont {Jin}},\ and\
  \bibinfo {author} {\bibfnamefont {H.-F.}\ \bibnamefont {Yu}},\ }\href@noop {}
  {\bibfield  {journal} {\bibinfo  {journal} {arXiv:2407.20338}\ } (\bibinfo
  {year} {2024})}\BibitemShut {NoStop}%
\bibitem [{\citenamefont {Touzard}\ \emph {et~al.}(2019)\citenamefont
  {Touzard}, \citenamefont {Kou}, \citenamefont {Frattini}, \citenamefont
  {Sivak}, \citenamefont {Puri}, \citenamefont {Grimm}, \citenamefont
  {Frunzio}, \citenamefont {Shankar},\ and\ \citenamefont
  {Devoret}}]{touzard19}%
  \BibitemOpen
  \bibfield  {author} {\bibinfo {author} {\bibfnamefont {S.}~\bibnamefont
  {Touzard}}, \bibinfo {author} {\bibfnamefont {A.}~\bibnamefont {Kou}},
  \bibinfo {author} {\bibfnamefont {N.~E.}\ \bibnamefont {Frattini}}, \bibinfo
  {author} {\bibfnamefont {V.~V.}\ \bibnamefont {Sivak}}, \bibinfo {author}
  {\bibfnamefont {S.}~\bibnamefont {Puri}}, \bibinfo {author} {\bibfnamefont
  {A.}~\bibnamefont {Grimm}}, \bibinfo {author} {\bibfnamefont
  {L.}~\bibnamefont {Frunzio}}, \bibinfo {author} {\bibfnamefont
  {S.}~\bibnamefont {Shankar}},\ and\ \bibinfo {author} {\bibfnamefont {M.~H.}\
  \bibnamefont {Devoret}},\ }\href@noop {} {\bibfield  {journal} {\bibinfo
  {journal} {Phys. Rev. Lett.}\ }\textbf {\bibinfo {volume} {122}},\ \bibinfo
  {pages} {080502} (\bibinfo {year} {2019})}\BibitemShut {NoStop}%
\bibitem [{\citenamefont {Solano}\ \emph {et~al.}(2003)\citenamefont {Solano},
  \citenamefont {Agarwal},\ and\ \citenamefont {Walther}}]{solano03}%
  \BibitemOpen
  \bibfield  {author} {\bibinfo {author} {\bibfnamefont {E.}~\bibnamefont
  {Solano}}, \bibinfo {author} {\bibfnamefont {G.~S.}\ \bibnamefont
  {Agarwal}},\ and\ \bibinfo {author} {\bibfnamefont {H.}~\bibnamefont
  {Walther}},\ }\href@noop {} {\bibfield  {journal} {\bibinfo  {journal} {Phys.
  Rev. Lett.}\ }\textbf {\bibinfo {volume} {90}},\ \bibinfo {pages} {027903}
  (\bibinfo {year} {2003})}\BibitemShut {NoStop}%
\bibitem [{\citenamefont {Ayyash}\ \emph
  {et~al.}(2024{\natexlab{a}})\citenamefont {Ayyash}, \citenamefont {Xu},\ and\
  \citenamefont {Mariantoni}}]{ayyash24}%
  \BibitemOpen
  \bibfield  {author} {\bibinfo {author} {\bibfnamefont {M.}~\bibnamefont
  {Ayyash}}, \bibinfo {author} {\bibfnamefont {X.}~\bibnamefont {Xu}},\ and\
  \bibinfo {author} {\bibfnamefont {M.}~\bibnamefont {Mariantoni}},\
  }\href@noop {} {\bibfield  {journal} {\bibinfo  {journal} {Phys. Rev. A}\
  }\textbf {\bibinfo {volume} {109}},\ \bibinfo {pages} {023703} (\bibinfo
  {year} {2024}{\natexlab{a}})}\BibitemShut {NoStop}%
\bibitem [{\citenamefont {Ayyash}\ \emph
  {et~al.}(2024{\natexlab{b}})\citenamefont {Ayyash}, \citenamefont {Xu},
  \citenamefont {Ashhab},\ and\ \citenamefont {Mariantoni}}]{ayyash24a}%
  \BibitemOpen
  \bibfield  {author} {\bibinfo {author} {\bibfnamefont {M.}~\bibnamefont
  {Ayyash}}, \bibinfo {author} {\bibfnamefont {X.}~\bibnamefont {Xu}}, \bibinfo
  {author} {\bibfnamefont {S.}~\bibnamefont {Ashhab}},\ and\ \bibinfo {author}
  {\bibfnamefont {M.}~\bibnamefont {Mariantoni}},\ }\href@noop {} {\bibfield
  {journal} {\bibinfo  {journal} {Phys. Rev. A}\ }\textbf {\bibinfo {volume}
  {110}},\ \bibinfo {pages} {053711} (\bibinfo {year}
  {2024}{\natexlab{b}})}\BibitemShut {NoStop}%
\bibitem [{\citenamefont {Ayyash}(2025)}]{ayyash25}%
  \BibitemOpen
  \bibfield  {author} {\bibinfo {author} {\bibfnamefont {M.}~\bibnamefont
  {Ayyash}},\ }\href@noop {} {\bibfield  {journal} {\bibinfo  {journal}
  {arXiv:2503.15941}\ } (\bibinfo {year} {2025})}\BibitemShut {NoStop}%
\bibitem [{Note1()}]{Note1}%
  \BibitemOpen
  \bibinfo {note} {Technically, $\epsilon $ and $\Delta _j$ are negatives of
  the corresponding detunings since they are negative when the drive is
  blue-detuned. However, throughout the paper, we will abuse the terminology
  and refer to them simply as detunings.}\BibitemShut {Stop}%
\bibitem [{\citenamefont {Bruzewicz}\ \emph {et~al.}(2019)\citenamefont
  {Bruzewicz}, \citenamefont {Chiaverini}, \citenamefont {McConnell},\ and\
  \citenamefont {Sage}}]{bruzewicz19}%
  \BibitemOpen
  \bibfield  {author} {\bibinfo {author} {\bibfnamefont {C.~D.}\ \bibnamefont
  {Bruzewicz}}, \bibinfo {author} {\bibfnamefont {J.}~\bibnamefont
  {Chiaverini}}, \bibinfo {author} {\bibfnamefont {R.}~\bibnamefont
  {McConnell}},\ and\ \bibinfo {author} {\bibfnamefont {J.~M.}\ \bibnamefont
  {Sage}},\ }\href@noop {} {\bibfield  {journal} {\bibinfo  {journal} {Appl.
  Phys. Rev.}\ }\textbf {\bibinfo {volume} {6}},\ \bibinfo {pages} {021314}
  (\bibinfo {year} {2019})}\BibitemShut {NoStop}%
\bibitem [{\citenamefont {Leibfried}\ \emph {et~al.}(2003)\citenamefont
  {Leibfried}, \citenamefont {DeMarco}, \citenamefont {Meyer}, \citenamefont
  {Lucas}, \citenamefont {Barrett}, \citenamefont {Britton}, \citenamefont
  {Itano}, \citenamefont {Jelenkovic}, \citenamefont {Langer}, \citenamefont
  {Rosenband},\ and\ \citenamefont {Wineland}}]{leibfried03}%
  \BibitemOpen
  \bibfield  {author} {\bibinfo {author} {\bibfnamefont {D.}~\bibnamefont
  {Leibfried}}, \bibinfo {author} {\bibfnamefont {B.}~\bibnamefont {DeMarco}},
  \bibinfo {author} {\bibfnamefont {V.}~\bibnamefont {Meyer}}, \bibinfo
  {author} {\bibfnamefont {D.}~\bibnamefont {Lucas}}, \bibinfo {author}
  {\bibfnamefont {M.}~\bibnamefont {Barrett}}, \bibinfo {author} {\bibfnamefont
  {J.}~\bibnamefont {Britton}}, \bibinfo {author} {\bibfnamefont {W.~M.}\
  \bibnamefont {Itano}}, \bibinfo {author} {\bibfnamefont {B.}~\bibnamefont
  {Jelenkovic}}, \bibinfo {author} {\bibfnamefont {C.}~\bibnamefont {Langer}},
  \bibinfo {author} {\bibfnamefont {T.}~\bibnamefont {Rosenband}},\ and\
  \bibinfo {author} {\bibfnamefont {D.~J.}\ \bibnamefont {Wineland}},\
  }\href@noop {} {\bibfield  {journal} {\bibinfo  {journal} {Nature (London)}\
  }\textbf {\bibinfo {volume} {422}},\ \bibinfo {pages} {412} (\bibinfo {year}
  {2003})}\BibitemShut {NoStop}%
\bibitem [{\citenamefont {Kim}\ \emph {et~al.}(2022)\citenamefont {Kim},
  \citenamefont {Morvan}, \citenamefont {Nguyen}, \citenamefont {Naik},
  \citenamefont {J{\"u}nger}, \citenamefont {Chen}, \citenamefont {Kreikebaum},
  \citenamefont {Santiago},\ and\ \citenamefont {Siddiqi}}]{kim22}%
  \BibitemOpen
  \bibfield  {author} {\bibinfo {author} {\bibfnamefont {Y.}~\bibnamefont
  {Kim}}, \bibinfo {author} {\bibfnamefont {A.}~\bibnamefont {Morvan}},
  \bibinfo {author} {\bibfnamefont {L.~B.}\ \bibnamefont {Nguyen}}, \bibinfo
  {author} {\bibfnamefont {R.~K.}\ \bibnamefont {Naik}}, \bibinfo {author}
  {\bibfnamefont {C.}~\bibnamefont {J{\"u}nger}}, \bibinfo {author}
  {\bibfnamefont {L.}~\bibnamefont {Chen}}, \bibinfo {author} {\bibfnamefont
  {J.~M.}\ \bibnamefont {Kreikebaum}}, \bibinfo {author} {\bibfnamefont
  {D.~I.}\ \bibnamefont {Santiago}},\ and\ \bibinfo {author} {\bibfnamefont
  {I.}~\bibnamefont {Siddiqi}},\ }\href@noop {} {\bibfield  {journal} {\bibinfo
   {journal} {Nat. Phys.}\ }\textbf {\bibinfo {volume} {18}},\ \bibinfo {pages}
  {783} (\bibinfo {year} {2022})}\BibitemShut {NoStop}%
\bibitem [{\citenamefont {Jandura}\ \emph {et~al.}(2024)\citenamefont
  {Jandura}, \citenamefont {Srivastava}, \citenamefont {Pecorari},
  \citenamefont {Brennen},\ and\ \citenamefont {Pupillo}}]{jandura24}%
  \BibitemOpen
  \bibfield  {author} {\bibinfo {author} {\bibfnamefont {S.}~\bibnamefont
  {Jandura}}, \bibinfo {author} {\bibfnamefont {V.}~\bibnamefont {Srivastava}},
  \bibinfo {author} {\bibfnamefont {L.}~\bibnamefont {Pecorari}}, \bibinfo
  {author} {\bibfnamefont {G.~K.}\ \bibnamefont {Brennen}},\ and\ \bibinfo
  {author} {\bibfnamefont {G.}~\bibnamefont {Pupillo}},\ }\href@noop {}
  {\bibfield  {journal} {\bibinfo  {journal} {Phys. Rev. A}\ }\textbf {\bibinfo
  {volume} {110}},\ \bibinfo {pages} {062610} (\bibinfo {year}
  {2024})}\BibitemShut {NoStop}%
\bibitem [{\citenamefont {Barends}\ \emph {et~al.}(2019)\citenamefont
  {Barends}, \citenamefont {Quintana}, \citenamefont {Petukhov}, \citenamefont
  {Chen}, \citenamefont {Kafri}, \citenamefont {Kechedzhi}, \citenamefont
  {Collins}, \citenamefont {Naaman}, \citenamefont {Boixo}, \citenamefont
  {Arute}, \citenamefont {Arya}, \citenamefont {Buell}, \citenamefont
  {Burkett}, \citenamefont {Chen}, \citenamefont {Chiaro}, \citenamefont
  {Dunsworth}, \citenamefont {Foxen}, \citenamefont {Fowler}, \citenamefont
  {Gidney}, \citenamefont {Giustina}, \citenamefont {Graff}, \citenamefont
  {Huang}, \citenamefont {Jeffrey}, \citenamefont {Kelly}, \citenamefont
  {Klimov}, \citenamefont {Kostritsa}, \citenamefont {Landhuis}, \citenamefont
  {Lucero}, \citenamefont {McEwen}, \citenamefont {Megrant}, \citenamefont
  {Mi}, \citenamefont {Mutus}, \citenamefont {Neeley}, \citenamefont {Neill},
  \citenamefont {Ostby}, \citenamefont {Roushan}, \citenamefont {Sank},
  \citenamefont {Satzinger}, \citenamefont {Vainsencher}, \citenamefont
  {White}, \citenamefont {Yao}, \citenamefont {Yeh}, \citenamefont {Zalcman},
  \citenamefont {Neven}, \citenamefont {Smelyanskiy},\ and\ \citenamefont
  {Martinis}}]{barends19}%
  \BibitemOpen
  \bibfield  {author} {\bibinfo {author} {\bibfnamefont {R.}~\bibnamefont
  {Barends}}, \bibinfo {author} {\bibfnamefont {C.~M.}\ \bibnamefont
  {Quintana}}, \bibinfo {author} {\bibfnamefont {A.~G.}\ \bibnamefont
  {Petukhov}}, \bibinfo {author} {\bibfnamefont {Y.}~\bibnamefont {Chen}},
  \bibinfo {author} {\bibfnamefont {D.}~\bibnamefont {Kafri}}, \bibinfo
  {author} {\bibfnamefont {K.}~\bibnamefont {Kechedzhi}}, \bibinfo {author}
  {\bibfnamefont {R.}~\bibnamefont {Collins}}, \bibinfo {author} {\bibfnamefont
  {O.}~\bibnamefont {Naaman}}, \bibinfo {author} {\bibfnamefont
  {S.}~\bibnamefont {Boixo}}, \bibinfo {author} {\bibfnamefont
  {F.}~\bibnamefont {Arute}}, \bibinfo {author} {\bibfnamefont
  {K.}~\bibnamefont {Arya}}, \bibinfo {author} {\bibfnamefont {D.}~\bibnamefont
  {Buell}}, \bibinfo {author} {\bibfnamefont {B.}~\bibnamefont {Burkett}},
  \bibinfo {author} {\bibfnamefont {Z.}~\bibnamefont {Chen}}, \bibinfo {author}
  {\bibfnamefont {B.}~\bibnamefont {Chiaro}}, \bibinfo {author} {\bibfnamefont
  {A.}~\bibnamefont {Dunsworth}}, \bibinfo {author} {\bibfnamefont
  {B.}~\bibnamefont {Foxen}}, \bibinfo {author} {\bibfnamefont
  {A.}~\bibnamefont {Fowler}}, \bibinfo {author} {\bibfnamefont
  {C.}~\bibnamefont {Gidney}}, \bibinfo {author} {\bibfnamefont
  {M.}~\bibnamefont {Giustina}}, \bibinfo {author} {\bibfnamefont
  {R.}~\bibnamefont {Graff}}, \bibinfo {author} {\bibfnamefont
  {T.}~\bibnamefont {Huang}}, \bibinfo {author} {\bibfnamefont
  {E.}~\bibnamefont {Jeffrey}}, \bibinfo {author} {\bibfnamefont
  {J.}~\bibnamefont {Kelly}}, \bibinfo {author} {\bibfnamefont {P.~V.}\
  \bibnamefont {Klimov}}, \bibinfo {author} {\bibfnamefont {F.}~\bibnamefont
  {Kostritsa}}, \bibinfo {author} {\bibfnamefont {D.}~\bibnamefont {Landhuis}},
  \bibinfo {author} {\bibfnamefont {E.}~\bibnamefont {Lucero}}, \bibinfo
  {author} {\bibfnamefont {M.}~\bibnamefont {McEwen}}, \bibinfo {author}
  {\bibfnamefont {A.}~\bibnamefont {Megrant}}, \bibinfo {author} {\bibfnamefont
  {X.}~\bibnamefont {Mi}}, \bibinfo {author} {\bibfnamefont {J.}~\bibnamefont
  {Mutus}}, \bibinfo {author} {\bibfnamefont {M.}~\bibnamefont {Neeley}},
  \bibinfo {author} {\bibfnamefont {C.}~\bibnamefont {Neill}}, \bibinfo
  {author} {\bibfnamefont {E.}~\bibnamefont {Ostby}}, \bibinfo {author}
  {\bibfnamefont {P.}~\bibnamefont {Roushan}}, \bibinfo {author} {\bibfnamefont
  {D.}~\bibnamefont {Sank}}, \bibinfo {author} {\bibfnamefont {K.~J.}\
  \bibnamefont {Satzinger}}, \bibinfo {author} {\bibfnamefont {A.}~\bibnamefont
  {Vainsencher}}, \bibinfo {author} {\bibfnamefont {T.}~\bibnamefont {White}},
  \bibinfo {author} {\bibfnamefont {J.}~\bibnamefont {Yao}}, \bibinfo {author}
  {\bibfnamefont {P.}~\bibnamefont {Yeh}}, \bibinfo {author} {\bibfnamefont
  {A.}~\bibnamefont {Zalcman}}, \bibinfo {author} {\bibfnamefont
  {H.}~\bibnamefont {Neven}}, \bibinfo {author} {\bibfnamefont {V.~N.}\
  \bibnamefont {Smelyanskiy}},\ and\ \bibinfo {author} {\bibfnamefont {J.~M.}\
  \bibnamefont {Martinis}},\ }\href@noop {} {\bibfield  {journal} {\bibinfo
  {journal} {Phys. Rev. Lett.}\ }\textbf {\bibinfo {volume} {123}},\ \bibinfo
  {pages} {210501} (\bibinfo {year} {2019})}\BibitemShut {NoStop}%
\bibitem [{\citenamefont {Manovitz}\ \emph {et~al.}(2017)\citenamefont
  {Manovitz}, \citenamefont {Rotem}, \citenamefont {Shaniv}, \citenamefont
  {Cohen}, \citenamefont {Shapira}, \citenamefont {Akerman}, \citenamefont
  {Retzker},\ and\ \citenamefont {Ozeri}}]{manovitz17}%
  \BibitemOpen
  \bibfield  {author} {\bibinfo {author} {\bibfnamefont {T.}~\bibnamefont
  {Manovitz}}, \bibinfo {author} {\bibfnamefont {A.}~\bibnamefont {Rotem}},
  \bibinfo {author} {\bibfnamefont {R.}~\bibnamefont {Shaniv}}, \bibinfo
  {author} {\bibfnamefont {I.}~\bibnamefont {Cohen}}, \bibinfo {author}
  {\bibfnamefont {Y.}~\bibnamefont {Shapira}}, \bibinfo {author} {\bibfnamefont
  {N.}~\bibnamefont {Akerman}}, \bibinfo {author} {\bibfnamefont
  {A.}~\bibnamefont {Retzker}},\ and\ \bibinfo {author} {\bibfnamefont
  {R.}~\bibnamefont {Ozeri}},\ }\href@noop {} {\bibfield  {journal} {\bibinfo
  {journal} {Phys. Rev. Lett.}\ }\textbf {\bibinfo {volume} {119}},\ \bibinfo
  {pages} {220505} (\bibinfo {year} {2017})}\BibitemShut {NoStop}%
\bibitem [{\citenamefont {Mizrahi}(2013)}]{mizrahi2013ultrafast}%
  \BibitemOpen
  \bibfield  {author} {\bibinfo {author} {\bibfnamefont {J.~A.}\ \bibnamefont
  {Mizrahi}},\ }\emph {\bibinfo {title} {Ultrafast control of spin and motion
  in trapped ions}},\ \href@noop {} {Ph.D. thesis},\ \bibinfo  {school}
  {University of Maryland, College Park} (\bibinfo {year} {2013})\BibitemShut
  {NoStop}%
\bibitem [{\citenamefont {Garc{\'\i}a-Ripoll}\ \emph
  {et~al.}(2003)\citenamefont {Garc{\'\i}a-Ripoll}, \citenamefont {Zoller},\
  and\ \citenamefont {Cirac}}]{garcia2003speed}%
  \BibitemOpen
  \bibfield  {author} {\bibinfo {author} {\bibfnamefont {J.~J.}\ \bibnamefont
  {Garc{\'\i}a-Ripoll}}, \bibinfo {author} {\bibfnamefont {P.}~\bibnamefont
  {Zoller}},\ and\ \bibinfo {author} {\bibfnamefont {J.~I.}\ \bibnamefont
  {Cirac}},\ }\href@noop {} {\bibfield  {journal} {\bibinfo  {journal} {Phys.
  Rev. Lett.}\ }\textbf {\bibinfo {volume} {91}},\ \bibinfo {pages} {157901}
  (\bibinfo {year} {2003})}\BibitemShut {NoStop}%
\bibitem [{\citenamefont {Carr}\ and\ \citenamefont {Purcell}(1954)}]{carr54}%
  \BibitemOpen
  \bibfield  {author} {\bibinfo {author} {\bibfnamefont {H.~Y.}\ \bibnamefont
  {Carr}}\ and\ \bibinfo {author} {\bibfnamefont {E.~M.}\ \bibnamefont
  {Purcell}},\ }\href@noop {} {\bibfield  {journal} {\bibinfo  {journal} {Phys.
  Rev.}\ }\textbf {\bibinfo {volume} {94}},\ \bibinfo {pages} {630} (\bibinfo
  {year} {1954})}\BibitemShut {NoStop}%
\bibitem [{\citenamefont {Meiboom}\ and\ \citenamefont
  {Gill}(1958)}]{meiboom58}%
  \BibitemOpen
  \bibfield  {author} {\bibinfo {author} {\bibfnamefont {S.}~\bibnamefont
  {Meiboom}}\ and\ \bibinfo {author} {\bibfnamefont {D.}~\bibnamefont {Gill}},\
  }\href@noop {} {\bibfield  {journal} {\bibinfo  {journal} {Rev. Sci.
  Instrum.}\ }\textbf {\bibinfo {volume} {29}},\ \bibinfo {pages} {688}
  (\bibinfo {year} {1958})}\BibitemShut {NoStop}%
\bibitem [{\citenamefont {Uhrig}(2007)}]{uhrig07}%
  \BibitemOpen
  \bibfield  {author} {\bibinfo {author} {\bibfnamefont {G.~S.}\ \bibnamefont
  {Uhrig}},\ }\href@noop {} {\bibfield  {journal} {\bibinfo  {journal} {Phys.
  Rev. Lett.}\ }\textbf {\bibinfo {volume} {98}},\ \bibinfo {pages} {100504}
  (\bibinfo {year} {2007})}\BibitemShut {NoStop}%
\bibitem [{\citenamefont {Golan}\ \emph {et~al.}(2025)\citenamefont {Golan}
  \emph {et~al.}}]{golan25}%
  \BibitemOpen
  \bibfield  {author} {\bibinfo {author} {\bibfnamefont {O.}~\bibnamefont
  {Golan}} \emph {et~al.},\ }\href@noop {} {\bibfield  {journal} {\bibinfo
  {journal} {in preparation}\ } (\bibinfo {year} {2025})}\BibitemShut {NoStop}%
\bibitem [{\citenamefont {Shapira}\ \emph {et~al.}(2020)\citenamefont
  {Shapira}, \citenamefont {Shaniv}, \citenamefont {Manovitz}, \citenamefont
  {Akerman}, \citenamefont {Peleg}, \citenamefont {Gazit}, \citenamefont
  {Ozeri},\ and\ \citenamefont {Stern}}]{shapira20}%
  \BibitemOpen
  \bibfield  {author} {\bibinfo {author} {\bibfnamefont {Y.}~\bibnamefont
  {Shapira}}, \bibinfo {author} {\bibfnamefont {R.}~\bibnamefont {Shaniv}},
  \bibinfo {author} {\bibfnamefont {T.}~\bibnamefont {Manovitz}}, \bibinfo
  {author} {\bibfnamefont {N.}~\bibnamefont {Akerman}}, \bibinfo {author}
  {\bibfnamefont {L.}~\bibnamefont {Peleg}}, \bibinfo {author} {\bibfnamefont
  {L.}~\bibnamefont {Gazit}}, \bibinfo {author} {\bibfnamefont
  {R.}~\bibnamefont {Ozeri}},\ and\ \bibinfo {author} {\bibfnamefont
  {A.}~\bibnamefont {Stern}},\ }\href@noop {} {\bibfield  {journal} {\bibinfo
  {journal} {Phys. Rev. A}\ }\textbf {\bibinfo {volume} {101}},\ \bibinfo
  {pages} {032330} (\bibinfo {year} {2020})}\BibitemShut {NoStop}%
\bibitem [{\citenamefont {Zhang}\ \emph {et~al.}(2022)\citenamefont {Zhang},
  \citenamefont {Srinivasan}, \citenamefont {Sundaresan}, \citenamefont
  {Bogorin}, \citenamefont {Martin}, \citenamefont {Hertzberg}, \citenamefont
  {Timmerwilke}, \citenamefont {Pritchett}, \citenamefont {Yau}, \citenamefont
  {Wang}, \citenamefont {Landers}, \citenamefont {Lewandowski}, \citenamefont
  {Narasgond}, \citenamefont {Rosenblatt}, \citenamefont {Keefe}, \citenamefont
  {Lauer}, \citenamefont {Rothwell}, \citenamefont {McClure}, \citenamefont
  {Dial}, \citenamefont {Orcutt}, \citenamefont {Brink},\ and\ \citenamefont
  {Chow}}]{zhang22i}%
  \BibitemOpen
  \bibfield  {author} {\bibinfo {author} {\bibfnamefont {E.~J.}\ \bibnamefont
  {Zhang}}, \bibinfo {author} {\bibfnamefont {S.}~\bibnamefont {Srinivasan}},
  \bibinfo {author} {\bibfnamefont {N.}~\bibnamefont {Sundaresan}}, \bibinfo
  {author} {\bibfnamefont {D.~F.}\ \bibnamefont {Bogorin}}, \bibinfo {author}
  {\bibfnamefont {Y.}~\bibnamefont {Martin}}, \bibinfo {author} {\bibfnamefont
  {J.~B.}\ \bibnamefont {Hertzberg}}, \bibinfo {author} {\bibfnamefont
  {J.}~\bibnamefont {Timmerwilke}}, \bibinfo {author} {\bibfnamefont {E.~J.}\
  \bibnamefont {Pritchett}}, \bibinfo {author} {\bibfnamefont {J.-B.}\
  \bibnamefont {Yau}}, \bibinfo {author} {\bibfnamefont {C.}~\bibnamefont
  {Wang}}, \bibinfo {author} {\bibfnamefont {W.}~\bibnamefont {Landers}},
  \bibinfo {author} {\bibfnamefont {E.~P.}\ \bibnamefont {Lewandowski}},
  \bibinfo {author} {\bibfnamefont {A.}~\bibnamefont {Narasgond}}, \bibinfo
  {author} {\bibfnamefont {S.}~\bibnamefont {Rosenblatt}}, \bibinfo {author}
  {\bibfnamefont {G.~A.}\ \bibnamefont {Keefe}}, \bibinfo {author}
  {\bibfnamefont {I.}~\bibnamefont {Lauer}}, \bibinfo {author} {\bibfnamefont
  {M.~B.}\ \bibnamefont {Rothwell}}, \bibinfo {author} {\bibfnamefont {D.~T.}\
  \bibnamefont {McClure}}, \bibinfo {author} {\bibfnamefont {O.~E.}\
  \bibnamefont {Dial}}, \bibinfo {author} {\bibfnamefont {J.~S.}\ \bibnamefont
  {Orcutt}}, \bibinfo {author} {\bibfnamefont {M.}~\bibnamefont {Brink}},\ and\
  \bibinfo {author} {\bibfnamefont {J.~M.}\ \bibnamefont {Chow}},\ }\href@noop
  {} {\bibfield  {journal} {\bibinfo  {journal} {Sci. Adv.}\ }\textbf {\bibinfo
  {volume} {8}},\ \bibinfo {pages} {eabi6690} (\bibinfo {year}
  {2022})}\BibitemShut {NoStop}%
\bibitem [{\citenamefont {Grzesiak}\ \emph {et~al.}(2020)\citenamefont
  {Grzesiak}, \citenamefont {Bl{\"u}mel}, \citenamefont {Wright}, \citenamefont
  {Beck}, \citenamefont {Pisenti}, \citenamefont {Li}, \citenamefont {Chaplin},
  \citenamefont {Amini}, \citenamefont {Debnath}, \citenamefont {Chen},\ and\
  \citenamefont {Nam}}]{grzesiak20}%
  \BibitemOpen
  \bibfield  {author} {\bibinfo {author} {\bibfnamefont {N.}~\bibnamefont
  {Grzesiak}}, \bibinfo {author} {\bibfnamefont {R.}~\bibnamefont
  {Bl{\"u}mel}}, \bibinfo {author} {\bibfnamefont {K.}~\bibnamefont {Wright}},
  \bibinfo {author} {\bibfnamefont {K.~M.}\ \bibnamefont {Beck}}, \bibinfo
  {author} {\bibfnamefont {N.~C.}\ \bibnamefont {Pisenti}}, \bibinfo {author}
  {\bibfnamefont {M.}~\bibnamefont {Li}}, \bibinfo {author} {\bibfnamefont
  {V.}~\bibnamefont {Chaplin}}, \bibinfo {author} {\bibfnamefont {J.~M.}\
  \bibnamefont {Amini}}, \bibinfo {author} {\bibfnamefont {S.}~\bibnamefont
  {Debnath}}, \bibinfo {author} {\bibfnamefont {J.-S.}\ \bibnamefont {Chen}},\
  and\ \bibinfo {author} {\bibfnamefont {Y.}~\bibnamefont {Nam}},\ }\href@noop
  {} {\bibfield  {journal} {\bibinfo  {journal} {Nat. Commun.}\ }\textbf
  {\bibinfo {volume} {11}},\ \bibinfo {pages} {2963} (\bibinfo {year}
  {2020})}\BibitemShut {NoStop}%
\bibitem [{\citenamefont {Dai}\ \emph {et~al.}(2026)\citenamefont {Dai},
  \citenamefont {Hazra}, \citenamefont {Weiss}, \citenamefont {Kurilovich},
  \citenamefont {Connolly}, \citenamefont {Babla}, \citenamefont {Singh},
  \citenamefont {Joshi}, \citenamefont {Ding}, \citenamefont {Parakh},
  \citenamefont {Venkatraman}, \citenamefont {Xiao}, \citenamefont {Frunzio},\
  and\ \citenamefont {Devoret}}]{dai26}%
  \BibitemOpen
  \bibfield  {author} {\bibinfo {author} {\bibfnamefont {W.}~\bibnamefont
  {Dai}}, \bibinfo {author} {\bibfnamefont {S.}~\bibnamefont {Hazra}}, \bibinfo
  {author} {\bibfnamefont {D.~K.}\ \bibnamefont {Weiss}}, \bibinfo {author}
  {\bibfnamefont {P.~D.}\ \bibnamefont {Kurilovich}}, \bibinfo {author}
  {\bibfnamefont {T.}~\bibnamefont {Connolly}}, \bibinfo {author}
  {\bibfnamefont {H.~K.}\ \bibnamefont {Babla}}, \bibinfo {author}
  {\bibfnamefont {S.}~\bibnamefont {Singh}}, \bibinfo {author} {\bibfnamefont
  {V.~R.}\ \bibnamefont {Joshi}}, \bibinfo {author} {\bibfnamefont {A.~Z.}\
  \bibnamefont {Ding}}, \bibinfo {author} {\bibfnamefont {P.~D.}\ \bibnamefont
  {Parakh}}, \bibinfo {author} {\bibfnamefont {J.}~\bibnamefont {Venkatraman}},
  \bibinfo {author} {\bibfnamefont {X.}~\bibnamefont {Xiao}}, \bibinfo {author}
  {\bibfnamefont {L.}~\bibnamefont {Frunzio}},\ and\ \bibinfo {author}
  {\bibfnamefont {M.~H.}\ \bibnamefont {Devoret}},\ }\href@noop {} {\bibfield
  {journal} {\bibinfo  {journal} {Phys. Rev. X}\ }\textbf {\bibinfo {volume}
  {16}},\ \bibinfo {pages} {011011} (\bibinfo {year} {2026})}\BibitemShut
  {NoStop}%
\bibitem [{\citenamefont {Mercurio}\ \emph {et~al.}(2025)\citenamefont
  {Mercurio}, \citenamefont {Huang}, \citenamefont {Cai}, \citenamefont {Chen},
  \citenamefont {Savona},\ and\ \citenamefont {Nori}}]{QuantumToolbox.jl2025}%
  \BibitemOpen
  \bibfield  {author} {\bibinfo {author} {\bibfnamefont {A.}~\bibnamefont
  {Mercurio}}, \bibinfo {author} {\bibfnamefont {Y.-T.}\ \bibnamefont {Huang}},
  \bibinfo {author} {\bibfnamefont {L.-X.}\ \bibnamefont {Cai}}, \bibinfo
  {author} {\bibfnamefont {Y.-N.}\ \bibnamefont {Chen}}, \bibinfo {author}
  {\bibfnamefont {V.}~\bibnamefont {Savona}},\ and\ \bibinfo {author}
  {\bibfnamefont {F.}~\bibnamefont {Nori}},\ }\href@noop {} {\bibfield
  {journal} {\bibinfo  {journal} {{Quantum}}\ }\textbf {\bibinfo {volume}
  {9}},\ \bibinfo {pages} {1866} (\bibinfo {year} {2025})}\BibitemShut
  {NoStop}%
\bibitem [{Note2()}]{Note2}%
  \BibitemOpen
  \bibinfo {note} {The Nuttall window is defined as $w(n) = a_0 - a_1 \cos
  (2\pi n/N_s) +a_2 \cos (4\pi n/N_s) - a_3 \cos (6\pi n/N_s)$ with $a_0 =
  0.355768$, $a_1=0.487396$, $a_2=0.144232$, and $a_3=0.012604$. $N_s$ is the
  number of time steps that the window is broken into.}\BibitemShut {Stop}%
\bibitem [{\citenamefont {G{\"o}ppl}\ \emph {et~al.}(2008)\citenamefont
  {G{\"o}ppl}, \citenamefont {Fragner}, \citenamefont {Baur}, \citenamefont
  {Bianchetti}, \citenamefont {Filipp}, \citenamefont {Fink}, \citenamefont
  {Leek}, \citenamefont {Puebla}, \citenamefont {Steffen},\ and\ \citenamefont
  {Wallraff}}]{goppl08a}%
  \BibitemOpen
  \bibfield  {author} {\bibinfo {author} {\bibfnamefont {M.}~\bibnamefont
  {G{\"o}ppl}}, \bibinfo {author} {\bibfnamefont {A.}~\bibnamefont {Fragner}},
  \bibinfo {author} {\bibfnamefont {M.}~\bibnamefont {Baur}}, \bibinfo {author}
  {\bibfnamefont {R.}~\bibnamefont {Bianchetti}}, \bibinfo {author}
  {\bibfnamefont {S.}~\bibnamefont {Filipp}}, \bibinfo {author} {\bibfnamefont
  {J.~M.}\ \bibnamefont {Fink}}, \bibinfo {author} {\bibfnamefont {P.~J.}\
  \bibnamefont {Leek}}, \bibinfo {author} {\bibfnamefont {G.}~\bibnamefont
  {Puebla}}, \bibinfo {author} {\bibfnamefont {L.}~\bibnamefont {Steffen}},\
  and\ \bibinfo {author} {\bibfnamefont {A.}~\bibnamefont {Wallraff}},\
  }\href@noop {} {\bibfield  {journal} {\bibinfo  {journal} {J. Appl. Phys.}\
  }\textbf {\bibinfo {volume} {104}},\ \bibinfo {pages} {113904} (\bibinfo
  {year} {2008})}\BibitemShut {NoStop}%
\bibitem [{\citenamefont {Bose}(2007)}]{bose07}%
  \BibitemOpen
  \bibfield  {author} {\bibinfo {author} {\bibfnamefont {S.}~\bibnamefont
  {Bose}},\ }\href@noop {} {\bibfield  {journal} {\bibinfo  {journal} {Contemp.
  Phys.}\ }\textbf {\bibinfo {volume} {48}},\ \bibinfo {pages} {13} (\bibinfo
  {year} {2007})}\BibitemShut {NoStop}%
\bibitem [{\citenamefont {Yao}\ \emph {et~al.}(2011)\citenamefont {Yao},
  \citenamefont {Jiang}, \citenamefont {Gorshkov}, \citenamefont {Gong},
  \citenamefont {Zhai}, \citenamefont {Duan},\ and\ \citenamefont
  {Lukin}}]{yao11}%
  \BibitemOpen
  \bibfield  {author} {\bibinfo {author} {\bibfnamefont {N.~Y.}\ \bibnamefont
  {Yao}}, \bibinfo {author} {\bibfnamefont {L.}~\bibnamefont {Jiang}}, \bibinfo
  {author} {\bibfnamefont {A.~V.}\ \bibnamefont {Gorshkov}}, \bibinfo {author}
  {\bibfnamefont {Z.~X.}\ \bibnamefont {Gong}}, \bibinfo {author}
  {\bibfnamefont {A.}~\bibnamefont {Zhai}}, \bibinfo {author} {\bibfnamefont
  {L.~M.}\ \bibnamefont {Duan}},\ and\ \bibinfo {author} {\bibfnamefont
  {M.~D.}\ \bibnamefont {Lukin}},\ }\href@noop {} {\bibfield  {journal}
  {\bibinfo  {journal} {Phys. Rev. Lett.}\ }\textbf {\bibinfo {volume} {106}},\
  \bibinfo {pages} {040505} (\bibinfo {year} {2011})}\BibitemShut {NoStop}%
\bibitem [{\citenamefont {Lu}\ \emph {et~al.}(2019)\citenamefont {Lu},
  \citenamefont {Zhang}, \citenamefont {Zhang}, \citenamefont {Chen},
  \citenamefont {Shen}, \citenamefont {Zhang}, \citenamefont {Zhang},\ and\
  \citenamefont {Kim}}]{lu19}%
  \BibitemOpen
  \bibfield  {author} {\bibinfo {author} {\bibfnamefont {Y.}~\bibnamefont
  {Lu}}, \bibinfo {author} {\bibfnamefont {S.}~\bibnamefont {Zhang}}, \bibinfo
  {author} {\bibfnamefont {K.}~\bibnamefont {Zhang}}, \bibinfo {author}
  {\bibfnamefont {W.}~\bibnamefont {Chen}}, \bibinfo {author} {\bibfnamefont
  {Y.}~\bibnamefont {Shen}}, \bibinfo {author} {\bibfnamefont {J.}~\bibnamefont
  {Zhang}}, \bibinfo {author} {\bibfnamefont {J.-N.}\ \bibnamefont {Zhang}},\
  and\ \bibinfo {author} {\bibfnamefont {K.}~\bibnamefont {Kim}},\ }\href@noop
  {} {\bibfield  {journal} {\bibinfo  {journal} {Nature (London)}\ }\textbf
  {\bibinfo {volume} {572}},\ \bibinfo {pages} {363} (\bibinfo {year}
  {2019})}\BibitemShut {NoStop}%
\bibitem [{\citenamefont {Leibfried}\ and\ \citenamefont {{and
  Wineland}}(2018)}]{leibfried18}%
  \BibitemOpen
  \bibfield  {author} {\bibinfo {author} {\bibfnamefont {D.}~\bibnamefont
  {Leibfried}}\ and\ \bibinfo {author} {\bibfnamefont {D.~J.}\ \bibnamefont
  {{and Wineland}}},\ }\href@noop {} {\bibfield  {journal} {\bibinfo  {journal}
  {J. Mod. Opt.}\ }\textbf {\bibinfo {volume} {65}},\ \bibinfo {pages} {774}
  (\bibinfo {year} {2018})}\BibitemShut {NoStop}%
\bibitem [{\citenamefont {Burd}\ \emph {et~al.}(2021)\citenamefont {Burd},
  \citenamefont {Srinivas}, \citenamefont {Knaack}, \citenamefont {Ge},
  \citenamefont {Wilson}, \citenamefont {Wineland}, \citenamefont {Leibfried},
  \citenamefont {Bollinger}, \citenamefont {Allcock},\ and\ \citenamefont
  {Slichter}}]{burd21}%
  \BibitemOpen
  \bibfield  {author} {\bibinfo {author} {\bibfnamefont {S.~C.}\ \bibnamefont
  {Burd}}, \bibinfo {author} {\bibfnamefont {R.}~\bibnamefont {Srinivas}},
  \bibinfo {author} {\bibfnamefont {H.~M.}\ \bibnamefont {Knaack}}, \bibinfo
  {author} {\bibfnamefont {W.}~\bibnamefont {Ge}}, \bibinfo {author}
  {\bibfnamefont {A.~C.}\ \bibnamefont {Wilson}}, \bibinfo {author}
  {\bibfnamefont {D.~J.}\ \bibnamefont {Wineland}}, \bibinfo {author}
  {\bibfnamefont {D.}~\bibnamefont {Leibfried}}, \bibinfo {author}
  {\bibfnamefont {J.~J.}\ \bibnamefont {Bollinger}}, \bibinfo {author}
  {\bibfnamefont {D.~T.~C.}\ \bibnamefont {Allcock}},\ and\ \bibinfo {author}
  {\bibfnamefont {D.~H.}\ \bibnamefont {Slichter}},\ }\href@noop {} {\bibfield
  {journal} {\bibinfo  {journal} {Nat. Phys.}\ }\textbf {\bibinfo {volume}
  {17}},\ \bibinfo {pages} {898} (\bibinfo {year} {2021})}\BibitemShut
  {NoStop}%
\bibitem [{\citenamefont {Ali~Ahmed}\ \emph {et~al.}(2013)\citenamefont
  {Ali~Ahmed}, \citenamefont {\'Alvarez},\ and\ \citenamefont
  {Suter}}]{ali-ahmed13}%
  \BibitemOpen
  \bibfield  {author} {\bibinfo {author} {\bibfnamefont {M.~A.}\ \bibnamefont
  {Ali~Ahmed}}, \bibinfo {author} {\bibfnamefont {G.~A.}\ \bibnamefont
  {\'Alvarez}},\ and\ \bibinfo {author} {\bibfnamefont {D.}~\bibnamefont
  {Suter}},\ }\href@noop {} {\bibfield  {journal} {\bibinfo  {journal} {Phys.
  Rev. A}\ }\textbf {\bibinfo {volume} {87}},\ \bibinfo {pages} {042309}
  (\bibinfo {year} {2013})}\BibitemShut {NoStop}%
\bibitem [{\citenamefont {Solgun}\ and\ \citenamefont
  {Srinivasan}(2022)}]{solgun22}%
  \BibitemOpen
  \bibfield  {author} {\bibinfo {author} {\bibfnamefont {F.}~\bibnamefont
  {Solgun}}\ and\ \bibinfo {author} {\bibfnamefont {S.}~\bibnamefont
  {Srinivasan}},\ }\href@noop {} {\bibfield  {journal} {\bibinfo  {journal}
  {Phys. Rev. Applied}\ }\textbf {\bibinfo {volume} {18}},\ \bibinfo {pages}
  {044025} (\bibinfo {year} {2022})}\BibitemShut {NoStop}%
\bibitem [{\citenamefont {Bravyi}\ \emph {et~al.}(2011)\citenamefont {Bravyi},
  \citenamefont {DiVincenzo},\ and\ \citenamefont {Loss}}]{bravyi11b}%
  \BibitemOpen
  \bibfield  {author} {\bibinfo {author} {\bibfnamefont {S.}~\bibnamefont
  {Bravyi}}, \bibinfo {author} {\bibfnamefont {D.~P.}\ \bibnamefont
  {DiVincenzo}},\ and\ \bibinfo {author} {\bibfnamefont {D.}~\bibnamefont
  {Loss}},\ }\href@noop {} {\bibfield  {journal} {\bibinfo  {journal} {Ann.
  Phys.}\ }\textbf {\bibinfo {volume} {326}},\ \bibinfo {pages} {2793}
  (\bibinfo {year} {2011})}\BibitemShut {NoStop}%
\bibitem [{\citenamefont {Mundada}\ \emph {et~al.}(2019)\citenamefont
  {Mundada}, \citenamefont {Zhang}, \citenamefont {Hazard},\ and\ \citenamefont
  {Houck}}]{mundada19}%
  \BibitemOpen
  \bibfield  {author} {\bibinfo {author} {\bibfnamefont {P.}~\bibnamefont
  {Mundada}}, \bibinfo {author} {\bibfnamefont {G.}~\bibnamefont {Zhang}},
  \bibinfo {author} {\bibfnamefont {T.}~\bibnamefont {Hazard}},\ and\ \bibinfo
  {author} {\bibfnamefont {A.}~\bibnamefont {Houck}},\ }\href@noop {}
  {\bibfield  {journal} {\bibinfo  {journal} {Phys. Rev. Applied}\ }\textbf
  {\bibinfo {volume} {12}},\ \bibinfo {pages} {054023} (\bibinfo {year}
  {2019})}\BibitemShut {NoStop}%
\bibitem [{\citenamefont {Heunisch}\ \emph {et~al.}(2023)\citenamefont
  {Heunisch}, \citenamefont {Eichler},\ and\ \citenamefont
  {Hartmann}}]{heunisch23a}%
  \BibitemOpen
  \bibfield  {author} {\bibinfo {author} {\bibfnamefont {L.}~\bibnamefont
  {Heunisch}}, \bibinfo {author} {\bibfnamefont {C.}~\bibnamefont {Eichler}},\
  and\ \bibinfo {author} {\bibfnamefont {M.~J.}\ \bibnamefont {Hartmann}},\
  }\href@noop {} {\bibfield  {journal} {\bibinfo  {journal} {Phys. Rev. Appl.}\
  }\textbf {\bibinfo {volume} {20}},\ \bibinfo {pages} {064037} (\bibinfo
  {year} {2023})}\BibitemShut {NoStop}%
\end{thebibliography}

\end{document}